\documentclass[12pt, captions=tableabove]{article}
\usepackage[a4paper, total={6.5in, 9in}]{geometry}
\usepackage{makecell}
\usepackage{threeparttable}
\usepackage{setspace}
\usepackage{caption}
\usepackage{physics}
\usepackage{subcaption}
\usepackage{amsmath}
\usepackage{bm}
\usepackage{graphicx} 
\usepackage[authoryear]{natbib}
\usepackage{xcolor}
\usepackage{authblk}
\usepackage{float}
\usepackage{comment}
\usepackage{array}
\usepackage{hyperref}

\hypersetup{colorlinks=true,linkcolor=blue,anchorcolor=blue,citecolor=blue,filecolor=blue,urlcolor=blue,bookmarksnumbered=true,pdfview=FitB} 

\title{Analyzing Zero-Inflated Clustered Longitudinal Ordinal Outcomes Using GEE-Type Models With an Application to Dental Fluorosis Studies}

\author[1]{Shoumi Sarkar}
\author[2]{Anish Mukherjee}
\author[2]{Jeremy T. Gaskins}
\author[3,4]{Steven Levy}
\author[1]{Peihua Qiu}
\author[1]{Somnath Datta\footnote{To whom correspondence should be addressed (\textcolor{blue}{\href{mailto:somnath.datta@ufl.edu}{somnath.datta@ufl.edu}})}}

\affil[1]{Department of Biostatistics, University of Florida, Gainesville, Florida}
\affil[2]{Department of Bioinformatics \& Biostatistics, University of Louisville, Louisville, Kentucky}
\affil[3]{Department of Preventive and Community Dentistry, University of Iowa, Iowa City, Iowa}
\affil[4]{Department of Epidemiology, University of Iowa, Iowa City, Iowa}

\date{}

\begin{document}

\maketitle

\begin{abstract}
Motivated by the Iowa Fluoride Study (IFS), which tracked fluoride intake and dental outcomes from childhood through young adulthood (ages 9, 13, 17, and 23), we analyze dental fluorosis—a condition caused by excessive fluoride exposure during enamel formation. In this context, fluorosis scores across tooth surfaces present as zero-inflated, clustered, and longitudinal ordinal outcomes, prompting the development of a unified modeling framework. Leveraging generalized estimating equations (GEEs), we construct separate models for the presence and severity of fluorosis and propose a combined model that links these components through shared covariates. To improve estimation efficiency and borrowing strength across timepoints, we incorporate James–Stein shrinkage estimators. We compare several working correlation structures, including a data-driven jackknifed structure, and perform model selection via rank aggregation. Simulation studies validate the finite-sample performance of the proposed models, and a bootstrap-based power analysis further confirms the validity of the testing procedure. In our analysis of the IFS data, early-life total daily fluoride intake, average home water fluoride concentration, and specific teeth and zones emerge as significant risk factors for dental fluorosis, while the maxillary lateral incisors and zones closer to the gum show protective effects across different ages. These findings reveal novel age-specific associations between early-life exposures and the progression of dental fluorosis through early adulthood.
\end{abstract}

\begin{center}
\end{center}

\section{Introduction}
\label{ch4:section:Introduction}

This paper is motivated by the Iowa Fluoride Study (IFS) \citep{warren2001prevalence, levy2001patterns, levy2003patterns, marshall2003dental, marshall2005roles, warren2002dental, marshall2005diet, marshall2004associations, curtis2020decline, yazdanbakhsh2024community}, a prospective longitudinal cohort study that began in the early 1990s with the recruitment of mothers of newborns from 8 postpartum wards across Iowa. The IFS was designed to examine how fluoride intake, diet, and other factors influence dental outcomes over time. While data collection began shortly after birth, the earliest dental fluorosis assessments of the permanent teeth were conducted at about age 9. In this paper, we focus on four key time points — ages 9, 13, 17, and 23 — where fluorosis data are available and consistently measured across participants. 
This analysis should be viewed in the context of the dual role of fluoride: while fluoride exposures, including fluoridated tap water, are known to reduce the risk of dental caries, excessive intake during early childhood can lead to dental fluorosis, a condition characterized by white striations on the surface of teeth.

The data collected in this study exhibit several unique features. The repeated assessments over time give it a longitudinal structure. Each participant, with their own genetic background and lifestyle habits, serves as an independent cluster. Within these clusters, caries count outcomes are recorded across all available teeth. For fluorosis, ordinal scores ranging from 0 to 3, based on the Fluorosis Risk Index (FRI) \citep{warren2021measurement}, are scored at four zones of each permanent tooth:  three zones on the buccal surface, and a fourth zone that is incisal for anterior teeth or occlusal for posterior teeth. An FRI score of 0 indicates no visible signs of fluorosis, while higher scores reflect increasing severity. This structure naturally leads to nested data: zones within teeth, and teeth within individuals. Furthermore, because most people do not develop caries or fluorosis, these outcomes exhibit zero inflation, manifesting itself as an excess of zero counts for caries and FRI scores of 0 for fluorosis. Although the fluorosis outcome is ordinal, we refer to the lowest category (corresponding to no fluorosis) as the ``zero" category and describe the inflation of this level as zero inflation. This results in a complex, multi-level data structure that is zero-inflated, clustered, and longitudinal, featuring count outcomes for caries and ordinal responses for fluorosis, alongside continuous and categorical covariates.
 Standard statistical methods are inadequate for analyzing such data, motivating the development of new methodologies that can accommodate these intricacies.
 
Previous studies using IFS data have focused primarily on modeling caries count outcomes \citep{choo2016marginal, choo2018bayesian, choo2018analyzing, kang2021longitudinal, kang2021analyzing, mukherjee2024modeling}. An exception was \citet{kang2023analyzing}, who addressed fluorosis ordinal outcomes via a Bayesian framework using data collected up to age 17—specifically at ages 9, 13, and 17. The inclusion of age 23 in the most recent wave offers an opportunity to refine our understanding of the development of dental fluorosis at critical stages of dental maturation. In addition, there is a need to develop flexible modeling frameworks for such data within the frequentist paradigm, which remains more familiar and widely adopted among clinicians and public health practitioners. This is particularly important in contexts such as the IFS, where the findings have direct relevance to dental practitioners and can inform decision-making in real-world clinical settings.
Motivated by these considerations, we propose a novel frequentist framework for modeling zero-inflated, clustered, longitudinal ordinal outcomes, aimed at analyzing the fluorosis data from the IFS across all available ages—9, 13, 17, and 23. Although the methodological approaches described in this paper are illustrated through the analysis of dental fluorosis data, the proposed framework is broadly applicable to similarly structured data involving zero-inflated, longitudinal, and clustered ordinal outcomes across diverse fields such as education research, epidemiology, health services research, developmental psychology, environmental sciences, labor economics, and so on.

While there are methods in the current literature that address individual characteristics - such as the ordinal nature of outcomes or zero-inflation - of such complexly structured data, there remains a need for unified approaches that can accommodate their full range of features. For instance, cumulative logit models within a generalized linear modeling (GLM) framework — such as the proportional odds model — are commonly used for ordinal responses \citep{mccullagh1980regression}. However, standard GLMs do not account for within-subject correlation arising from repeated measurements over time, nor do they capture hierarchical nesting (e.g., zones within teeth, and teeth within individuals). Mixed-effects extensions of GLMs \citep{clayton1996generalized} allow for random effects to address clustering, while hurdle \citep{cragg1971some} and zero-inflated models \citep{lambert1992zero, greene1994accounting} have been proposed to handle excess zeros in count data. Mixture models have also been developed to accommodate zero inflation in ordinal data by incorporating a binary component that distinguishes between zero-inflated and non-zero-inflated subpopulations \citep{kelley2008zero}. However, such methods are not readily applicable when ordinal outcomes exhibit both multilevel and longitudinal structure. This gap motivates the development of a specialized modeling framework.

In this paper, we introduce a suite of modeling schemes tailored to the complex features of the data described above. To account for the multilevel clustering inherent in the data, we utilize Generalized Estimating Equations (GEEs) \citep{liang1986longitudinal} for estimation. GEEs, a quasi-likelihood approach that requires only the specification of the first two moments, yield consistent and robust estimates even when the working correlation structure is mis-specified. First, motivated by the hurdle model—which assumes that all zeroes arise from a distinct data-generating mechanism—we propose age-specific models that separately handle the zero-inflated outcome category and the non-zero categories. In the context of the IFS fluorosis dataset, where the unit of analysis is the tooth zone, this decomposition allows us to investigate the age-specific effects of dietary and non-dietary predictors on two distinct aspects of the outcome: (i) the presence or absence of fluorosis, and (ii) the severity of fluorosis, conditional on its presence. We refer to these models as the \emph{presence model}, which models whether the fluorosis outcome falls in the lowest category (i.e., ``zero''), and the \emph{severity model}, which models the ordinal fluorosis score given that it is non-zero. The separation of the model into presence and severity components is motivated by substantial zero inflation in the data: the proportion of surfaces with no fluorosis is much higher than what would be expected under a standard ordinal regression model. Including the zero category within a conventional proportional odds framework would likely lead to underestimation of the true frequency of zero responses. By contrast, modeling the presence of fluorosis separately offers a more parsimonious and interpretable solution, with its own parameter set, and better accounts for excessive zeroes. Next, we introduce a unified modeling framework that links the two submodels through shared parameters, with the goal of improving estimation efficiency. This \emph{combined model} enables simultaneous estimation of the presence and severity components under parametric constraints, leveraging the full dataset rather than fitting separate models to disjoint subsets. Finally, we adopt an empirical Bayes approach \citep{robbins1964empirical} to ``borrow strength'' across different ages. By incorporating information from models fitted at other ages, this strategy enhances the stability and overall accuracy of parameter estimates in age-specific models. Together, these techniques yield a comprehensive and flexible modeling framework capable of addressing the full range of complexities in the IFS fluorosis data. In our analysis, early life factors, such as total daily fluoride intake and average fluoride concentration in home water, and spatial factors, such as involvement of the central maxillary incisors and zones closer to the tip of the tooth, were significantly associated with the risk of fluorosis.

The remainder of the paper is organized as follows. Section~\ref{ch4:section: Methods and Modeling} details our statistical methods, including options for correlation structures and a model selection strategy. Section~\ref{ch4:section:Real data analysis} presents the IFS data analysis, highlighting associations of fluoride exposure, and dietary and non-dietary factors with fluorosis risk and severity, along with selected optimal models. Section~\ref{ch4:section:Simulation studies} reports simulation results evaluating method performance across varying sample sizes. We conclude with a discussion in Section~\ref{section:Discussion}.


\section{Methods and Modeling}
\label{ch4:section: Methods and Modeling}

\subsection{Notation}
\label{ch4:subsection:Notation}
The following notation is presented with the FRI score outcome obtained under the IFS in mind, though the proposed methods are general and equally applicable to similarly structured multilevel outcome in other contexts. Thus, using the IFS data as an example, let $Y_{tijk}$ denote the ordinal fluorosis outcome, represented by the Fluorosis Risk Index (FRI) score, measured at time-point $t$ on the $k$-th zone within the $j$-th tooth of the $i$-th individual. In this study, measurements were taken at four time-points corresponding to participant ages 9, 13, 17, and 23; thus, we define $T = 4$ and index the time-points as $t = 1, \dots, T$. The FRI scores are recorded on an ordinal scale with four levels: 0 (no or negligible fluorosis), 1 (questionable fluorosis), 2 (definitive mild fluorosis), and 3 (severe fluorosis). More generally, $Y_{tijk}$ takes on $L+1$ ordered levels, $0 < 1 < \ldots < L$, where $Y_{tijk} = 0$ denotes the lowest severity category and $Y_{tijk} = L$ the highest. Let $i = 1, \dots, N$ index individuals in the cohort. For each individual $i$ at time $t$, we consider $J$ teeth, indexed by $j = 1, \dots, J$. Within each tooth $j$ of individual $i$ at time $t$, FRI scores are measured at $K$ zones measured from root to tip, indexed by $k = 1, \dots, K$. In the IFS data,
$K=4$, corresponding to the cervical third (C), middle third (M), and incisal third (I) of the buccal surface, along with the incisal edge (O), measured on each of the four maxillary incisors. Ideally, a complete dataset would consist of $N \times T \times J \times K$ observations. However, some $(i, t, j, k)$ combinations can be missing in practice due to incomplete data collection or dropout.

Based on $Y_{tijk}$, we further derive the following outcomes: $W_{P,tijk} = I[Y_{tijk}>0]$, which denotes presence or absence of fluorosis of the tooth, serving as the response for a ``presence'' model; and, when $Y_{tijk}>0$, we define $W_{S,tijk}$ as the integer value of $Y_{tijk}$, denoting the severity of fluorosis when it is present - this serves as the response for a ``severity'' model. Details of presence and severity models are presented in Sections \ref{ch4:subsection:Presence Model} and \ref{ch4:subsection:Severity Model}. 

The multilevel structure of $Y_{tijk}$ induces correlations among observations within the same individual, tooth, and zone. For a pair of FRI scores $Y_{tijk}$ and $Y_{ti'j'k'}$ measured at the same time-point $t$ for the same individual $i$, we define $\rho_{P,t,jk,j'k'}$ as the correlation between outcomes at locations $(j,k)$ and $(j',k')$ under the presence model (i.e., between $W_{P,tijk}$ and $W_{P,tij'k'}$). Similarly, $\rho_{S,t,jk,j'k'}$ denotes the corresponding correlation under the severity model (i.e., between $W_{S,tijk}$ and $W_{S,tij'k'}$). The specification of these correlation parameters is provided in Section~\ref{ch4:subsection:Correlation parameters}.

For the presence model at time-point $t$, the linear predictor is defined using an intercept $\alpha_{P,t}$ and a vector of regression coefficients $\boldsymbol{\beta}_{P,t} = (\beta_{P,t,1}, \beta_{P,t,2}, \dots, \beta_{P,t,q})^\top$. Similarly, for the severity model, the linear predictor includes category-specific intercepts $\alpha_{S,t,1|2}, \dots, \alpha_{S,t,L-1|L}$ and a corresponding vector of regression coefficients $\boldsymbol{\beta}_{S,t} = (\beta_{S,t,1}, \beta_{S,t,2}, \dots, \beta_{S,t,q})^\top$. Here, bold notation denotes vectors or matrices. In the combined model introduced in Section~\ref{Estimation of presence and severity parameters when the parameters are functionally related}, the regression coefficients for the severity component are functionally linked to those of the presence component via a time-specific scalar $\gamma_t$, such that $\beta_{S,t,i} = \gamma_t \beta_{P,t,i}$ for the $i$-th covariate. This constraint imposes proportionality between covariate effects across the presence and severity components, facilitating dimensionality reduction while leveraging all available data. A more general formulation that relaxes this assumption is also discussed in that section.


\subsubsection{Generalized Estimating Equations (GEEs)}
\label{ch4:subsection:Generalized Estimating Equations (GEEs)}
 Generalized estimating equations proposed by \cite{liang1986longitudinal} extend the GLM framework to analyze longitudinal and clustered data. GEEs are designed to handle correlated observations. This makes them particularly useful in situations where repeated measures are collected on the same subject, where observations in these subject-specific ``clusters'' are not necessarily independent.  GEEs use quasi-likelihood methods, focusing on the first two moments in addition to a working correlation structure, rather than fully specifying the joint distribution of the data. Parameter estimates are obtained numerically via iterative methods, such as the Newton-Raphson method. In a setting with repeated measures on $N$ subjects, the GEE, $\boldsymbol{\psi}(\boldsymbol{\beta})=0$, is based on the generalized estimating function:
\begin{equation*}
\boldsymbol{\psi}(\boldsymbol{\beta}) = \sum_{i=1}^{N} \boldsymbol{D_i}^T \boldsymbol{V_i}^{-1} (\boldsymbol{Y_i} - \boldsymbol{\mu_i})
\end{equation*}
where \(\boldsymbol{\beta}\) are parameters,
\(\boldsymbol{Y_i}\) is the vector of responses for subject \(i\), $\boldsymbol{X_i}$ denotes covariates for subject $i$, and $\boldsymbol{\mu_i} = E(\boldsymbol{Y_i}) = g(\boldsymbol{X_i'\beta})$, where $g(.)$ is a link function, is the mean response. The quantity \(\boldsymbol{D_i} = \frac{\partial \boldsymbol{\mu_i}}{\partial \boldsymbol{\beta}}\) is the matrix of derivatives, and \(\boldsymbol{V_i}\) is the working covariance matrix. A working correlation structure $\boldsymbol{V_i}$ is specified for repeated measurements within cluster $i$. GEEs provide consistent parameter estimates despite the mis-specification of $\boldsymbol{V_i}$; this robustness is valuable when the true correlation structure is unknown or complex.

\subsubsection{Presence Model}
\label{ch4:subsection:Presence Model}
To address the excess of zeroes in the ordinal FRI response, we model them separately using the derived binary outcome variable $W_{P,tijk}:= I[Y_{tijk}>0]$, which indicates the presence of fluorosis. With $\mu_{P,tijk} := {P(W_{P,tijk}=0)}$, we specify McCullagh's model \citep{mccullagh1980regression} for fluorosis \emph{presence}:
\begin{align}
\mu_{P,tijk} = F(\alpha_{P,t} + \boldsymbol{x'_{P,tijk}} \boldsymbol{\beta_{P,t}}),
    \label{ch4:eqn:McCullagh-presence}
\end{align}
where $F(\cdot)$ is a known, fixed cumulative distribution function (CDF). Different choices of $F(\cdot)$ correspond to different link functions: for example, \( F(t) = \Phi(t) \), where $\Phi(t)$ is the standard normal CDF, yields the probit link, while \( F(t) = 1 - \exp(-\exp(t)) \) corresponds to the complementary log-log link. The choice of $F(\cdot)$ determines the link function and, consequently, the form of the relationship between the predictors and the response.  The choice of F can be guided by considerations of interpretability. In this paper, we illustrate the case where $F(\cdot)$ is the logistic CDF, i.e., \( F(t) = \frac{\exp(t)}{1 + \exp(t)} \), which corresponds to a logistic regression model when the outcomes are independent. Accordingly, we have
\begin{equation}
 {P(W_{P,tijk}=1)} = \frac{\exp{(\alpha_{P,t} + \boldsymbol{x'_{P,tijk}} \boldsymbol{\beta_{P,t}})}}{1+\exp{(\alpha_{P,t} + \boldsymbol{x'_{P,tijk}} \boldsymbol{\beta_{P,t}})}},
 \label{ch4:eqn:presence_prob_expression}
\end{equation}
which is equivalent to
\begin{equation*}
     \log\bigg(\frac{P(W_{P,tijk}=0)}{P(W_{P,tijk}=1)}\bigg) = \alpha_{P,t} + \boldsymbol{x'_{P,tijk}} \boldsymbol{\beta_{P,t}}.
    \label{ch4:eqn:model-presence-separate}
\end{equation*}
Specific to each time-point $t=1, \ldots, T$, estimates for $\alpha_{P,t}$ and $\boldsymbol{\beta_{P,t}}$ are obtained by solving the GEE $\boldsymbol{\psi_{P,t}}(\alpha_{P,t}, \boldsymbol{\beta_{P,t}}) = \boldsymbol{0}$, where
\begin{equation}
    \boldsymbol{\psi_{P,t}}(\alpha_{P,t}, \boldsymbol{\beta_{P,t}}) := \sum_{i_p=1}^{N_{p,t}} \boldsymbol{D_{p,i_p,t}^{T}} \boldsymbol{V_{p,i_p,t}^{-1}} (\boldsymbol{W_{P,i_p, t} - \mu_{p,i_p, t}}), \text{ } t=1, \ldots T
    \label{ch4:eqn:presence_GEF}
\end{equation}
Here, $\boldsymbol{W_{P,i_p,t}}$ denotes the vector of $W_{P,tijk}$ for fixed $t$ and $i$, stacked over the levels $j=1, \ldots, J$ and $k=1, \ldots, K$. For the presence model, this is the probability given in (\ref{ch4:eqn:presence_prob_expression}). $\boldsymbol{\mu_{P,i_p,t} = E(W_{P,i_p,t}})$, which is also a vector of $\mu_{P,tijk}$ for fixed $t$ and $i$, stacked over the levels of $j=1, \ldots, J$ and $k=1, \ldots, K$. $\boldsymbol{V_{p,i_p,t} = A^{1/2}_{p,i_p,t} R_{p,i_p,t} A^{1/2}_{p,i_p,t}}$, where $\boldsymbol{R_{p,i_p,t}}$ denotes the working cluster correlation matrix of cluster $i$ at time of dental visit $t$ and $\boldsymbol{A_{p,i_p,t}}$ denotes $\boldsymbol{Var(W_{P,i_p,t}})$, the variance-covariance matrix of $\boldsymbol{W_{P,i_p,t}}$. The entries of $\boldsymbol{R_{p, i_p, t}}$ are specific to the chosen working correlation structure; additional details on these entries and the choices of working correlation structures are provided in Section \ref{ch4:subsection:Correlation parameters}. Additional details on the GEE components can be found in Section 2 of the Supplementary File. 

The presence GEE, denoted by $\psi_{P,t}(\boldsymbol{\alpha}_{P,t}, \boldsymbol{\beta}_{P,t}) = 0$, corresponding to the estimating function in Equation~(\ref{ch4:eqn:presence_GEF}) can be solved numerically using iterative methods such as the Newton–Raphson algorithm. However, convergence of the standard Newton–Raphson method can be unstable when estimating high-dimensional parameter vectors. To address this, we adopt a modified Newton–Raphson approach as described by \citet{lauritzen2009newtonraphson}, which improves convergence by introducing a scaling parameter $\kappa$ to reduce the step size and a quadratic adjustment term to prevent overly large updates when the estimating function has a large magnitude. In our implementation, a value of $\kappa = 0.25$ was found to be sufficient to ensure convergence. The iterates from the modified Newton-Raphson method for the parameters
 $\big(\alpha_{P,t}, \boldsymbol{\beta_{P,t}}\big)^T$ corresponding to the presence model at time-point $t$ are given by
\begin{align*}
    \bigg( \alpha^{(h)}_{P,t}, \boldsymbol{\beta_{P,t}^{(h+1)}} \bigg)^T = &\bigg( \alpha^{(h)}_{P,t}, \boldsymbol{\beta_{P,t}^{(h)}} \bigg)^T + \nonumber\\
    &\kappa \bigg( \sum_{i_p=1}^{N_{p,t}} \boldsymbol{D_{p,i_p,t}^{(h)T} V_{p,i_p,t}^{(h)-1}D_{p,i_p,t}^{(h)}} + \psi^{(h)}_{P,t} \psi^{(h)T}_{P,t}
    \bigg)^{-1} \psi^{(h)}_{P,t}
\label{ch4:eqn:NR_beta_iterate_presence_GEE}
\end{align*}

\subsubsection{Severity Model}
\label{ch4:subsection:Severity Model}
With the excessive zeroes in the outcome addressed through the presence model, we now focus on modeling the non-zero outcomes, which capture the extent or severity of dental fluorosis given that it is present. This is leads to the \emph{severity model}, where the outcome of interest, $W_{S,tijk}$, is the same as the integer value of $Y_{tijk}$ in cases when $Y_{tijk} > 0$. This model employs a proportional odds framework to model the probabilities of belonging to different levels of severity of fluorosis. Mathematically, this can be expressed as
\begin{equation}
    \log\bigg(\frac{P(W_{S,tijk}\leq l)}{P(W_{S,tijk} > l)}\bigg) = \alpha_{S,t,l|l+1} + \boldsymbol{x'_{S,tijk}} \boldsymbol{\beta_{S,t}},
    \label{ch4:eqn:model-severity-separate}
\end{equation}
which leads to
\begin{align}
    \mu_{S,tijk,l} :&= P(W_{S,tijk}\leq l) \nonumber \\
    &= \frac{\exp(\alpha_{S,t,l|l+1} + \boldsymbol{x'_{S,tijk}} \boldsymbol{\beta_{S,t}})}{1+\exp(\alpha_{S,t,l|l+1} + \boldsymbol{x'_{S,tijk}} \boldsymbol{\beta_{S,t}})}
    \label{ch4:eqn:mu_severity-separate}
\end{align}
Estimates for the parameters $\alpha_{S,t,1|2}$, $\ldots$, $ \alpha_{S,t,L-1|L},$ and $\boldsymbol{\beta_{S,t}}$ of the severity model at time-point $t$ are obtained by solving the severity GEE, $\boldsymbol{\psi_{S,t}}(\alpha_{S,t,1|2}, \ldots, \alpha_{S,t,L-1|L}, \\ \boldsymbol{\beta_{S,t}})=0$, arising from the estimating function
\begin{equation}
    \boldsymbol{\psi_{S,t}}(\alpha_{S,t,1|2}, \ldots, \alpha_{S,t,L-1|L}, \boldsymbol{\beta_{S,t}}) := \sum_{i=1}^{N_s} \boldsymbol{D_{S,it}^{T}} \boldsymbol{V_{S,it}^{-1}} (\boldsymbol{Z_{S,it} - \pi_{S,it}}), \text{  } t=1,\ldots,T,
    \label{ch4:eqn:severity-separate_GEF}
\end{equation}
where 
\begin{itemize}
    \item $\boldsymbol{Z_{S,it} := (Z_{S,t,111}, \ldots, Z_{S,tijk}, \ldots, Z_{S,t,NJK}})^{T}$, 
    \item $\boldsymbol{Z_{S,tijk}}:=(Z_{S,tijk,1}, Z_{S,tijk,2}, Z_{S,tijk,3})^{T}$, 
    \item $Z_{S,tijk,l} := I[W_{S,tijk}=l]$, $l=1, \ldots, L$, 
    \item $\boldsymbol{\pi_{S,it}} := \boldsymbol{E[Z_{S,it}]} = (\pi_{S,t,111}, \ldots, \pi_{S,tijk}, \ldots, \pi_{S,t,NJK})^{T}$, 
    \item $\boldsymbol{\pi_{S,tijk}} := (\pi_{S,tijk,1}, \pi_{S,tijk,2}, \pi_{S,tijk,3})^{T}$, 
    \item $\pi_{S,tijk,l} := P[W_{S,tijk}=l] = \mu_{S,tijk,l+1} - \mu_{S,tijk,l}$, $l=1, \ldots, L-1$,
    \item $\boldsymbol{V_{S,it} = A^{1/2}_{S,it} R_{S,it} A^{1/2}_{S,it}}$, where $\boldsymbol{R_{S,it}}$ denotes the working cluster correlation matrix of cluster $i$ at time-point $t$, and
    \item $\boldsymbol{A_{S,i_p,t}}$ denotes $\boldsymbol{Var(Z_{S,it})}$, the variance-covariance matrix of $\boldsymbol{Z_{S,it}}$.
\end{itemize}

The entries of $\boldsymbol{R_{S, it}}$, specific to the chosen working correlation structure, are detailed in Section \ref{ch4:subsection:Correlation parameters}; additional details on the GEE components can be found in Section S2 of the Supplementary File. The severity GEE is solved using the modified Newton-Raphson method, with updated iterates given by 
\begin{align*}
    \bigg( \alpha^{(h+1)}_{S,t,1|2}, \ldots, \alpha^{(h+1)}_{S,t,L-1|L}, \boldsymbol{\beta^{(h+1)}_{S,t}} \bigg)^T = &\bigg( \alpha^{(h)}_{S,t,1|2}, \ldots, \alpha^{(h)}_{S,t,L-1|L}, \boldsymbol{\beta^{(h)}_{S,t}} \bigg)^T \\ 
    &+ \kappa \bigg( \sum_{i_s=1}^{N_{S,t}} \boldsymbol{D_{s,i_s,t}^{(h)T} V_{s,i_s,t}^{(h)-1}D_{s,i_s,t}^{(h)}} + \psi^{(h)}_{S,t} \psi^{(h)T}_{S,t}
    \bigg)^{-1} \psi^{(h)}_{S,t}
    \label{ch4:eqn:NR_beta_iterate_severity_GEE}
\end{align*}
\subsubsection{Estimation of Presence and Severity Parameters when the Parameters are Functionally Related}
\label{Estimation of presence and severity parameters when the parameters are functionally related}
The presence and severity models described thus far are estimated on disjoint subsets of the data: the presence model uses all observations based on binary presence/absence responses, while the severity model is fit only to observations with non-zero FRI scores. To utilize the full information in the dataset, we outline a joint modeling framework that links the presence and severity components.

To link the $g$-th coefficients ${\beta}_{P,t}^{g}$ and ${\beta}_{S,t}^{g}$ from the presence and severity models, respectively, at time point $t$ (for $g = 1, \ldots, q$) into a unified framework, we propose a functional relationship of the form ${\beta}_{S,t}^{g} = h({\beta}_{P,t}^{g}; \gamma_t)$, where $h(\cdot)$ is a known parametric function. Using the estimates $\hat{\beta}_{P,t}^{g}$ and $\hat{\beta}_{S,t}^{g}$ from the separately fitted models, the time-specific scalar $\gamma_t$ can be estimated by minimizing the squared loss function $\sum_{g=1}^{q} \big( \hat{\beta}_{S,t}^{g} - h(\hat{\beta}_{P,t}^{g}; \gamma_t) \big)^2$. That is,
\begin{equation}
  \hat{\gamma}_t = argmin_{\gamma_t} \sum_{g=1}^{q} \big( \hat{\beta}_{S,t}^{g} - h(\hat{\beta}_{P,t}^{g}; \gamma_{t})\big)^2.
  \label{ch4:eqn:argmin}
\end{equation}
For example,
\begin{equation}
    h(\beta_{P,t}^{g}; \gamma_{t}) = \gamma_{t} \beta_{P,t}^{g}
    \label{ch4:eqn:example_h}
\end{equation}
implies that at time-point $t$, the direction of each predictor’s effect on severity mirrors its effect on presence, scaled by \( \gamma_t \). For example, a predictor that increases the probability of fluorosis presence is assumed to also increase severity (or decrease both), with the magnitude of this effect governed by the time-specific scaling factor. While this is a simplifying assumption - since the relationship between presence and severity may vary across predictors - it provides a simple, intuitively appealing structure for jointly modeling the two components. For simplicity, we will explore this choice of the parametric function $h(.)$ extensively in this paper. With this choice, following (\ref{ch4:eqn:argmin}), $\hat{\gamma}_t = \frac{\sum_{g=1}^{q} \hat{\beta}_{S,t}^{g} \hat{\beta}_{P,t}^{g}}{\sum_{g=1}^{q} {\hat{\beta}_{P,t}^{g2} }}$. When $\hat{\beta}_{P,t}^{g}$ is near constant in $t$, the estimate of ${\gamma}_{t}$ simplifies to
\begin{equation}
    \hat{\gamma}_t = \frac{\sum_{g=1}^{q} \hat{\beta}_{S,t}^{g}}{\sum_{g=1}^{q} \hat{\beta}_{P,t}^{g}},
    \label{ch4:eqn:gammaratio}
\end{equation}
the ratio of the sum of the coefficient estimates in the separate severity model to their corresponding estimates in their presence counterpart, representing the factor by which the severity coefficients are magnified overall compared to their presence counterparts. 

Additionally, under the parameterization (\ref{ch4:eqn:example_h}), we have the following simultaneous models for the presence and severity pieces:
\begin{equation}
 \log\bigg(\frac{P(W_{P,tijk}=0)}{P(W_{P,tijk}\neq0)}\bigg) = \alpha_{P,t} + \boldsymbol{x'_{P,tijk}} \boldsymbol{\beta_{P,t}}
 \label{ch4:eqn:comb-piece-pres}
\end{equation}
\begin{equation}
    \log\bigg(\frac{P(W_{S,tijk}\leq l)}{P(W_{S,tijk} > l)}\bigg) = \alpha_{S,t,l|l+1} + \boldsymbol{x'_{S,tijk}} (\gamma_t \boldsymbol{\beta_{P,t}}).
     \label{ch4:model:comb-piece-sev}
\end{equation}
The estimation problem (\ref{ch4:eqn:comb-piece-pres})-(\ref{ch4:model:comb-piece-sev}) would reduce to solving only for the intercepts and the presence model slopes $\boldsymbol{\beta}_{P,t}$ if $\gamma_t$ were known. We leverage this to propose a two-step estimation approach. In the first step, the parameter $\gamma_t$ is pre-estimated from separately obtained estimates of the presence and severity model coefficients at time $t$, and subsequently held fixed. Specifically, $\hat{\gamma}_t$ is computed as 
$\hat{\gamma}_t = \frac{\sum_{g=1}^{q} \hat{\beta}_{S,t,g}}{\sum_{g=1}^{q} \hat{\beta}_{P,t,g}}$, following (\ref{ch4:eqn:gammaratio}).

Additionally, since the intercepts $\alpha_{P,t}, \alpha_{S,t,1|2}, \dots, \alpha_{S,t,L-1|L}$ are ancillary to the slope parameters, they are estimated from the separate model components: $\alpha_{P,t}$ from the presence model in Equation~\eqref{ch4:eqn:McCullagh-presence}, and $\alpha_{S,t,1|2}, \dots, \alpha_{S,t,L-1|L}$ from the severity model in Equation~\eqref{ch4:eqn:model-severity-separate}, and then held fixed.

In the second step, only the presence model slopes $\beta_{P,t,1}, \dots, \beta_{P,t,q}$, which appear in both components (\ref{ch4:eqn:comb-piece-pres}) and (\ref{ch4:model:comb-piece-sev}) of the combined model, remain to be estimated. These parameters are obtained iteratively by solving the estimating equation $\psi_t(\boldsymbol{\beta_{P,t}})=0$ derived from the estimating function in~\eqref{ch4:eqn:combined-GEF} detailed below; estimates from the first step remain fixed during this iteration. Once estimates $\boldsymbol{\hat{\beta}}_{P,t}$ are obtained, the severity model coefficients for the combined scheme are derived as $\boldsymbol{\hat{\beta}}_{S,t} = \hat{\gamma}_t \boldsymbol{\hat{\beta}}_{P,t}$.

The parameters in $\boldsymbol{\beta_{P,t}}$ is derived by solving the combined estimating equation $\psi_t(\boldsymbol{\beta_{P,t}}) = 0$, where
\begin{equation}
    \psi_t(\boldsymbol{\beta_{P,t}}) := \psi_{p,t}(\boldsymbol{\beta_{P,t}}) + \psi_{S,t}( \boldsymbol{\beta_{P,t}}),
\label{ch4:eqn:combined-GEF}
\end{equation}
where
\begin{itemize}
    \item $\boldsymbol{\psi_{P,t}}(\boldsymbol{\beta_{P,t}}) := \boldsymbol{\psi_{P,t}}(\alpha_{P,t}, \boldsymbol{\beta_{P,t}})$, of the same form as the generalized estimating function for presence in (\ref{ch4:eqn:presence_GEF}),
    
    \item $\boldsymbol{\psi_{S,t}}(\boldsymbol{\beta_{P,t}}) : = \boldsymbol{\psi_{S,t}}(\alpha_{S,t,1|2}, \ldots, \alpha_{S,t,L-1|L}, \hat{\gamma}_t \boldsymbol{\beta_{P,t}})$, that is, the generalized estimation function (\ref{ch4:eqn:severity-separate_GEF}) with the substitution of the $\boldsymbol{\beta_{S,t}}$ term with $ \hat{\gamma}_t \boldsymbol{\beta_{P,t}}$.
\end{itemize}

 This combined estimating function $\psi_t(\boldsymbol{\beta_{P,t}})=0$ arising from the estimating function (\ref{ch4:eqn:combined-GEF}) is solved through the modified Newton-Raphson method, updating $\boldsymbol{\beta_{P,t}}$ for $t=1, \ldots, T$ iteratively until convergence:
\begin{align*}
    \boldsymbol{\beta_{P,t}^{(h+1)}} = \boldsymbol{\beta_{P,t}^{(h)}} + \kappa \bigg( &\sum_{i_p=1}^{N_{p,t}} \boldsymbol{D_{p,i_p,t}^{(h)T} V_{p,i_p,t}^{(h)-1}D_{p,i_p,t}^{(h)}} \nonumber\\
    &+ \sum_{i_s=1}^{N_s} \boldsymbol{D_{S,i_s,t}^{(h)T} V_{S,i_s,t}^{(h)-1}D_{S,i_s,t}^{(h)}} + \psi_t(\boldsymbol{\beta_{P,t}}^{(h)}) \psi_t^T(\boldsymbol{\beta_{P,t}}^{(h)}) \bigg)^{-1} \psi_{t}(\boldsymbol{\beta_{P,t}}^{(h)})
    \label{ch4:eqn:NR_beta_iterate_combined_GEE}
\end{align*}

\subsection{Correlation Parameters}
\label{ch4:subsection:Correlation parameters}

The clustering of the outcome induces correlations at each hierarchical level. In our time-specific models, consider two outcomes $Y_{tijk}$ and $Y_{ti'j'k'}$ measured at the same time point $t$. Since individuals are treated as independent sampling units, outcomes are uncorrelated when $i \neq i'$. However, when $i = i'$, correlation arises due to shared individual-level characteristics such as genetics and lifestyle. In particular, outcomes from different teeth ($j \neq j'$) of the same individual are likely to be correlated. Furthermore, surfaces $k$ and $k'$ from the same individual—and especially those on the same tooth ($j = j'$)—are subject to stronger correlation due to shared biological environment during early tooth development and mineralization.

To construct working correlation structures for our GEEs, we explicitly account for the multilevel dependencies inherent in the data. Unlike standard GEE applications that model repeated measures at a single level—typically over time—our setting involves spatially indexed outcomes observed at a single time point, requiring more nuanced handling. Conventional structures like exchangeable or AR(1) are inadequate when used in isolation. In our application, the AR(1) structure is defined over the spatial ordering of teeth rather than time, based on the standard dental numbering scheme, and reflects the assumption that adjacent teeth within the mouth exhibit stronger correlation. To address the complexity of these dependencies, we build composite working correlation matrices by combining structures across hierarchical levels using the Kronecker product. This approach enables us to model within-individual dependencies while capturing correlations across teeth and surfaces. 

We define correlation parameters separately for the presence and severity models at time point $t$. Let $\rho_{P,t,jk,j'k'}$ denote the correlation between presence outcomes $W_{P,tijk}$ and $W_{P,ti'j'k'}$, and $\rho_{S,t,jk,j'k'}$ denote the correlation between severity outcomes $W_{S,tijk}$ and $W_{S,ti'j'k'}$. These are constructed using combinations of standard structures—independence, exchangeable, AR(1), and jackknife—at the appropriate hierarchical level.

\paragraph{Standard Correlation Structures:}
\begin{itemize}
    \item \textbf{Independence}:
    \[
    \rho_{P,t,jk,j'k'} = 
    \begin{cases}
    1, & j = j',\; k = k', \\
    0, & \text{otherwise}
    \end{cases}
    \]
    \[
    \rho_{S,t,jk,j'k'} =
    \begin{cases}
    1, & j = j',\; k = k', \\
    0, & \text{otherwise}
    \end{cases}
    \]

    \item \textbf{Exchangeable} (with parameters \(\rho_{P,t,\text{exch}},\; \rho_{S,t,\text{exch}}\)):
    \[
    \rho_{P,t,jk,j'k'} = 
    \begin{cases}
    1, & j = j',\; k = k', \\
    \rho_{P,t,\text{exch}}, & \text{otherwise}
    \end{cases}
    \]     \[
    \rho_{S,t,jk,j'k'} = 
    \begin{cases}
    1, & j = j',\; k = k', \\
    \rho_{S,t,\text{exch}}, & \text{otherwise}
    \end{cases}
    \]

    \item \textbf{AR(1)} (with parameters \(\rho_{P,t,\text{AR(1)}},\; \rho_{S,t,\text{AR(1)}}\)):
    \[
    \rho_{P,t,jk,j'k'} = 
    \begin{cases}
    1, & j = j', \\
    \rho_{P,t,\text{AR(1)}}^{|j-j'|}, & \text{otherwise}
    \end{cases}
    \]     \[
    \rho_{S,t,jk,j'k'} = 
    \begin{cases}
    1, & j = j', \\
    \rho_{S,t,\text{AR(1)}}^{|j-j'|}, & \text{otherwise}
    \end{cases}
    \]

    \item \textbf{Jackknife}: With $\rho_{P,t,jk,j'k',\text{jackknife}}$ and $\rho_{S,t,jk,j'k',\text{jackknife}}$ estimated from the data,
    \[
    \rho_{P,t,jk,j'k'} = 
    \begin{cases}
    1, & j = j',\; k = k', \\
    \rho_{P,t,jk,j'k',\text{jackknife}}, & \text{otherwise}
    \end{cases}
    \]     \[
    \rho_{S,t,jk,j'k'} = 
    \begin{cases}
    1, & j = j',\; k = k', \\
    \rho_{S,t,jk,j'k',\text{jackknife}}, & \text{otherwise}
    \end{cases}
    \]
\end{itemize}

\paragraph{Severity Model Indicator Correlation:} As discussed in Section \ref{ch4:subsection:Severity Model}, the GEE model for the severity outcome $W_{S,tijk}$ is based on the transformed vector of indicator variables \(\boldsymbol{Z}_{S,tijk} = (Z_{S,tijk,1}, Z_{S,tijk,2}, Z_{S,tijk,3})\), with
\[
\boldsymbol{Z}_{S,tijk} \sim \text{Multinomial}(n=1, \pi_{S,tijk,1}, \pi_{S,tijk,2}, \pi_{S,tijk,3}).
\]
The covariances between indicators are:
\[
\text{Cov}(Z_{S,tijk,l}, Z_{S,tijk,l'}) = 
\begin{cases}
-\pi_{S,tijk,l} \pi_{S,tijk,l'}, & l \neq l', \\
\pi_{S,tijk,l}(1 - \pi_{S,tijk,l}), & l = l'.
\end{cases}
\]
The corresponding correlations are:
\[
\theta_{tijk,l,l'} := \text{Cor}(Z_{S,tijk,l}, Z_{S,tijk,l'}) = 
\begin{cases}
1, & l = l', \\
\frac{\text{Cov}(Z_{S,tijk,l}, Z_{S,tijk,l'})}{\sqrt{\text{Var}(Z_{S,tijk,l}) \text{Var}(Z_{S,tijk,l'})}}, & l \neq l'.
\end{cases}
\]

\paragraph{Composite Working Correlation Structure:}
The full composite correlation for the severity model is given by:
\[
\rho_{S,t,jkl,j'k'l'} := \text{Cor}(Z_{S,tijk,l}, Z_{S,tijk,l'}) =
\begin{cases}
\rho_{S,t,jk,j'k'} \cdot \theta_{tijk,l,l'}, & i = i',\; j = j',\; k = k', \\
\rho_{S,t,jk,j'k'}, & \text{otherwise}.
\end{cases}
\]

Alternatively, this structure can be constructed using matrix operations. Let $\boldsymbol{B}_{S,tijk}$ be the within-observation correlation matrix of the indicator variables for observation $(i,t,j,k)$:
\begin{equation*}
    \boldsymbol{B}_{S,tijk} =
\begin{pmatrix}
1 & \theta_{tijk,1,2} & \theta_{tijk,1,3} \\
\theta_{tijk,2,1} & 1 & \theta_{tijk,2,3} \\
\theta_{tijk,3,1} & \theta_{tijk,3,2} & 1
\end{pmatrix}.
\end{equation*}
Stacking these across all observations within cluster $(i,t)$, define:
\begin{equation*}
    \boldsymbol{B}_{S,it} =
\begin{pmatrix}
\boldsymbol{B}_{S,it11} & \boldsymbol{0} & \ldots & \boldsymbol{0} \\
\boldsymbol{0} & \boldsymbol{B}_{S,it12} & \ldots & \boldsymbol{0} \\
\vdots & \vdots & \ddots & \vdots \\
\boldsymbol{0} & \boldsymbol{0} & \ldots & \boldsymbol{B}_{S,tijk}
\end{pmatrix}.
\end{equation*}
Let $\boldsymbol{\Phi}_{S,it}$ be the between-observation correlation matrix of dimension $JK \times JK$ (e.g., exchangeable, AR(1), etc.) for the severity model at time $t$. For instance, for the exchangeable correlation structure,
\begin{equation*}
  \boldsymbol{\Phi_{S,it}} = 
\begin{pmatrix}
1 & \rho_{S,t,exch} & \ldots & \rho_{S,t,exch}\\
\rho_{S,t,exch} & 1 & \ldots & \rho_{S,t,exch}\\
\vdots & \vdots & \ddots & \vdots\\
\rho_{S,t,exch} & \rho_{S,t,exch} & \ldots & 1\\
\end{pmatrix}
\end{equation*}
Then the full composite correlation matrix is:
\begin{equation*}
    \boldsymbol{R}_{S,it} = (\boldsymbol{\Phi}_{S,it} \otimes \boldsymbol{J}_{n_i}) \circ \boldsymbol{B}_{S,it},
\end{equation*}
where $\otimes$ denotes the Kronecker product, $\circ$ denotes the Hadamard (elementwise) product, and $\boldsymbol{J}_{n_i}$ is the $n_i \times n_i$ matrix of ones. The composite correlation matrix for presence is derived similarly. Since the outcomes remain univariate under the presence model and do not require expansion into multiple indicator variables, the analogous $\boldsymbol{B}_{P,tijk}$ term is equal to $1$. As a result, the working correlation structure simplifies to:
\begin{equation*}
    \boldsymbol{R}_{P,it} = \boldsymbol{\Phi}_{P,it} \otimes \boldsymbol{J}_{n_i},
\end{equation*}
where $\boldsymbol{\Phi}_{P,it}$ represents the cluster-level correlation matrix for presence at time $t$, and $\boldsymbol{J}_{n_i}$ denotes an $n_i \times n_i$ matrix of ones.

\subsubsection{Estimation of Correlation Parameters}
\label{ch4:subsubsection:Estimation of correlation parameters}
The correlation parameters for the exchangeable and AR(1) cases are estimated according to the following schemes proposed by \cite{liang1986longitudinal}. Suppose we have
$$r_{P,tijk} = \frac{w_{P,tijk} - \mu_{P,tijk}}{\sqrt{Var(W_{P,tijk})}},$$
 and 
$$r_{S,tijk,l} = \frac{Z_{S,tijk,l} - \pi_{S,tijk,l}}{\sqrt{Var(Z_{S,tijk,l})}}.$$ 
We define
\begin{equation}
\label{ch4:eqn:phi_P}
\phi_{P,t} = \frac{1}{\sum_{i=1}^{N} \sum_{j,j'=1}^{J_{p,t_i}} \sum_{k,k'=1}^{K_{p,t_i}} 1} \sum_{i=1}^{N} \sum_{j,j'=1}^{J_{p,t_i}} \sum_{k,k'=1}^{K_{p,t_i}} \hat{r}^2_{P,tijk}
\end{equation}
and
\begin{equation}
\label{ch4:eqn:phi_S}
{\phi}_{S,t} = \frac{1}{\sum_{i=1}^{N} \sum_{l=1}^{L} \sum_{j=1}^{J_{S,ti}} \sum_{\substack{k=1}}^{K_{S,ti}} 1}
 \sum_{i=1}^{N} \sum_{l=1}^{L} \sum_{j=1}^{J_{S,ti}} \sum_{\substack{k=1}}^{K_{S,ti}} \hat{r}^2_{S,tijk,l}.
 \end{equation} 
 Then, the exchangeable correlation parameter under the presence model at time-point $t$ is given by
  \begin{equation*}
      \hat{\rho}_{P,t,exch} = \frac{N_{P,t,exch}}{D_{P,t,exch}},
  \end{equation*}
  with
  \begin{equation*}
      N_{P,t,exch} = {\phi}_{P,t} 
  \sum_{i=1}^{N} \sum_{\substack{j,j'=1 \\ j \neq j'}}^{J_{P,ti}} \sum_{\substack{k,k'=1 \\ k \neq k}}^{K_{P,ti}} \hat{r}_{P,tijk} \hat{r}_{P,tij'k'}.
  \end{equation*}
  and
  \begin{equation*}
      D_{P,t,exch} ={\sum_{i=1}^{N} \sum_{j,j'=1}^{J_{P,ti}} \sum_{\substack{k,k'=1 \\ j \neq j, k \neq k}}^{K_{P,ti}} 1}.
  \end{equation*}

Similarly, under the severity model, the exchangeable correlation parameter can be estimated as
\begin{equation*}
\hat{\rho}_{S,t,exch} = \frac{N_{S,t,exch}}{D_{S,t,exch}}      
\end{equation*}
where
  \begin{equation*}
N_{S,t,exch} =
  {\phi}_{S,t}
 \sum_{i=1}^{N} \sum_{l,l'=1}^{L} \sum_{j,j'=1}^{J_{S,ti}} \sum_{\substack{k,k'=1 \\ l \neq l', j \neq j', k \neq k}}^{K_{S,ti}} \hat{r}_{S,tijk,l} \hat{r}_{S,tij'k',l'}
\end{equation*}
and
\begin{equation*}
    D_{S,t,exch} = {\sum_{i=1}^{N} \sum_{l,l'=1}^{L} \sum_{j,j'=1}^{J_{S,ti}} \sum_{\substack{k,k'=1 \\ l \neq l', j \neq j', k \neq k}}^{K_{S,ti}} 1}
\end{equation*}
 For the AR(1) cluster correlation structure, the correlation parameter $\rho_{P,t,AR(1)}$ is estimated from the slope of the regression of $\log(\hat{r}_{P,tijk} \hat{r}_{P,tij'k'})$ on 
 $|j-j'|$.
 Likewise, $\rho_{S,t,AR(1)}$ is estimated from the slope of the regression of $\log(\hat{r}_{S,tijk,l} \hat{r}_{S,tij'k',l})$ on 
 $|j-j'|$.
 
As a data-driven approach to estimating the correlation parameters, we adopt a jackknife resampling technique. This method systematically excludes one individual at a time and recomputes the correlation estimate based on the remaining data, allowing for reduced bias and improved stability in the estimation of the correlation structure. For the jackknifed correlation parameters, the estimation is carried out as follows:

 \begin{equation*}
   \hat{\rho}_{P,t,jk,j'k', jackknife (-i)} = \frac{N_{P,t,jk,j'k', jackknife (-i)}}{D_{P,t,jk,j'k', jackknife (-i)}} , 
 \end{equation*}
 with
 \begin{equation*}
      {N_{P,t,jk,j'k', jackknife (-i)}} =
    {\phi}_{P,t,-i}
 \sum_{\substack{i^*, i^{**}=1 \\ i^*, i^{**} \neq i}}^{N} \sum_{j,j'=1}^{J_{P,ti}} \sum_{k,k'=1}^{K_{P,ti}} 
 \hat{r}_{P,t{i^*}jk} \hat{r}_{P,t{i^{**}}j'k'}  ,
 \end{equation*}
 and
 \begin{equation*}
     {D_{P,t,jk,j'k', jackknife (-i)}} = {\sum_{\substack{i^*, i^{**}=1 \\ i^*, i^{**} \neq i}}^{N} \sum_{j,j'=1}^{J_{P,ti}} \sum_{k,k'=1}^{K_{P,ti}} 1 },
 \end{equation*}
 where $\phi_{P,t,-i}$ corresponds to the quantity in (\ref{ch4:eqn:phi_P}) computed leaving out cluster $i$.
Then,
\begin{equation*}
    \hat{\rho}_{P,t,jk,j'k', jackknife} = \frac{1}{N} \sum_{i=1}^{N} \hat{\rho}_{P,t,jk,j'k', jackknife (-i)}.
\end{equation*}
Likewise, 
\begin{equation*}
    \hat{\rho}_{S,t,jk,j'k', jackknife} = \frac{1}{N} \sum_{i=1}^{N} \hat{\rho}_{S,t,jk,j'k', jackknife (-i)},
\end{equation*}
where
\begin{equation*}
    \hat{\rho}_{S,t,jk,j'k', jackknife (-i)} = \frac{N_{S,t,jk,j'k', jackknife (-i)}}{D_{S,t,jk,j'k', jackknife (-i)}}, 
\end{equation*}
with
\begin{equation*}
    N_{S,t,jk,j'k', jackknife (-i)} = 
    {\phi}_{S,t,-i}
    \sum_{\substack{i^*, i^{**}=1 i^*, i^{**} \neq i}}^{N} 
    \sum_{l,l'=1}^{L-1} \sum_{j,j'=1}^{J_{S,ti}} \sum_{k,k'=1}^{K_{S,ti}} 
    \hat{r}_{S,t{i^*}jk,l} \hat{r}_{S,t{i^{**}}j'k',l'},
\end{equation*}
and
\begin{equation*}
D_{S,t,jk,j'k', jackknife (-i)} = {\sum_{\substack{i^*, i^{**}=1 \\ i^*, i^{**} \neq i}}^{N} \sum_{l,l'=1}^{L-1} \sum_{j,j'=1}^{J_{S,ti}} \sum_{k,k'=1}^{K_{S,ti}} 1}
\end{equation*}
with $\phi_{S,t,-i}$ corresponding to the quantity in (\ref{ch4:eqn:phi_S}) computed leaving out cluster $i$.

\subsection{An Empirical Bayes Adjustment}
\label{ch4:subsection:An Empirical Bayes Adjustment}
The models described thus far are specific to a given time point $t$, where $t = 1, \ldots, T$. That is, each time-specific model estimates effects based solely on the subset of data corresponding to that particular time point. While this approach is useful for isolating time-specific associations, it does not fully leverage the longitudinal structure inherent in the data. Ideally, we would like these estimates to borrow information across time points, enabling a more coherent understanding of temporal trends and capturing potential longitudinal effects. Across dental visits at times $t = 1, \ldots, T$, the proposed models yield time-specific estimates $\beta_{P,t,g}$ for the presence component and $\beta_{S,t,g}$ for the severity component, where $g = 1, \ldots, q$ indexes the covariates. That is, each coefficient captures the age-specific effect of a given protective or risk factor on dental fluorosis. To ``borrow strength'' across the available time points, we apply an empirical Bayes adjustment \citep{robbins1964empirical} using the James-Stein shrinkage estimator \citep{james1992estimation}. This approach possesses a desirable large-sample property: it yields a set of refined estimators whose average mean squared error (MSE) is lower than that of the corresponding original estimators.

The calculation of James-Stein estimates requires the standard deviations of the estimates to be known; equivalently, the estimates must be standardized. Thus, we perform shrinkage on standardized coefficient estimates. As an illustration, we first outline how the James-Stein estimates are obtained from the presence model.

Let the vector $\boldsymbol{\hat{\beta}_{P,g}}$ denote the collection of time-specific coefficient estimates for the $g$-th predictor across time points $t = 1, \ldots, T$:
\[
\boldsymbol{\hat{\beta}_{P,g}} = (\hat{\beta}_{P,1,g}, \hat{\beta}_{P,2,g}, \ldots, \hat{\beta}_{P,T,g}) \quad \text{for } g = 1, \ldots, q.
\]
Let $\mathrm{SD}(\hat{\beta}_{P,t,g})$ denote the standard deviation of $\hat{\beta}_{P,t,g}$, obtained via jackknife resampling over clusters. Define the standardized estimates as:
\[
\boldsymbol{\hat{\beta}^{*}_{P,g}} = \left( \frac{\hat{\beta}_{P,1,g}}{\mathrm{SD}(\hat{\beta}_{P,1,g})}, \ldots, \frac{\hat{\beta}_{P,T,g}}{\mathrm{SD}(\hat{\beta}_{P,T,g})} \right).
\]
The James-Stein estimator $\boldsymbol{\hat{\beta}^{* \mathrm{JS}}_{P,g}}$ of the standardized presence estimates $\boldsymbol{\hat{\beta}^{*}_{P,g}}$ is given by:
\begin{equation*}
\boldsymbol{\hat{\beta}^{* \mathrm{JS}}_{P,g}} = \left( 1 - \frac{T - 2}{\|\boldsymbol{\hat{\beta}^{*}_{P,g}}\|^2} \right) \boldsymbol{\hat{\beta}^{*}_{P,g}},
\end{equation*}
where $\|\cdot\|$ denotes the Euclidean norm.
Similarly, for the severity model, define:
\[
\boldsymbol{\hat{\beta}_{S,g}} = (\hat{\beta}_{S,1,g}, \hat{\beta}_{S,2,g}, \ldots, \hat{\beta}_{S,T,g}),
\]
and the corresponding standardized vector:
\[
\boldsymbol{\hat{\beta}^{*}_{S,g}} = \left( \frac{\hat{\beta}_{S,1,g}}{\mathrm{SD}(\hat{\beta}_{S,1,g})}, \ldots, \frac{\hat{\beta}_{S,T,g}}{\mathrm{SD}(\hat{\beta}_{S,T,g})} \right).
\]
Then the James-Stein estimator for the severity component is:
\begin{equation*}
    \boldsymbol{\hat{\beta}^{* \mathrm{JS}}_{S,g}} = \left( 1 - \frac{T - 2}{\|\boldsymbol{\hat{\beta}^{*}_{S,g}}\|^2} \right) \boldsymbol{\hat{\beta}^{*}_{S,g}}.
\end{equation*}
The direction and magnitude of the James-Stein estimates reflect the effect of each predictor on dental fluorosis at a given level: positive values indicate protective effects, while negative values suggest increased risk. We report the James-Stein adjusted standardized estimates along with their 95\% bootstrapped confidence intervals (CIs), with bootstrapping performed at the level of the individual. A CI that excludes zero is interpreted as evidence of a statistically significant effect, while the sign of the point estimate indicates whether the predictor is protective against dental fluorosis or associated with increased risk.

\subsection{Model Selection via Rank Aggregation}
\label{ch4:subsection:Model selection via rank aggregation}

To determine which working correlation structures yield the most precise estimates in our models, we employ a data-driven strategy based on rank aggregation \citep{pihur2009rankaggreg}. Traditional model selection tools such as the Akaike Information Criterion (AIC) \citep{akaike1987factor} are not directly applicable to GEE models, as GEEs do not rely on full likelihood specification. Although the Quasi-likelihood Information Criterion (QIC) \citep{pan2001akaike} can be used to compare nested GEE models and assess different working correlation structures, it is limited in its ability to compare models across differing outcome distributions.

To enable broader and more flexible comparisons, we adopt a precision-based ranking approach. For each age $t \in \{9, 13, 17, 23\}$ and for each candidate working correlation structure, we evaluate the bootstrapped mean squared errors (MSEs) of all predictors in the presence model, with bootstrapping performed at the level of the individual (details on computing the bootstrapped MSE is provided in Section \ref{ch4:section:Results}). For each predictor, the correlation structures are ranked from lowest MSE (indicating higher precision) to highest MSE (lower precision). These predictor-specific ranked lists—one for each of the $q$ predictors—represent the relative performance of different correlation structures.

To summarize this information across predictors and identify an overall preferred correlation structure, we apply a rank aggregation method as proposed by \citet{pihur2009rankaggreg}. Let $L_1, L_2, \ldots, L_q$ denote the ordered lists corresponding to the $q$ predictors. Our objective is to find a consensus ranking $\delta^*$ that best represents the information contained in these input lists. Formally, we define $\delta^*$ as the permutation that minimizes a weighted sum of distances, $\Phi(\delta) = \sum_{i=1}^{q} w_i \, d(\delta, L_i)$, to the individual ranked lists:
\begin{equation*}
    \delta^* := \arg\min_{\delta} \; \Phi(\delta),
\end{equation*}
where:
\begin{itemize}
    \item $w_i$ is the weight assigned to list $L_i$ (we assume equal weights, i.e., $w_i = 1$ for all $i$),
    \item $d(\cdot, \cdot)$ is a distance function that quantifies the dissimilarity between two ranked lists, such as the Kendall tau or Spearman footrule distance.
\end{itemize}

This optimization problem is solved using a Cross-Entropy Monte Carlo algorithm, which efficiently explores the space of permutations to identify the consensus list $\delta^*$. This list reflects the overall ranking of working correlation structures that yield the most precise estimates across predictors at a given age. Subsequently, the consensus rankings from each age are themselves aggregated—again using rank aggregation—to produce a final, age-agnostic summary ranking. This final list identifies the working correlation structures that most consistently result in precise estimates across the different time-specific presence models.

While this illustration is based on the presence model, we apply the same procedure to the severity model as well as to the combined presence-and-severity framework. Depending on the analysis objective, the rank aggregation framework can also be used flexibly to examine the relative performance of a specific correlation structure across modeling components—for example, to assess whether a given structure yields more precise estimates in the presence model compared to the severity model. Such comparisons are not feasible using traditional criteria like QIC.

\section{Analysis of the Iowa Fluoride Study Data}
\label{ch4:section:Real data analysis}

\subsection{An Overview of the Data}
\label{ch4:subsection:An overview of the IFS data}
One of the aims of the Iowa Fluoride Study (IFS) is to investigate the relationships between dental fluorosis and various dietary and non-dietary factors in a cohort of Iowa children. Dental fluorosis, a condition affecting tooth enamel, occurs due to excessive fluoride intake during tooth development in early childhood, though it only becomes visible years later after the teeth have erupted \citep{hong2006fluoride}. This study specifically focuses on the maxillary (upper jaw) incisors, which are among the first teeth to erupt and are most clinically relevant due to their visibility and esthetic importance. Given their early eruption, the analysis uses early-life exposures to fluoride and other factors — aggregated over the ages 0 to 5 — as predictors for all outcome ages. Additional details on these predictors are provided in Section S1 of the Supplementary File.

\begin{figure}
    \centering
    \includegraphics[width=1\linewidth]{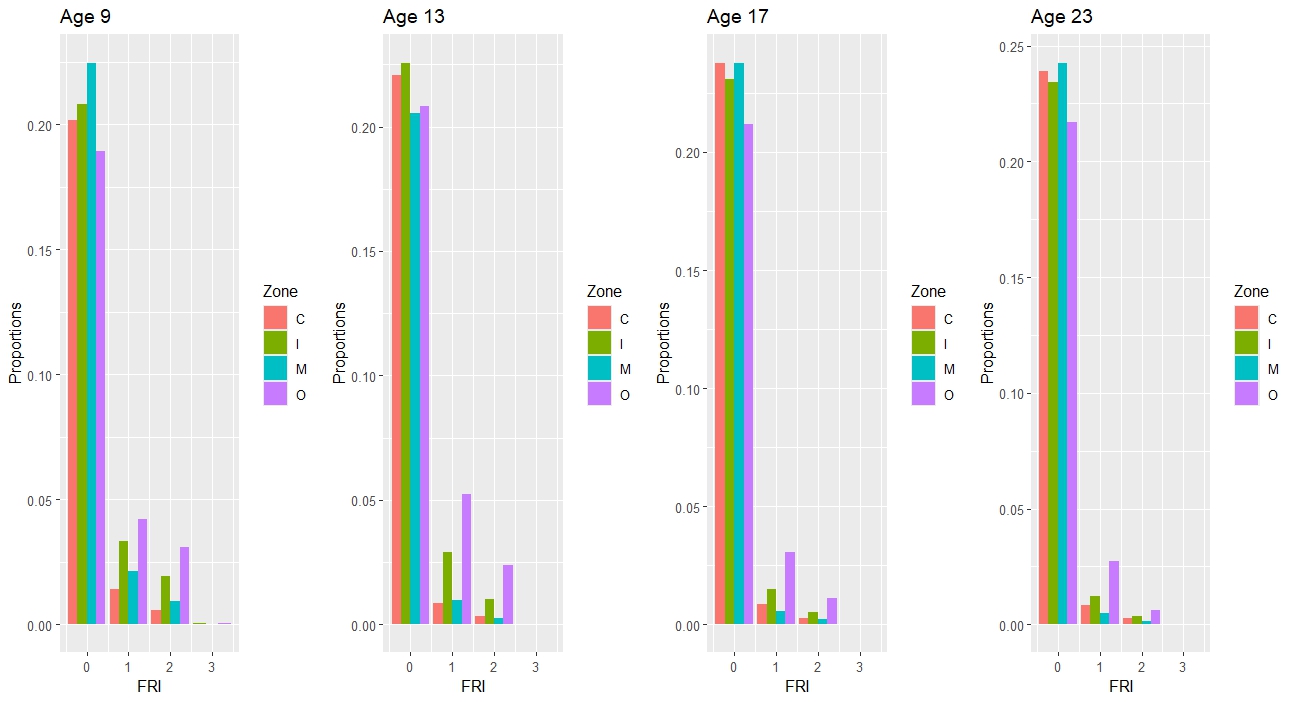}
    \caption{FRI score distributions across ages 9, 13, 17, and 23. The FRI categories include: 0=no fluorosis; 1=less than 50\% of the zone covered by white striations; 2=more than 50\% of the zone covered by white striations; and 3=substantial pitting, staining and/or deformity, considered over the following zones (from root to tip): cervical third (C), middle third (M), and incisal third (I) of the buccal surface, and incisal edge (E). Bar heights represent the relative frequency of each FRI score–zone combination, calculated as the proportion of the total number of observations at that age.
}
    \label{ch4:fig:FRI_distribution}
\end{figure}

The Fluorosis Risk Index (FRI) was used to assess fluorosis severity of the permanent teeth, categorizing it into four levels for each specific zone on each tooth: (0) no fluorosis; (1) less than 50\% of the zone covered by white striations; (2) more than 50\% of the zone covered by white striations; and (3) substantial pitting, staining, and/or deformity. FRI scores were collected at four time-points, corresponding to the ages of 9, 13, 17, and 23. For each tooth, scores were recorded for each of the four zones: C (cervical third, nearest the gum), M (middle third), I (incisal third), and E (incisal edge, tip of the tooth). The distributions of FRI scores across different ages is shown in Figure \ref{ch4:fig:FRI_distribution}. There is a clear predominance of observations in the lowest (no) fluorosis category (no fluorosis) across all zones, with this preponderance becoming more pronounced with increasing age.

The dataset includes 606 children with available fluorosis outcomes at one or more of these time-points and complete covariate information at age 5, resulting in a total of 21,407 zone observations. The cohort consists of 425 children assessed at age 9, 419 at age 13, 341 at age 17, and 253 at age 23. The analysis incorporates covariates of categorical variables for tooth location (tooth 7 upper right, lateral incisor as reference; 8 right central; 9 left central; 10 left lateral), and tooth zone (C, M, I, or E, with zone C as the reference). Additionally, three continuous covariates measure fluoride exposures from home tap water, professional dental fluoride treatments, and all combined fluoride sources. Other predictors considered include patient age at the time of the dental examination, intake of sugary beverages, tooth brushing frequency, and dental visit frequency. Additional details on these predictors are provided in Section S1 of the Supplementary File.

\subsection{Results}
\label{ch4:section:Results}
For brevity, we present estimation results corresponding to the best-performing correlation structures — identified via rank aggregation (outlined in Section \ref{ch4:subsection:Model selection via rank aggregation}) — in the main text of this paper. The AR(1) working correlation structure emerged as the top choice for the separate presence modeling, while jackknifing consistently performed the best across the correlation choices in the separate severity and combined models. 
Estimates from both separate and combined presence and severity models are shown in Tables~\ref{ch4:table:sep:pres:ar1}–\ref{ch4:table:comb:sev:jackknifed}, with additional correlation structures and their rankings provided in Section S5 of the Supplementary File. We fit models of the form A.$c.t$, B.$c.t$, or C.$c_P.c_S.t$, reflecting the modeling scheme (A: separate presence model; B: separate severity model; C: combined modeling of presence and severity), the assumed correlation structure for each separate model ($c$), or for the presence ($c_P$) and severity ($c_S$) components of the combined model. The index $t$ denotes the time-points of dental assessments, with $t=1,2,3,4$ corresponding to ages 9, 13, 17, and 23, respectively. Correlation structure codes are defined as follows: 1 for independence, 2 for exchangeable, 3 for AR(1), and 4 for jackknifing. For example, model A.1.3 denotes a separate presence model using an exchangeable working correlation structure for age 17. Similarly, model C.2.2.4 refers to a combined model with exchangeable correlation structures for both presence and severity components, applied to age 23. We report separate estimates for the presence and severity components of the combined model to facilitate direct comparison with their respective counterparts from the piecewise models. Additionally, although the combined models are fit jointly, the standardization and James-Stein shrinkage steps are performed separately for each component, resulting in distinct sets of estimates. 

We illustrate our findings using model A.3.4 (Table~\ref{ch4:table:sep:pres:ar1}), the separate presence model that employs the AR(1) working correlation structure. The estimated effect of  {Dental\_age} is $1.444$, with its positive sign indicating an association with decreased fluorosis risk. To update this estimate to its James-Stein counterpart—borrowing strength from the corresponding estimates in models A.3.1, A.3.2, and A.3.3 at the other ages — the first step involves standardizing the estimate. We compute its standard deviation (SD) using jackknife resampling by systematically leaving out one cluster at a time and evaluating the variability across the resulting leave-one-cluster-out estimates. This yields a standardized estimate of $0.648$. The standard error of this value is obtained via cluster-level bootstrapping with a resample size of $B = 500$ \citep{davison1997bootstrap}, where individuals (clusters) are resampled with replacement—the choice of $B$ deemed sufficient for our purposes \citep{choo2016marginal}. The standard deviation of the resulting bootstrapped standardized estimates yields an SE of $0.687$. Next, using the standardized estimates of {Dental\_age} from age-specific presence models with the same working correlation structure — A.3.1, A.3.2, A.3.3, and A.3.4 corresponding to ages 9, 13, 17, and 23 — we apply James-Stein shrinkage to obtain the updated estimate. For model A.3.4, the James-Stein updated standardized estimate is $0.215$ with an SE of $0.718$. Using these bootstrapped James-Stein estimates, we compute a quantile-based 95\% confidence interval (CI). For  {Dental\_age}, this results in a 95\% CI of $(-0.805,\ 1.722)$, which includes zero and thus indicates a non-significant effect at the 0.05 level. In contrast, for  {Avg\_homeppm}, the 95\% CI is $(-2.793,\ -0.221)$, indicating a statistically significant negative association at the 0.05 level and identifying it as a risk factor for the presence of dental fluorosis at age 23. It is important to note that the James-Stein point estimates and their corresponding confidence intervals (CIs) are obtained through different procedures. The James-Stein point estimate is calculated using the full dataset, while the 95\% CI is derived from empirical quantiles based on a leave-one-cluster-out bootstrap. Consequently, it is possible — and not unexpected — for the point estimate to fall outside the associated bootstrap CI, due to both sampling variability and the shrinkage effect introduced by the James-Stein procedure. For covariates whose CIs do not overlap with zero (e.g., entirely negative or entirely positive, indicating significance of the covariate at the 0.05 level), we consider the direction of the effect to be more reliably reflected by the CI than by the point estimate alone. Therefore, we interpret covariate effects primarily based on the James-Stein intervals, which tend to exhibit improved stability and reduced variance compared to the original estimates.

In addition to computing the standard error (SE) of the James-Stein estimator, we also estimate its bias and, consequently, its mean squared error (MSE). The MSE serves as a criterion for model comparison through the rank aggregation procedure described in Section \ref{ch4:subsection:Model selection via rank aggregation}. Both the SE and bias are obtained using a cluster-level bootstrap approach with a resample size of $B=500$. To estimate the bias, we treat the James-Stein estimate based on the full relevant dataset as the "true" parameter value. The bias for each bootstrap sample is computed as the difference between its James-Stein estimate and this reference value. The average of these differences is reported as the bias. Finally, the MSE is calculated using the estimated bias and SE.

For severity models, the interpretation follows a similar logic: negative values indicate a greater risk of more severe fluorosis, while positive values correspond to reduced severity. These interpretations extend analogously to the presence and severity components of the combined models. Across different working correlation structures, parameter estimates were generally similar, consistent with the known robustness of GEEs to mis-specification of the working correlation structure. Nevertheless, to identify the best-performing structure, we perform model selection using the rank aggregation approach described in Section~\ref{ch4:subsection:Model selection via rank aggregation}.

Significant risk effects for fluorosis presence are observed with {Avg\_homeppm},  {Tooth10}, and {ZoneE}, with risk increasing progressively from the surface closest to the gum, {ZoneC}, to the tip of the tooth, {ZoneE} (Table \ref{ch4:table:sep:pres:ar1}). In contrast, {Tooth10} demonstrates a protective effect across its zones. Though not significant at the 0.05 level, {Dental\_age} emerges as a protective factor (Table \ref{ch4:table:sep:pres:ar1}). From the rank aggregation results (presented in the Supplementary File), the AR(1) working correlation structure yields more precise estimates overall for the presence models; the jackknifed cluster correlation structure performs well for the severity model as well as both components of the combined modeling.

These findings are further supported by the presence component of the combined model (C.4.4.4, 
Table \ref{ch4:table:comb:pres:jackknifed}), which yield results consistent with the separate presence models (
A.3.4). Significant risk factors again include  {Tooth8},  {Tooth9},  {ZoneM},  {ZoneI}, and  {ZoneE}, along with  {Avg\_homeppm}.  {Tooth10} continues to demonstrate protective effects across its surfaces, particularly at ages 9 and 13. For these models, the jackknifed working correlation structure performed the best overall (detailed in Section S5 of the Supplementary File).

Turning to the severity models, results from the separate specification (B.4.4, 
Table \ref{ch4:table:sep:sev:jackknifed}) show that covariate effects generally align in direction with those observed in the presence models (
A.3.4). However, statistical significance is less frequently observed—likely due to the reduced sample size, as these models are fit only on the subset of individuals with non-zero Fluorosis Risk Index (FRI) scores. Notable exceptions include ZoneI and ZoneE, which consistently show significant risk effects, as well as SugarAddedBeverageOzPerDay, which emerges as significant at later ages (17 and 23). The jackknifed working correlation structure was the optimal correlation structure for a majority of the models - age 9, age 17, and age 23 (detailed in the Supplementary File). 

The severity component of the combined models (
C.4.4.4, 
Table \ref{ch4:table:comb:sev:jackknifed}) leverages the full IFS dataset, thereby offering improved statistical power compared to the subset-based separate models. This enhancement enables the detection of additional significant effects. 
BrushingFrequencyPerDay and Avg\_homeppm are significant risk factors at younger ages (9 and 13), with BrushingFrequencyPerDay transitioning to a protective effect at ages 17 and 23. Tooth8, ZoneC, ZoneI, and ZoneE continue to show significant risk effects, with risk intensifying progressively from the gum line to the tip of the tooth. Again, the jackknifed working correlation structure performed the best for these models (detailed in the Supplementary File).

\section{Simulation Studies}
\label{ch4:section:Simulation studies}

We conducted simulation studies to evaluate the finite-sample performance of the proposed methodology. We begin by outlining the data generation process and the parameter settings used. We then assess the quality of the point estimates in terms of bias, variance, and mean squared error (MSE) to evaluate estimation accuracy. In addition, we examine whether the large-sample property of reduced average MSE is observed for the cluster sizes considered. As a proof of concept, we also evaluate the empirical size of our significance testing procedure—defined as the proportion of times the bootstrapped confidence interval excludes zero—and compare it to the nominal significance level.


\subsection{Data Generation}
\label{ch4:subsection:Simulation-Data-Generation}

We considered three choices for the number of clusters (individuals): $N = 30$, $50$, and $200$, each with a fixed cluster size of $n_{it} = 16$, corresponding to the full set of tooth-zone combinations (teeth 7–10 and zones C, M, I, E). For notational simplicity, we indexed these combinations using a single index $r$, ordered lexicographically, rather than explicitly using tooth ($j$) and zone ($k$) subscripts.

Simulation covariates were obtained by resampling $N$ unique individuals from the real IFS data, restricted to those with complete information across all 16 tooth-zone combinations. Eight predictors were used: two continuous variables (Dental\_age and Avg\_homeppm, and binary indicators for teeth and zones (Tooth8, Tooth9, Tooth10, ZoneM, ZoneI, and ZoneE).

To generate correlated ordinal outcomes, we adopted a latent variable approach in which each ordinal response arises from an underlying multivariate normal distribution. For the presence component, we generated a latent vector $\bm{L}{p,t,i} \sim \mathrm{MVN}{n_{it}}(0, \boldsymbol{\Sigma}_p)$, where the correlation matrix $\Sigma_p$ reflects one of three working structures: independence (identity matrix), exchangeable, or jackknifed. The specific choices of these structures and parameters are described in Section~\ref{ch4:subsection:Simulation-Parameters}. Each latent value was first transformed into a uniform random variable using the standard normal CDF, and then thresholded based on the estimated probabilities of belonging to each ordinal class (arising from \ref{ch4:eqn:presence_prob_expression}). These probabilities were derived from model parameters obtained from the age-specific combined models C.2.2.1–C.2.2.4, based on the analysis of the IFS data
, resulting in binary presence indicators.

A similar latent-to-ordinal process was used to simulate severity responses. Latent variables $\bm{L}{s,t,i}$ were drawn from $\mathrm{MVN}{n_{it}}(0, \boldsymbol{\Sigma_s})$ using the specified correlation structures. These were mapped to ordinal severity levels (1, 2, or 3) using cumulative probabilities from the severity model (arising from \ref{ch4:eqn:mu_severity-separate}). The overall FRI score for each observation was then constructed by combining the presence and severity indicators: observations with presence equal to 1 were assigned a severity level, while those with presence equal to 0 received an FRI score of 0.

\subsection{Parameter Choices}
\label{ch4:subsection:Simulation-Parameters}

We simulate data under four distinct combinations of working correlation structures for the presence and severity components.

\begin{enumerate}
    \item Presence: independence; Severity: exchangeable ($\rho_s = 0.6$)
    \item Presence: exchangeable ($\rho_p = 0.6$); Severity: independence
    \item Presence: exchangeable ($\rho_p = 0.3$); Severity: exchangeable ($\rho_s = 0.8$)
    \item Presence: jackknifed; Severity: jackknifed
\end{enumerate}

These structures were chosen to balance methodological relevance with computational feasibility. AR(1) structures, while conceptually appealing, were excluded due to the complexity of parameter estimation. Jackknifed structures are also computationally intensive, but we retained them to represent a fully data-driven approach to estimating within-cluster dependence. For brevity and to maintain consistency with the real data analysis, we present simulation results only for the final combination using jackknifed cluster correlations. Complete results for all correlation structure combinations are provided in Section S4 of the Supplementary File.

To control runtime, we generate 100 Monte Carlo datasets per age and cluster size, which is sufficient for finite-sample evaluation \citep{choo2016marginal}. Latent variables are sampled from multivariate normal distributions with the specified correlation structure. However, the nonlinear transformation used to simulate ordinal outcomes alters the magnitude of these correlations: while structured forms, like exchangeable correlation, maintain their general pattern, the precise ``true'' parameter values are not preserved.

To address this challenge, true parameter values for bias calculation are approximated using estimates from a large-scale simulation with $N = 10{,}000$ clusters under the independence structure. This simplification is reasonable, as GEE estimators remain consistent and asymptotically efficient even under mis-specified working correlation structures.

We denote simulation configurations using the notation A$_s$.c.t, B$_s$.c.t, and C$_s$.c$_P$.c$_S$.t. Here, A and B refer to presence and severity models, while C indicates the combined model. The subscript $s$ designates simulation. The indices $c$, $c_P$, and $c_S$ indicate the assumed working correlation: $1$ = independence, $2$ = exchangeable, $3$ = jackknifed. The final index $t = 1, 2, 3, 4$ corresponds to ages 9, 13, 17, and 23, respectively. In our setup, the same correlation structure is typically used for both data generation and model fitting. 

The case of jackknifed correlation ($c = 3$) is particularly interesting. In this setting, latent variables are generated using an unstructured correlation matrix estimated from a fixed reference dataset. However, the nonlinear transformation to ordinal outcomes distorts the underlying dependence, rendering the original correlation unrecoverable. As a result, the working correlation structure must be re-estimated from each simulated dataset via a fresh jackknife procedure. This setup closely mirrors real-world scenarios, where the true within-cluster correlation is unknown and must be inferred from the data itself.

\subsection{Estimation Performance}
\label{ch4:subsection:BiasVarianceMSE}

We evaluate estimator performance using bias, SE, and mean squared error (MSE), computed based on 100 simulation replicates for each cluster size $N$ and age. As expected, all three metrics decrease as $N$ increases, reflecting consistency and improved finite-sample performance. To limit the number of tables while still presenting results from each model type (presence, severity, combined presence, combined severity), we focus on age 23, corresponding to the jackknifed working correlation structures 
(Table \ref{ch4:table:sim:raw:jackknifed}). Results for additional working correlations are presented in Section S6 of the Supplementary File. 

As an example, consider Table~\ref{ch4:table:sim:raw:jackknifed}, which presents simulation results under data generated using jackknifed correlation structures for both the presence and severity components. Results are shown for the separate presence model (A$_s$.3.4), separate severity model (B$_s$.3.4), and the corresponding components of the combined model (C$_s$.3.3.4), across increasing cluster sizes $N = 30$, $50$, and $200$. At a small sample size of $N = 30$, several variables - particularly in the severity models - exhibit elevated bias, high SEs, and inflated MSEs. However, these metrics stabilize rapidly as $N$ increases. By $N = 200$, the estimates are substantially more precise, and both SE and MSE are markedly reduced, indicating strong finite-sample improvement.

Importantly, when comparing the separate and combined models at $N = 200$, we observe that MSE either remains unchanged or decreases across nearly all covariates. This trend holds for both the presence and severity components. For instance, MSE values in the combined presence model are comparable to those in the separate model, suggesting that joint modeling does not compromise precision. In terms of MSE, the combined severity model frequently outperforms the separate severity model, such as for Dental\_age and Avg\_homeppm.

\subsection{Effect of James-Stein Shrinkage on Average MSE}
\label{ch4:subsection:JamesStein-MSE}
In this section, we examine the behavior of standardized estimates and their James–Stein counterparts under increasing sample sizes. Within each model type (e.g., presence-only or combined presence), the MSE of standardized estimates generally decreases with $N$, reflecting improved estimation stability. Comparing across model types—such as the separate severity model B$_s$.3.4 (Table~\ref{ch4:table:sim:std:sev:jackknifed}) versus the combined severity model C$_s$.3.3.4 (Table~\ref{ch4:table:sim:std:comb:sev:jackknifed})—we observe that the average MSEs of the standardized estimates are often higher in the combined model compared to the separate models, which suggests that the benefits of combining information may not be fully realized in the standardized setting. In contrast, the direct (unstandardized) estimates from the combined models consistently show reduced MSEs, indicating that the combined framework does improve estimation performance on the original scale.

These discrepancies in the standardized setting can be attributed to two key factors. First, due to the complex nature of the combined model, the true underlying parameters in the simulation are not directly recoverable (discussed in Section \ref{ch4:subsection:Simulation-Parameters}). Although we can specify certain regression parameters for the latent variables in the simulation, their proportionality — a key aspect of the combined framework — is not guaranteed to hold after the required transformations are applied. This limitation is not unique to our chosen formulation; simulation setups employing alternative functional forms to link model components would encounter similar challenges. In addition, the theoretical justification for James–Stein shrinkage relies on assumptions of known (or unit) standard deviations and that the parameter vector follows an exact multivariate normal distribution. In practice, the standard deviations are estimated from data, and the multivariate normality assumption holds only asymptotically. Thus, the classical James–Stein theory may not apply directly to a finite-sample context. 

Notwithstanding the complexities described above, the standardized James-Stein estimates consistently outperform their unshrunken counterparts in terms of MSE, consistent with the theoretical MSE dominance of James–Stein shrinkage in large samples. Although the reduction in MSE is an asymptotic property and may not always materialize at the sample sizes we examine, the observed trends provide empirical support for the use of shrinkage in improving estimation stability (Tables \ref{ch4:table:sim:std:pres:jackknifed}-\ref{ch4:table:sim:std:comb:sev:jackknifed} and additional tables in Section S6.2 of the Supplementary File).  

\subsection{Empirical Size Assessment of Bootstrap-Based Significance Tests}
\label{ch4:subsection:Power}

As a proof of concept, we conduct a power study to assess the empirical performance of our significance testing procedure. Due to the high computational cost of these simulations, we restrict our analysis to a smaller cluster size ($N = 30$) and focus on a simplified model that includes one continuous covariate, Avg\_homeppm, and binary indicators for Tooth and Zone. An independence working correlation structure is assumed within clusters.

For each of the 100 Monte Carlo replicates, we generate data under the assumption that Avg\_homeppm has no effect (i.e., true coefficient equals zero), allowing us to evaluate the empirical size of the test. We then fit the model and compute a 95\% confidence interval for the estimated effect of Avg\_homeppm using cluster-level bootstrapping with $B = 10{,}000$ replicates. If the resulting confidence interval includes zero, the covariate is deemed non-significant at the 5\% level. Although estimated parameters generally differ from the values used during data generation (as noted in Section~\ref{ch4:subsection:BiasVarianceMSE}), they are expected to align at zero when the true effect is null. Specifically, the covariate Avg\_homeppm was intentionally assigned a true coefficient of zero—i.e., no effect—in the data generation process. Consequently, its estimated effect should be approximately zero, allowing the corresponding confidence intervals to appropriately reflect the absence of an association.

In our analysis, the empirical size is 0.052 for the presence model and 0.049 for the severity model, both closely aligned with the nominal 0.05 level. For the combined model, which jointly models presence and severity, the empirical size is slightly elevated at 0.061. 
While the combined models show a modest inflation in type I error — possibly due to being overpowered — this result is based on a small sample size ($N = 30$).

We also conduct power analyses under alternative hypotheses, where Avg\_homeppm is assumed to have a non-zero effect. The power to reject the null hypothesis is evaluated across multiple effect sizes. Due to transformation steps involved in the data generation and modeling process, the actual parameter effect size in the fitted model does not directly correspond to the values specified during simulation. To address this discrepancy, we compute the mean and standard deviation of the estimated coefficients across 100 Monte Carlo replicates and report standardized effect sizes for each alternative scenario. As a result of these transformations, the power is not strictly monotonic in the specified beta values, but in all cases considered under the alternative, the empirical power substantially exceeds the nominal level of 0.05, as expected (Table \ref{ch4:table:alt_H1}).

\section{Discussion}
\label{section:Discussion}
This study developed a unified GEE-based modeling framework for analyzing fluorosis outcomes, including both presence and severity components, using separate and combined modeling approaches. The framework accommodates zero-inflated ordinal outcomes in clustered longitudinal data and allows for flexible, data-driven correlation structures using jackknife estimation where possible. Rank aggregation was employed for flexible model selection, and James–Stein shrinkage estimators were incorporated to borrow strength across different ages. 
Our empirical size analysis confirmed that the separate models maintain appropriate type I error control at small sample sizes (e.g., N = 30). The combined model, which we explored using a simplified functional relationship with approximation (Equation~\ref{ch4:eqn:gammaratio}), may benefit from further refinement. Potential improvements include identifying more appropriate functional forms that do not rely on approximation.

From a clinical standpoint, we identified several important risk and protective factors for fluorosis presence and severity. The models captured significant risk effects for Total\_mgFPerDay, Avg\_homeppm, Tooth8, Tooth9, ZoneM, ZoneI, and ZoneE, with risk increasing as we move from the root to the tip of the tooth (ZoneC to ZoneE). In contrast, Tooth10 consistently exhibited a protective effect across its zones. 

In the separate severity models, effects were directionally aligned with those from the separate presence models, though statistical significance was less frequent due to smaller sample size — limited to individuals with non-zero FRI scores. The severity component of the combined models utilized the full dataset, increasing statistical power and enabling the detection of additional effects. Specifically, BrushingFrequencyPerDay and Avg\_homeppm were significant at early ages (9 and 13), with BrushingFrequencyPerDay shifting to a protective factor at older ages (17 and 23). Tooth8 and all tooth zones consistently showed significant risk effects, with risk intensifying as we move from ZoneC to ZoneE. 

It is important to note that we report both separate and combined model results to leverage their respective strengths. The separate models focus exclusively on one aspect of the outcome — presence or severity — allowing for tailored inference within that component. In contrast, the combined model counterpart benefits from the full dataset, potentially yielding more stable and precise estimates due to the larger information base. When the estimated effects from the separate and combined models align in direction, we recommend using the combined model estimates, as they are informed by a richer data structure. However, in cases where the effect directions diverge, we prioritize the separate model estimates for directional inference, as they are directly targeted to that specific component and less prone to dilution or averaging effects across components.

From a biological and developmental perspective, the protective effect of Tooth10 (and reference-level Tooth7) at ages 9 and 13 likely reflects differences in enamel formation and fluoride exposure windows. Lateral incisors (Teeth 7 and 10) begin enamel formation and mineralization several months after central incisors (Teeth 8 and 9), typically starting around 10–12 months of age and completing by age 4–5. This delayed development results in reduced fluoride exposure during the most vulnerable periods of mineralization \citep{sheoran2023study}. 


The observed association between SugarAddedBeverageOzPerDay and increased fluorosis severity in older ages highlights the cumulative effect of fluoride from sugar-added beverages. Many of these are processed with fluoridated water, and chronic consumption can lead to enamel porosity and staining. Prior research shows that beverage intake in infancy contributes significantly to fluorosis risk \citep{marshall2004associations}.

The significant association of BrushingFrequencyPerDay with dental fluorosis at younger ages aligns with known patterns: excessive brushing in high-fluoride environments in early childhood may increase fluorosis risk \citep{mascarenhas1998fluorosis}.

The progressive increase in risk from ZoneC to ZoneE is consistent with enamel zone-specific vulnerability and is supported by previous studies \citep{kang2023analyzing} that highlight higher fluoride uptake on moving from Zones C to E.

Although not statistically significant in some models, greater dental visit frequency (Prop\_DentalAppt) and total fluoride intake from multiple sources (Total\_mgFPerDay) remain of clinical interest - Prop\_DentalAppt is associated with more frequent professional fluoride application \citep{marinho2002fluoride}, and Total\_mgFPerDay is an established contributor to fluorosis risk \citep{marshall2004associations}.

In addition, we gain insights into the progression of dental fluorosis from early childhood (age 9) through young adulthood (age 23). Fluoride ingestion during early life leads to structural changes in enamel that may not be immediately visible but emerge as permanent teeth erupt. While these changes originate during tooth development, their associations with behavioral and environmental predictors may vary across age. In our presence models (Table~\ref{ch4:table:sep:pres:ar1}), variables such as BrushingFrequencyPerDay, total\_mgFPerDay, and SugarAddedBeverageOzPerDay exhibit their strongest effects at age 23, suggesting that their influence becomes more detectable or pronounced in later developmental stages. Similarly, in the severity models (Table~\ref{ch4:table:sep:sev:jackknifed}), SugarAddedBeverageOzPerDay and Prop\_DentalAppt show their largest estimated effects at age 23.

From a methodological standpoint, the James–Stein estimates demonstrated reductions in average MSE compared to their unshrunken counterparts, consistent with the estimator's large-sample properties. Our power analysis showed strong alignment with the nominal 0.05 level for the separate presence and severity models at a small cluster size ($N = 30$). The combined model exhibited a slightly elevated empirical size of 0.061. Future work could investigate the performance of the combined model under larger $N$ and consider alternative, potentially more appropriate, functional relationships linking the presence and severity components. Despite these limitations, the models successfully identified biologically plausible and clinically relevant protective and risk factors for dental fluorosis.

\section{Acknowledgments}
This research has been partially supported by NIH grant R03DE030502.

\section{Supporting information}
Codes for the analyses and simulations presented in this paper can be found at \\ \texttt{https://github.com/shoumisarkar/IFS-fluorosis/}.

\newpage 

\begin{table}[ht]
\centering
\caption{ Estimates from models A.3.1-A.3.4, the separate presence models with the AR(1) cluster correlation structure.}
\label{ch4:table:sep:pres:ar1}
\begin{threeparttable}
\centering
\begin{subtable}{\linewidth}
\centering
\caption{Model A.3.1 (age 9)}
\scalebox{0.55}{
\begin{tabular}{rrrrrrrrl}
Variable & Estimate & \makecell{Standardized\\Estimate} & \makecell{SE\\(Standardized\\Estimate)} & \makecell{James-Stein\\Estimator} & \makecell{Bias\\(James-Stein\\Estimator)} & \makecell{SE\\(James-Stein\\Estimator)} & \makecell{MSE\\(James-Stein\\Estimator)} & \makecell{95\% CI\\(James-Stein\\Estimator)} \\
  \hline
dental\_age & -0.161 & -0.601 & 0.836 & -0.199 &  0.117 & 0.727 & 0.741 & (-1.077,  0.873) \\ 
  Total\_mgFPerDay & -0.121 & -0.654 & 0.863 & -0.220 & -0.053 & 0.668 & 0.671 & (-1.637,  1.013) \\ 
  SugarAddedBeverageOzPerDay & -0.005 & -0.490 & 1.023 & -0.339 &  0.052 & 0.788 & 0.791 & (-2.140,  0.933) \\ 
  BrushingFrequencyPerDay & -0.049 & -0.336 & 0.853 &  1.124 & -1.137 & 0.540 & 1.832 & (-1.232,  1.267) \\ 
  Avg\_homeppm & -0.613 & -3.230 & 0.850 & -2.916 &  0.638 & 0.844 & 1.251 & (-3.884, -0.747)$^{*-}$ \\ 
  Prop\_DentAppt &  0.055 &  0.116 & 0.863 & -0.851 &  1.019 & 0.669 & 1.706 & (-1.048,  1.887) \\ 
  Prop\_FluorideTreatment & -0.063 & -0.077 & 0.871 &  0.087 & -0.090 & 0.538 & 0.546 & (-1.307,  1.132) \\ 
  Tooth8 & -0.319 & -3.999 & 1.293 & -3.566 &  1.164 & 1.339 & 2.693 & (-4.578,  0.001) \\ 
  Tooth9 & -0.331 & -4.154 & 1.258 & -3.709 &  1.266 & 1.343 & 2.947 & (-4.705,  0.042) \\ 
  Tooth10 &  0.241 &  3.511 & 1.010 &  3.104 & -0.827 & 1.037 & 1.721 & ( 0.403,  4.398)$^{*+}$ \\ 
  ZoneM & -1.608 & -3.244 & 2.219 & -2.680 &  0.220 & 2.207 & 2.255 & (-6.982,  1.405) \\ 
  ZoneI & -1.635 & -3.078 & 1.960 & -2.852 &  0.764 & 1.967 & 2.551 & (-6.649,  0.402) \\ 
  ZoneE & -1.701 & -3.284 & 2.024 & -3.138 &  0.834 & 1.944 & 2.639 & (-5.842,  0.040) \\ 
   \hline
\end{tabular}
}
\end{subtable}
\begin{subtable}{\linewidth}
\centering
\caption{Model A.3.2 (age 13)}
\scalebox{0.55}{
\begin{tabular}{rrrrrrrrl}
Variable & Estimate & \makecell{Standardized\\Estimate} & \makecell{SE\\(Standardized\\Estimate)} & \makecell{James-Stein\\Estimator} & \makecell{Bias\\(James-Stein\\Estimator)} & \makecell{SE\\(James-Stein\\Estimator)} & \makecell{MSE\\(James-Stein\\Estimator)} & \makecell{95\% CI\\(James-Stein\\Estimator)} \\
  \hline
dental\_age &  0.346 &  0.780 & 0.765 &  0.258 &  0.041 & 0.500 & 0.502 & ( -0.549,  1.367) \\ 
  Total\_mgFPerDay & -0.122 & -0.713 & 0.924 & -0.240 & -0.089 & 0.618 & 0.625 & ( -1.778,  0.799) \\ 
  SugarAddedBeverageOzPerDay &  0.011 &  1.306 & 0.888 &  0.903 & -0.110 & 0.715 & 0.727 & ( -0.297,  2.313) \\ 
  BrushingFrequencyPerDay &  0.038 &  0.253 & 0.922 & -0.846 &  0.947 & 0.560 & 1.458 & ( -0.954,  1.205) \\ 
  Avg\_homeppm & -0.579 & -2.374 & 1.035 & -2.143 &  0.017 & 1.009 & 1.009 & ( -4.038, -0.180)$^{*-}$ \\ 
  Prop\_DentAppt &  0.051 &  0.071 & 1.010 & -0.524 &  0.529 & 0.811 & 1.091 & ( -1.901,  1.522) \\ 
  Prop\_FluorideTreatment & -0.116 & -0.110 & 0.823 &  0.123 & -0.154 & 0.681 & 0.705 & ( -1.334,  1.221) \\ 
  Tooth8 & -0.097 & -0.795 & 0.931 & -0.709 & -0.023 & 0.826 & 0.826 & ( -2.291,  0.619) \\ 
  Tooth9 &  0.014 &  0.107 & 0.880 &  0.096 & -0.057 & 0.738 & 0.741 & ( -1.196,  1.398) \\ 
  Tooth10 &  0.178 &  2.147 & 0.986 &  1.898 & -0.083 & 0.930 & 0.937 & (  0.426,  3.790)$^{*+}$ \\ 
  ZoneM &  7.772 &  0.175 & 0.738 &  0.145 & -0.268 & 0.703 & 0.775 & ( -2.355,  0.599) \\ 
  ZoneI & -1.764 & -4.204 & 3.121 & -3.895 & -0.287 & 3.263 & 3.345 & (-10.845,  0.088) \\ 
  ZoneE & -2.136 & -5.287 & 2.112 & -5.051 &  0.032 & 2.061 & 2.062 & ( -9.359, -0.950)$^{*-}$ \\ 
   \hline
\end{tabular}
}
\end{subtable}
\begin{subtable}{\linewidth}
\centering
\caption{Model A.3.3 (age 17)}
\scalebox{0.55}{
\begin{tabular}{rrrrrrrrl}
Variable & Estimate & \makecell{Standardized\\Estimate} & \makecell{SE\\(Standardized\\Estimate)} & \makecell{James-Stein\\Estimator} & \makecell{Bias\\(James-Stein\\Estimator)} & \makecell{SE\\(James-Stein\\Estimator)} & \makecell{MSE\\(James-Stein\\Estimator)} & \makecell{95\% CI\\(James-Stein\\Estimator)} \\
  \hline
dental\_age &  0.673 &  1.264 & 0.764 &  0.418 & -0.004 & 0.612 &  0.612 & ( -0.802,  1.735) \\ 
  Total\_mgFPerDay & -0.182 & -0.754 & 0.729 & -0.254 & -0.132 & 0.570 &  0.587 & ( -1.662,  0.529) \\ 
  SugarAddedBeverageOzPerDay &  0.019 &  1.751 & 0.798 &  1.210 & -0.404 & 0.768 &  0.931 & ( -0.436,  2.399) \\ 
  BrushingFrequencyPerDay & -0.054 & -0.264 & 0.797 &  0.883 & -0.880 & 0.502 &  1.276 & ( -0.956,  1.233) \\ 
  Avg\_homeppm & -0.616 & -1.469 & 0.706 & -1.326 &  0.135 & 0.646 &  0.664 & ( -2.579, -0.184)$^{*-}$ \\ 
  Prop\_DentAppt &  0.485 &  0.470 & 0.885 & -3.459 &  3.610 & 0.707 & 13.742 & ( -1.069,  2.033) \\ 
  Prop\_FluorideTreatment & -1.411 & -0.961 & 0.868 &  1.081 & -1.269 & 0.645 &  2.255 & ( -1.384,  1.230) \\ 
  Tooth8 & -0.109 & -0.214 & 1.100 & -0.191 & -0.722 & 1.068 &  1.589 & ( -3.715,  0.406) \\ 
  Tooth9 & -0.019 & -0.042 & 0.950 & -0.037 & -0.485 & 0.860 &  1.095 & ( -2.650,  1.053) \\ 
  Tooth10 &  0.010 &  0.035 & 0.709 &  0.031 &  0.261 & 0.610 &  0.678 & ( -1.086,  1.168) \\ 
  ZoneM &  4.504 &  0.977 & 0.665 &  0.807 & -0.869 & 1.012 &  1.768 & ( -2.434,  1.234) \\ 
  ZoneI &  1.081 &  0.017 & 1.569 &  0.016 & -0.846 & 1.527 &  2.244 & ( -4.693,  0.219) \\ 
  ZoneE & -2.921 & -2.463 & 3.322 & -2.353 & -0.573 & 3.186 &  3.515 & (-10.021,  0.143) \\ 
   \hline
\end{tabular}
}
\end{subtable}
\begin{subtable}{\linewidth}
\centering
\caption{Model A.3.4 (age 23)}
\scalebox{0.55}{
\begin{tabular}{rrrrrrrrl}
Variable & Estimate & \makecell{Standardized\\Estimate} & \makecell{SE\\(Standardized\\Estimate)} & \makecell{James-Stein\\Estimator} & \makecell{Bias\\(James-Stein\\Estimator)} & \makecell{SE\\(James-Stein\\Estimator)} & \makecell{MSE\\(James-Stein\\Estimator)} & \makecell{95\% CI\\(James-Stein\\Estimator)} \\
  \hline
dental\_age &  1.444 &  0.648 & 0.687 &  0.215 &  0.174 & 0.718 & 0.749 & (-0.805,  1.722) \\ 
  Total\_mgFPerDay & -0.424 & -1.230 & 0.910 & -0.414 & -0.420 & 0.940 & 1.117 & (-2.688,  0.721) \\ 
  SugarAddedBeverageOzPerDay & -0.020 & -1.210 & 0.785 & -0.836 &  0.152 & 0.739 & 0.762 & (-2.497,  0.390) \\ 
  BrushingFrequencyPerDay & -0.306 & -0.462 & 0.842 &  1.548 & -1.767 & 0.600 & 3.721 & (-1.791,  0.755) \\ 
  Avg\_homeppm & -0.888 & -1.522 & 0.851 & -1.374 & -0.026 & 0.788 & 0.789 & (-2.793, -0.221)$^{*-}$ \\ 
  Prop\_DentAppt & -0.004 & -0.002 & 0.890 &  0.016 &  0.022 & 0.792 & 0.793 & (-1.511,  1.841) \\ 
  Prop\_FluorideTreatment &  0.011 &  0.005 & 0.546 & -0.006 &  0.026 & 0.412 & 0.412 & (-0.798,  0.782) \\ 
  Tooth8 & -0.730 & -1.353 & 0.842 & -1.207 &  0.090 & 0.854 & 0.862 & (-2.982,  0.041) \\ 
  Tooth9 & -0.703 & -1.196 & 0.829 & -1.068 &  0.078 & 0.823 & 0.829 & (-2.742,  0.077) \\ 
  Tooth10 &  0.107 &  0.541 & 0.645 &  0.479 & -0.197 & 0.542 & 0.580 & (-0.836,  1.304) \\ 
  ZoneM &  1.968 &  0.031 & 0.586 &  0.025 & -0.153 & 0.902 & 0.926 & (-2.042,  1.116) \\ 
  ZoneI & -2.022 & -0.234 & 1.070 & -0.217 & -0.442 & 1.045 & 1.240 & (-3.709,  0.530) \\ 
  ZoneE &  4.761 &  0.091 & 1.146 &  0.087 & -0.515 & 1.118 & 1.384 & (-4.111,  0.199) \\ 
   \hline
\end{tabular}
}
\end{subtable}
\begin{tablenotes}[para,flushleft]
\scriptsize
\item Superscripts $*+$ and $*-$ denote significant protective and risk effects at the 5\% significance level, respectively.
\end{tablenotes}
\end{threeparttable}
\end{table}


\addtocounter{table}{-1} 

\begin{table}[ht]
\centering
\caption{ Estimates from models B.4.1-B.4.4, the separate severity models with the jackknifed cluster correlation structure.}
\label{ch4:table:sep:sev:jackknifed}
\begin{threeparttable}
\centering
\begin{subtable}{\linewidth}
\centering
\caption{Model B.4.1 (age 9)}
\scalebox{0.55}{
\begin{tabular}{rrrrrrrrl}
Variable & Estimate & \makecell{Standardized\\Estimate} & \makecell{SE\\(Standardized\\Estimate)} & \makecell{James-Stein\\Estimator} & \makecell{Bias\\(James-Stein\\Estimator)} & \makecell{SE\\(James-Stein\\Estimator)} & \makecell{MSE\\(James-Stein\\Estimator)} & \makecell{95\% CI\\(James-Stein\\Estimator)} \\
  \hline
dental\_age & -0.494 & -0.375 & 0.874 &  3.280 & -3.302 & 0.660 & 11.566 & (-1.705, 1.284) \\ 
  Total\_mgFPerDay & -0.598 & -0.639 & 1.175 &  1.130 & -1.340 & 0.931 &  2.726 & (-2.849, 1.225) \\ 
  SugarAddedBeverageOzPerDay & -0.004 & -0.110 & 0.913 & -0.059 & -0.427 & 0.714 &  0.896 & (-2.051, 0.809) \\ 
  BrushingFrequencyPerDay &  0.259 &  0.343 & 0.951 & -0.532 &  0.723 & 0.817 &  1.340 & (-1.558, 1.458) \\ 
  Avg\_homeppm & -0.106 & -0.160 & 0.860 &  1.292 & -1.567 & 0.841 &  3.296 & (-2.356, 0.989) \\ 
  Prop\_DentAppt & -1.152 & -0.731 & 0.967 &  0.039 & -0.564 & 0.798 &  1.115 & (-2.112, 0.831) \\ 
  Prop\_FluorideTreatment &  0.560 &  0.136 & 0.827 & -0.173 &  0.335 & 0.539 &  0.652 & (-0.878, 1.029) \\ 
  Tooth8 &  1.101 &  1.200 & 1.071 & -0.264 &  0.497 & 1.022 &  1.269 & (-1.124, 1.950) \\ 
  Tooth9 &  0.022 &  0.039 & 1.109 & -0.192 &  0.024 & 0.936 &  0.937 & (-2.860, 1.181) \\ 
  Tooth10 & -0.448 & -0.333 & 1.201 &  0.271 & -0.683 & 0.907 &  1.374 & (-2.427, 1.203) \\ 
  ZoneM & -0.708 & -0.773 & 1.420 &  0.996 & -1.035 & 1.331 &  2.402 & (-3.741, 2.755) \\ 
  ZoneI & -1.161 & -1.144 & 1.049 & -0.483 & -0.681 & 1.045 &  1.508 & (-3.726, 0.169) \\ 
  ZoneE & -1.278 & -1.437 & 1.444 & -0.392 & -1.545 & 1.797 &  4.185 & (-5.544, 1.399) \\ 
   \hline
\end{tabular}
}
\end{subtable}
\begin{subtable}{\linewidth}
\centering
\caption{Model B.4.2 (age 13)}
\scalebox{0.55}{
\begin{tabular}{rrrrrrrrl}
Variable & Estimate & \makecell{Standardized\\Estimate} & \makecell{SE\\(Standardized\\Estimate)} & \makecell{James-Stein\\Estimator} & \makecell{Bias\\(James-Stein\\Estimator)} & \makecell{SE\\(James-Stein\\Estimator)} & \makecell{MSE\\(James-Stein\\Estimator)} & \makecell{95\% CI\\(James-Stein\\Estimator)} \\
  \hline
dental\_age &  1.212 &  0.198 & 0.512 & -1.729 &  1.463 & 0.829 & 2.970 & (-2.167, 1.136) \\ 
  Total\_mgFPerDay & -0.051 & -0.096 & 0.657 &  0.170 & -0.215 & 0.499 & 0.546 & (-0.808, 0.794) \\ 
  SugarAddedBeverageOzPerDay & -0.015 & -0.090 & 0.815 & -0.048 & -0.185 & 0.480 & 0.514 & (-1.554, 0.743) \\ 
  BrushingFrequencyPerDay & -0.096 & -0.059 & 0.713 &  0.092 & -0.318 & 0.742 & 0.843 & (-1.783, 1.105) \\ 
  Avg\_homeppm & -0.441 & -0.296 & 0.986 &  2.393 & -2.500 & 0.892 & 7.142 & (-1.524, 1.486) \\ 
  Prop\_DentAppt & -2.556 & -0.664 & 0.929 &  0.035 & -0.448 & 0.820 & 1.020 & (-2.138, 0.711) \\ 
  Prop\_FluorideTreatment &  2.171 &  0.209 & 0.645 & -0.266 &  0.255 & 1.483 & 1.549 & (-1.455, 1.469) \\ 
  Tooth8 & -0.011 & -0.011 & 0.750 &  0.002 & -0.056 & 0.750 & 0.754 & (-1.674, 1.129) \\ 
  Tooth9 &  0.206 &  0.294 & 0.723 & -1.449 &  1.318 & 0.544 & 2.283 & (-1.243, 0.653) \\ 
  Tooth10 & -1.322 & -0.906 & 0.770 &  0.739 & -1.155 & 0.838 & 2.172 & (-2.235, 0.924) \\ 
  ZoneM & -0.509 & -0.516 & 0.769 &  0.664 & -0.821 & 0.931 & 1.606 & (-1.954, 0.822) \\ 
  ZoneI & -1.709 & -1.321 & 1.033 & -0.558 & -0.432 & 0.830 & 1.016 & (-2.831, 0.222) \\ 
  ZoneE &  0.157 &  0.058 & 0.687 &  0.016 &  0.113 & 0.536 & 0.549 & (-0.618, 1.689) \\ 
   \hline
\end{tabular}
}
\end{subtable}
\begin{subtable}{\linewidth}
\centering
\caption{Model B.4.3 (age 17)}
\scalebox{0.55}{
\begin{tabular}{rrrrrrrrl}
Variable & Estimate & \makecell{Standardized\\Estimate} & \makecell{SE\\(Standardized\\Estimate)} & \makecell{James-Stein\\Estimator} & \makecell{Bias\\(James-Stein\\Estimator)} & \makecell{SE\\(James-Stein\\Estimator)} & \makecell{MSE\\(James-Stein\\Estimator)} & \makecell{95\% CI\\(James-Stein\\Estimator)} \\
  \hline
dental\_age &  0.141 &  0.038 & 0.559 & -0.335 &  0.384 & 0.583 & 0.730 & (-1.234, 1.196) \\ 
  Total\_mgFPerDay & -1.159 & -0.550 & 0.796 &  0.972 & -1.064 & 0.696 & 1.828 & (-1.726, 1.250) \\ 
  SugarAddedBeverageOzPerDay & -0.053 & -0.669 & 1.023 & -0.359 & -0.699 & 0.960 & 1.449 & (-3.132, 0.240) \\ 
  BrushingFrequencyPerDay &  0.489 &  0.804 & 0.671 & -1.249 &  1.359 & 0.548 & 2.394 & (-1.212, 1.156) \\ 
  Avg\_homeppm &  0.076 &  0.030 & 0.789 & -0.242 &  0.235 & 0.701 & 0.756 & (-1.312, 1.501) \\ 
  Prop\_DentAppt & -1.704 & -0.346 & 0.872 &  0.018 & -0.070 & 0.650 & 0.655 & (-1.244, 1.146) \\ 
  Prop\_FluorideTreatment & -4.013 & -0.897 & 0.832 &  1.143 & -1.311 & 0.962 & 2.679 & (-2.264, 2.158) \\ 
  Tooth8 & -0.217 & -0.258 & 0.708 &  0.057 & -0.146 & 0.741 & 0.762 & (-1.701, 1.049) \\ 
  Tooth9 &  0.171 &  0.118 & 1.051 & -0.584 &  0.502 & 0.875 & 1.127 & (-1.819, 0.751) \\ 
  Tooth10 & -0.936 & -0.385 & 0.960 &  0.314 & -0.451 & 0.751 & 0.955 & (-1.642, 1.486) \\ 
  ZoneM & -0.107 & -0.063 & 1.219 &  0.081 & -0.199 & 0.562 & 0.602 & (-1.907, 0.510) \\ 
  ZoneI & -0.214 & -0.059 & 1.882 & -0.025 & -0.525 & 1.032 & 1.307 & (-3.309, 0.738) \\ 
  ZoneE & -1.669 & -0.825 & 1.836 & -0.225 & -0.657 & 1.013 & 1.445 & (-3.826, 0.397) \\ 
   \hline
\end{tabular}
}
\end{subtable}
\begin{subtable}{\linewidth}
\centering
\caption{Model B.4.4 (age 23)}
\scalebox{0.55}{
\begin{tabular}{rrrrrrrrl}
Variable & Estimate & \makecell{Standardized\\Estimate} & \makecell{SE\\(Standardized\\Estimate)} & \makecell{James-Stein\\Estimator} & \makecell{Bias\\(James-Stein\\Estimator)} & \makecell{SE\\(James-Stein\\Estimator)} & \makecell{MSE\\(James-Stein\\Estimator)} & \makecell{95\% CI\\(James-Stein\\Estimator)} \\
  \hline
dental\_age & -0.672 & -0.155 & 0.440 &  1.357 & -1.405 & 0.387 & 2.360 & (-0.818, 0.411) \\ 
  Total\_mgFPerDay & -0.031 & -0.048 & 0.533 &  0.085 & -0.174 & 0.368 & 0.398 & (-0.878, 0.444) \\ 
  SugarAddedBeverageOzPerDay & -0.060 & -1.961 & 1.263 & -1.051 & -0.005 & 1.196 & 1.197 & (-3.607, 0.194) \\ 
  BrushingFrequencyPerDay & -0.127 & -0.125 & 0.667 &  0.194 & -0.274 & 0.589 & 0.664 & (-1.168, 0.684) \\ 
  Avg\_homeppm &  0.381 &  0.326 & 0.653 & -2.633 &  2.598 & 0.519 & 7.269 & (-1.104, 0.936) \\ 
  Prop\_DentAppt &  2.873 &  0.897 & 0.749 & -0.047 &  0.512 & 0.633 & 0.895 & (-0.298, 1.866) \\ 
  Prop\_FluorideTreatment & -0.704 & -0.116 & 0.548 &  0.147 & -0.090 & 0.484 & 0.492 & (-0.596, 0.715) \\ 
  Tooth8 & -0.587 & -0.364 & 0.461 &  0.080 & -0.147 & 0.397 & 0.418 & (-0.860, 0.951) \\ 
  Tooth9 & -0.487 & -0.485 & 0.766 &  2.393 & -2.465 & 0.733 & 6.808 & (-1.314, 1.670) \\ 
  Tooth10 &  0.340 &  0.146 & 0.650 & -0.119 &  0.240 & 0.540 & 0.597 & (-0.673, 1.183) \\ 
  ZoneM &  0.194 &  0.080 & 1.135 & -0.103 & -0.045 & 0.614 & 0.616 & (-1.473, 0.564) \\ 
  ZoneI &  0.903 &  0.637 & 1.099 &  0.269 & -0.146 & 0.955 & 0.976 & (-2.669, 2.092) \\ 
  ZoneE &  0.054 &  0.030 & 1.203 &  0.008 & -0.365 & 0.857 & 0.990 & (-2.937, 0.414) \\ 
   \hline
\end{tabular}
}
\end{subtable}
\begin{tablenotes}[para,flushleft]
\scriptsize
\item Superscripts $*+$ and $*-$ denote significant protective and risk effects at the 5\% significance level, respectively.
\end{tablenotes}
\end{threeparttable}
\end{table}


\addtocounter{table}{-1} 

\begin{table}[ht]
\centering
\caption{ Presence estimates from models C.4.4.1-C.4.4.4, the combined models with the jackknifed and jackknifed presence and severity cluster correlation structures respectively. }
\label{ch4:table:comb:pres:jackknifed}
\begin{threeparttable}
\centering
\begin{subtable}{\linewidth}
\centering
\caption{Model C.4.4.1 (age 9)}
\scalebox{0.55}{
\begin{tabular}{rrrrrrrrl}
Variable & Estimate & \makecell{Standardized\\Estimate} & \makecell{SE\\(Standardized\\Estimate)} & \makecell{James-Stein\\Estimator} & \makecell{Bias\\(James-Stein\\Estimator)} & \makecell{SE\\(James-Stein\\Estimator)} & \makecell{MSE\\(James-Stein\\Estimator)} & \makecell{95\% CI\\(James-Stein\\Estimator)} \\
  \hline
dental\_age & -0.303 & -1.031 & 0.670 & -0.812 &  0.629 & 0.620 & 1.016 & (-0.605,  0.478) \\ 
  Total\_mgFPerDay & -0.095 & -0.362 & 1.245 & -0.272 &  1.098 & 1.101 & 2.307 & ( 0.186,  2.002)$^{*-}$ \\ 
  SugarAddedBeverageOzPerDay & -0.006 & -0.420 & 0.826 & -0.230 & -0.422 & 1.176 & 1.354 & (-1.908,  0.031) \\ 
  BrushingFrequencyPerDay & -0.072 & -0.375 & 0.819 & -0.007 &  0.160 & 0.333 & 0.359 & (-0.110,  0.502) \\ 
  Avg\_homeppm & -0.657 & -3.127 & 1.611 & -2.846 &  0.563 & 0.423 & 0.740 & (-2.544, -1.831)$^{*-}$ \\ 
  Prop\_DentAppt & -0.191 & -0.189 & 0.894 &  0.407 & -0.284 & 0.209 & 0.289 & (-0.010,  0.347) \\ 
  Prop\_FluorideTreatment &  0.386 &  0.271 & 0.909 & -1.067 &  1.338 & 0.950 & 2.741 & (-0.692,  1.093) \\ 
  Tooth8 & -0.228 & -0.482 & 1.959 & -0.313 & -0.088 & 0.432 & 0.440 & (-0.855, -0.067)$^{*-}$ \\ 
  Tooth9 & -0.139 & -0.167 & 2.172 & -0.088 &  0.036 & 0.110 & 0.111 & (-0.122,  0.066) \\ 
  Tooth10 &  0.370 &  1.265 & 1.230 &  0.040 &  1.199 & 0.522 & 1.960 & ( 0.888,  1.796)$^{*+}$ \\ 
  ZoneM & -0.845 & -2.003 & 1.740 & -1.254 & -1.339 & 1.731 & 3.524 & (-3.912, -0.770)$^{*-}$ \\ 
  ZoneI & -1.738 & -3.231 & 3.912 & -3.044 & -2.215 & 3.554 & 8.458 & (-7.859, -1.495)$^{*-}$ \\ 
  ZoneE & -2.164 & -4.765 & 5.224 & -4.656 & -2.290 & 3.832 & 9.076 & (-9.233, -2.850)$^{*-}$ \\ 
   \hline
\end{tabular}
}
\end{subtable}
\begin{subtable}{\linewidth}
\centering
\caption{Model C.4.4.2 (age 13)}
\scalebox{0.55}{
\begin{tabular}{rrrrrrrrl}
Variable & Estimate & \makecell{Standardized\\Estimate} & \makecell{SE\\(Standardized\\Estimate)} & \makecell{James-Stein\\Estimator} & \makecell{Bias\\(James-Stein\\Estimator)} & \makecell{SE\\(James-Stein\\Estimator)} & \makecell{MSE\\(James-Stein\\Estimator)} & \makecell{95\% CI\\(James-Stein\\Estimator)} \\
  \hline
dental\_age &  0.548 &  1.001 & 0.663 &  0.789 & -0.253 & 0.537 & 0.601 & ( 0.004,  1.019)$^{*+}$ \\ 
  Total\_mgFPerDay &  0.013 &  0.009 & 0.618 &  0.007 &  0.223 & 0.388 & 0.437 & (-0.034,  0.643) \\ 
  SugarAddedBeverageOzPerDay &  0.005 &  0.234 & 0.752 &  0.128 & -0.121 & 0.335 & 0.349 & (-0.237,  0.362) \\ 
  BrushingFrequencyPerDay & -0.016 & -0.022 & 0.755 &  0.000 &  0.121 & 0.344 & 0.359 & (-0.131,  0.485) \\ 
  Avg\_homeppm & -0.621 & -0.974 & 1.169 & -0.886 & -0.084 & 0.823 & 0.831 & (-1.780, -0.219)$^{*-}$ \\ 
  Prop\_DentAppt & -0.167 & -0.049 & 0.679 &  0.106 &  0.161 & 0.733 & 0.759 & (-0.272,  1.043) \\ 
  Prop\_FluorideTreatment &  0.410 &  0.080 & 0.640 & -0.315 & -0.334 & 1.611 & 1.722 & (-2.368,  0.423) \\ 
  Tooth8 & -0.183 & -0.193 & 0.865 & -0.126 & -0.157 & 0.133 & 0.158 & (-0.391, -0.144)$^{*-}$ \\ 
  Tooth9 & -0.099 & -0.216 & 0.606 & -0.113 & -0.030 & 0.060 & 0.061 & (-0.208, -0.105)$^{*-}$ \\ 
  Tooth10 &  0.124 &  0.144 & 0.733 &  0.005 &  0.036 & 0.082 & 0.083 & (-0.010,  0.128) \\ 
  ZoneM & -0.406 & -0.732 & 0.928 & -0.458 & -0.815 & 0.553 & 1.218 & (-1.859, -0.867)$^{*-}$ \\ 
  ZoneI & -1.387 & -2.711 & 2.294 & -2.555 & -0.687 & 2.416 & 2.888 & (-5.773, -1.355)$^{*-}$ \\ 
  ZoneE & -2.123 & -2.228 & 3.830 & -2.177 & -1.624 & 1.165 & 3.802 & (-4.907, -2.692)$^{*-}$ \\ 
   \hline
\end{tabular}
}
\end{subtable}
\begin{subtable}{\linewidth}
\centering
\caption{Model C.4.4.3 (age 17)}
\scalebox{0.55}{
\begin{tabular}{rrrrrrrrl}
Variable & Estimate & \makecell{Standardized\\Estimate} & \makecell{SE\\(Standardized\\Estimate)} & \makecell{James-Stein\\Estimator} & \makecell{Bias\\(James-Stein\\Estimator)} & \makecell{SE\\(James-Stein\\Estimator)} & \makecell{MSE\\(James-Stein\\Estimator)} & \makecell{95\% CI\\(James-Stein\\Estimator)} \\
  \hline
dental\_age &  0.770 &  2.257 & 0.726 &  1.779 & -0.639 & 0.812 & 1.220 & ( 0.420,  1.952)$^{*+}$ \\ 
  Total\_mgFPerDay & -0.151 & -1.276 & 1.004 & -0.961 &  1.201 & 0.483 & 1.925 & (-0.122,  0.750) \\ 
  SugarAddedBeverageOzPerDay &  0.015 &  1.473 & 0.802 &  0.805 & -0.183 & 0.485 & 0.519 & ( 0.188,  1.105)$^{*+}$ \\ 
  BrushingFrequencyPerDay &  0.191 &  1.247 & 0.780 &  0.023 & -0.613 & 0.587 & 0.964 & (-1.067,  0.019) \\ 
  Avg\_homeppm & -0.764 & -2.724 & 0.724 & -2.479 &  0.530 & 0.760 & 1.041 & (-2.693, -1.251)$^{*-}$ \\ 
  Prop\_DentAppt & -3.465 & -0.740 & 0.900 &  1.591 & -2.157 & 0.796 & 5.448 & (-1.408,  0.021) \\ 
  Prop\_FluorideTreatment &  2.026 &  0.462 & 1.204 & -1.820 &  2.361 & 1.243 & 6.820 & (-0.581,  1.772) \\ 
  Tooth8 & -0.099 & -0.166 & 0.899 & -0.108 & -0.376 & 0.182 & 0.323 & (-0.677, -0.357)$^{*-}$ \\ 
  Tooth9 &  0.046 &  0.104 & 0.751 &  0.054 &  0.016 & 0.077 & 0.077 & ( 0.022,  0.152)$^{*+}$ \\ 
  Tooth10 & -0.209 & -0.580 & 0.903 & -0.018 &  0.587 & 0.659 & 1.004 & ( 0.110,  1.270)$^{*-}$ \\ 
  ZoneM & -0.352 & -0.824 & 1.078 & -0.516 & -0.846 & 0.735 & 1.450 & (-2.079, -0.685)$^{*-}$ \\ 
  ZoneI & -1.741 & -3.655 & 2.156 & -3.444 & -0.411 & 2.160 & 2.329 & (-6.165, -2.520)$^{*-}$ \\ 
  ZoneE & -2.374 & -7.216 & 2.374 & -7.050 &  1.145 & 2.900 & 4.211 & (-9.005, -4.188)$^{*-}$ \\ 
   \hline
\end{tabular}
}
\end{subtable}
\begin{subtable}{\linewidth}
\centering
\caption{Model C.4.4.4 (age 23)}
\scalebox{0.55}{
\begin{tabular}{rrrrrrrrl}
Variable & Estimate & \makecell{Standardized\\Estimate} & \makecell{SE\\(Standardized\\Estimate)} & \makecell{James-Stein\\Estimator} & \makecell{Bias\\(James-Stein\\Estimator)} & \makecell{SE\\(James-Stein\\Estimator)} & \makecell{MSE\\(James-Stein\\Estimator)} & \makecell{95\% CI\\(James-Stein\\Estimator)} \\
  \hline
dental\_age &  0.720 &  1.510 & 0.757 &  1.190 &  0.208 & 0.362 & 0.406 & ( 1.067,  1.754)$^{*+}$ \\ 
  Total\_mgFPerDay & -0.343 & -2.516 & 1.326 & -1.894 &  1.951 & 0.792 & 4.597 & (-0.717,  0.786) \\ 
  SugarAddedBeverageOzPerDay & -0.014 & -1.418 & 0.795 & -0.775 & -0.009 & 0.942 & 0.943 & (-1.787, -0.128)$^{*-}$ \\ 
  BrushingFrequencyPerDay & -0.177 & -0.584 & 0.720 & -0.011 &  0.197 & 0.308 & 0.347 & (-0.076,  0.500) \\ 
  Avg\_homeppm & -0.986 & -2.028 & 0.849 & -1.846 &  0.590 & 0.793 & 1.141 & (-2.010, -0.504)$^{*-}$ \\ 
  Prop\_DentAppt & -0.274 & -0.222 & 0.922 &  0.478 & -0.291 & 0.715 & 0.799 & (-0.384,  0.932) \\ 
  Prop\_FluorideTreatment &  0.479 &  0.334 & 0.617 & -1.316 &  0.335 & 0.699 & 0.811 & (-1.727, -0.549)$^{*+}$ \\ 
  Tooth8 & -0.512 & -2.328 & 1.164 & -1.513 &  0.870 & 0.156 & 0.913 & (-0.759, -0.478)$^{*-}$ \\ 
  Tooth9 & -0.463 & -2.032 & 1.032 & -1.068 &  0.494 & 0.454 & 0.699 & (-1.041, -0.195)$^{*-}$ \\ 
  Tooth10 &  0.057 &  0.331 & 0.585 &  0.010 & -0.032 & 0.234 & 0.235 & (-0.233,  0.210) \\ 
  ZoneM & -0.570 & -0.344 & 2.065 & -0.215 & -0.049 & 0.560 & 0.562 & (-0.862,  0.095) \\ 
  ZoneI & -2.318 & -1.881 & 5.314 & -1.772 &  0.030 & 1.715 & 1.716 & (-3.574, -0.725)$^{*-}$ \\ 
  ZoneE & -3.156 & -2.719 & 6.847 & -2.657 & -0.025 & 2.453 & 2.453 & (-5.302, -1.251)$^{*-}$ \\ 
   \hline
\end{tabular}
}
\end{subtable}
\begin{tablenotes}[para,flushleft]
\scriptsize
\item Superscripts $*+$ and $*-$ denote significant protective and risk effects at the 5\% significance level, respectively.
\end{tablenotes}
\end{threeparttable}
\end{table}

\addtocounter{table}{-1} 

\begin{table}[ht]
\centering
\caption{ Severity estimates from models C.4.4.1-C.4.4.4, the combined models with the jackknifed and jackknifed presence and severity cluster correlation structures respectively. }
\label{ch4:table:comb:sev:jackknifed}
\begin{threeparttable}
\centering
\begin{subtable}{\linewidth}
\centering
\caption{Model C.4.4.1 (age 9)}
\scalebox{0.55}{
\begin{tabular}{rrrrrrrrl}
Variable & Estimate & \makecell{Standardized\\Estimate} & \makecell{SE\\(Standardized\\Estimate)} & \makecell{James-Stein\\Estimator} & \makecell{Bias\\(James-Stein\\Estimator)} & \makecell{SE\\(James-Stein\\Estimator)} & \makecell{MSE\\(James-Stein\\Estimator)} & \makecell{95\% CI\\(James-Stein\\Estimator)} \\
  \hline
dental\_age & -0.208 & -0.596 & 0.523 & -0.345 &  0.224 & 0.650 & 0.700 & (-0.804,  0.383) \\ 
  Total\_mgFPerDay & -0.065 & -0.394 & 0.733 &  0.038 &  0.501 & 2.031 & 2.282 & (-1.179,  2.616) \\ 
  SugarAddedBeverageOzPerDay & -0.004 & -0.295 & 0.644 &  0.007 & -0.170 & 0.209 & 0.238 & (-0.387, -0.019)$^{*-}$ \\ 
  BrushingFrequencyPerDay & -0.050 & -0.376 & 0.519 &  0.088 &  0.702 & 0.611 & 1.103 & ( 0.136,  1.166)$^{*-}$ \\ 
  Avg\_homeppm & -0.453 & -0.725 & 0.759 & -0.604 &  0.234 & 0.155 & 0.209 & (-0.500, -0.211)$^{*-}$ \\ 
  Prop\_DentAppt & -0.132 & -0.151 & 0.517 &  0.376 & -0.607 & 0.528 & 0.896 & (-0.790,  0.154) \\ 
  Prop\_FluorideTreatment &  0.266 &  0.206 & 0.600 & -1.160 &  1.082 & 0.522 & 1.692 & (-0.411,  0.480) \\ 
  Tooth8 & -0.157 & -1.024 & 0.939 &  0.728 & -0.982 & 0.500 & 1.464 & (-0.773,  0.149) \\ 
  Tooth9 & -0.096 & -0.212 & 1.005 &  2.584 & -2.437 & 1.001 & 6.941 & (-0.454,  1.217) \\ 
  Tooth10 &  0.255 &  0.510 & 0.565 & -1.236 &  0.742 & 0.282 & 0.832 & (-0.793, -0.288)$^{*+}$ \\ 
  ZoneM & -0.582 & -0.574 & 1.064 &  0.569 & -0.595 & 0.637 & 0.990 & (-0.442,  0.654) \\ 
  ZoneI & -1.198 & -0.640 & 1.100 & -0.441 &  0.147 & 0.455 & 0.477 & (-0.688,  0.167) \\ 
  ZoneE & -1.491 & -0.681 & 1.335 & -0.566 &  0.155 & 0.442 & 0.466 & (-0.854, -0.020)$^{*-}$ \\ 
   \hline
\end{tabular}
}
\end{subtable}
\begin{subtable}{\linewidth}
\centering
\caption{Model C.4.4.2 (age 13)}
\scalebox{0.55}{
\begin{tabular}{rrrrrrrrl}
Variable & Estimate & \makecell{Standardized\\Estimate} & \makecell{SE\\(Standardized\\Estimate)} & \makecell{James-Stein\\Estimator} & \makecell{Bias\\(James-Stein\\Estimator)} & \makecell{SE\\(James-Stein\\Estimator)} & \makecell{MSE\\(James-Stein\\Estimator)} & \makecell{95\% CI\\(James-Stein\\Estimator)} \\
  \hline
dental\_age &  0.404 &  0.298 & 0.433 &  0.173 & -0.292 & 0.099 & 0.184 & (-0.215, -0.027)$^{*+}$ \\ 
  Total\_mgFPerDay &  0.010 &  0.006 & 0.371 & -0.001 &  0.221 & 0.619 & 0.668 & (-0.263,  0.869) \\ 
  SugarAddedBeverageOzPerDay &  0.004 &  0.148 & 0.379 & -0.003 & -0.039 & 0.250 & 0.251 & (-0.305,  0.146) \\ 
  BrushingFrequencyPerDay & -0.012 & -0.016 & 0.374 &  0.004 & -0.300 & 0.536 & 0.626 & (-0.847,  0.152) \\ 
  Avg\_homeppm & -0.458 & -0.342 & 0.605 & -0.285 &  0.210 & 0.127 & 0.171 & (-0.197,  0.044) \\ 
  Prop\_DentAppt & -0.123 & -0.035 & 0.356 &  0.088 & -0.707 & 1.104 & 1.603 & (-1.793,  0.153) \\ 
  Prop\_FluorideTreatment &  0.302 &  0.054 & 0.375 & -0.304 &  1.236 & 1.693 & 3.222 & (-0.119,  2.743) \\ 
  Tooth8 & -0.135 & -0.106 & 0.791 &  0.076 & -0.327 & 0.203 & 0.310 & (-0.400, -0.036)$^{*-}$ \\ 
  Tooth9 & -0.073 & -0.130 & 0.546 &  1.584 & -1.712 & 0.672 & 3.602 & (-0.692,  0.561) \\ 
  Tooth10 &  0.091 &  0.116 & 0.409 & -0.281 &  0.261 & 0.023 & 0.091 & (-0.043,  0.000)$^{*+}$ \\ 
  ZoneM & -0.299 & -0.343 & 0.645 &  0.340 & -0.644 & 0.249 & 0.663 & (-0.536, -0.063)$^{*-}$ \\ 
  ZoneI & -1.023 & -0.326 & 0.858 & -0.225 &  0.030 & 0.192 & 0.193 & (-0.399, -0.084)$^{*-}$ \\ 
  ZoneE & -1.565 & -0.339 & 0.908 & -0.282 &  0.137 & 0.175 & 0.193 & (-0.326, -0.004)$^{*-}$ \\ 
   \hline
\end{tabular}
}
\end{subtable}
\begin{subtable}{\linewidth}
\centering
\caption{Model C.4.4.3 (age 17)}
\scalebox{0.55}{
\begin{tabular}{rrrrrrrrl}
Variable & Estimate & \makecell{Standardized\\Estimate} & \makecell{SE\\(Standardized\\Estimate)} & \makecell{James-Stein\\Estimator} & \makecell{Bias\\(James-Stein\\Estimator)} & \makecell{SE\\(James-Stein\\Estimator)} & \makecell{MSE\\(James-Stein\\Estimator)} & \makecell{95\% CI\\(James-Stein\\Estimator)} \\
  \hline
dental\_age &  1.039 &  2.056 & 0.405 &  1.192 & -1.590 & 0.230 &  2.759 & (-0.560, -0.154)$^{*+}$ \\ 
  Total\_mgFPerDay & -0.203 & -1.256 & 0.678 &  0.121 &  0.497 & 0.512 &  0.759 & ( 0.074,  0.984)$^{*-}$ \\ 
  SugarAddedBeverageOzPerDay &  0.021 &  1.327 & 0.837 & -0.030 & -0.470 & 0.570 &  0.791 & (-1.052,  0.031) \\ 
  BrushingFrequencyPerDay &  0.257 &  1.198 & 0.560 & -0.280 & -0.843 & 0.217 &  0.928 & (-1.342, -0.933)$^{*+}$ \\ 
  Avg\_homeppm & -1.031 & -3.352 & 0.694 & -2.792 &  2.329 & 0.155 &  5.580 & (-0.620, -0.331)$^{*-}$ \\ 
  Prop\_DentAppt & -4.674 & -0.726 & 0.698 &  1.812 & -2.192 & 1.553 &  6.358 & (-2.040,  0.584) \\ 
  Prop\_FluorideTreatment &  2.733 &  0.468 & 0.692 & -2.639 &  3.510 & 0.921 & 13.243 & ( 0.156,  1.838)$^{*+}$ \\ 
  Tooth8 & -0.134 & -0.180 & 0.719 &  0.128 & -0.163 & 0.589 &  0.615 & (-0.526,  0.571) \\ 
  Tooth9 &  0.063 &  0.105 & 0.514 & -1.281 &  0.965 & 0.381 &  1.311 & (-0.697,  0.022) \\ 
  Tooth10 & -0.282 & -0.513 & 0.579 &  1.243 & -1.464 & 0.118 &  2.261 & (-0.342, -0.122)$^{*-}$ \\ 
  ZoneM & -0.475 & -0.714 & 0.652 &  0.707 & -0.873 & 0.873 &  1.634 & (-0.980,  0.677) \\ 
  ZoneI & -2.348 & -2.420 & 0.869 & -1.670 &  1.328 & 0.526 &  2.289 & (-0.792,  0.193) \\ 
  ZoneE & -3.203 & -3.354 & 0.828 & -2.791 &  2.361 & 0.372 &  5.947 & (-0.680, -0.033)$^{*-}$ \\ 
   \hline
\end{tabular}
}
\end{subtable}
\begin{subtable}{\linewidth}
\centering
\caption{Model C.4.4.4 (age 23)}
\scalebox{0.55}{
\begin{tabular}{rrrrrrrrl}
Variable & Estimate & \makecell{Standardized\\Estimate} & \makecell{SE\\(Standardized\\Estimate)} & \makecell{James-Stein\\Estimator} & \makecell{Bias\\(James-Stein\\Estimator)} & \makecell{SE\\(James-Stein\\Estimator)} & \makecell{MSE\\(James-Stein\\Estimator)} & \makecell{95\% CI\\(James-Stein\\Estimator)} \\
  \hline
dental\_age & -0.187 & -0.300 & 0.656 & -0.174 &  0.459 & 0.258 & 0.469 & ( 0.050,  0.538)$^{*-}$ \\ 
  Total\_mgFPerDay &  0.089 &  0.301 & 0.901 & -0.029 & -0.746 & 0.656 & 1.212 & (-1.309, -0.096)$^{*+}$ \\ 
  SugarAddedBeverageOzPerDay &  0.004 &  0.294 & 0.500 & -0.007 & -0.067 & 0.183 & 0.188 & (-0.235,  0.111) \\ 
  BrushingFrequencyPerDay &  0.046 &  0.209 & 0.565 & -0.049 & -0.433 & 0.187 & 0.375 & (-0.680, -0.344)$^{*+}$ \\ 
  Avg\_homeppm &  0.256 &  0.275 & 0.896 &  0.229 & -0.061 & 0.103 & 0.106 & ( 0.063,  0.255)$^{*+}$ \\ 
  Prop\_DentAppt &  0.071 &  0.145 & 0.561 & -0.363 &  0.820 & 0.305 & 0.977 & ( 0.146,  0.718)$^{*+}$ \\ 
  Prop\_FluorideTreatment & -0.124 & -0.193 & 0.369 &  1.089 & -0.584 & 0.935 & 1.276 & (-0.152,  1.500) \\ 
  Tooth8 &  0.133 &  0.278 & 0.969 & -0.198 &  0.127 & 0.300 & 0.316 & (-0.391,  0.126) \\ 
  Tooth9 &  0.120 &  0.281 & 0.934 & -3.417 &  2.339 & 1.256 & 6.728 & (-2.338,  0.030) \\ 
  Tooth10 & -0.015 & -0.219 & 0.397 &  0.530 & -0.632 & 0.246 & 0.645 & (-0.358,  0.094) \\ 
  ZoneM &  0.148 &  0.219 & 1.493 & -0.217 &  0.310 & 0.089 & 0.185 & (-0.002,  0.155) \\ 
  ZoneI &  0.601 &  0.279 & 1.867 &  0.193 & -0.108 & 0.169 & 0.181 & (-0.091,  0.224) \\ 
  ZoneE &  0.818 &  0.282 & 1.863 &  0.235 & -0.106 & 0.134 & 0.145 & ( 0.007,  0.261)$^{*+}$ \\ 
   \hline
\end{tabular}
}
\end{subtable}
\begin{tablenotes}[para,flushleft]
\scriptsize
\item Superscripts $*+$ and $*-$ denote significant protective and risk effects at the 5\% significance level, respectively.
\end{tablenotes}
\end{threeparttable}
\end{table}

\addtocounter{table}{-1} 


\begin{table}[ht]
\centering
\caption{ Simulation results for age 23 data generated with jackknifed and jackknifed correlation structures for presence and severity respectively - properties of direct estimates. }
\label{ch4:table:sim:raw:jackknifed}
\begin{threeparttable}
\centering
\begin{subtable}{\linewidth}
\centering
\caption{Presence model A$_s$.3.4}
\scalebox{0.65}{
\begin{tabular}{l|cccc|cccc|cccc}
{} & \multicolumn{4}{c|}{\textbf{N=30}} & \multicolumn{4}{c|}{\textbf{N=50}} & \multicolumn{4}{c}{\textbf{N=200}} \\ \hline
Variable & Estimate & Bias & SE & MSE & Estimate & Bias & SE & MSE & Estimate & Bias & SE & MSE \\ 
  \hline
dental\_age & -0.286 & 0.015 & 0.642 & 0.412 & -0.250 & 0.051 & 0.274 & 0.078 & -0.305 & -0.004 & 0.114 & 0.013 \\ 
  Avg\_homeppm & -0.634 & 0.049 & 0.313 & 0.100 & -0.667 & 0.016 & 0.258 & 0.067 & -0.674 & 0.008 & 0.136 & 0.018 \\ 
  Tooth8 & -0.228 & -0.009 & 0.316 & 0.100 & -0.245 & -0.026 & 0.257 & 0.067 & -0.199 & 0.020 & 0.116 & 0.014 \\ 
  Tooth9 & -0.179 & -0.089 & 0.340 & 0.124 & -0.167 & -0.077 & 0.234 & 0.061 & -0.119 & -0.030 & 0.124 & 0.016 \\ 
  Tooth10 & 0.366 & -0.056 & 0.364 & 0.136 & 0.349 & -0.073 & 0.287 & 0.088 & 0.411 & -0.011 & 0.144 & 0.021 \\ 
  ZoneM & -0.917 & -0.149 & 0.494 & 0.267 & -0.769 & -0.001 & 0.350 & 0.123 & -0.890 & -0.121 & 0.186 & 0.049 \\ 
  ZoneI & -1.788 & -0.041 & 0.443 & 0.198 & -1.672 & 0.075 & 0.313 & 0.103 & -1.793 & -0.046 & 0.171 & 0.031 \\ 
  ZoneE & -2.237 & -0.080 & 0.458 & 0.216 & -2.130 & 0.026 & 0.312 & 0.098 & -2.196 & -0.040 & 0.173 & 0.032 \\ 
   \hline
\end{tabular}
}
\end{subtable}
\begin{subtable}{\linewidth}
\centering
\caption{Severity model B$_s$.3.4}
\scalebox{0.65}{
\begin{tabular}{l|cccc|cccc|cccc}
{} & \multicolumn{4}{c|}{\textbf{N=30}} & \multicolumn{4}{c|}{\textbf{N=50}} & \multicolumn{4}{c}{\textbf{N=200}} \\ \hline
Variable & Estimate & Bias & SE & MSE & Estimate & Bias & SE & MSE & Estimate & Bias & SE & MSE \\ 
  \hline
dental\_age & 0.119 & 0.227 & 17.992 & 323.752 & -0.646 & -0.538 & 3.404 & 11.880 & -0.309 & -0.202 & 1.786 & 3.231 \\ 
  Avg\_homeppm & -0.457 & -0.228 & 0.804 & 0.698 & -0.509 & -0.280 & 0.470 & 0.299 & -0.457 & -0.229 & 0.265 & 0.123 \\ 
  Tooth8 & -0.154 & -0.140 & 0.617 & 0.401 & -0.225 & -0.211 & 0.449 & 0.246 & -0.143 & -0.129 & 0.352 & 0.141 \\ 
  Tooth9 & -0.093 & -0.185 & 0.591 & 0.383 & -0.103 & -0.195 & 0.421 & 0.215 & -0.060 & -0.152 & 0.188 & 0.058 \\ 
  Tooth10 & 0.337 & 0.112 & 0.721 & 0.532 & 0.338 & 0.113 & 0.457 & 0.221 & 0.291 & 0.066 & 0.389 & 0.156 \\ 
  ZoneM & -4.979 & -4.855 & 12.140 & 170.954 & -2.672 & -2.548 & 8.360 & 76.378 & -0.445 & -0.321 & 0.485 & 0.338 \\ 
  ZoneI & -6.164 & -5.628 & 12.528 & 188.624 & -3.370 & -2.833 & 8.552 & 81.159 & -1.086 & -0.550 & 0.501 & 0.553 \\ 
  ZoneE & -6.445 & -5.813 & 12.577 & 191.986 & -3.755 & -3.123 & 8.563 & 83.087 & -1.421 & -0.789 & 0.531 & 0.905 \\ 
   \hline
\end{tabular}
}
\end{subtable}
\begin{subtable}{\linewidth}
\centering
\caption{Presence piece of model C$_s$.3.3.4}
\scalebox{0.65}{
\begin{tabular}{l|cccc|cccc|cccc}
{} & \multicolumn{4}{c|}{\textbf{N=30}} & \multicolumn{4}{c|}{\textbf{N=50}} & \multicolumn{4}{c}{\textbf{N=200}} \\ \hline
Variable & Estimate & Bias & SE & MSE & Estimate & Bias & SE & MSE & Estimate & Bias & SE & MSE \\ 
  \hline
dental\_age & -0.274 & 0.027 & 0.505 & 0.255 & -0.253 & 0.048 & 0.255 & 0.067 & -0.297 & 0.004 & 0.108 & 0.012 \\ 
  Avg\_homeppm & -0.614 & 0.069 & 0.314 & 0.104 & -0.677 & 0.006 & 0.244 & 0.059 & -0.672 & 0.011 & 0.135 & 0.018 \\ 
  Tooth8 & -0.214 & 0.005 & 0.310 & 0.096 & -0.226 & -0.007 & 0.241 & 0.058 & -0.199 & 0.020 & 0.116 & 0.014 \\ 
  Tooth9 & -0.208 & -0.119 & 0.286 & 0.096 & -0.163 & -0.074 & 0.223 & 0.055 & -0.118 & -0.029 & 0.126 & 0.017 \\ 
  Tooth10 & 0.339 & -0.083 & 0.362 & 0.138 & 0.346 & -0.076 & 0.292 & 0.091 & 0.415 & -0.006 & 0.145 & 0.021 \\ 
  ZoneM & -0.851 & -0.082 & 0.479 & 0.236 & -0.764 & 0.005 & 0.348 & 0.121 & -0.895 & -0.126 & 0.186 & 0.051 \\ 
  ZoneI & -1.689 & 0.057 & 0.405 & 0.167 & -1.659 & 0.087 & 0.287 & 0.090 & -1.802 & -0.055 & 0.171 & 0.032 \\ 
  ZoneE & -2.087 & 0.070 & 0.389 & 0.157 & -2.079 & 0.077 & 0.288 & 0.089 & -2.199 & -0.042 & 0.177 & 0.033 \\ 
   \hline
\end{tabular}
}
\end{subtable}
\begin{subtable}{\linewidth}
\centering
\caption{Severity piece of model C$_s$.3.3.4}
\scalebox{0.65}{
\begin{tabular}{l|cccc|cccc|cccc}
{} & \multicolumn{4}{c|}{\textbf{N=30}} & \multicolumn{4}{c|}{\textbf{N=50}} & \multicolumn{4}{c}{\textbf{N=200}} \\ \hline
Variable & Estimate & Bias & SE & MSE & Estimate & Bias & SE & MSE & Estimate & Bias & SE & MSE \\ 
  \hline
dental\_age & -0.632 & -0.524 & 3.493 & 12.472 & -0.083 & 0.025 & 1.808 & 3.271 & -0.189 & -0.081 & 0.144 & 0.027 \\ 
  Avg\_homeppm & -1.614 & -1.386 & 5.244 & 29.417 & -1.612 & -1.383 & 4.476 & 21.946 & -0.435 & -0.206 & 0.376 & 0.184 \\ 
  Tooth8 & -0.257 & -0.243 & 3.088 & 9.595 & -0.304 & -0.289 & 1.436 & 2.145 & -0.124 & -0.110 & 0.133 & 0.030 \\ 
  Tooth9 & -0.459 & -0.551 & 2.076 & 4.615 & -0.330 & -0.422 & 1.182 & 1.576 & -0.060 & -0.152 & 0.092 & 0.032 \\ 
  Tooth10 & 1.475 & 1.251 & 7.819 & 62.698 & 0.854 & 0.630 & 2.677 & 7.563 & 0.267 & 0.042 & 0.194 & 0.039 \\ 
  ZoneM & -2.818 & -2.694 & 7.919 & 69.973 & -1.600 & -1.476 & 4.061 & 18.672 & -0.568 & -0.443 & 0.392 & 0.350 \\ 
  ZoneI & -4.766 & -4.229 & 14.150 & 218.121 & -3.792 & -3.255 & 9.285 & 96.813 & -1.140 & -0.604 & 0.791 & 0.991 \\ 
  ZoneE & -6.034 & -5.402 & 17.295 & 328.289 & -4.665 & -4.033 & 11.317 & 144.340 & -1.404 & -0.773 & 1.018 & 1.632 \\ 
   \hline
\end{tabular}
}
\end{subtable}
\end{threeparttable}
\end{table}


\addtocounter{table}{-1} 

\begin{table}[ht]
\centering
\caption{ Simulation results for age 23 data generated with jackknifed and jackknifed correlation structures for presence and severity respectively - properties of standardized estimates arising from the presence model A$_s$.3.4. }
\label{ch4:table:sim:std:pres:jackknifed}
\begin{threeparttable}
\centering
\begin{subtable}{\linewidth}
\centering
\caption{N=30}
\scalebox{0.65}{
\begin{tabular}{lcccc|cccc}
Variable & \makecell{Standardized\\Estimate} & \makecell{Bias\\(Standardized\\Estimate)} & \makecell{SE\\(Standardized\\Estimate)} & \makecell{MSE\\(Standardized\\Estimate)} & \makecell{Standardized\\James-Stein\\Estimate} & \makecell{Bias\\(Standardized\\James-Stein\\Estimate)} & \makecell{SE\\(Standardized\\James-Stein\\Estimate)} & \makecell{MSE\\(Standardized\\James-Stein\\Estimate)} \\ 
  \hline
Dental\_age & -0.150 & 0.037 & 0.228 & 0.053 & -0.089 & 0.099 & 0.210 & 0.054 \\ 
  Avg\_homeppm & -0.346 & 0.036 & 0.178 & 0.033 & -0.306 & 0.076 & 0.170 & 0.035 \\ 
  Tooth8 & -0.125 & 0.000 & 0.179 & 0.032 & -0.072 & 0.053 & 0.159 & 0.080 \\ 
  Tooth9 & -0.098 & -0.047 & 0.199 & 0.042 & -0.055 & -0.004 & 0.150 & 0.023 \\ 
  Tooth10 & 0.184 & -0.038 & 0.187 & 0.036 & 0.118 & -0.104 & 0.154 & 0.035 \\ 
  ZoneM & -0.337 & -0.023 & 0.164 & 0.027 & -0.289 & 0.024 & 0.157 & 0.025 \\ 
  ZoneI & -0.727 & 0.034 & 0.144 & 0.022 & -0.706 & 0.055 & 0.146 & 0.024 \\ 
  ZoneE & -0.924 & 0.038 & 0.164 & 0.029 & -0.91 & 0.052 & 0.162 & 0.029 \\ 
  \hline Average MSE &  &  &  & 0.034 &  &  &  & 0.032 \\ 
   \hline
\end{tabular}
}
\end{subtable}
\begin{subtable}{\linewidth}
\centering
\caption{N=50}
\scalebox{0.65}{
\begin{tabular}{lcccc|cccc}
Variable & \makecell{Standardized\\Estimate} & \makecell{Bias\\(Standardized\\Estimate)} & \makecell{SE\\(Standardized\\Estimate)} & \makecell{MSE\\(Standardized\\Estimate)} & \makecell{Standardized\\James-Stein\\Estimate} & \makecell{Bias\\(Standardized\\James-Stein\\Estimate)} & \makecell{SE\\(Standardized\\James-Stein\\Estimate)} & \makecell{MSE\\(Standardized\\James-Stein\\Estimate)} \\ 
  \hline
Dental\_age & -0.161 & 0.027 & 0.188 & 0.036 & -0.116 & 0.072 & 0.183 & 0.039 \\ 
  Avg\_homeppm & -0.373 & 0.009 & 0.168 & 0.080 & -0.349 & 0.033 & 0.165 & 0.080 \\ 
  Tooth8 & -0.135 & -0.010 & 0.141 & 0.020 & -0.087 & 0.037 & 0.118 & 0.015 \\ 
  Tooth9 & -0.089 & -0.038 & 0.125 & 0.017 & -0.044 & 0.007 & 0.109 & 0.012 \\ 
  Tooth10 & 0.179 & -0.043 & 0.144 & 0.023 & 0.148 & -0.073 & 0.126 & 0.021 \\ 
  ZoneM & -0.304 & 0.009 & 0.128 & 0.016 & -0.279 & 0.034 & 0.120 & 0.016 \\ 
  ZoneI & -0.726 & 0.035 & 0.113 & 0.014 & -0.715 & 0.047 & 0.111 & 0.015 \\ 
  ZoneE & -0.943 & 0.019 & 0.108 & 0.012 & -0.933 & 0.030 & 0.108 & 0.013 \\ 
  \hline Average MSE &  &  &  & 0.021 &  &  &  & 0.020 \\ 
   \hline
\end{tabular}
}
\end{subtable}
\begin{subtable}{\linewidth}
\centering
\caption{N=200}
\scalebox{0.65}{
\begin{tabular}{lcccc|cccc}
Variable & \makecell{Standardized\\Estimate} & \makecell{Bias\\(Standardized\\Estimate)} & \makecell{SE\\(Standardized\\Estimate)} & \makecell{MSE\\(Standardized\\Estimate)} & \makecell{Standardized\\James-Stein\\Estimate} & \makecell{Bias\\(Standardized\\James-Stein\\Estimate)} & \makecell{SE\\(Standardized\\James-Stein\\Estimate)} & \makecell{MSE\\(Standardized\\James-Stein\\Estimate)} \\ 
  \hline
Dental\_age & -0.195 & -0.007 & 0.082 & 0.007 & -0.172 & 0.016 & 0.078 & 0.006 \\ 
  Avg\_homeppm & -0.385 & -0.003 & 0.083 & 0.007 & -0.382 & 0.000 & 0.080 & 0.006 \\ 
  Tooth8 & -0.112 & 0.012 & 0.066 & 0.005 & -0.095 & 0.030 & 0.062 & 0.005 \\ 
  Tooth9 & -0.067 & -0.017 & 0.070 & 0.005 & -0.048 & 0.002 & 0.057 & 0.003 \\ 
  Tooth10 & 0.217 & -0.005 & 0.077 & 0.006 & 0.204 & -0.018 & 0.076 & 0.006 \\ 
  ZoneM & -0.354 & -0.040 & 0.069 & 0.006 & -0.341 & -0.027 & 0.068 & 0.005 \\ 
  ZoneI & -0.759 & 0.002 & 0.055 & 0.003 & -0.753 & 0.008 & 0.055 & 0.003 \\ 
  ZoneE & -0.950 & 0.012 & 0.064 & 0.004 & -0.946 & 0.017 & 0.065 & 0.004 \\ 
  \hline Average MSE &  &  &  & 0.005 &  &  &  & 0.005 \\ 
   \hline
\end{tabular}
}
\end{subtable}
\end{threeparttable}
\end{table}


\addtocounter{table}{-1} 

\begin{table}[ht]
\centering
\caption{ Simulation results for age 23 data generated with jackknifed and jackknifed correlation structures for presence and severity respectively - properties of standardized estimates arising from the severity model B$_s$.3.4. }
\label{ch4:table:sim:std:sev:jackknifed}
\begin{threeparttable}
\centering
\begin{subtable}{\linewidth}
\centering
\caption{N=30}
\scalebox{0.65}{
\begin{tabular}{lcccc|cccc}
Variable & \makecell{Standardized\\Estimate} & \makecell{Bias\\(Standardized\\Estimate)} & \makecell{SE\\(Standardized\\Estimate)} & \makecell{MSE\\(Standardized\\Estimate)} & \makecell{Standardized\\James-Stein\\Estimate} & \makecell{Bias\\(Standardized\\James-Stein\\Estimate)} & \makecell{SE\\(Standardized\\James-Stein\\Estimate)} & \makecell{MSE\\(Standardized\\James-Stein\\Estimate)} \\ 
  \hline
Dental\_age & -0.015 & 0.025 & 0.070 & 0.006 & 0.049 & 0.090 & 1.229 & 1.518 \\ 
  Avg\_homeppm & -0.103 & -0.035 & 0.168 & 0.029 & -0.057 & 0.012 & 0.137 & 0.019 \\ 
  Tooth8 & -0.046 & -0.042 & 0.168 & 0.030 & -0.046 & -0.042 & 0.122 & 0.017 \\ 
  Tooth9 & -0.027 & -0.055 & 0.154 & 0.027 & 0.011 & -0.017 & 0.124 & 0.016 \\ 
  Tooth10 & 0.074 & 0.014 & 0.160 & 0.026 & 0.044 & -0.016 & 0.120 & 0.015 \\ 
  ZoneM & -0.147 & -0.123 & 0.511 & 0.276 & -0.064 & -0.040 & 0.737 & 0.544 \\ 
  ZoneI & -0.242 & -0.130 & 0.584 & 0.358 & -0.204 & -0.093 & 0.605 & 0.375 \\ 
  ZoneE & -0.282 & -0.146 & 0.645 & 0.438 & -0.239 & -0.104 & 0.669 & 0.458 \\ 
  \hline Average MSE &  &  &  & 0.149 &  &  &  & 0.370 \\ 
   \hline
\end{tabular}
}
\end{subtable}
\begin{subtable}{\linewidth}
\centering
\caption{N=50}
\scalebox{0.65}{
\begin{tabular}{lcccc|cccc}
Variable & \makecell{Standardized\\Estimate} & \makecell{Bias\\(Standardized\\Estimate)} & \makecell{SE\\(Standardized\\Estimate)} & \makecell{MSE\\(Standardized\\Estimate)} & \makecell{Standardized\\James-Stein\\Estimate} & \makecell{Bias\\(Standardized\\James-Stein\\Estimate)} & \makecell{SE\\(Standardized\\James-Stein\\Estimate)} & \makecell{MSE\\(Standardized\\James-Stein\\Estimate)} \\ 
  \hline
Dental\_age & -0.017 & 0.023 & 0.090 & 0.009 & 0.073 & 0.114 & 1.281 & 1.654 \\ 
  Avg\_homeppm & -0.129 & -0.060 & 0.124 & 0.019 & -0.091 & -0.022 & 0.097 & 0.010 \\ 
  Tooth8 & -0.068 & -0.064 & 0.134 & 0.022 & -0.040 & -0.035 & 0.096 & 0.010 \\ 
  Tooth9 & -0.029 & -0.056 & 0.122 & 0.018 & -0.010 & -0.038 & 0.085 & 0.009 \\ 
  Tooth10 & 0.088 & 0.080 & 0.118 & 0.015 & 0.038 & -0.022 & 0.087 & 0.008 \\ 
  ZoneM & -0.091 & -0.067 & 0.342 & 0.122 & -0.048 & -0.024 & 0.337 & 0.114 \\ 
  ZoneI & -0.218 & -0.106 & 0.665 & 0.454 & -0.191 & -0.079 & 0.667 & 0.451 \\ 
  ZoneE & -0.294 & -0.158 & 0.780 & 0.633 & -0.256 & -0.120 & 0.787 & 0.633 \\ 
  \hline Average MSE &  &  &  & 0.162 &  &  &  & 0.361 \\ 
   \hline
\end{tabular}
}
\end{subtable}
\begin{subtable}{\linewidth}
\centering
\caption{N=200}
\scalebox{0.65}{
\begin{tabular}{lcccc|cccc}
Variable & \makecell{Standardized\\Estimate} & \makecell{Bias\\(Standardized\\Estimate)} & \makecell{SE\\(Standardized\\Estimate)} & \makecell{MSE\\(Standardized\\Estimate)} & \makecell{Standardized\\James-Stein\\Estimate} & \makecell{Bias\\(Standardized\\James-Stein\\Estimate)} & \makecell{SE\\(Standardized\\James-Stein\\Estimate)} & \makecell{MSE\\(Standardized\\James-Stein\\Estimate)} \\ 
  \hline
Dental\_age & -0.045 & -0.005 & 0.067 & 0.004 & -0.024 & 0.016 & 0.113 & 0.013 \\ 
  Avg\_homeppm & -0.134 & -0.065 & 0.079 & 0.010 & -0.119 & -0.050 & 0.072 & 0.008 \\ 
  Tooth8 & -0.046 & -0.042 & 0.113 & 0.015 & -0.049 & -0.044 & 0.095 & 0.011 \\ 
  Tooth9 & -0.019 & -0.046 & 0.058 & 0.006 & -0.006 & -0.034 & 0.036 & 0.002 \\ 
  Tooth10 & 0.082 & 0.022 & 0.111 & 0.013 & 0.068 & 0.008 & 0.096 & 0.009 \\ 
  ZoneM & -0.079 & -0.055 & 0.086 & 0.010 & -0.067 & -0.044 & 0.075 & 0.008 \\ 
  ZoneI & -0.207 & -0.095 & 0.089 & 0.017 & -0.196 & -0.084 & 0.090 & 0.015 \\ 
  ZoneE & -0.275 & -0.140 & 0.092 & 0.080 & -0.266 & -0.131 & 0.093 & 0.026 \\ 
  \hline Average MSE &  &  &  & 0.013 &  &  &  & 0.012 \\ 
   \hline
\end{tabular}
}
\end{subtable}
\end{threeparttable}
\end{table}


\addtocounter{table}{-1} 

\begin{table}[ht]
\centering
\caption{ Simulation results for age 23 data generated with jackknifed and jackknifed correlation structures for presence and severity respectively - properties of standardized estimates arising from the presence piece of the combined model C$_s$.3.3.4. }
\label{ch4:table:sim:std:comb:pres:jackknifed}
\begin{threeparttable}
\centering
\begin{subtable}{\linewidth}
\centering
\caption{N=30}
\scalebox{0.65}{
\begin{tabular}{lcccc|cccc}
Variable & \makecell{Standardized\\Estimate} & \makecell{Bias\\(Standardized\\Estimate)} & \makecell{SE\\(Standardized\\Estimate)} & \makecell{MSE\\(Standardized\\Estimate)} & \makecell{Standardized\\James-Stein\\Estimate} & \makecell{Bias\\(Standardized\\James-Stein\\Estimate)} & \makecell{SE\\(Standardized\\James-Stein\\Estimate)} & \makecell{MSE\\(Standardized\\James-Stein\\Estimate)} \\ 
  \hline
Dental\_age & 1.743 & 1.930 & 1.331 & 5.498 & 1.909 & 2.097 & 1.524 & 6.72 \\ 
  Avg\_homeppm & 2.405 & 2.787 & 4.826 & 31.053 & 2.693 & 3.075 & 5.757 & 42.603 \\ 
  Tooth8 & 3.935 & 4.059 & 3.932 & 31.94 & 4.124 & 4.248 & 4.656 & 39.729 \\ 
  Tooth9 & -0.154 & -0.104 & 0.275 & 0.087 & -0.212 & -0.161 & 0.235 & 0.081 \\ 
  Tooth10 & -0.32 & -0.542 & 0.175 & 0.325 & -0.262 & -0.484 & 0.166 & 0.262 \\ 
  ZoneM & -0.089 & 0.224 & 0.132 & 0.068 & -0.005 & 0.308 & 0.128 & 0.111 \\ 
  ZoneI & -0.092 & 0.669 & 0.134 & 0.466 & -0.026 & 0.735 & 0.084 & 0.548 \\ 
  ZoneE & 0.150 & 1.113 & 0.170 & 1.267 & 0.070 & 1.033 & 0.135 & 1.084 \\ 
  \hline Average MSE &  &  &  & 8.838 &  &  &  & 11.392 \\ 
   \hline
\end{tabular}
}
\end{subtable}
\begin{subtable}{\linewidth}
\centering
\caption{N=50}
\scalebox{0.65}{
\begin{tabular}{lcccc|cccc}
Variable & \makecell{Standardized\\Estimate} & \makecell{Bias\\(Standardized\\Estimate)} & \makecell{SE\\(Standardized\\Estimate)} & \makecell{MSE\\(Standardized\\Estimate)} & \makecell{Standardized\\James-Stein\\Estimate} & \makecell{Bias\\(Standardized\\James-Stein\\Estimate)} & \makecell{SE\\(Standardized\\James-Stein\\Estimate)} & \makecell{MSE\\(Standardized\\James-Stein\\Estimate)} \\ 
  \hline
Dental\_age & 1.897 & 2.085 & 1.026 & 5.399 & 1.978 & 2.166 & 0.994 & 5.678 \\ 
  Avg\_homeppm & 0.97 & 1.352 & 0.868 & 2.58 & 0.907 & 1.289 & 0.555 & 1.968 \\ 
  Tooth8 & 3.239 & 3.364 & 1.392 & 13.25 & 3.248 & 3.373 & 0.96 & 12.298 \\ 
  Tooth9 & -0.141 & -0.091 & 0.139 & 0.027 & -0.139 & -0.089 & 0.124 & 0.023 \\ 
  Tooth10 & -0.357 & -0.579 & 0.131 & 0.353 & -0.351 & -0.573 & 0.126 & 0.344 \\ 
  ZoneM & -0.092 & 0.221 & 0.104 & 0.060 & -0.046 & 0.267 & 0.102 & 0.082 \\ 
  ZoneI & -0.070 & 0.691 & 0.101 & 0.488 & 0.006 & 0.768 & 0.103 & 0.600 \\ 
  ZoneE & 0.161 & 1.124 & 0.142 & 1.283 & 0.139 & 1.101 & 0.127 & 1.228 \\ 
  \hline Average MSE &  &  &  & 2.930 &  &  &  & 2.778 \\ 
   \hline
\end{tabular}
}
\end{subtable}
\begin{subtable}{\linewidth}
\centering
\caption{N=200}
\scalebox{0.65}{
\begin{tabular}{lcccc|cccc}
Variable & \makecell{Standardized\\Estimate} & \makecell{Bias\\(Standardized\\Estimate)} & \makecell{SE\\(Standardized\\Estimate)} & \makecell{MSE\\(Standardized\\Estimate)} & \makecell{Standardized\\James-Stein\\Estimate} & \makecell{Bias\\(Standardized\\James-Stein\\Estimate)} & \makecell{SE\\(Standardized\\James-Stein\\Estimate)} & \makecell{MSE\\(Standardized\\James-Stein\\Estimate)} \\ 
  \hline
Dental\_age & 2.153 & 2.341 & 0.568 & 5.803 & 2.144 & 2.332 & 0.55 & 5.74 \\ 
  Avg\_homeppm & 0.855 & 1.237 & 0.361 & 1.66 & 0.89 & 1.272 & 0.327 & 1.725 \\ 
  Tooth8 & 3.515 & 3.64 & 0.552 & 13.553 & 3.579 & 3.704 & 0.56 & 14.035 \\ 
  Tooth9 & -0.167 & -0.117 & 0.063 & 0.018 & -0.155 & -0.105 & 0.060 & 0.015 \\ 
  Tooth10 & -0.357 & -0.579 & 0.074 & 0.340 & -0.357 & -0.579 & 0.070 & 0.340 \\ 
  ZoneM & -0.080 & 0.233 & 0.048 & 0.057 & -0.059 & 0.254 & 0.045 & 0.067 \\ 
  ZoneI & -0.050 & 0.711 & 0.054 & 0.509 & -0.029 & 0.732 & 0.041 & 0.537 \\ 
  ZoneE & 0.181 & 1.143 & 0.064 & 1.31 & 0.168 & 1.130 & 0.064 & 1.281 \\ 
  \hline Average MSE &  &  &  & 2.906 &  &  &  & 2.968 \\ 
   \hline
\end{tabular}
}
\end{subtable}
\end{threeparttable}
\end{table}


\addtocounter{table}{-1} 

\begin{table}[ht]
\centering
\caption{ Simulation results for age 23 data generated with jackknifed and jackknifed correlation structures for presence and severity respectively - properties of standardized estimates arising from the severity piece of the combined model C$_s$.3.3.4. }
\label{ch4:table:sim:std:comb:sev:jackknifed}
\begin{threeparttable}
\centering
\begin{subtable}{\linewidth}
\centering
\caption{N=30}
\scalebox{0.65}{
\begin{tabular}{lcccc|cccc}
Variable & \makecell{Standardized\\Estimate} & \makecell{Bias\\(Standardized\\Estimate)} & \makecell{SE\\(Standardized\\Estimate)} & \makecell{MSE\\(Standardized\\Estimate)} & \makecell{Standardized\\James-Stein\\Estimate} & \makecell{Bias\\(Standardized\\James-Stein\\Estimate)} & \makecell{SE\\(Standardized\\James-Stein\\Estimate)} & \makecell{MSE\\(Standardized\\James-Stein\\Estimate)} \\ 
  \hline
Dental\_age & 7.523 & 7.563 & 27.399 & 807.936 & 2.272 & 2.313 & 7.807 & 66.303 \\ 
  Avg\_homeppm & 12.887 & 12.956 & 30.842 & 1119.054 & 8.118 & 8.187 & 18.114 & 395.156 \\ 
  Tooth8 & -0.122 & -0.118 & 0.258 & 0.080 & -0.176 & -0.172 & 0.196 & 0.068 \\ 
  Tooth9 & -0.177 & -0.205 & 0.350 & 0.165 & -0.255 & -0.282 & 0.266 & 0.151 \\ 
  Tooth10 & -0.069 & -0.129 & 0.187 & 0.052 & -0.043 & -0.103 & 0.151 & 0.034 \\ 
  ZoneM & -0.048 & -0.024 & 0.137 & 0.019 & -0.005 & 0.018 & 0.075 & 0.006 \\ 
  ZoneI & 0.071 & 0.183 & 0.213 & 0.079 & 0.054 & 0.166 & 0.133 & 0.045 \\ 
  ZoneE & -0.225 & -0.089 & 0.319 & 0.110 & -0.337 & -0.202 & 0.201 & 0.081 \\ 
  \hline Average MSE &  &  &  & 240.937 &  &  &  & 57.730 \\ 
   \hline
\end{tabular}
}
\end{subtable}
\begin{subtable}{\linewidth}
\centering
\caption{N=50}
\scalebox{0.65}{
\begin{tabular}{lcccc|cccc}
Variable & \makecell{Standardized\\Estimate} & \makecell{Bias\\(Standardized\\Estimate)} & \makecell{SE\\(Standardized\\Estimate)} & \makecell{MSE\\(Standardized\\Estimate)} & \makecell{Standardized\\James-Stein\\Estimate} & \makecell{Bias\\(Standardized\\James-Stein\\Estimate)} & \makecell{SE\\(Standardized\\James-Stein\\Estimate)} & \makecell{MSE\\(Standardized\\James-Stein\\Estimate)} \\ 
  \hline
Dental\_age & 3.445 & 3.486 & 13.226 & 187.089 & 2.175 & 2.215 & 4.674 & 26.753 \\ 
  Avg\_homeppm & 12.850 & 12.919 & 42.408 & 1965.331 & 10.778 & 10.847 & 30.399 & 1041.784 \\ 
  Tooth8 & -0.108 & -0.104 & 0.176 & 0.042 & -0.110 & -0.106 & 0.165 & 0.038 \\ 
  Tooth9 & -0.29 & -0.318 & 0.271 & 0.174 & -0.285 & -0.313 & 0.263 & 0.167 \\ 
  Tooth10 & -0.082 & -0.142 & 0.154 & 0.044 & -0.049 & -0.109 & 0.136 & 0.030 \\ 
  ZoneM & -0.055 & -0.031 & 0.102 & 0.011 & 0.018 & 0.042 & 0.102 & 0.012 \\ 
  ZoneI & 0.119 & 0.231 & 0.176 & 0.084 & 0.111 & 0.222 & 0.146 & 0.071 \\ 
  ZoneE & -0.248 & -0.113 & 0.265 & 0.083 & -0.214 & -0.078 & 0.250 & 0.069 \\ 
  \hline Average MSE &  &  &  & 269.107 &  &  &  & 133.616 \\ 
   \hline
\end{tabular}
}
\end{subtable}
\begin{subtable}{\linewidth}
\centering
\caption{N=200}
\scalebox{0.65}{
\begin{tabular}{lcccc|cccc}
Variable & \makecell{Standardized\\Estimate} & \makecell{Bias\\(Standardized\\Estimate)} & \makecell{SE\\(Standardized\\Estimate)} & \makecell{MSE\\(Standardized\\Estimate)} & \makecell{Standardized\\James-Stein\\Estimate} & \makecell{Bias\\(Standardized\\James-Stein\\Estimate)} & \makecell{SE\\(Standardized\\James-Stein\\Estimate)} & \makecell{MSE\\(Standardized\\James-Stein\\Estimate)} \\ 
  \hline
Dental\_age & 3.435 & 3.475 & 12.041 & 157.07 & 4.059 & 4.099 & 14.246 & 219.742 \\ 
  Avg\_homeppm & 12.430 & 12.499 & 47.212 & 2385.171 & 14.618 & 14.687 & 55.917 & 3342.376 \\ 
  Tooth8 & -0.162 & -0.158 & 0.085 & 0.032 & -0.148 & -0.143 & 0.087 & 0.080 \\ 
  Tooth9 & -0.350 & -0.378 & 0.164 & 0.170 & -0.342 & -0.369 & 0.185 & 0.171 \\ 
  Tooth10 & -0.094 & -0.154 & 0.080 & 0.030 & -0.071 & -0.131 & 0.079 & 0.023 \\ 
  ZoneM & -0.041 & -0.017 & 0.056 & 0.003 & -0.020 & 0.003 & 0.040 & 0.002 \\ 
  ZoneI & 0.167 & 0.278 & 0.089 & 0.085 & 0.150 & 0.262 & 0.092 & 0.077 \\ 
  ZoneE & -0.364 & -0.229 & 0.144 & 0.073 & -0.341 & -0.206 & 0.160 & 0.068 \\ 
  \hline Average MSE &  &  &  & 317.829 &  &  &  & 445.311 \\ 
   \hline
\end{tabular}
}
\end{subtable}
\end{threeparttable}
\end{table}


\addtocounter{table}{-1}

\begin{table}[]
\caption{Power and standardized effects ($\beta^*$) under varying true parameter values ($\beta$) for {Avg\_homeppm}, including both null and non-null cases. When $\beta = 0$, the standardized effect remains zero after transformation. For non-zero $\beta$, $\beta^*$ is computed from Monte Carlo estimates.}
    \label{ch4:table:alt_H1}
\centering
\begin{tabular}{r|rr|rr|rr|rr}
{} & \multicolumn{2}{c|}{\textbf{Presence}} & \multicolumn{2}{c|}{\textbf{Severity}} & \multicolumn{2}{c|}{\textbf{\makecell{Combined\\Presence}}} & \multicolumn{2}{c}{\textbf{\makecell{Combined\\Severity}}} \\ \hline
$\beta$ & $\beta^*$ & Power & $\beta^*$ & Power & $\beta^*$ & Power & $\beta^*$ & Power \\
\hline
-2 & -6.623 & 1.000 & -0.021 & 0.083 & -6.530 & 1.000 & -0.302 & 1.000 \\
-1 & -3.258 & 0.919 &  0.117 & 0.086 & -3.212 & 0.915 & -0.414 & 0.915 \\
 0 &  0.000 & 0.052 &  0.000 & 0.049 &  0.000 & 0.061 &  0.000 & 0.061 \\
 1 &  2.205 & 0.571 & -0.271 & 0.284 &  2.170 & 0.573 &  0.094 & 0.573 \\
 2 &  2.863 & 0.889 & -0.360 & 0.181 &  3.212 & 0.929 & -0.091 & 0.929 \\
\hline
\end{tabular}
\end{table}


\clearpage
\bibliography{bib}
\end{document}


\maketitle

\label{section:Appendix}
\section{Covariates in Iowa Fluoride Study (IFS) data}
\subsection{Categorical predictors}
\label{appendC:Categorical predictors}
\begin{itemize}
    \item Tooth: tooth locations 7,8,9,10, coded by the universal numbering system. Tooth 7 (upper right, lateral incisor) is treated as the reference level. Teeth 8 and 9 are the two maxillary central incisors (right and left, respectively), and tooth 10 is the upper left, lateral incisor.
    \item Zone: Tooth zones include the cervical third (C), middle third (M), incisal third (I) of the buccal surface, and the incisal edge (E), measured on each of the four maxillary incisors (reference: zone C).
\end{itemize}
\subsection{Continuous Covariates}
    \begin{itemize}
        \item Dental\_age: patient age at dental examination. It is the child’s age in years at the time of the dental examination associated with measurement occasion t. In practice, the means of actual
dental examination ages are slightly higher than the scheduled ages of 9, 13 and 17, and age is centered by the scheduled time. For instance, if the child’s dental appointment is 2 months after the scheduled time, the dental age would be reported as $2/12 = 0.17$.
        \item Total\_mgFPerDay:  total daily fluoride intake (mg) determined by combining fluoride intake from water, other beverages and selected foods, ingested toothpaste, and dietary fluoride supplements before permanent teeth eruption. The value was computed through an AUC (area under the curve) trapezoidal method using all available
data measured approximately every 6 months for ages of 0-5.
        \item SugarAddedBeverageOzPerDay: daily sugar-added beverage intake (oz.) was computed by an AUC trapezoidal method using all available data measured approximately every 6 months for ages of
0-5.
        \item BrushingFrequencyPerDay:  average of the tooth brushing frequencies per day was
reported approximately every 6 months for ages of 0-5 and combined using an AUC
trapezoidal approach.
        \item Avg\_homeppm: fluoride exposure from home tap water. It is the average home tap water fluoride level (ppm) reported for all returned
questionnaires approximately every 6 months for ages of 0-5, combined through a
trapezoidal AUC method.
        \item Prop\_DentalAppt:  proportion of times a dental visit was reported for approximately 6-month intervals for ages of 0-5.
        \item Prop\_FluorideTreatment: proportion of times a professional dental fluoride treatment was reported in approximately 6-month intervals for ages of 0-5.
    \end{itemize}

\section{Components of the Presence GEE}
\label{appendC: Components of the presence GEE}
\allowdisplaybreaks
\begin{align}
    \mu_{P,tijk} &= \frac{\exp(\alpha + \boldsymbol{x_{P,tijk}'\beta})}{1+\exp(\alpha + \boldsymbol{x_{P,tijk}'\beta})} \\
    \mu_{P,tijk} &= \big( \mu_{P,i, 111}, \mu_{P,i, 112}, \cdots, \mu_{P,i, T J K} \big)^T \\
    \pdv{\mu_{P,tijk}}{\alpha} &= \frac{\exp(\alpha + \boldsymbol{x_{P,tijk}'\beta})}{1+\exp(\alpha + \boldsymbol{x_{P,tijk}'\beta})} - \Bigg( \frac{\exp(\alpha + \boldsymbol{x_{P,tijk}'\beta})}{1+\exp(\alpha + \boldsymbol{x_{P,tijk}'\beta})}\Bigg)^2 = \mu_{P,tijk}(1-\mu_{P,tijk}) \\ \pdv{\mu_{P,tijk}}{\beta_g} &= \mu_{P,tijk}(1-\mu_{P,tijk}) x_{tijk, g} \text{, where  } g=1, \cdots q \\
    \pdv{\mu_{i}}{\alpha} &= \bigg( \pdv{\mu_{i, 111}}{\alpha}, \pdv{\mu_{i, 112}}{\alpha},  \cdots, \pdv{\mu_{i, T J K}}{\alpha}  \bigg)^T \\
    \pdv{\mu_{ti}}{\beta_g} &= \bigg( \pdv{\mu_{ti11}}{\beta_g}, \pdv{\mu_{ti12}}{\beta_g},  \cdots, \pdv{\mu_{TiJK}}{\beta_g}  \bigg)^T, \text{for  } g=1, \cdots, q  \\
    \boldsymbol{Y_{ti}} &= (Y_{ti11}, Y_{ti12}, \cdots, Y_{TiJK})^T, 
    \label{C-7}
\end{align}
  where the indicators $Y_{P,tijk}=1$ if the corresponding FRI Score is non-zero (in \ref{C-7}).
  
\section{Components of the Severity GEE}
\label{appendC: Components of the severity GEE}
We present the derivations corresponding to $L=3$, as is the case for the real IFS data. Setting $\gamma=1$ corresponds to the case of separate modeling severity piece. 

\begin{align}
\mu_{S,{S,tijk},1} &= \frac{\exp(\alpha_{1} + \gamma(\boldsymbol{x_{S,tijk}'\beta}))}{1+\exp(\alpha_{1} + \gamma(\boldsymbol{x_{S,tijk}'\beta}))}\\
\mu_{S,tijk,2} &= \frac{\exp(\alpha_{2} + \gamma(\boldsymbol{x_{S,tijk}'\beta}))}{1+\exp(\alpha_{2} + \gamma(\boldsymbol{x_{S,tijk}'\beta}))} - \frac{\exp(\alpha_{1} + \gamma(\boldsymbol{x_{S,tijk}'\beta}))}{1+\exp(\alpha_{1} + \gamma(\boldsymbol{x_{S,tijk}'\beta}))}\\
\mu_{S,tijk,3} &= 1-\frac{\exp(\alpha_{2} + \gamma(\boldsymbol{x_{S,tijk}'\beta}))}{1+\exp(\alpha_{2} + \gamma(\boldsymbol{x_{S,tijk}'\beta}))}\\
 \mu_{S,ti,1} &= \big( \mu_{S,ti11,1} , \mu_{S,ti12,1}, \cdots \mu_{S,tiJK,1} \big)^{T}\\
  \mu_{S,ti,2} &= \big( \mu_{S,ti11,2} , \mu_{S,ti12,2}, \cdots \mu_{S,tiJK,2} \big)^{T}\\
   \mu_{S,ti,3} &= \big( \mu_{S,ti11,3} , \mu_{S,ti12,3}, \cdots \mu_{S,tiJK,3} \big)^{T}\\
\mu_{S,ti} &= \big( \mu_{S,ti,1}^T, \mu_{S,ti,2}^T, \mu_{S,ti,3}^T\big)^T \\
\pdv{\mu_{S,{tijk},1}}{\alpha_{1}} &= \mu_{S,{tijk}, 1}(1-\mu_{S,{tijk}, 1})\\
\pdv{\mu_{S,{tijk},2}}{\alpha_{1}} &= -\mu_{S,{tijk}, 1}(1-\mu_{S,{tijk}, 1}) \\
\pdv{\mu_{S,{tijk},3}}{\alpha_1} &= 0 \\
\pdv{\mu_{S,{tijk},1}}{\alpha_{2}} &= 0 \\
\pdv{\mu_{S,{tijk},2}}{\alpha_{2}} &= \mu_{S,{tijk},2}(1-\mu_{S,{tijk},2}) \\
\pdv{\mu_{S,{tijk},3}}{\alpha_{2}} &= -\mu_{S,{tijk},2}(1-\mu_{S,{tijk},2}) \\
\pdv{\mu_{S,{tijk},1}}{\beta_g} &= \mu_{S,{tijk},1}(1-\mu_{S,{tijk},1}) x_{tijk,g} \text{ for } g=1, \cdots, q \\
\pdv{\mu_{S,{tijk},2}}{\beta_g} &= \bigg(\mu_{S,{tijk},2}(1-\mu_{S,{tijk},2}) - \mu_{S,{tijk},1}(1-\mu_{S,{tijk},1})\bigg) x_{tijk,g} \text{ for } g=1, \cdots, q \\
\pdv{\mu_{S,{tijk},3}}{\beta_g} &= - \mu_{S,{tijk},2}(1-\mu_{S,{tijk},2}) x_{tijk,g} \text{ for } g=1, \cdots, q \\
\pdv{\mu_{S,ti}}{\alpha_{1}} &= \bigg( \pdv{\mu_{S,ti11}}{\alpha_{1}}, \pdv{\mu_{S,ti12}}{\alpha_{1}}, \cdots \pdv{\mu_{S,tiJK}}{\alpha_{1}} \bigg)^T \\
\pdv{\mu_{S,ti}}{\alpha_{2}} &= \bigg( \pdv{\mu_{S,ti11}}{\alpha_{2}}, \pdv{\mu_{S,ti12}}{\alpha_{2}}, \cdots \pdv{\mu_{S,tiJK}}{\alpha_{2}} \bigg)^T \\
\pdv{\mu_{S,ti}}{\beta_g} &= \bigg( \pdv{\mu_{S,ti11}}{\beta_g}, \pdv{\mu_{S,ti12}}{\beta_g}, \cdots \pdv{\mu_{S,tiJK}}{\beta_g} \bigg)^T \text{  for } g=1, \cdots, q \\
\boldsymbol{Y_{tijk}} &= (I(Y_{tijk}=1), I(Y_{tijk}=2), I(Y_{tijk}=3))^T 
\end{align}

\begin{align}
    \boldsymbol{Y_{i}} &= 
\begin{pmatrix}
\boldsymbol{Y_{i 111}^T}\\[6pt]
\boldsymbol{Y_{i 112}^T}\\[6pt]
\vdots \\[6pt]
\boldsymbol{Y_{i T J K}^T}
\end{pmatrix}
\end{align}

\section{Additional IFS Data Analysis Results}

Results corresponding to additional choices of cluster correlation structures are presented below.




\begin{table}[H]
\centering
\caption{ Estimates from models A.1.1-A.1.4, the separate presence models with the independence cluster correlation structure }
\label{ch4:table:sep:pres:independence}
\begin{threeparttable}
\centering
\begin{subtable}{\linewidth}
\centering
\caption{Model A.1.1 (age 9)}
        \scalebox{0.55}{
\begin{tabular}{rrrrrrrrl}
Variable & Estimate & \makecell{Standardized\\Estimate} & \makecell{SE\\(Standardized\\Estimate)} & \makecell{James-Stein\\Estimator} & \makecell{Bias\\(James-Stein\\Estimator)} & \makecell{SE\\(James-Stein\\Estimator)} & \makecell{MSE\\(James-Stein\\Estimator)} & \makecell{95\% CI\\(James-Stein\\Estimator)} \\
  \hline
dental\_age & -0.291 &  -1.034 & 0.891 &  -0.702 &  0.119 & 0.731 & 0.745 & ( -2.121,   0.631) \\ 
  Total\_mgFPerDay & -0.109 &  -0.550 & 0.968 &  -0.408 & -0.029 & 0.798 & 0.799 & ( -2.165,   0.937) \\ 
  SugarAddedBeverageOzPerDay & -0.006 &  -0.527 & 1.061 &  -0.344 &  0.017 & 0.853 & 0.854 & ( -2.108,   1.152) \\ 
  BrushingFrequencyPerDay & -0.098 &  -0.644 & 0.884 &   1.019 & -1.326 & 0.619 & 2.377 & ( -1.555,   0.894) \\ 
  Avg\_homeppm & -0.631 &  -2.985 & 0.780 &  -2.784 & -0.195 & 0.771 & 0.809 & ( -4.664,  -1.166)$^{*-}$ \\ 
  Prop\_DentAppt & -0.027 &  -0.054 & 0.909 &   1.591 & -1.511 & 0.619 & 2.903 & ( -1.055,   1.505) \\ 
  Prop\_FluorideTreatment &  0.126 &   0.148 & 0.956 &  -0.182 &  0.222 & 0.653 & 0.703 & ( -1.468,   1.225) \\ 
  Tooth8 & -0.409 &  -5.159 & 0.950 &  -4.980 & -0.013 & 0.912 & 0.913 & ( -6.654,  -3.289)$^{*-}$ \\ 
  Tooth9 & -0.407 &  -5.076 & 0.927 &  -4.844 & -0.052 & 0.897 & 0.899 & ( -6.554,  -3.292)$^{*-}$ \\ 
  Tooth10 &  0.239 &   3.338 & 0.935 &   2.951 & -0.041 & 0.963 & 0.964 & (  1.152,   4.568)$^{*+}$ \\ 
  ZoneM & -0.648 &  -5.783 & 0.884 &  -5.561 & -0.210 & 0.881 & 0.925 & ( -7.343,  -4.019)$^{*-}$ \\ 
  ZoneI & -1.553 & -11.772 & 0.690 & -11.689 & -0.194 & 0.682 & 0.720 & (-13.132, -10.575)$^{*-}$ \\ 
  ZoneE & -2.003 & -13.854 & 0.685 & -13.797 & -0.213 & 0.702 & 0.748 & (-15.263, -12.736)$^{*-}$ \\ 
   \hline
\end{tabular}
}
\end{subtable}
\begin{subtable}{\linewidth}
\centering
\caption{Model A.1.2 (age 13)}
\scalebox{0.55}{
\begin{tabular}{rrrrrrrrl}
Variable & Estimate & \makecell{Standardized\\Estimate} & \makecell{SE\\(Standardized\\Estimate)} & \makecell{James-Stein\\Estimator} & \makecell{Bias\\(James-Stein\\Estimator)} & \makecell{SE\\(James-Stein\\Estimator)} & \makecell{MSE\\(James-Stein\\Estimator)} & \makecell{95\% CI\\(James-Stein\\Estimator)} \\
  \hline
dental\_age &  0.577 &   1.191 & 0.812 &   0.809 & -0.101 & 0.656 & 0.666 & ( -0.618,   1.811) \\ 
  Total\_mgFPerDay &  0.043 &   0.249 & 0.971 &   0.185 &  0.009 & 0.726 & 0.726 & ( -1.163,   1.836) \\ 
  SugarAddedBeverageOzPerDay &  0.004 &   0.498 & 0.966 &   0.325 &  0.026 & 0.735 & 0.736 & ( -1.100,   1.669) \\ 
  BrushingFrequencyPerDay & -0.011 &  -0.066 & 0.965 &   0.104 & -0.169 & 0.589 & 0.617 & ( -1.156,   1.168) \\ 
  Avg\_homeppm & -0.627 &  -3.022 & 1.072 &  -2.818 &  0.043 & 1.042 & 1.044 & ( -4.758,  -0.764)$^{*-}$ \\ 
  Prop\_DentAppt &  0.068 &   0.089 & 1.067 &  -2.608 &  2.679 & 0.746 & 7.921 & ( -1.520,   1.854) \\ 
  Prop\_FluorideTreatment &  0.112 &   0.102 & 0.889 &  -0.126 &  0.165 & 0.595 & 0.623 & ( -1.124,   1.339) \\ 
  Tooth8 & -0.200 &  -2.341 & 1.053 &  -2.260 &  0.024 & 1.035 & 1.036 & ( -3.952,   0.023) \\ 
  Tooth9 & -0.135 &  -1.514 & 1.003 &  -1.445 &  0.035 & 0.997 & 0.999 & ( -3.120,   0.680) \\ 
  Tooth10 &  0.163 &   2.136 & 0.938 &   1.888 &  0.179 & 0.904 & 0.936 & (  0.561,   3.832)$^{*+}$ \\ 
  ZoneM & -0.420 &  -3.265 & 1.076 &  -3.139 & -0.095 & 1.021 & 1.030 & ( -5.127,  -1.148)$^{*-}$ \\ 
  ZoneI & -1.419 &  -8.763 & 0.778 &  -8.701 & -0.049 & 0.746 & 0.749 & (-10.118,  -7.524)$^{*-}$ \\ 
  ZoneE & -2.175 & -13.008 & 0.739 & -12.954 & -0.066 & 0.717 & 0.722 & (-14.490, -11.856)$^{*-}$ \\ 
   \hline
\end{tabular}
}
\end{subtable}
\begin{subtable}{\linewidth}
\centering
\caption{Model A.1.3 (age 17)}
\scalebox{0.55}{
\begin{tabular}{rrrrrrrrl}
Variable & Estimate & \makecell{Standardized\\Estimate} & \makecell{SE\\(Standardized\\Estimate)} & \makecell{James-Stein\\Estimator} & \makecell{Bias\\(James-Stein\\Estimator)} & \makecell{SE\\(James-Stein\\Estimator)} & \makecell{MSE\\(James-Stein\\Estimator)} & \makecell{95\% CI\\(James-Stein\\Estimator)} \\
  \hline
dental\_age &  0.587 &  1.205 & 0.827 &  0.818 & -0.021 & 0.722 & 0.722 & ( -0.505,  2.331) \\ 
  Total\_mgFPerDay & -0.116 & -0.625 & 0.962 & -0.463 & -0.006 & 0.811 & 0.811 & ( -2.401,  0.950) \\ 
  SugarAddedBeverageOzPerDay &  0.015 &  1.695 & 0.888 &  1.107 &  0.125 & 0.831 & 0.847 & ( -0.087,  2.697) \\ 
  BrushingFrequencyPerDay &  0.000 & -0.002 & 1.011 &  0.004 &  0.050 & 0.633 & 0.635 & ( -1.063,  1.591) \\ 
  Avg\_homeppm & -0.745 & -2.292 & 0.824 & -2.138 & -0.010 & 0.804 & 0.804 & ( -3.665, -0.721)$^{*-}$ \\ 
  Prop\_DentAppt & -0.040 & -0.043 & 1.116 &  1.272 & -1.291 & 0.771 & 2.438 & ( -1.584,  1.402) \\ 
  Prop\_FluorideTreatment & -1.182 & -0.841 & 1.133 &  1.035 & -1.429 & 0.842 & 2.886 & ( -2.117,  1.431) \\ 
  Tooth8 & -0.351 & -3.854 & 1.005 & -3.720 & -0.175 & 0.993 & 1.023 & ( -5.480, -2.148)$^{*-}$ \\ 
  Tooth9 & -0.244 & -2.588 & 1.021 & -2.470 & -0.080 & 0.967 & 0.974 & ( -4.571, -0.750)$^{*-}$ \\ 
  Tooth10 &  0.096 &  0.996 & 0.917 &  0.880 &  0.071 & 0.858 & 0.863 & ( -0.388,  2.470) \\ 
  ZoneM & -0.466 & -2.199 & 0.882 & -2.115 &  0.000 & 0.846 & 0.846 & ( -4.012, -0.854)$^{*-}$ \\ 
  ZoneI & -1.820 & -7.057 & 0.662 & -7.008 & -0.118 & 0.694 & 0.708 & ( -8.527, -5.825)$^{*-}$ \\ 
  ZoneE & -2.550 & -9.618 & 0.830 & -9.578 & -0.159 & 0.852 & 0.878 & (-11.245, -8.015)$^{*-}$ \\ 
   \hline
\end{tabular}
}
\end{subtable}
\begin{subtable}{\linewidth}
\centering
\caption{Model A.1.4 (age 23)}
\scalebox{0.55}{
\begin{tabular}{rrrrrrrrl}
Variable & Estimate & \makecell{Standardized\\Estimate} & \makecell{SE\\(Standardized\\Estimate)} & \makecell{James-Stein\\Estimator} & \makecell{Bias\\(James-Stein\\Estimator)} & \makecell{SE\\(James-Stein\\Estimator)} & \makecell{MSE\\(James-Stein\\Estimator)} & \makecell{95\% CI\\(James-Stein\\Estimator)} \\
  \hline
dental\_age &  0.716 &  1.516 & 0.767 &  1.030 &  0.115 & 0.719 &  0.733 & ( -0.140,  2.788) \\ 
  Total\_mgFPerDay & -0.331 & -2.641 & 0.934 & -1.957 &  0.099 & 0.984 &  0.994 & ( -3.442,  0.147) \\ 
  SugarAddedBeverageOzPerDay & -0.013 & -1.537 & 0.950 & -1.004 & -0.222 & 0.864 &  0.914 & ( -3.029,  0.231) \\ 
  BrushingFrequencyPerDay & -0.144 & -0.596 & 1.002 &  0.944 & -1.332 & 0.714 &  2.488 & ( -2.001,  1.057) \\ 
  Avg\_homeppm & -0.878 & -2.520 & 0.759 & -2.350 & -0.028 & 0.736 &  0.737 & ( -3.713, -0.931)$^{*-}$ \\ 
  Prop\_DentAppt & -0.281 & -0.231 & 1.107 &  6.763 & -6.747 & 0.756 & 46.284 & ( -1.672,  1.337) \\ 
  Prop\_FluorideTreatment &  0.555 &  0.397 & 0.767 & -0.488 &  0.692 & 0.523 &  1.002 & ( -0.759,  1.275) \\ 
  Tooth8 & -0.477 & -3.266 & 0.990 & -3.153 & -0.105 & 0.979 &  0.990 & ( -5.077, -1.653)$^{*-}$ \\ 
  Tooth9 & -0.432 & -2.986 & 0.927 & -2.849 & -0.015 & 0.917 &  0.917 & ( -4.439, -1.327)$^{*-}$ \\ 
  Tooth10 &  0.076 &  0.745 & 1.028 &  0.658 &  0.014 & 0.933 &  0.933 & ( -1.078,  2.333) \\ 
  ZoneM & -0.904 & -1.737 & 0.741 & -1.670 & -0.006 & 0.708 &  0.708 & ( -3.015, -0.050)$^{*-}$ \\ 
  ZoneI & -2.538 & -4.211 & 1.685 & -4.181 & -0.370 & 1.638 &  1.775 & ( -7.360, -0.131)$^{*-}$ \\ 
  ZoneE & -3.361 & -5.631 & 2.446 & -5.608 & -0.780 & 2.387 &  2.996 & (-10.570, -0.156)$^{*-}$ \\ 
   \hline
\end{tabular}
}
\end{subtable}
\begin{tablenotes}[para,flushleft]
\scriptsize
\item Superscripts $*+$ and $*-$ denote significant protective and risk effects at the 5\% significance level, respectively.
\end{tablenotes}
\end{threeparttable}
\end{table}

\addtocounter{table}{-1}

\begin{table}[H]
\centering
\caption{ Estimates from models A.2.1-A.2.4, the separate presence models with the exchangeable cluster correlation structure }
\label{ch4:table:sep:pres:exchangeable}
\begin{threeparttable}
\centering
\begin{subtable}{\linewidth}
\centering
\caption{Model A.2.1 (age 9)}
\scalebox{0.55}{
\begin{tabular}{rrrrrrrrl}
Variable & Estimate & \makecell{Standardized\\Estimate} & \makecell{SE\\(Standardized\\Estimate)} & \makecell{James-Stein\\Estimator} & \makecell{Bias\\(James-Stein\\Estimator)} & \makecell{SE\\(James-Stein\\Estimator)} & \makecell{MSE\\(James-Stein\\Estimator)} & \makecell{95\% CI\\(James-Stein\\Estimator)} \\
  \hline
dental\_age & -0.291 &  -1.034 & 0.891 &  -0.703 &  0.119 & 0.731 & 0.745 & ( -2.122,   0.630) \\ 
  Total\_mgFPerDay & -0.109 &  -0.550 & 0.968 &  -0.408 & -0.029 & 0.798 & 0.799 & ( -2.165,   0.937) \\ 
  SugarAddedBeverageOzPerDay & -0.006 &  -0.526 & 1.061 &  -0.344 &  0.017 & 0.853 & 0.854 & ( -2.107,   1.153) \\ 
  BrushingFrequencyPerDay & -0.098 &  -0.643 & 0.884 &   1.020 & -1.326 & 0.619 & 2.378 & ( -1.554,   0.894) \\ 
  Avg\_homeppm & -0.631 &  -2.985 & 0.780 &  -2.784 & -0.194 & 0.771 & 0.809 & ( -4.664,  -1.164)$^{*-}$ \\ 
  Prop\_DentAppt & -0.027 &  -0.054 & 0.909 &   1.590 & -1.510 & 0.619 & 2.901 & ( -1.055,   1.505) \\ 
  Prop\_FluorideTreatment &  0.127 &   0.149 & 0.956 &  -0.183 &  0.223 & 0.653 & 0.703 & ( -1.468,   1.226) \\ 
  Tooth8 & -0.409 &  -5.158 & 0.950 &  -4.979 & -0.013 & 0.912 & 0.912 & ( -6.653,  -3.289)$^{*-}$ \\ 
  Tooth9 & -0.407 &  -5.075 & 0.927 &  -4.842 & -0.052 & 0.896 & 0.899 & ( -6.552,  -3.291)$^{*-}$ \\ 
  Tooth10 &  0.239 &   3.339 & 0.935 &   2.952 & -0.041 & 0.963 & 0.965 & (  1.155,   4.571)$^{*+}$ \\ 
  ZoneM & -0.648 &  -5.784 & 0.884 &  -5.561 & -0.210 & 0.882 & 0.926 & ( -7.345,  -4.019)$^{*-}$ \\ 
  ZoneI & -1.553 & -11.771 & 0.690 & -11.687 & -0.194 & 0.682 & 0.720 & (-13.131, -10.573)$^{*-}$ \\ 
  ZoneE & -2.003 & -13.853 & 0.685 & -13.796 & -0.213 & 0.703 & 0.748 & (-15.264, -12.734)$^{*-}$ \\ 
   \hline
\end{tabular}
}
\end{subtable}
\begin{subtable}{\linewidth}
\centering
\caption{Model A.2.2 (age 13)}
\scalebox{0.55}{
\begin{tabular}{rrrrrrrrl}
Variable & Estimate & \makecell{Standardized\\Estimate} & \makecell{SE\\(Standardized\\Estimate)} & \makecell{James-Stein\\Estimator} & \makecell{Bias\\(James-Stein\\Estimator)} & \makecell{SE\\(James-Stein\\Estimator)} & \makecell{MSE\\(James-Stein\\Estimator)} & \makecell{95\% CI\\(James-Stein\\Estimator)} \\
  \hline
dental\_age &  0.578 &   1.191 & 0.812 &   0.810 & -0.102 & 0.656 & 0.666 & ( -0.617,   1.812) \\ 
  Total\_mgFPerDay &  0.043 &   0.252 & 0.971 &   0.187 &  0.010 & 0.727 & 0.727 & ( -1.159,   1.845) \\ 
  SugarAddedBeverageOzPerDay &  0.004 &   0.496 & 0.966 &   0.324 &  0.025 & 0.735 & 0.736 & ( -1.101,   1.668) \\ 
  BrushingFrequencyPerDay & -0.011 &  -0.066 & 0.965 &   0.105 & -0.170 & 0.589 & 0.618 & ( -1.157,   1.168) \\ 
  Avg\_homeppm & -0.627 &  -3.022 & 1.072 &  -2.818 &  0.042 & 1.042 & 1.044 & ( -4.758,  -0.766)$^{*-}$ \\ 
  Prop\_DentAppt &  0.068 &   0.089 & 1.067 &  -2.601 &  2.671 & 0.746 & 7.880 & ( -1.521,   1.855) \\ 
  Prop\_FluorideTreatment &  0.113 &   0.103 & 0.889 &  -0.126 &  0.166 & 0.595 & 0.623 & ( -1.123,   1.340) \\ 
  Tooth8 & -0.200 &  -2.340 & 1.053 &  -2.258 &  0.024 & 1.035 & 1.036 & ( -3.949,   0.025) \\ 
  Tooth9 & -0.135 &  -1.514 & 1.003 &  -1.444 &  0.035 & 0.997 & 0.999 & ( -3.120,   0.681) \\ 
  Tooth10 &  0.163 &   2.137 & 0.938 &   1.889 &  0.179 & 0.904 & 0.936 & (  0.561,   3.836)$^{*+}$ \\ 
  ZoneM & -0.420 &  -3.263 & 1.076 &  -3.137 & -0.095 & 1.022 & 1.031 & ( -5.124,  -1.144)$^{*-}$ \\ 
  ZoneI & -1.419 &  -8.762 & 0.779 &  -8.700 & -0.049 & 0.747 & 0.749 & (-10.118,  -7.520)$^{*-}$ \\ 
  ZoneE & -2.175 & -13.007 & 0.739 & -12.954 & -0.066 & 0.717 & 0.722 & (-14.488, -11.854)$^{*-}$ \\ 
   \hline
\end{tabular}
}
\end{subtable}
\begin{subtable}{\linewidth}
\centering
\caption{Model A.2.3 (age 17)}
\scalebox{0.55}{
\begin{tabular}{rrrrrrrrl}
Variable & Estimate & \makecell{Standardized\\Estimate} & \makecell{SE\\(Standardized\\Estimate)} & \makecell{James-Stein\\Estimator} & \makecell{Bias\\(James-Stein\\Estimator)} & \makecell{SE\\(James-Stein\\Estimator)} & \makecell{MSE\\(James-Stein\\Estimator)} & \makecell{95\% CI\\(James-Stein\\Estimator)} \\
  \hline
dental\_age &  0.587 &  1.205 & 0.827 &  0.819 & -0.021 & 0.722 & 0.722 & ( -0.504,  2.331) \\ 
  Total\_mgFPerDay & -0.116 & -0.624 & 0.962 & -0.463 & -0.006 & 0.810 & 0.810 & ( -2.400,  0.949) \\ 
  SugarAddedBeverageOzPerDay &  0.015 &  1.695 & 0.888 &  1.107 &  0.125 & 0.831 & 0.847 & ( -0.087,  2.697) \\ 
  BrushingFrequencyPerDay &  0.000 & -0.001 & 1.011 &  0.002 &  0.053 & 0.633 & 0.635 & ( -1.060,  1.592) \\ 
  Avg\_homeppm & -0.745 & -2.295 & 0.824 & -2.140 & -0.010 & 0.804 & 0.804 & ( -3.667, -0.723)$^{*-}$ \\ 
  Prop\_DentAppt & -0.041 & -0.044 & 1.116 &  1.295 & -1.315 & 0.771 & 2.500 & ( -1.585,  1.401) \\ 
  Prop\_FluorideTreatment & -1.183 & -0.841 & 1.133 &  1.033 & -1.428 & 0.843 & 2.883 & ( -2.116,  1.431) \\ 
  Tooth8 & -0.351 & -3.853 & 1.004 & -3.719 & -0.175 & 0.993 & 1.023 & ( -5.480, -2.146)$^{*-}$ \\ 
  Tooth9 & -0.243 & -2.587 & 1.021 & -2.469 & -0.080 & 0.967 & 0.974 & ( -4.570, -0.749)$^{*-}$ \\ 
  Tooth10 &  0.097 &  0.997 & 0.917 &  0.881 &  0.071 & 0.858 & 0.863 & ( -0.386,  2.471) \\ 
  ZoneM & -0.466 & -2.198 & 0.883 & -2.113 &  0.000 & 0.846 & 0.846 & ( -4.013, -0.853)$^{*-}$ \\ 
  ZoneI & -1.820 & -7.057 & 0.662 & -7.007 & -0.118 & 0.694 & 0.708 & ( -8.526, -5.825)$^{*-}$ \\ 
  ZoneE & -2.550 & -9.618 & 0.830 & -9.578 & -0.160 & 0.852 & 0.878 & (-11.245, -8.016)$^{*-}$ \\ 
   \hline
\end{tabular}
}
\end{subtable}
\begin{subtable}{\linewidth}
\centering
\caption{Model A.2.4 (age 23)}
\scalebox{0.55}{
\begin{tabular}{rrrrrrrrl}
Variable & Estimate & \makecell{Standardized\\Estimate} & \makecell{SE\\(Standardized\\Estimate)} & \makecell{James-Stein\\Estimator} & \makecell{Bias\\(James-Stein\\Estimator)} & \makecell{SE\\(James-Stein\\Estimator)} & \makecell{MSE\\(James-Stein\\Estimator)} & \makecell{95\% CI\\(James-Stein\\Estimator)} \\
  \hline
dental\_age &  0.716 &  1.517 & 0.767 &  1.031 &  0.115 & 0.719 &  0.733 & ( -0.139,  2.789) \\ 
  Total\_mgFPerDay & -0.331 & -2.640 & 0.934 & -1.957 &  0.099 & 0.984 &  0.994 & ( -3.442,  0.148) \\ 
  SugarAddedBeverageOzPerDay & -0.013 & -1.538 & 0.950 & -1.004 & -0.223 & 0.864 &  0.914 & ( -3.030,  0.231) \\ 
  BrushingFrequencyPerDay & -0.144 & -0.596 & 1.002 &  0.945 & -1.333 & 0.714 &  2.490 & ( -2.001,  1.059) \\ 
  Avg\_homeppm & -0.878 & -2.520 & 0.759 & -2.351 & -0.028 & 0.736 &  0.737 & ( -3.713, -0.931)$^{*-}$ \\ 
  Prop\_DentAppt & -0.282 & -0.231 & 1.108 &  6.762 & -6.747 & 0.756 & 46.271 & ( -1.672,  1.336) \\ 
  Prop\_FluorideTreatment &  0.555 &  0.397 & 0.767 & -0.487 &  0.691 & 0.523 &  1.001 & ( -0.759,  1.274) \\ 
  Tooth8 & -0.476 & -3.266 & 0.991 & -3.153 & -0.105 & 0.980 &  0.991 & ( -5.076, -1.652)$^{*-}$ \\ 
  Tooth9 & -0.432 & -2.986 & 0.927 & -2.849 & -0.015 & 0.917 &  0.917 & ( -4.439, -1.326)$^{*-}$ \\ 
  Tooth10 &  0.076 &  0.745 & 1.028 &  0.659 &  0.014 & 0.933 &  0.933 & ( -1.078,  2.334) \\ 
  ZoneM & -0.904 & -1.737 & 0.739 & -1.670 & -0.006 & 0.707 &  0.707 & ( -3.015, -0.056)$^{*-}$ \\ 
  ZoneI & -2.538 & -4.211 & 1.682 & -4.181 & -0.370 & 1.637 &  1.774 & ( -7.359, -0.141)$^{*-}$ \\ 
  ZoneE & -3.361 & -5.631 & 2.442 & -5.608 & -0.780 & 2.385 &  2.994 & (-10.572, -0.169)$^{*-}$ \\ 
   \hline
\end{tabular}
}
\end{subtable}
\begin{tablenotes}[para,flushleft]
\scriptsize
\item Superscripts $*+$ and $*-$ denote significant protective and risk effects at the 5\% significance level, respectively.
\end{tablenotes}
\end{threeparttable}
\end{table}

\addtocounter{table}{-1}

\begin{table}[H]
\centering
\caption{ Estimates from models A.4.1-A.4.4, the separate presence models with the jackknifed cluster correlation structure }
\label{ch4:table:sep:pres:jackknifed}
\begin{threeparttable}
\centering
\begin{subtable}{\linewidth}
\centering
\caption{Model A.4.1 (age 9)}
\scalebox{0.55}{
\begin{tabular}{rrrrrrrrl}
Variable & Estimate & \makecell{Standardized\\Estimate} & \makecell{SE\\(Standardized\\Estimate)} & \makecell{James-Stein\\Estimator} & \makecell{Bias\\(James-Stein\\Estimator)} & \makecell{SE\\(James-Stein\\Estimator)} & \makecell{MSE\\(James-Stein\\Estimator)} & \makecell{95\% CI\\(James-Stein\\Estimator)} \\
  \hline
dental\_age & -0.291 &  -1.033 & 0.907 &  -0.702 &  0.158 & 0.693 & 0.718 & ( -2.077,   0.633) \\ 
  Total\_mgFPerDay & -0.109 &  -0.549 & 0.971 &  -0.407 & -0.057 & 0.814 & 0.817 & ( -2.179,   1.002) \\ 
  SugarAddedBeverageOzPerDay & -0.006 &  -0.530 & 1.072 &  -0.346 &  0.078 & 0.858 & 0.864 & ( -2.228,   1.168) \\ 
  BrushingFrequencyPerDay & -0.098 &  -0.645 & 0.904 &   1.014 & -1.342 & 0.636 & 2.436 & ( -1.577,   0.907) \\ 
  Avg\_homeppm & -0.631 &  -2.987 & 0.788 &  -2.786 & -0.222 & 0.800 & 0.850 & ( -4.739,  -1.162)$^{*-}$ \\ 
  Prop\_DentAppt & -0.027 &  -0.054 & 0.906 &   1.573 & -1.525 & 0.586 & 2.912 & ( -1.164,   1.423) \\ 
  Prop\_FluorideTreatment &  0.125 &   0.147 & 0.952 &  -0.182 &  0.218 & 0.673 & 0.721 & ( -1.538,   1.265) \\ 
  Tooth8 & -0.408 &  -5.150 & 0.955 &  -4.971 & -0.043 & 0.904 & 0.905 & ( -6.647,  -3.296)$^{*-}$ \\ 
  Tooth9 & -0.406 &  -5.069 & 0.935 &  -4.837 & -0.067 & 0.895 & 0.900 & ( -6.541,  -3.222)$^{*-}$ \\ 
  Tooth10 &  0.239 &   3.338 & 0.918 &   2.950 &  0.002 & 0.960 & 0.960 & (  1.107,   4.591)$^{*+}$ \\ 
  ZoneM & -0.648 &  -5.779 & 0.842 &  -5.556 & -0.182 & 0.880 & 0.913 & ( -7.335,  -3.952)$^{*-}$ \\ 
  ZoneI & -1.553 & -11.767 & 0.677 & -11.684 & -0.203 & 0.700 & 0.741 & (-13.205, -10.558)$^{*-}$ \\ 
  ZoneE & -2.002 & -13.851 & 0.692 & -13.794 & -0.205 & 0.729 & 0.771 & (-15.311, -12.722)$^{*-}$ \\ 
   \hline
\end{tabular}
}
\end{subtable}
\begin{subtable}{\linewidth}
\centering
\caption{Model A.4.2 (age 13)}
\scalebox{0.55}{
\begin{tabular}{rrrrrrrrl}
Variable & Estimate & \makecell{Standardized\\Estimate} & \makecell{SE\\(Standardized\\Estimate)} & \makecell{James-Stein\\Estimator} & \makecell{Bias\\(James-Stein\\Estimator)} & \makecell{SE\\(James-Stein\\Estimator)} & \makecell{MSE\\(James-Stein\\Estimator)} & \makecell{95\% CI\\(James-Stein\\Estimator)} \\
  \hline
dental\_age &  0.576 &   1.188 & 0.815 &   0.807 & -0.112 & 0.637 & 0.649 & ( -0.539,   1.840) \\ 
  Total\_mgFPerDay &  0.043 &   0.249 & 0.971 &   0.184 &  0.096 & 0.700 & 0.709 & ( -0.978,   1.849) \\ 
  SugarAddedBeverageOzPerDay &  0.004 &   0.497 & 0.969 &   0.325 & -0.060 & 0.702 & 0.705 & ( -1.139,   1.613) \\ 
  BrushingFrequencyPerDay & -0.011 &  -0.068 & 0.988 &   0.107 & -0.178 & 0.599 & 0.631 & ( -1.174,   1.109) \\ 
  Avg\_homeppm & -0.627 &  -3.022 & 1.080 &  -2.818 & -0.029 & 1.032 & 1.033 & ( -4.770,  -1.198)$^{*-}$ \\ 
  Prop\_DentAppt &  0.069 &   0.090 & 1.091 &  -2.634 &  2.705 & 0.766 & 8.082 & ( -1.626,   1.930) \\ 
  Prop\_FluorideTreatment &  0.110 &   0.101 & 0.885 &  -0.124 &  0.187 & 0.572 & 0.607 & ( -1.075,   1.383) \\ 
  Tooth8 & -0.200 &  -2.336 & 1.057 &  -2.254 & -0.016 & 1.003 & 1.003 & ( -3.956,  -0.007)$^{*-}$ \\ 
  Tooth9 & -0.135 &  -1.511 & 0.991 &  -1.441 &  0.003 & 0.954 & 0.954 & ( -3.122,   0.429) \\ 
  Tooth10 &  0.163 &   2.135 & 0.949 &   1.887 &  0.154 & 0.917 & 0.941 & (  0.541,   3.733)$^{*+}$ \\ 
  ZoneM & -0.420 &  -3.265 & 1.052 &  -3.139 & -0.129 & 0.996 & 1.013 & ( -5.144,  -1.333)$^{*-}$ \\ 
  ZoneI & -1.418 &  -8.764 & 0.750 &  -8.702 & -0.050 & 0.685 & 0.687 & (-10.049,  -7.697)$^{*-}$ \\ 
  ZoneE & -2.174 & -13.004 & 0.759 & -12.950 & -0.036 & 0.690 & 0.692 & (-14.345, -11.820)$^{*-}$ \\ 
   \hline
\end{tabular}
}
\end{subtable}
\begin{subtable}{\linewidth}
\centering
\caption{Model A.4.3 (age 17)}
\scalebox{0.55}{
\begin{tabular}{rrrrrrrrl}
Variable & Estimate & \makecell{Standardized\\Estimate} & \makecell{SE\\(Standardized\\Estimate)} & \makecell{James-Stein\\Estimator} & \makecell{Bias\\(James-Stein\\Estimator)} & \makecell{SE\\(James-Stein\\Estimator)} & \makecell{MSE\\(James-Stein\\Estimator)} & \makecell{95\% CI\\(James-Stein\\Estimator)} \\
  \hline
dental\_age &  0.587 &  1.205 & 0.843 &  0.818 &  0.020 & 0.748 & 0.749 & ( -0.508,  2.340) \\ 
  Total\_mgFPerDay & -0.117 & -0.627 & 0.956 & -0.465 & -0.008 & 0.799 & 0.799 & ( -2.481,  0.858) \\ 
  SugarAddedBeverageOzPerDay &  0.015 &  1.696 & 0.869 &  1.108 &  0.085 & 0.827 & 0.834 & ( -0.089,  2.733) \\ 
  BrushingFrequencyPerDay & -0.001 & -0.006 & 1.013 &  0.009 &  0.062 & 0.659 & 0.663 & ( -1.106,  1.622) \\ 
  Avg\_homeppm & -0.745 & -2.292 & 0.848 & -2.137 & -0.015 & 0.821 & 0.821 & ( -3.731, -0.702)$^{*-}$ \\ 
  Prop\_DentAppt & -0.040 & -0.043 & 1.123 &  1.270 & -1.270 & 0.788 & 2.400 & ( -1.613,  1.466) \\ 
  Prop\_FluorideTreatment & -1.181 & -0.840 & 1.148 &  1.038 & -1.399 & 0.868 & 2.827 & ( -2.165,  1.483) \\ 
  Tooth8 & -0.350 & -3.853 & 0.980 & -3.719 & -0.203 & 0.972 & 1.013 & ( -5.471, -2.343)$^{*-}$ \\ 
  Tooth9 & -0.243 & -2.584 & 1.033 & -2.466 & -0.113 & 0.968 & 0.981 & ( -4.692, -0.760)$^{*-}$ \\ 
  Tooth10 &  0.095 &  0.981 & 0.939 &  0.867 &  0.114 & 0.882 & 0.895 & ( -0.433,  2.454) \\ 
  ZoneM & -0.469 & -2.209 & 0.881 & -2.124 & -0.032 & 0.856 & 0.857 & ( -4.188, -0.859)$^{*-}$ \\ 
  ZoneI & -1.818 & -7.052 & 0.639 & -7.002 & -0.128 & 0.662 & 0.678 & ( -8.381, -5.954)$^{*-}$ \\ 
  ZoneE & -2.549 & -9.613 & 0.801 & -9.573 & -0.171 & 0.810 & 0.840 & (-11.177, -8.046)$^{*-}$ \\ 
   \hline
\end{tabular}
}
\end{subtable}
\begin{subtable}{\linewidth}
\centering
\caption{Model A.4.4 (age 23)}
\scalebox{0.55}{
\begin{tabular}{rrrrrrrrl}
Variable & Estimate & \makecell{Standardized\\Estimate} & \makecell{SE\\(Standardized\\Estimate)} & \makecell{James-Stein\\Estimator} & \makecell{Bias\\(James-Stein\\Estimator)} & \makecell{SE\\(James-Stein\\Estimator)} & \makecell{MSE\\(James-Stein\\Estimator)} & \makecell{95\% CI\\(James-Stein\\Estimator)} \\
  \hline
dental\_age &  0.715 &  1.516 & 0.774 &  1.030 &  0.124 & 0.747 &  0.763 & ( -0.146,  2.809) \\ 
  Total\_mgFPerDay & -0.331 & -2.641 & 0.933 & -1.958 &  0.102 & 0.974 &  0.984 & ( -3.467, -0.038)$^{*-}$ \\ 
  SugarAddedBeverageOzPerDay & -0.013 & -1.537 & 0.924 & -1.004 & -0.192 & 0.852 &  0.889 & ( -2.812,  0.242) \\ 
  BrushingFrequencyPerDay & -0.144 & -0.597 & 1.005 &  0.940 & -1.338 & 0.737 &  2.526 & ( -2.108,  1.091) \\ 
  Avg\_homeppm & -0.878 & -2.521 & 0.752 & -2.351 & -0.022 & 0.734 &  0.735 & ( -3.654, -0.881)$^{*-}$ \\ 
  Prop\_DentAppt & -0.281 & -0.231 & 1.071 &  6.757 & -6.765 & 0.758 & 46.529 & ( -1.757,  1.330) \\ 
  Prop\_FluorideTreatment &  0.555 &  0.396 & 0.767 & -0.490 &  0.688 & 0.534 &  1.008 & ( -0.772,  1.288) \\ 
  Tooth8 & -0.476 & -3.269 & 0.995 & -3.155 & -0.069 & 0.988 &  0.993 & ( -5.069, -1.634)$^{*-}$ \\ 
  Tooth9 & -0.432 & -2.984 & 0.920 & -2.847 &  0.023 & 0.914 &  0.914 & ( -4.334, -1.314)$^{*-}$ \\ 
  Tooth10 &  0.076 &  0.746 & 1.049 &  0.660 & -0.042 & 0.931 &  0.933 & ( -1.155,  2.218) \\ 
  ZoneM & -0.903 & -1.737 & 0.766 & -1.670 &  0.032 & 0.737 &  0.738 & ( -3.075, -0.050)$^{*-}$ \\ 
  ZoneI & -2.538 & -4.205 & 1.754 & -4.175 & -0.405 & 1.745 &  1.910 & ( -7.376, -0.129)$^{*-}$ \\ 
  ZoneE & -3.360 & -5.622 & 2.534 & -5.599 & -0.855 & 2.530 &  3.261 & (-10.686, -0.153)$^{*-}$ \\ 
   \hline
\end{tabular}
}
\end{subtable}
\begin{tablenotes}[para,flushleft]
\scriptsize
\item Superscripts $*+$ and $*-$ denote significant protective and risk effects at the 5\% significance level, respectively.
\end{tablenotes}
\end{threeparttable}
\end{table}



\addtocounter{table}{-1}

\begin{table}[H]
\centering
\caption{ Estimates from models B.1.1-B.1.4, the separate severity models with the independence cluster correlation structure }
\label{ch4:table:sep:sev:independence}
\begin{threeparttable}
\centering
\begin{subtable}{\linewidth}
\centering
\caption{Model B.1.1 (age 9)}
\scalebox{0.55}{
\begin{tabular}{rrrrrrrrl}
Variable & Estimate & \makecell{Standardized\\Estimate} & \makecell{SE\\(Standardized\\Estimate)} & \makecell{James-Stein\\Estimator} & \makecell{Bias\\(James-Stein\\Estimator)} & \makecell{SE\\(James-Stein\\Estimator)} & \makecell{MSE\\(James-Stein\\Estimator)} & \makecell{95\% CI\\(James-Stein\\Estimator)} \\
  \hline
dental\_age & -0.493 & -0.374 & 0.959 &  4.563 & -4.638 & 0.853 & 22.364 & (-2.233, 1.320) \\ 
  Total\_mgFPerDay & -0.599 & -0.642 & 1.157 &  1.086 & -1.103 & 1.141 &  2.356 & (-2.375, 2.470) \\ 
  SugarAddedBeverageOzPerDay & -0.004 & -0.109 & 0.898 &  0.205 & -0.556 & 0.720 &  1.029 & (-1.652, 0.991) \\ 
  BrushingFrequencyPerDay &  0.258 &  0.341 & 0.967 & -0.498 &  0.718 & 0.830 &  1.345 & (-1.576, 2.076) \\ 
  Avg\_homeppm & -0.105 & -0.158 & 0.852 &  4.937 & -5.134 & 0.789 & 27.149 & (-1.939, 0.948) \\ 
  Prop\_DentAppt & -1.152 & -0.730 & 0.950 &  0.414 & -0.927 & 0.868 &  1.728 & (-2.518, 0.828) \\ 
  Prop\_FluorideTreatment &  0.556 &  0.135 & 0.834 & -0.202 &  0.347 & 0.645 &  0.765 & (-1.465, 1.230) \\ 
  Tooth8 &  1.101 &  1.200 & 1.057 & -0.309 &  0.639 & 0.984 &  1.392 & (-0.845, 2.219) \\ 
  Tooth9 &  0.022 &  0.038 & 1.110 & -0.515 &  0.015 & 1.182 &  1.182 & (-3.443, 1.125) \\ 
  Tooth10 & -0.446 & -0.331 & 1.244 &  1.995 & -2.477 & 1.050 &  7.186 & (-2.847, 1.261) \\ 
  ZoneM & -0.710 & -0.775 & 1.417 &  1.669 & -1.900 & 1.339 &  4.947 & (-3.464, 2.246) \\ 
  ZoneI & -1.162 & -1.144 & 1.076 & -0.127 & -0.975 & 1.242 &  2.192 & (-3.459, 0.764) \\ 
  ZoneE & -1.277 & -1.437 & 1.424 & -0.412 & -1.668 & 1.695 &  4.477 & (-5.544, 0.709) \\ 
   \hline
\end{tabular}
}
\end{subtable}
\begin{subtable}{\linewidth}
\centering
\caption{Model B.1.2 (age 13)}
\scalebox{0.55}{
\begin{tabular}{rrrrrrrrl}
Variable & Estimate & \makecell{Standardized\\Estimate} & \makecell{SE\\(Standardized\\Estimate)} & \makecell{James-Stein\\Estimator} & \makecell{Bias\\(James-Stein\\Estimator)} & \makecell{SE\\(James-Stein\\Estimator)} & \makecell{MSE\\(James-Stein\\Estimator)} & \makecell{95\% CI\\(James-Stein\\Estimator)} \\
  \hline
dental\_age &  1.393 &  0.045 & 0.553 & -0.549 &  0.386 & 0.806 &  0.955 & (-2.223, 1.160) \\ 
  Total\_mgFPerDay & -0.039 & -0.013 & 0.699 &  0.023 & -0.030 & 0.470 &  0.471 & (-0.722, 0.808) \\ 
  SugarAddedBeverageOzPerDay & -0.013 & -0.055 & 0.828 &  0.104 & -0.394 & 0.726 &  0.881 & (-1.640, 0.813) \\ 
  BrushingFrequencyPerDay & -0.147 & -0.010 & 0.742 &  0.014 & -0.290 & 0.767 &  0.851 & (-1.924, 0.972) \\ 
  Avg\_homeppm & -0.410 & -0.131 & 1.054 &  4.081 & -4.256 & 1.138 & 19.248 & (-3.273, 1.590) \\ 
  Prop\_DentAppt & -2.567 & -0.214 & 0.876 &  0.122 & -0.550 & 0.721 &  1.024 & (-2.076, 0.738) \\ 
  Prop\_FluorideTreatment &  2.227 &  0.063 & 0.698 & -0.094 &  0.183 & 1.252 &  1.286 & (-1.567, 1.659) \\ 
  Tooth8 & -0.053 & -0.007 & 0.799 &  0.002 & -0.097 & 0.681 &  0.690 & (-1.956, 1.104) \\ 
  Tooth9 &  0.197 &  0.283 & 0.774 & -3.810 &  3.642 & 0.847 & 14.108 & (-3.040, 1.097) \\ 
  Tooth10 & -1.283 & -0.091 & 0.773 &  0.549 & -0.923 & 0.911 &  1.762 & (-2.521, 1.318) \\ 
  ZoneM & -0.543 & -0.147 & 0.727 &  0.317 & -0.432 & 1.009 &  1.196 & (-1.730, 1.438) \\ 
  ZoneI & -1.709 & -0.921 & 1.073 & -0.102 & -0.819 & 1.084 &  1.755 & (-3.285, 0.590) \\ 
  ZoneE &  0.197 &  0.023 & 0.699 &  0.006 &  0.088 & 0.558 &  0.566 & (-0.914, 1.631) \\ 
   \hline
\end{tabular}
}
\end{subtable}
\begin{subtable}{\linewidth}
\centering
\caption{Model B.1.3 (age 17)}
\scalebox{0.55}{
\begin{tabular}{rrrrrrrrl}
Variable & Estimate & \makecell{Standardized\\Estimate} & \makecell{SE\\(Standardized\\Estimate)} & \makecell{James-Stein\\Estimator} & \makecell{Bias\\(James-Stein\\Estimator)} & \makecell{SE\\(James-Stein\\Estimator)} & \makecell{MSE\\(James-Stein\\Estimator)} & \makecell{95\% CI\\(James-Stein\\Estimator)} \\
  \hline
dental\_age &  0.137 &  0.037 & 0.567 & -0.457 &  0.469 & 0.701 & 0.920 & (-0.956, 1.152) \\ 
  Total\_mgFPerDay & -1.153 & -0.552 & 0.830 &  0.935 & -1.101 & 0.718 & 1.930 & (-1.815, 1.179) \\ 
  SugarAddedBeverageOzPerDay & -0.053 & -0.720 & 1.039 &  1.352 & -2.239 & 1.079 & 6.091 & (-3.237, 0.820) \\ 
  BrushingFrequencyPerDay &  0.489 &  0.832 & 0.710 & -1.213 &  1.300 & 0.522 & 2.211 & (-1.162, 1.205) \\ 
  Avg\_homeppm &  0.070 &  0.028 & 0.731 & -0.869 &  0.836 & 0.669 & 1.368 & (-1.326, 1.985) \\ 
  Prop\_DentAppt & -1.711 & -0.349 & 0.825 &  0.198 & -0.174 & 0.633 & 0.664 & (-1.164, 1.317) \\ 
  Prop\_FluorideTreatment & -4.012 & -0.877 & 0.868 &  1.315 & -1.548 & 0.953 & 3.348 & (-2.444, 1.755) \\ 
  Tooth8 & -0.215 & -0.324 & 0.703 &  0.084 & -0.194 & 0.635 & 0.673 & (-1.821, 0.930) \\ 
  Tooth9 &  0.173 &  0.123 & 0.978 & -1.657 &  1.630 & 0.933 & 3.588 & (-1.845, 1.725) \\ 
  Tooth10 & -0.936 & -0.396 & 0.902 &  2.392 & -2.525 & 0.764 & 7.140 & (-1.804, 1.435) \\ 
  ZoneM & -0.111 & -0.065 & 3.557 &  0.141 & -0.560 & 2.496 & 2.810 & (-2.524, 0.872) \\ 
  ZoneI & -0.207 & -0.057 & 4.084 & -0.006 & -1.015 & 4.185 & 5.214 & (-5.049, 0.738) \\ 
  ZoneE & -1.671 & -0.850 & 3.747 & -0.243 & -0.874 & 3.092 & 3.855 & (-4.531, 0.258) \\ 
   \hline
\end{tabular}
}
\end{subtable}
\begin{subtable}{\linewidth}
\centering
\caption{Model B.1.4 (age 23)}
\scalebox{0.55}{
\begin{tabular}{rrrrrrrrl}
Variable & Estimate & \makecell{Standardized\\Estimate} & \makecell{SE\\(Standardized\\Estimate)} & \makecell{James-Stein\\Estimator} & \makecell{Bias\\(James-Stein\\Estimator)} & \makecell{SE\\(James-Stein\\Estimator)} & \makecell{MSE\\(James-Stein\\Estimator)} & \makecell{95\% CI\\(James-Stein\\Estimator)} \\
  \hline
dental\_age & -0.340 & -0.090 & 0.457 &  1.099 & -1.109 & 0.343 &  1.572 & (-0.800, 0.527) \\ 
  Total\_mgFPerDay & -0.244 & -0.160 & 0.544 &  0.271 & -0.276 & 0.562 &  0.638 & (-0.948, 0.942) \\ 
  SugarAddedBeverageOzPerDay & -0.090 & -0.402 & 1.560 &  0.754 & -1.785 & 1.587 &  4.775 & (-4.127, 0.709) \\ 
  BrushingFrequencyPerDay &  0.149 &  0.074 & 0.632 & -0.108 &  0.055 & 0.486 &  0.489 & (-1.128, 0.666) \\ 
  Avg\_homeppm &  0.222 &  0.138 & 0.638 & -4.313 &  4.264 & 0.591 & 18.768 & (-1.387, 0.904) \\ 
  Prop\_DentAppt &  3.210 &  0.759 & 0.767 & -0.430 &  0.925 & 0.720 &  1.576 & (-0.379, 2.169) \\ 
  Prop\_FluorideTreatment & -0.625 & -0.091 & 0.550 &  0.137 & -0.081 & 0.962 &  0.968 & (-1.201, 0.998) \\ 
  Tooth8 & -1.177 & -0.213 & 0.461 &  0.055 & -0.075 & 0.462 &  0.467 & (-0.891, 1.018) \\ 
  Tooth9 & -1.321 & -0.204 & 0.772 &  2.738 & -2.829 & 0.791 &  8.797 & (-1.620, 1.866) \\ 
  Tooth10 &  0.220 &  0.098 & 0.623 & -0.591 &  0.665 & 0.472 &  0.915 & (-0.676, 1.314) \\ 
  ZoneM & -0.767 & -0.085 & 1.139 &  0.183 & -0.403 & 1.048 &  1.210 & (-2.778, 0.631) \\ 
  ZoneI &  1.627 &  0.298 & 1.063 &  0.033 &  0.002 & 0.924 &  0.924 & (-2.889, 1.245) \\ 
  ZoneE & -1.216 & -0.128 & 1.201 & -0.037 & -0.378 & 1.089 &  1.232 & (-3.630, 0.441) \\ 
   \hline
\end{tabular}
}
\end{subtable}
\begin{tablenotes}[para,flushleft]
\scriptsize
\item Superscripts $*+$ and $*-$ denote significant protective and risk effects at the 5\% significance level, respectively.
\end{tablenotes}
\end{threeparttable}
\end{table}

\addtocounter{table}{-1} 

\begin{table}[H]
\centering
\caption{ Estimates from models B.2.1-B.2.4, the separate severity models with the exchangeable cluster correlation structure }
\label{ch4:table:sep:sev:exchangeable}
\begin{threeparttable}
\centering
\begin{subtable}{\linewidth}
\centering
\caption{Model B.2.1 (age 9)}
\scalebox{0.55}{
\begin{tabular}{rrrrrrrrl}
Variable & Estimate & \makecell{Standardized\\Estimate} & \makecell{SE\\(Standardized\\Estimate)} & \makecell{James-Stein\\Estimator} & \makecell{Bias\\(James-Stein\\Estimator)} & \makecell{SE\\(James-Stein\\Estimator)} & \makecell{MSE\\(James-Stein\\Estimator)} & \makecell{95\% CI\\(James-Stein\\Estimator)} \\
  \hline
dental\_age & -0.492 & -0.374 & 0.973 &  3.370 & -3.441 & 0.883 & 12.723 & (-2.253, 1.416) \\ 
  Total\_mgFPerDay & -0.598 & -0.641 & 1.160 &  1.119 & -1.145 & 1.343 &  2.653 & (-2.477, 2.471) \\ 
  SugarAddedBeverageOzPerDay & -0.004 & -0.110 & 0.898 & -0.057 & -0.268 & 0.725 &  0.797 & (-1.600, 1.147) \\ 
  BrushingFrequencyPerDay &  0.258 &  0.341 & 0.965 & -0.607 &  0.833 & 0.804 &  1.497 & (-1.605, 2.081) \\ 
  Avg\_homeppm & -0.105 & -0.159 & 0.838 &  1.634 & -1.844 & 0.783 &  4.183 & (-1.919, 1.106) \\ 
  Prop\_DentAppt & -1.153 & -0.731 & 0.949 &  0.247 & -0.755 & 0.869 &  1.440 & (-2.500, 0.777) \\ 
  Prop\_FluorideTreatment &  0.563 &  0.137 & 0.833 & -0.188 &  0.309 & 0.678 &  0.774 & (-1.677, 1.303) \\ 
  Tooth8 &  1.101 &  1.200 & 1.057 & -0.302 &  0.629 & 0.986 &  1.382 & (-0.958, 2.221) \\ 
  Tooth9 &  0.022 &  0.039 & 1.110 & -0.189 & -0.270 & 1.134 &  1.207 & (-3.316, 1.133) \\ 
  Tooth10 & -0.447 & -0.331 & 1.243 &  0.418 & -0.797 & 1.158 &  1.794 & (-2.790, 1.678) \\ 
  ZoneM & -0.709 & -0.775 & 1.415 &  1.239 & -1.491 & 1.320 &  3.542 & (-3.477, 2.289) \\ 
  ZoneI & -1.162 & -1.145 & 1.081 & -0.058 & -1.048 & 1.224 &  2.322 & (-3.647, 0.352) \\ 
  ZoneE & -1.277 & -1.436 & 1.424 & -0.411 & -1.674 & 1.688 &  4.490 & (-5.549, 0.564) \\ 
   \hline
\end{tabular}
}
\end{subtable}
\begin{subtable}{\linewidth}
\centering
\caption{Model B.2.2 (age 13)}
\scalebox{0.55}{
\begin{tabular}{rrrrrrrrl}
Variable & Estimate & \makecell{Standardized\\Estimate} & \makecell{SE\\(Standardized\\Estimate)} & \makecell{James-Stein\\Estimator} & \makecell{Bias\\(James-Stein\\Estimator)} & \makecell{SE\\(James-Stein\\Estimator)} & \makecell{MSE\\(James-Stein\\Estimator)} & \makecell{95\% CI\\(James-Stein\\Estimator)} \\
  \hline
dental\_age &  1.401 &  0.215 & 0.559 & -1.940 &  1.810 & 0.783 & 4.061 & (-1.760, 1.238) \\ 
  Total\_mgFPerDay & -0.030 & -0.025 & 0.674 &  0.044 & -0.052 & 0.384 & 0.387 & (-0.705, 0.800) \\ 
  SugarAddedBeverageOzPerDay & -0.013 & -0.049 & 0.881 & -0.025 & -0.318 & 0.765 & 0.866 & (-2.324, 0.708) \\ 
  BrushingFrequencyPerDay & -0.177 & -0.104 & 0.749 &  0.185 & -0.458 & 0.752 & 0.962 & (-1.959, 0.968) \\ 
  Avg\_homeppm & -0.400 & -0.260 & 1.101 &  2.664 & -2.883 & 1.152 & 9.462 & (-3.330, 1.623) \\ 
  Prop\_DentAppt & -2.544 & -0.247 & 0.939 &  0.083 & -0.487 & 0.772 & 1.009 & (-2.031, 1.143) \\ 
  Prop\_FluorideTreatment &  2.240 &  0.087 & 0.673 & -0.120 &  0.195 & 1.229 & 1.267 & (-1.652, 1.666) \\ 
  Tooth8 & -0.065 & -0.043 & 0.806 &  0.011 & -0.079 & 0.698 & 0.704 & (-1.621, 1.155) \\ 
  Tooth9 &  0.198 &  0.094 & 0.772 & -0.453 &  0.281 & 0.881 & 0.960 & (-3.128, 1.296) \\ 
  Tooth10 & -1.253 & -0.779 & 0.765 &  0.984 & -1.336 & 0.911 & 2.697 & (-2.389, 1.658) \\ 
  ZoneM & -0.546 & -0.291 & 0.735 &  0.466 & -0.613 & 0.905 & 1.282 & (-1.890, 1.391) \\ 
  ZoneI & -1.721 & -0.580 & 1.034 & -0.029 & -0.876 & 1.059 & 1.826 & (-3.317, 0.506) \\ 
  ZoneE &  0.226 &  0.080 & 0.700 &  0.023 &  0.065 & 0.565 & 0.569 & (-0.922, 1.602) \\ 
   \hline
\end{tabular}
}
\end{subtable}
\begin{subtable}{\linewidth}
\centering
\caption{Model B.2.3 (age 17)}
\scalebox{0.55}{
\begin{tabular}{rrrrrrrrl}
Variable & Estimate & \makecell{Standardized\\Estimate} & \makecell{SE\\(Standardized\\Estimate)} & \makecell{James-Stein\\Estimator} & \makecell{Bias\\(James-Stein\\Estimator)} & \makecell{SE\\(James-Stein\\Estimator)} & \makecell{MSE\\(James-Stein\\Estimator)} & \makecell{95\% CI\\(James-Stein\\Estimator)} \\
  \hline
dental\_age &  0.136 &  0.036 & 0.570 & -0.328 &  0.350 & 0.698 & 0.821 & (-1.092, 1.062) \\ 
  Total\_mgFPerDay & -1.154 & -0.554 & 0.838 &  0.968 & -1.155 & 0.742 & 2.075 & (-1.848, 1.201) \\ 
  SugarAddedBeverageOzPerDay & -0.053 & -0.682 & 1.049 & -0.357 & -0.549 & 1.086 & 1.388 & (-3.238, 0.886) \\ 
  BrushingFrequencyPerDay &  0.489 &  0.762 & 0.722 & -1.358 &  1.441 & 0.527 & 2.602 & (-1.371, 1.215) \\ 
  Avg\_homeppm &  0.072 &  0.029 & 0.740 & -0.293 &  0.265 & 0.693 & 0.763 & (-1.362, 2.059) \\ 
  Prop\_DentAppt & -1.710 & -0.344 & 0.829 &  0.116 & -0.100 & 0.641 & 0.651 & (-1.059, 1.337) \\ 
  Prop\_FluorideTreatment & -4.011 & -0.861 & 0.862 &  1.187 & -1.434 & 0.900 & 2.957 & (-2.446, 1.514) \\ 
  Tooth8 & -0.215 & -0.285 & 0.706 &  0.072 & -0.178 & 0.631 & 0.662 & (-1.758, 0.845) \\ 
  Tooth9 &  0.173 &  0.119 & 0.996 & -0.572 &  0.580 & 0.846 & 1.182 & (-1.766, 1.709) \\ 
  Tooth10 & -0.937 & -0.390 & 0.893 &  0.492 & -0.590 & 0.736 & 1.084 & (-1.641, 1.424) \\ 
  ZoneM & -0.112 & -0.064 & 3.566 &  0.102 & -0.551 & 2.542 & 2.845 & (-2.480, 0.597) \\ 
  ZoneI & -0.208 & -0.057 & 4.079 & -0.003 & -1.022 & 4.173 & 5.218 & (-5.163, 0.746) \\ 
  ZoneE & -1.670 & -0.856 & 3.510 & -0.245 & -0.927 & 3.187 & 4.047 & (-4.614, 0.186) \\ 
   \hline
\end{tabular}
}
\end{subtable}
\begin{subtable}{\linewidth}
\centering
\caption{Model B.2.4 (age 23)}
\scalebox{0.55}{
\begin{tabular}{rrrrrrrrl}
Variable & Estimate & \makecell{Standardized\\Estimate} & \makecell{SE\\(Standardized\\Estimate)} & \makecell{James-Stein\\Estimator} & \makecell{Bias\\(James-Stein\\Estimator)} & \makecell{SE\\(James-Stein\\Estimator)} & \makecell{MSE\\(James-Stein\\Estimator)} & \makecell{95\% CI\\(James-Stein\\Estimator)} \\
  \hline
dental\_age & -0.416 & -0.110 & 0.448 &  0.995 & -1.020 & 0.447 & 1.486 & (-0.749, 0.533) \\ 
  Total\_mgFPerDay & -0.046 & -0.097 & 0.524 &  0.169 & -0.210 & 0.417 & 0.461 & (-0.852, 1.011) \\ 
  SugarAddedBeverageOzPerDay & -0.061 & -1.928 & 1.451 & -1.010 &  0.010 & 1.454 & 1.454 & (-4.296, 0.674) \\ 
  BrushingFrequencyPerDay & -0.107 & -0.109 & 0.649 &  0.194 & -0.244 & 0.489 & 0.549 & (-1.184, 0.614) \\ 
  Avg\_homeppm &  0.346 &  0.290 & 0.634 & -2.976 &  2.931 & 0.539 & 9.127 & (-1.244, 0.928) \\ 
  Prop\_DentAppt &  2.862 &  0.884 & 0.749 & -0.299 &  0.777 & 0.678 & 1.281 & (-0.440, 1.965) \\ 
  Prop\_FluorideTreatment & -1.428 & -0.271 & 0.548 &  0.374 & -0.324 & 0.894 & 0.999 & (-1.427, 0.854) \\ 
  Tooth8 & -0.410 & -0.273 & 0.474 &  0.069 & -0.087 & 0.477 & 0.484 & (-0.886, 1.319) \\ 
  Tooth9 & -0.462 & -0.565 & 0.755 &  2.725 & -2.855 & 0.723 & 8.872 & (-1.742, 1.588) \\ 
  Tooth10 &  0.278 &  0.125 & 0.645 & -0.158 &  0.265 & 0.516 & 0.586 & (-0.710, 1.721) \\ 
  ZoneM &  0.461 &  0.283 & 1.287 & -0.453 &  0.221 & 1.209 & 1.258 & (-2.795, 0.825) \\ 
  ZoneI &  0.956 &  0.675 & 1.176 &  0.034 & -0.015 & 1.034 & 1.034 & (-3.030, 1.270) \\ 
  ZoneE &  0.067 &  0.038 & 1.251 &  0.011 & -0.410 & 1.138 & 1.306 & (-3.641, 0.441) \\ 
   \hline
\end{tabular}
}
\end{subtable}
\begin{tablenotes}[para,flushleft]
\scriptsize
\item Superscripts $*+$ and $*-$ denote significant protective and risk effects at the 5\% significance level, respectively.
\end{tablenotes}
\end{threeparttable}
\end{table}

\addtocounter{table}{-1} 

\begin{table}[H]
\centering
\caption{ Estimates from models B.3.1-B.3.4, the separate severity models with the ar1 cluster correlation structure }
\label{ch4:table:sep:sev:ar1}
\begin{threeparttable}
\centering
\begin{subtable}{\linewidth}
\centering
\caption{Model B.3.1 (age 9)}
\scalebox{0.55}{
\begin{tabular}{rrrrrrrrl}
Variable & Estimate & \makecell{Standardized\\Estimate} & \makecell{SE\\(Standardized\\Estimate)} & \makecell{James-Stein\\Estimator} & \makecell{Bias\\(James-Stein\\Estimator)} & \makecell{SE\\(James-Stein\\Estimator)} & \makecell{MSE\\(James-Stein\\Estimator)} & \makecell{95\% CI\\(James-Stein\\Estimator)} \\
  \hline
dental\_age & -0.182 & -0.138 & 0.750 &  0.755 & -0.911 & 1.843 &  2.674 & (-2.905, 2.969) \\ 
  Total\_mgFPerDay & -0.668 & -0.695 & 1.133 &  1.888 & -1.737 & 1.253 &  4.272 & (-2.465, 2.656) \\ 
  SugarAddedBeverageOzPerDay & -0.010 & -0.314 & 0.861 &  1.046 & -1.293 & 0.771 &  2.442 & (-1.720, 1.198) \\ 
  BrushingFrequencyPerDay &  0.373 &  0.608 & 0.818 & -1.289 &  1.420 & 1.197 &  3.212 & (-1.874, 2.561) \\ 
  Avg\_homeppm & -0.077 & -0.130 & 0.837 &  2.781 & -2.937 & 1.110 &  9.739 & (-3.187, 1.672) \\ 
  Prop\_DentAppt & -1.114 & -0.687 & 0.991 &  0.958 & -1.184 & 1.050 &  2.452 & (-2.340, 1.984) \\ 
  Prop\_FluorideTreatment & -1.704 & -0.397 & 0.770 &  3.481 & -3.080 & 1.562 & 11.048 & (-1.974, 3.675) \\ 
  Tooth8 &  0.982 &  1.236 & 0.830 & -0.335 &  0.251 & 1.280 &  1.343 & (-2.352, 2.424) \\ 
  Tooth9 &  0.077 &  0.144 & 1.013 & -1.504 &  1.343 & 1.303 &  3.107 & (-3.082, 3.152) \\ 
  Tooth10 & -0.747 & -0.838 & 1.098 &  1.452 & -1.790 & 2.099 &  5.304 & (-4.178, 3.942) \\ 
  ZoneM & -0.644 & -0.704 & 1.364 &  2.102 & -1.994 & 1.727 &  5.704 & (-3.413, 3.937) \\ 
  ZoneI & -1.300 & -1.168 & 0.999 &  0.298 & -0.930 & 1.289 &  2.153 & (-3.140, 2.074) \\ 
  ZoneE & -1.128 & -1.025 & 1.558 &  0.187 & -2.006 & 1.980 &  6.002 & (-5.678, 1.532) \\ 
   \hline
\end{tabular}
}
\end{subtable}
\begin{subtable}{\linewidth}
\centering
\caption{Model B.3.2 (age 13)}
\scalebox{0.55}{
\begin{tabular}{rrrrrrrrl}
Variable & Estimate & \makecell{Standardized\\Estimate} & \makecell{SE\\(Standardized\\Estimate)} & \makecell{James-Stein\\Estimator} & \makecell{Bias\\(James-Stein\\Estimator)} & \makecell{SE\\(James-Stein\\Estimator)} & \makecell{MSE\\(James-Stein\\Estimator)} & \makecell{95\% CI\\(James-Stein\\Estimator)} \\
  \hline
dental\_age &  1.470 &  0.532 & 0.251 & -2.913 &  2.048 & 2.707 &  6.900 & (-4.368, 0.165) \\ 
  Total\_mgFPerDay &  0.125 &  0.117 & 0.362 & -0.319 &  0.154 & 0.553 &  0.576 & (-1.584, 0.540) \\ 
  SugarAddedBeverageOzPerDay & -0.006 & -0.109 & 0.341 &  0.363 & -0.373 & 0.280 &  0.419 & (-0.465, 0.687) \\ 
  BrushingFrequencyPerDay & -0.108 & -0.100 & 0.281 &  0.212 & -0.018 & 0.637 &  0.638 & (-0.568, 2.018) \\ 
  Avg\_homeppm & -0.450 & -0.219 & 0.490 &  4.685 & -4.468 & 0.693 & 20.659 & (-0.929, 1.948) \\ 
  Prop\_DentAppt & -1.597 & -0.279 & 0.459 &  0.389 & -0.430 & 0.420 &  0.604 & (-0.905, 0.449) \\ 
  Prop\_FluorideTreatment &  2.045 &  0.136 & 0.361 & -1.195 &  1.205 & 2.160 &  3.612 & (-1.538, 0.529) \\ 
  Tooth8 & -0.268 & -0.136 & 0.318 &  0.037 &  0.146 & 0.482 &  0.504 & (-0.329, 1.341) \\ 
  Tooth9 & -0.527 & -0.356 & 0.660 &  3.719 & -3.444 & 1.038 & 12.898 & (-1.093, 2.287) \\ 
  Tooth10 & -0.039 & -0.023 & 0.269 &  0.040 & -0.033 & 1.083 &  1.085 & (-1.407, 2.096) \\ 
  ZoneM & -0.139 & -0.040 & 0.331 &  0.118 &  0.013 & 0.909 &  0.909 & (-0.852, 2.620) \\ 
  ZoneI & -0.504 & -0.472 & 0.592 &  0.120 & -0.083 & 0.642 &  0.648 & (-0.923, 2.024) \\ 
  ZoneE & -0.451 & -0.196 & 0.430 &  0.036 & -0.110 & 0.402 &  0.414 & (-0.844, 0.966) \\ 
   \hline
\end{tabular}
}
\end{subtable}
\begin{subtable}{\linewidth}
\centering
\caption{Model B.3.3 (age 17)}
\scalebox{0.55}{
\begin{tabular}{rrrrrrrrl}
Variable & Estimate & \makecell{Standardized\\Estimate} & \makecell{SE\\(Standardized\\Estimate)} & \makecell{James-Stein\\Estimator} & \makecell{Bias\\(James-Stein\\Estimator)} & \makecell{SE\\(James-Stein\\Estimator)} & \makecell{MSE\\(James-Stein\\Estimator)} & \makecell{95\% CI\\(James-Stein\\Estimator)} \\
  \hline
dental\_age &  0.586 &  0.043 & 0.530 & -0.238 & -0.148 & 1.824 &  1.846 & ( -2.224, 1.195) \\ 
  Total\_mgFPerDay & -0.566 & -0.161 & 0.752 &  0.437 & -0.362 & 0.819 &  0.950 & ( -1.271, 2.028) \\ 
  SugarAddedBeverageOzPerDay & -0.046 & -0.468 & 0.818 &  1.559 & -1.791 & 1.222 &  4.429 & ( -2.434, 2.460) \\ 
  BrushingFrequencyPerDay &  0.363 &  0.501 & 0.607 & -1.062 &  0.841 & 0.914 &  1.622 & ( -2.575, 1.390) \\ 
  Avg\_homeppm & -0.592 & -0.156 & 0.775 &  3.339 & -3.186 & 1.139 & 11.287 & ( -1.432, 3.683) \\ 
  Prop\_DentAppt & -1.980 & -0.338 & 0.792 &  0.471 & -0.405 & 0.741 &  0.905 & ( -1.481, 1.615) \\ 
  Prop\_FluorideTreatment & -1.805 & -0.141 & 0.751 &  1.238 & -1.117 & 1.616 &  2.864 & ( -2.491, 3.299) \\ 
  Tooth8 & -0.171 & -0.105 & 0.702 &  0.028 & -0.095 & 1.492 &  1.501 & ( -2.983, 1.966) \\ 
  Tooth9 & -0.057 & -0.058 & 0.799 &  0.611 & -0.480 & 0.814 &  1.044 & ( -1.606, 1.788) \\ 
  Tooth10 & -0.433 & -0.170 & 0.789 &  0.294 & -0.534 & 1.832 &  2.117 & ( -5.031, 3.512) \\ 
  ZoneM &  0.065 &  0.040 & 2.456 & -0.120 & -0.170 & 2.268 &  2.297 & ( -4.265, 1.588) \\ 
  ZoneI &  0.218 &  0.061 & 4.232 & -0.015 & -0.861 & 4.545 &  5.286 & ( -9.753, 1.603) \\ 
  ZoneE & -1.975 & -0.776 & 3.628 &  0.142 & -1.361 & 3.728 &  5.579 & (-15.715, 0.799) \\ 
   \hline
\end{tabular}
}
\end{subtable}
\begin{subtable}{\linewidth}
\centering
\caption{Model B.3.4 (age 23)}
\scalebox{0.55}{
\begin{tabular}{rrrrrrrrl}
Variable & Estimate & \makecell{Standardized\\Estimate} & \makecell{SE\\(Standardized\\Estimate)} & \makecell{James-Stein\\Estimator} & \makecell{Bias\\(James-Stein\\Estimator)} & \makecell{SE\\(James-Stein\\Estimator)} & \makecell{MSE\\(James-Stein\\Estimator)} & \makecell{95\% CI\\(James-Stein\\Estimator)} \\
  \hline
dental\_age & -0.980 & -0.071 & 0.483 &  0.388 & -0.492 & 1.105 & 1.347 & (-3.498, 2.204) \\ 
  Total\_mgFPerDay & -0.204 & -0.123 & 0.398 &  0.333 & -0.288 & 0.617 & 0.700 & (-1.441, 1.344) \\ 
  SugarAddedBeverageOzPerDay & -0.083 & -0.364 & 0.774 &  1.212 & -1.416 & 0.965 & 2.971 & (-2.429, 2.140) \\ 
  BrushingFrequencyPerDay & -0.308 & -0.098 & 0.512 &  0.208 & -0.457 & 1.021 & 1.229 & (-2.737, 1.193) \\ 
  Avg\_homeppm &  0.031 &  0.007 & 0.570 & -0.147 & -0.028 & 1.004 & 1.004 & (-2.689, 1.456) \\ 
  Prop\_DentAppt &  2.626 &  0.414 & 0.485 & -0.578 &  0.610 & 0.716 & 1.087 & (-1.538, 1.543) \\ 
  Prop\_FluorideTreatment &  4.919 &  0.093 & 0.439 & -0.818 &  0.753 & 1.121 & 1.688 & (-2.896, 1.579) \\ 
  Tooth8 & -0.621 & -0.130 & 0.409 &  0.035 &  0.284 & 1.155 & 1.236 & (-1.242, 4.078) \\ 
  Tooth9 & -0.495 & -0.155 & 0.406 &  1.622 & -1.513 & 0.865 & 3.153 & (-0.974, 1.493) \\ 
  Tooth10 &  0.037 &  0.005 & 0.406 & -0.009 &  0.003 & 1.711 & 1.711 & (-3.363, 4.095) \\ 
  ZoneM &  0.169 &  0.059 & 1.065 & -0.175 & -0.010 & 1.314 & 1.314 & (-3.778, 2.448) \\ 
  ZoneI &  0.182 &  0.042 & 0.918 & -0.011 & -0.123 & 0.916 & 0.931 & (-3.414, 0.981) \\ 
  ZoneE &  0.073 &  0.026 & 1.293 & -0.005 & -0.386 & 1.220 & 1.369 & (-3.584, 0.460) \\ 
   \hline
\end{tabular}
}
\end{subtable}
\begin{tablenotes}[para,flushleft]
\scriptsize
\item Superscripts $*+$ and $*-$ denote significant protective and risk effects at the 5\% significance level, respectively.
\end{tablenotes}
\end{threeparttable}
\end{table}



\addtocounter{table}{-1} 

\begin{table}[H]
\centering
\caption{ Presence estimates from models C.1.1.1-C.1.1.4, the combined models with the independence and independence presence and severity cluster correlation structures respectively. }
\label{ch4:table:comb:pres:independence}
\begin{threeparttable}
\centering
\begin{subtable}{\linewidth}
\centering
\caption{Model C.1.1.1 (age 9)}
\scalebox{0.55}{
\begin{tabular}{rrrrrrrrl}
Variable & Estimate & \makecell{Standardized\\Estimate} & \makecell{SE\\(Standardized\\Estimate)} & \makecell{James-Stein\\Estimator} & \makecell{Bias\\(James-Stein\\Estimator)} & \makecell{SE\\(James-Stein\\Estimator)} & \makecell{MSE\\(James-Stein\\Estimator)} & \makecell{95\% CI\\(James-Stein\\Estimator)} \\
  \hline
dental\_age & -0.303 & -1.031 & 0.849 & -0.641 &  0.137 & 0.702 &  0.721 & ( -2.036,  0.617) \\ 
  Total\_mgFPerDay & -0.095 & -0.361 & 0.837 & -0.108 & -0.057 & 0.769 &  0.772 & ( -2.153,  1.204) \\ 
  SugarAddedBeverageOzPerDay & -0.006 & -0.416 & 0.906 & -0.044 & -0.161 & 0.669 &  0.695 & ( -1.893,  1.014) \\ 
  BrushingFrequencyPerDay & -0.072 & -0.371 & 0.737 &  1.600 & -1.585 & 0.614 &  3.128 & ( -1.213,  1.426) \\ 
  Avg\_homeppm & -0.657 & -3.121 & 0.815 & -2.640 &  0.173 & 0.824 &  0.854 & ( -3.683, -0.671)$^{*-}$ \\ 
  Prop\_DentAppt & -0.192 & -0.189 & 0.770 &  2.145 & -1.985 & 0.898 &  4.837 & ( -1.063,  3.058) \\ 
  Prop\_FluorideTreatment &  0.388 &  0.271 & 0.862 & -2.016 &  2.114 & 0.802 &  5.272 & ( -1.298,  1.566) \\ 
  Tooth8 & -0.229 & -0.481 & 1.474 &  0.057 & -1.825 & 1.383 &  4.714 & ( -4.218,  0.034) \\ 
  Tooth9 & -0.139 & -0.166 & 1.637 &  0.037 & -1.464 & 1.578 &  3.722 & ( -4.745,  0.328) \\ 
  Tooth10 &  0.370 &  1.265 & 0.996 & -0.164 &  1.247 & 1.268 &  2.824 & ( -1.164,  3.100) \\ 
  ZoneM & -0.845 & -1.989 & 1.681 & -1.131 & -1.463 & 1.828 &  3.969 & ( -5.684, -0.081)$^{*-}$ \\ 
  ZoneI & -1.738 & -3.207 & 2.902 & -2.905 & -3.650 & 2.986 & 16.305 & (-11.057, -1.154)$^{*-}$ \\ 
  ZoneE & -2.165 & -4.733 & 3.435 & -4.492 & -4.016 & 3.492 & 19.621 & (-13.464, -2.026)$^{*-}$ \\ 
   \hline
\end{tabular}
}
\end{subtable}
\begin{subtable}{\linewidth}
\centering
\caption{Model C.1.1.2 (age 13)}
\scalebox{0.55}{
\begin{tabular}{rrrrrrrrl}
Variable & Estimate & \makecell{Standardized\\Estimate} & \makecell{SE\\(Standardized\\Estimate)} & \makecell{James-Stein\\Estimator} & \makecell{Bias\\(James-Stein\\Estimator)} & \makecell{SE\\(James-Stein\\Estimator)} & \makecell{MSE\\(James-Stein\\Estimator)} & \makecell{95\% CI\\(James-Stein\\Estimator)} \\
  \hline
dental\_age &  0.564 &  1.030 & 0.838 &  0.640 & -0.042 & 0.721 &  0.723 & ( -0.685,  1.862) \\ 
  Total\_mgFPerDay &  0.012 &  0.009 & 0.623 &  0.003 & -0.163 & 0.964 &  0.990 & ( -3.194,  1.420) \\ 
  SugarAddedBeverageOzPerDay &  0.006 &  0.283 & 0.702 &  0.030 &  0.108 & 0.572 &  0.584 & ( -0.522,  1.259) \\ 
  BrushingFrequencyPerDay & -0.009 & -0.013 & 0.739 &  0.054 & -0.059 & 0.434 &  0.437 & ( -0.862,  0.887) \\ 
  Avg\_homeppm & -0.650 & -1.231 & 1.124 & -1.041 & -0.332 & 1.063 &  1.174 & ( -3.890,  0.012) \\ 
  Prop\_DentAppt & -0.096 & -0.026 & 0.807 &  0.301 & -0.237 & 0.693 &  0.749 & ( -1.533,  1.567) \\ 
  Prop\_FluorideTreatment &  0.324 &  0.060 & 0.712 & -0.447 &  0.393 & 0.647 &  0.801 & ( -1.474,  1.263) \\ 
  Tooth8 & -0.180 & -0.202 & 0.797 &  0.024 & -0.781 & 0.746 &  1.356 & ( -2.571,  0.196) \\ 
  Tooth9 & -0.101 & -0.253 & 0.594 &  0.056 & -0.367 & 0.498 &  0.632 & ( -1.244,  0.349) \\ 
  Tooth10 &  0.136 &  0.171 & 0.744 & -0.022 &  0.418 & 0.719 &  0.893 & ( -0.783,  1.830) \\ 
  ZoneM & -0.403 & -0.687 & 1.144 & -0.391 & -0.913 & 1.179 &  2.013 & ( -3.738, -0.006)$^{*-}$ \\ 
  ZoneI & -1.386 & -2.691 & 2.734 & -2.438 & -1.976 & 2.853 &  6.759 & ( -8.835, -0.618)$^{*-}$ \\ 
  ZoneE & -2.130 & -2.190 & 4.245 & -2.078 & -4.939 & 4.417 & 28.815 & (-13.571, -1.055)$^{*-}$ \\ 
   \hline
\end{tabular}
}
\end{subtable}
\begin{subtable}{\linewidth}
\centering
\caption{Model C.1.1.3 (age 17)}
\scalebox{0.55}{
\begin{tabular}{rrrrrrrrl}
Variable & Estimate & \makecell{Standardized\\Estimate} & \makecell{SE\\(Standardized\\Estimate)} & \makecell{James-Stein\\Estimator} & \makecell{Bias\\(James-Stein\\Estimator)} & \makecell{SE\\(James-Stein\\Estimator)} & \makecell{MSE\\(James-Stein\\Estimator)} & \makecell{95\% CI\\(James-Stein\\Estimator)} \\
  \hline
dental\_age &  0.767 &  0.950 & 0.730 &  0.591 &  0.026 & 0.635 &  0.636 & ( -0.439,  1.992) \\ 
  Total\_mgFPerDay & -0.151 & -0.599 & 0.735 & -0.179 &  0.018 & 0.813 &  0.814 & ( -1.795,  1.773) \\ 
  SugarAddedBeverageOzPerDay &  0.015 &  0.702 & 0.854 &  0.074 &  0.545 & 0.895 &  1.193 & ( -0.898,  2.374) \\ 
  BrushingFrequencyPerDay &  0.191 &  0.379 & 0.775 & -1.633 &  1.661 & 0.620 &  3.378 & ( -1.351,  1.151) \\ 
  Avg\_homeppm & -0.764 & -1.078 & 0.925 & -0.912 & -0.493 & 0.890 &  1.134 & ( -3.323, -0.090)$^{*-}$ \\ 
  Prop\_DentAppt & -3.431 & -0.267 & 0.907 &  3.028 & -3.141 & 0.903 & 10.770 & ( -2.308,  1.747) \\ 
  Prop\_FluorideTreatment &  1.999 &  0.164 & 0.999 & -1.223 &  1.090 & 0.825 &  2.012 & ( -2.211,  1.576) \\ 
  Tooth8 & -0.099 & -0.063 & 1.145 &  0.007 & -1.241 & 1.097 &  2.636 & ( -3.977,  0.086) \\ 
  Tooth9 &  0.046 &  0.041 & 0.860 & -0.009 & -0.531 & 0.785 &  1.067 & ( -2.967,  0.150) \\ 
  Tooth10 & -0.212 & -0.226 & 0.725 &  0.029 &  0.168 & 0.541 &  0.569 & ( -0.710,  1.617) \\ 
  ZoneM & -0.350 & -0.368 & 0.969 & -0.209 & -0.583 & 0.944 &  1.284 & ( -3.092,  0.518) \\ 
  ZoneI & -1.745 & -1.571 & 2.302 & -1.423 & -2.013 & 2.323 &  6.374 & ( -7.623, -0.410)$^{*-}$ \\ 
  ZoneE & -2.378 & -3.107 & 2.933 & -2.949 & -2.409 & 2.978 &  8.783 & (-10.183, -1.222)$^{*-}$ \\ 
   \hline
\end{tabular}
}
\end{subtable}
\begin{subtable}{\linewidth}
\centering
\caption{Model C.1.1.4 (age 23)}
\scalebox{0.55}{
\begin{tabular}{rrrrrrrrl}
Variable & Estimate & \makecell{Standardized\\Estimate} & \makecell{SE\\(Standardized\\Estimate)} & \makecell{James-Stein\\Estimator} & \makecell{Bias\\(James-Stein\\Estimator)} & \makecell{SE\\(James-Stein\\Estimator)} & \makecell{MSE\\(James-Stein\\Estimator)} & \makecell{95\% CI\\(James-Stein\\Estimator)} \\
  \hline
dental\_age &  0.713 &  1.503 & 0.795 &  0.934 &  0.013 & 0.742 &  0.742 & ( -0.157,  2.793) \\ 
  Total\_mgFPerDay & -0.327 & -1.536 & 1.232 & -0.458 & -0.021 & 1.381 &  1.381 & ( -2.965,  2.392) \\ 
  SugarAddedBeverageOzPerDay & -0.013 & -1.221 & 0.946 & -0.129 & -0.461 & 0.968 &  1.180 & ( -2.134,  0.999) \\ 
  BrushingFrequencyPerDay & -0.140 & -0.308 & 2.201 &  1.330 & -1.600 & 2.228 &  4.788 & ( -1.648,  1.441) \\ 
  Avg\_homeppm & -0.860 & -0.747 & 0.884 & -0.632 & -0.609 & 0.776 &  1.147 & ( -2.835, -0.181)$^{*-}$ \\ 
  Prop\_DentAppt & -0.290 & -0.233 & 1.158 &  2.648 & -2.641 & 1.028 &  8.003 & ( -2.084,  2.015) \\ 
  Prop\_FluorideTreatment &  0.576 &  0.365 & 1.157 & -2.715 &  2.827 & 1.170 &  9.161 & ( -1.514,  1.447) \\ 
  Tooth8 & -0.475 & -1.230 & 1.791 &  0.146 & -1.579 & 1.798 &  4.292 & ( -4.451,  0.036) \\ 
  Tooth9 & -0.433 & -1.243 & 1.360 &  0.275 & -1.446 & 1.434 &  3.526 & ( -4.158,  0.520) \\ 
  Tooth10 &  0.074 &  0.301 & 0.619 & -0.039 &  0.212 & 0.506 &  0.551 & ( -0.800,  1.190) \\ 
  ZoneM & -0.938 & -0.265 & 2.229 & -0.151 & -0.527 & 2.048 &  2.325 & ( -4.189,  0.237) \\ 
  ZoneI & -2.567 & -1.100 & 5.277 & -0.996 & -1.753 & 4.782 &  7.855 & (-14.105, -0.421)$^{*-}$ \\ 
  ZoneE & -3.387 & -1.554 & 6.853 & -1.475 & -2.433 & 6.041 & 11.959 & (-18.095, -0.719)$^{*-}$ \\ 
   \hline
\end{tabular}
}
\end{subtable}
\begin{tablenotes}[para,flushleft]
\scriptsize
\item Superscripts $*+$ and $*-$ denote significant protective and risk effects at the 5\% significance level, respectively.
\end{tablenotes}
\end{threeparttable}
\end{table}

\addtocounter{table}{-1} 

\begin{table}[H]
\centering
\caption{ Presence estimates from models C.2.2.1-C.2.2.4, the combined models with the exchangeable and exchangeable presence and severity cluster correlation structures respectively. }
\label{ch4:table:comb:pres:exchangeable}
\begin{threeparttable}
\centering
\begin{subtable}{\linewidth}
\centering
\caption{Model C.2.2.1 (age 9)}
\scalebox{0.55}{
\begin{tabular}{rrrrrrrrl}
Variable & Estimate & \makecell{Standardized\\Estimate} & \makecell{SE\\(Standardized\\Estimate)} & \makecell{James-Stein\\Estimator} & \makecell{Bias\\(James-Stein\\Estimator)} & \makecell{SE\\(James-Stein\\Estimator)} & \makecell{MSE\\(James-Stein\\Estimator)} & \makecell{95\% CI\\(James-Stein\\Estimator)} \\
  \hline
dental\_age & -0.303 & -1.031 &  3.446 & -0.636 &  -0.590 &  2.825 &   3.172 & ( -4.927,  0.729) \\ 
  Total\_mgFPerDay & -0.095 & -0.363 &  1.756 & -0.178 &  -0.399 &  1.794 &   1.953 & ( -5.235,  1.502) \\ 
  SugarAddedBeverageOzPerDay & -0.006 & -0.418 &  2.116 & -0.083 &  -0.263 &  2.191 &   2.260 & ( -5.128,  1.437) \\ 
  BrushingFrequencyPerDay & -0.072 & -0.373 &  1.744 &  1.022 &  -1.289 &  1.864 &   3.527 & ( -3.512,  2.295) \\ 
  Avg\_homeppm & -0.657 & -3.117 &  3.919 & -2.696 &  -1.656 &  3.928 &   6.671 & (-14.641, -0.778)$^{*-}$ \\ 
  Prop\_DentAppt & -0.191 & -0.189 &  1.467 &  2.265 &  -2.505 &  1.537 &   7.813 & ( -4.562,  1.413) \\ 
  Prop\_FluorideTreatment &  0.388 &  0.272 &  2.068 & -2.303 &   2.810 &  2.015 &   9.908 & ( -1.536,  6.585) \\ 
  Tooth8 & -0.229 & -0.483 &  5.046 & -0.270 &  -3.560 &  5.633 &  18.308 & (-18.209,  0.158) \\ 
  Tooth9 & -0.140 & -0.167 &  4.897 & -0.065 &  -3.283 &  5.526 &  16.303 & (-16.802,  0.523) \\ 
  Tooth10 &  0.370 &  1.264 &  4.343 & -0.178 &   3.268 &  5.051 &  15.732 & ( -1.006, 20.771) \\ 
  ZoneM & -0.845 & -2.002 &  8.753 & -1.119 &  -4.931 &  9.812 &  34.125 & (-31.523, -0.220)$^{*-}$ \\ 
  ZoneI & -1.738 & -3.228 & 13.592 & -2.924 &  -9.750 & 15.118 & 110.178 & (-63.272, -1.364)$^{*-}$ \\ 
  ZoneE & -2.165 & -4.760 & 16.409 & -4.531 & -11.494 & 18.323 & 150.424 & (-74.635, -2.146)$^{*-}$ \\ 
   \hline
\end{tabular}
}
\end{subtable}
\begin{subtable}{\linewidth}
\centering
\caption{Model C.2.2.2 (age 13)}
\scalebox{0.55}{
\begin{tabular}{rrrrrrrrl}
Variable & Estimate & \makecell{Standardized\\Estimate} & \makecell{SE\\(Standardized\\Estimate)} & \makecell{James-Stein\\Estimator} & \makecell{Bias\\(James-Stein\\Estimator)} & \makecell{SE\\(James-Stein\\Estimator)} & \makecell{MSE\\(James-Stein\\Estimator)} & \makecell{95\% CI\\(James-Stein\\Estimator)} \\
  \hline
dental\_age &  0.568 &  0.946 &  3.520 &  0.584 &   1.591 &  3.681 &   6.212 & ( -0.724, 12.135) \\ 
  Total\_mgFPerDay &  0.012 &  0.006 &  2.993 &  0.003 &   0.508 &  2.991 &   3.249 & ( -2.202,  6.092) \\ 
  SugarAddedBeverageOzPerDay &  0.006 &  0.206 &  2.427 &  0.041 &  -0.006 &  1.820 &   1.820 & ( -4.266,  3.582) \\ 
  BrushingFrequencyPerDay & -0.008 & -0.008 &  2.395 &  0.023 &  -0.201 &  2.366 &   2.406 & ( -5.179,  5.825) \\ 
  Avg\_homeppm & -0.655 & -0.814 &  6.939 & -0.704 &  -3.767 &  6.704 &  20.897 & (-23.295,  0.061) \\ 
  Prop\_DentAppt & -0.077 & -0.016 &  3.574 &  0.191 &  -0.004 &  3.488 &   3.488 & ( -6.568,  6.479) \\ 
  Prop\_FluorideTreatment &  0.301 &  0.042 &  2.922 & -0.357 &   0.413 &  2.277 &   2.448 & ( -6.430,  5.155) \\ 
  Tooth8 & -0.178 & -0.141 &  3.317 & -0.079 &  -2.279 &  3.269 &   8.462 & (-13.412,  0.435) \\ 
  Tooth9 & -0.100 & -0.173 &  2.655 & -0.068 &  -1.083 &  2.798 &   3.970 & ( -9.280,  3.682) \\ 
  Tooth10 &  0.140 &  0.125 &  4.701 & -0.018 &   1.789 &  4.227 &   7.426 & ( -0.900, 10.460) \\ 
  ZoneM & -0.404 & -0.553 & 15.529 & -0.309 &  -5.220 & 17.853 &  45.104 & (-24.147, -0.080)$^{*-}$ \\ 
  ZoneI & -1.387 & -2.296 & 18.334 & -2.080 & -11.784 & 20.050 & 158.909 & (-59.093, -0.790)$^{*-}$ \\ 
  ZoneE & -2.132 & -1.726 & 28.621 & -1.643 & -19.111 & 30.966 & 396.195 & (-77.885, -1.386)$^{*-}$ \\ 
   \hline
\end{tabular}
}
\end{subtable}
\begin{subtable}{\linewidth}
\centering
\caption{Model C.2.2.3 (age 17)}
\scalebox{0.55}{
\begin{tabular}{rrrrrrrrl}
Variable & Estimate & \makecell{Standardized\\Estimate} & \makecell{SE\\(Standardized\\Estimate)} & \makecell{James-Stein\\Estimator} & \makecell{Bias\\(James-Stein\\Estimator)} & \makecell{SE\\(James-Stein\\Estimator)} & \makecell{MSE\\(James-Stein\\Estimator)} & \makecell{95\% CI\\(James-Stein\\Estimator)} \\
  \hline
dental\_age &  0.766 &  0.957 & 0.767 &  0.590 &  0.174 & 0.725 &  0.755 & ( -0.383,  2.261) \\ 
  Total\_mgFPerDay & -0.151 & -0.586 & 0.735 & -0.288 & -0.010 & 0.813 &  0.813 & ( -1.847,  1.579) \\ 
  SugarAddedBeverageOzPerDay &  0.015 &  0.676 & 0.848 &  0.135 &  0.642 & 0.875 &  1.288 & ( -0.900,  2.557) \\ 
  BrushingFrequencyPerDay &  0.192 &  0.356 & 0.801 & -0.974 &  1.087 & 0.566 &  1.748 & ( -0.985,  1.485) \\ 
  Avg\_homeppm & -0.768 & -1.097 & 0.934 & -0.949 & -0.563 & 0.941 &  1.258 & ( -3.494, -0.275)$^{*-}$ \\ 
  Prop\_DentAppt & -3.417 & -0.264 & 1.000 &  3.154 & -3.174 & 0.948 & 11.020 & ( -2.224,  1.873) \\ 
  Prop\_FluorideTreatment &  2.015 &  0.164 & 1.024 & -1.391 &  1.099 & 0.968 &  2.176 & ( -2.408,  1.643) \\ 
  Tooth8 & -0.096 & -0.060 & 1.116 & -0.034 & -1.372 & 1.097 &  2.979 & ( -4.049,  0.047) \\ 
  Tooth9 &  0.045 &  0.039 & 0.875 &  0.015 & -0.653 & 0.857 &  1.283 & ( -3.149,  0.194) \\ 
  Tooth10 & -0.212 & -0.223 & 0.807 &  0.031 &  0.258 & 0.703 &  0.769 & ( -0.661,  1.940) \\ 
  ZoneM & -0.348 & -0.358 & 1.091 & -0.200 & -0.735 & 1.101 &  1.641 & ( -3.195,  0.432) \\ 
  ZoneI & -1.747 & -1.513 & 2.481 & -1.371 & -2.242 & 2.563 &  7.590 & ( -7.736, -0.366)$^{*-}$ \\ 
  ZoneE & -2.377 & -3.030 & 3.205 & -2.885 & -2.690 & 3.344 & 10.578 & (-10.204, -1.201)$^{*-}$ \\ 
   \hline
\end{tabular}
}
\end{subtable}
\begin{subtable}{\linewidth}
\centering
\caption{Model C.2.2.4 (age 23)}
\scalebox{0.55}{
\begin{tabular}{rrrrrrrrl}
Variable & Estimate & \makecell{Standardized\\Estimate} & \makecell{SE\\(Standardized\\Estimate)} & \makecell{James-Stein\\Estimator} & \makecell{Bias\\(James-Stein\\Estimator)} & \makecell{SE\\(James-Stein\\Estimator)} & \makecell{MSE\\(James-Stein\\Estimator)} & \makecell{95\% CI\\(James-Stein\\Estimator)} \\
  \hline
dental\_age &  0.721 &  1.532 & 1.076 &  0.945 &  0.094 & 0.735 & 0.744 & ( -0.093,  2.692) \\ 
  Total\_mgFPerDay & -0.343 & -1.860 & 1.324 & -0.914 &  0.174 & 1.185 & 1.216 & ( -3.032,  0.649) \\ 
  SugarAddedBeverageOzPerDay & -0.014 & -1.351 & 1.368 & -0.270 & -0.397 & 0.953 & 1.111 & ( -2.366,  0.511) \\ 
  BrushingFrequencyPerDay & -0.176 & -0.519 & 2.224 &  1.421 & -1.839 & 2.297 & 5.679 & ( -1.673,  0.852) \\ 
  Avg\_homeppm & -0.984 & -1.794 & 1.048 & -1.551 &  0.264 & 0.815 & 0.885 & ( -2.943, -0.131)$^{*-}$ \\ 
  Prop\_DentAppt & -0.274 & -0.220 & 1.128 &  2.636 & -2.630 & 0.967 & 7.883 & ( -1.429,  1.920) \\ 
  Prop\_FluorideTreatment &  0.480 &  0.329 & 1.251 & -2.789 &  2.891 & 1.293 & 9.650 & ( -1.121,  1.450) \\ 
  Tooth8 & -0.511 & -2.067 & 2.248 & -1.155 & -0.376 & 1.870 & 2.011 & ( -4.549,  0.021) \\ 
  Tooth9 & -0.462 & -1.795 & 1.820 & -0.701 & -0.616 & 1.421 & 1.800 & ( -4.169,  0.132) \\ 
  Tooth10 &  0.058 &  0.302 & 0.807 & -0.043 &  0.208 & 0.520 & 0.563 & ( -0.638,  1.276) \\ 
  ZoneM & -0.580 & -0.309 & 3.317 & -0.173 & -0.638 & 2.305 & 2.712 & ( -5.512,  0.224) \\ 
  ZoneI & -2.323 & -1.812 & 5.449 & -1.641 & -1.276 & 4.807 & 6.434 & (-15.115, -0.388)$^{*-}$ \\ 
  ZoneE & -3.161 & -2.620 & 8.780 & -2.494 & -1.576 & 5.952 & 8.437 & (-18.680, -0.727)$^{*-}$ \\ 
   \hline
\end{tabular}
}
\end{subtable}
\begin{tablenotes}[para,flushleft]
\scriptsize
\item Superscripts $*+$ and $*-$ denote significant protective and risk effects at the 5\% significance level, respectively.
\end{tablenotes}
\end{threeparttable}
\end{table}

\addtocounter{table}{-1} 

\begin{table}[H]
\centering
\caption{ Presence estimates from models C.3.3.1-C.3.3.4, the combined models with the ar1 and ar1 presence and severity cluster correlation structures respectively. }
\label{ch4:table:comb:pres:ar1}
\begin{threeparttable}
\centering
\begin{subtable}{\linewidth}
\centering
\caption{Model C.3.3.1 (age 9)}
\scalebox{0.55}{
\begin{tabular}{rrrrrrrrl}
Variable & Estimate & \makecell{Standardized\\Estimate} & \makecell{SE\\(Standardized\\Estimate)} & \makecell{James-Stein\\Estimator} & \makecell{Bias\\(James-Stein\\Estimator)} & \makecell{SE\\(James-Stein\\Estimator)} & \makecell{MSE\\(James-Stein\\Estimator)} & \makecell{95\% CI\\(James-Stein\\Estimator)} \\
  \hline
dental\_age & -0.254 & -0.199 & 0.582 &   1.936 &  -1.710 & 0.914 &   3.837 & ( -1.320,  2.339) \\ 
  Total\_mgFPerDay & -0.145 & -0.146 & 0.480 &   0.415 &  -0.208 & 1.329 &   1.372 & ( -2.544,  3.270) \\ 
  SugarAddedBeverageOzPerDay & -0.008 & -0.176 & 0.722 &   1.833 &  -1.842 & 1.541 &   4.932 & ( -4.109,  2.268) \\ 
  BrushingFrequencyPerDay & -0.080 & -0.054 & 0.507 &   8.229 &  -8.393 & 3.728 &  74.175 & ( -8.671,  6.792) \\ 
  Avg\_homeppm & -0.556 & -0.528 & 0.960 &   1.797 &  -2.285 & 1.376 &   6.596 & ( -3.280,  1.296) \\ 
  Prop\_DentAppt & -0.189 & -0.062 & 0.505 &  11.470 & -11.142 & 1.896 & 126.035 & ( -3.213,  3.372) \\ 
  Prop\_FluorideTreatment &  0.152 &  0.032 & 0.482 &  -2.297 &   1.524 & 2.444 &   4.767 & ( -5.912,  3.020) \\ 
  Tooth8 & -0.378 & -0.170 & 1.055 &   1.499 &  -1.498 & 3.966 &   6.210 & ( -5.110,  6.255) \\ 
  Tooth9 & -0.263 & -0.099 & 1.057 &   1.837 &  -1.688 & 3.101 &   5.949 & ( -5.582,  6.386) \\ 
  Tooth10 &  0.124 &  0.080 & 0.650 & -14.906 &  13.123 & 2.492 & 174.704 & ( -6.944,  2.031) \\ 
  ZoneM & -0.690 & -0.201 & 1.289 &   8.703 &  -7.301 & 6.326 &  59.633 & ( -9.248, 15.692) \\ 
  ZoneI & -1.580 & -0.522 & 2.560 &   1.979 &  -3.180 & 3.173 &  13.288 & ( -9.817,  3.652) \\ 
  ZoneE & -1.997 & -0.915 & 2.961 &   0.720 &  -2.640 & 3.183 &  10.151 & (-12.498,  0.973) \\ 
   \hline
\end{tabular}
}
\end{subtable}
\begin{subtable}{\linewidth}
\centering
\caption{Model C.3.3.2 (age 13)}
\scalebox{0.55}{
\begin{tabular}{rrrrrrrrl}
Variable & Estimate & \makecell{Standardized\\Estimate} & \makecell{SE\\(Standardized\\Estimate)} & \makecell{James-Stein\\Estimator} & \makecell{Bias\\(James-Stein\\Estimator)} & \makecell{SE\\(James-Stein\\Estimator)} & \makecell{MSE\\(James-Stein\\Estimator)} & \makecell{95\% CI\\(James-Stein\\Estimator)} \\
  \hline
dental\_age &  0.269 &  0.076 & 0.368 & -0.736 &  0.383 & 0.683 &  0.829 & (-1.887, 0.722) \\ 
  Total\_mgFPerDay &  0.017 &  0.005 & 0.244 & -0.014 & -0.087 & 0.975 &  0.982 & (-2.556, 1.395) \\ 
  SugarAddedBeverageOzPerDay &  0.007 &  0.089 & 0.223 & -0.923 &  0.313 & 1.225 &  1.323 & (-3.893, 0.634) \\ 
  BrushingFrequencyPerDay & -0.052 & -0.025 & 0.186 &  3.843 & -4.659 & 2.431 & 24.141 & (-6.360, 2.332) \\ 
  Avg\_homeppm & -0.450 & -0.079 & 0.375 &  0.269 & -0.226 & 0.583 &  0.634 & (-0.623, 1.652) \\ 
  Prop\_DentAppt & -0.048 & -0.010 & 0.288 &  1.858 & -1.601 & 1.829 &  4.392 & (-3.088, 4.613) \\ 
  Prop\_FluorideTreatment & -0.034 & -0.004 & 0.228 &  0.295 & -0.064 & 1.925 &  1.929 & (-4.521, 3.686) \\ 
  Tooth8 & -0.202 & -0.097 & 0.200 &  0.855 & -0.443 & 0.938 &  1.135 & (-0.570, 2.546) \\ 
  Tooth9 & -0.116 & -0.065 & 0.163 &  1.208 & -0.498 & 2.098 &  2.346 & (-1.347, 7.200) \\ 
  Tooth10 &  0.065 &  0.030 & 0.290 & -5.527 &  4.170 & 2.255 & 19.643 & (-6.799, 0.715) \\ 
  ZoneM & 41.777 &  0.042 & 0.132 & -1.819 &  1.975 & 1.953 &  5.852 & (-1.796, 3.908) \\ 
  ZoneI & -1.796 & -0.170 & 0.454 &  0.646 & -0.646 & 0.456 &  0.874 & (-0.616, 1.110) \\ 
  ZoneE & -2.000 & -0.296 & 0.587 &  0.232 & -0.543 & 0.557 &  0.851 & (-1.260, 1.059) \\ 
   \hline
\end{tabular}
}
\end{subtable}
\begin{subtable}{\linewidth}
\centering
\caption{Model C.3.3.3 (age 17)}
\scalebox{0.55}{
\begin{tabular}{rrrrrrrrl}
Variable & Estimate & \makecell{Standardized\\Estimate} & \makecell{SE\\(Standardized\\Estimate)} & \makecell{James-Stein\\Estimator} & \makecell{Bias\\(James-Stein\\Estimator)} & \makecell{SE\\(James-Stein\\Estimator)} & \makecell{MSE\\(James-Stein\\Estimator)} & \makecell{95\% CI\\(James-Stein\\Estimator)} \\
  \hline
dental\_age &  0.559 &  0.239 & 0.356 & -2.324 &  1.695 & 1.057 &  3.931 & (-3.379,  0.510) \\ 
  Total\_mgFPerDay & -0.117 & -0.114 & 0.268 &  0.323 & -0.091 & 1.195 &  1.203 & (-1.125,  2.850) \\ 
  SugarAddedBeverageOzPerDay &  0.013 &  0.086 & 0.371 & -0.900 &  0.055 & 1.685 &  1.688 & (-4.320,  0.885) \\ 
  BrushingFrequencyPerDay & -0.001 & -0.001 & 0.334 &  0.097 & -0.295 & 2.880 &  2.966 & (-5.810,  4.731) \\ 
  Avg\_homeppm & -0.706 & -0.059 & 0.489 &  0.199 & -0.253 & 0.546 &  0.610 & (-1.540,  1.101) \\ 
  Prop\_DentAppt & -0.041 & -0.004 & 0.334 &  0.798 & -1.043 & 2.453 &  3.541 & (-5.269,  4.033) \\ 
  Prop\_FluorideTreatment & -1.184 & -0.145 & 0.365 & 10.271 & -9.530 & 2.530 & 93.345 & (-5.133,  5.038) \\ 
  Tooth8 & -0.362 & -0.085 & 0.601 &  0.745 &  0.169 & 2.030 &  2.059 & (-0.967,  4.454) \\ 
  Tooth9 & -0.231 & -0.078 & 0.446 &  1.450 & -0.627 & 1.791 &  2.184 & (-0.591,  4.913) \\ 
  Tooth10 &  0.112 &  0.032 & 0.164 & -5.871 &  5.347 & 2.097 & 30.687 & (-5.129,  3.040) \\ 
  ZoneM & -0.466 & -0.055 & 0.354 &  2.396 & -1.504 & 3.907 &  6.170 & (-2.016, 12.313) \\ 
  ZoneI & -1.822 & -0.103 & 0.928 &  0.389 & -0.427 & 1.086 &  1.269 & (-1.517,  1.118) \\ 
  ZoneE & -2.553 & -0.441 & 1.299 &  0.347 & -0.680 & 0.583 &  1.046 & (-1.715,  0.530) \\ 
   \hline
\end{tabular}
}
\end{subtable}
\begin{subtable}{\linewidth}
\centering
\caption{Model C.3.3.4 (age 23)}
\scalebox{0.55}{
\begin{tabular}{rrrrrrrrl}
Variable & Estimate & \makecell{Standardized\\Estimate} & \makecell{SE\\(Standardized\\Estimate)} & \makecell{James-Stein\\Estimator} & \makecell{Bias\\(James-Stein\\Estimator)} & \makecell{SE\\(James-Stein\\Estimator)} & \makecell{MSE\\(James-Stein\\Estimator)} & \makecell{95\% CI\\(James-Stein\\Estimator)} \\
  \hline
dental\_age &  0.712 &  0.290 & 0.687 & -2.819 &   2.308 & 1.230 &   6.558 & (-3.540, 1.408) \\ 
  Total\_mgFPerDay & -0.331 & -0.697 & 1.004 &  1.984 &  -0.960 & 1.790 &   2.712 & (-2.164, 5.175) \\ 
  SugarAddedBeverageOzPerDay & -0.013 & -0.359 & 0.687 &  3.738 &  -3.230 & 1.665 &  12.099 & (-2.510, 4.316) \\ 
  BrushingFrequencyPerDay & -0.148 & -0.097 & 0.366 & 14.891 & -14.342 & 2.237 & 207.940 & (-4.062, 4.178) \\ 
  Avg\_homeppm & -0.878 & -0.407 & 0.867 &  1.386 &  -1.607 & 0.863 &   3.444 & (-2.197, 1.171) \\ 
  Prop\_DentAppt & -0.341 & -0.082 & 0.624 & 15.234 & -14.606 & 2.424 & 215.770 & (-2.420, 5.040) \\ 
  Prop\_FluorideTreatment &  0.617 &  0.076 & 0.345 & -5.347 &   4.752 & 2.314 &  24.894 & (-4.827, 3.291) \\ 
  Tooth8 & -0.481 & -0.398 & 1.022 &  3.506 &  -2.449 & 2.622 &   8.620 & (-2.672, 6.522) \\ 
  Tooth9 & -0.449 & -0.287 & 0.823 &  5.317 &  -4.431 & 2.329 &  21.965 & (-2.119, 6.815) \\ 
  Tooth10 &  0.059 &  0.049 & 0.287 & -9.134 &   8.697 & 2.110 &  77.743 & (-4.504, 4.297) \\ 
  ZoneM & -0.908 & -0.001 & 1.119 &  0.061 &  -0.424 & 3.955 &   4.134 & (-6.223, 6.428) \\ 
  ZoneI & -2.534 & -0.325 & 3.198 &  1.231 &  -1.351 & 1.911 &   3.737 & (-3.935, 3.034) \\ 
  ZoneE & -3.337 & -0.012 & 4.492 &  0.009 &  -0.910 & 1.839 &   2.667 & (-5.240, 0.832) \\ 
   \hline
\end{tabular}
}
\end{subtable}
\begin{tablenotes}[para,flushleft]
\scriptsize
\item Superscripts $*+$ and $*-$ denote significant protective and risk effects at the 5\% significance level, respectively.
\end{tablenotes}
\end{threeparttable}
\end{table}



\addtocounter{table}{-1} 

\begin{table}[H]
\centering
\caption{ Severity estimates from models C.1.1.1-C.1.1.4, the combined models with the independence and independence presence and severity cluster correlation structures respectively. }
\label{table:comb:sev:independence}
\begin{threeparttable}
\centering
\begin{subtable}{\linewidth}
\centering
\caption{Model C.1.1.1 (age 9)}
\scalebox{0.55}{
\begin{tabular}{rrrrrrrrl}
Variable & Estimate & \makecell{Standardized\\Estimate} & \makecell{SE\\(Standardized\\Estimate)} & \makecell{James-Stein\\Estimator} & \makecell{Bias\\(James-Stein\\Estimator)} & \makecell{SE\\(James-Stein\\Estimator)} & \makecell{MSE\\(James-Stein\\Estimator)} & \makecell{95\% CI\\(James-Stein\\Estimator)} \\
  \hline
dental\_age & -0.209 & -0.595 & 0.519 &  0.705 & -0.323 & 0.968 &  1.072 & (-1.049, 2.494) \\ 
  Total\_mgFPerDay & -0.065 & -0.394 & 0.574 &  1.163 & -1.013 & 1.369 &  2.394 & (-3.397, 2.838) \\ 
  SugarAddedBeverageOzPerDay & -0.004 & -0.293 & 0.570 &  1.017 & -0.708 & 1.660 &  2.161 & (-1.949, 3.562) \\ 
  BrushingFrequencyPerDay & -0.049 & -0.373 & 0.450 &  2.361 & -2.012 & 1.668 &  5.715 & (-2.897, 3.560) \\ 
  Avg\_homeppm & -0.453 & -0.723 & 0.734 &  0.145 & -0.242 & 0.884 &  0.942 & (-1.611, 2.024) \\ 
  Prop\_DentAppt & -0.132 & -0.150 & 0.455 &  3.156 & -2.897 & 1.599 &  9.992 & (-3.267, 3.915) \\ 
  Prop\_FluorideTreatment &  0.267 &  0.205 & 0.545 & -5.440 &  5.265 & 1.506 & 29.222 & (-3.191, 2.639) \\ 
  Tooth8 & -0.158 & -1.022 & 0.900 &  0.905 & -1.116 & 1.239 &  2.484 & (-2.378, 2.844) \\ 
  Tooth9 & -0.096 & -0.211 & 0.907 &  6.468 & -6.642 & 1.626 & 45.736 & (-2.773, 2.577) \\ 
  Tooth10 &  0.255 &  0.509 & 0.546 & -2.623 &  2.010 & 1.703 &  5.745 & (-4.117, 2.253) \\ 
  ZoneM & -0.583 & -0.572 & 0.783 &  1.383 & -1.276 & 1.058 &  2.685 & (-2.062, 2.225) \\ 
  ZoneI & -1.198 & -0.637 & 0.879 &  0.295 & -0.599 & 0.997 &  1.356 & (-2.324, 2.038) \\ 
  ZoneE & -1.492 & -0.678 & 1.014 &  0.045 & -0.478 & 1.106 &  1.335 & (-3.125, 0.987) \\ 
   \hline
\end{tabular}
}
\end{subtable}
\begin{subtable}{\linewidth}
\centering
\caption{Model C.1.1.2 (age 13)}
\scalebox{0.55}{
\begin{tabular}{rrrrrrrrl}
Variable & Estimate & \makecell{Standardized\\Estimate} & \makecell{SE\\(Standardized\\Estimate)} & \makecell{James-Stein\\Estimator} & \makecell{Bias\\(James-Stein\\Estimator)} & \makecell{SE\\(James-Stein\\Estimator)} & \makecell{MSE\\(James-Stein\\Estimator)} & \makecell{95\% CI\\(James-Stein\\Estimator)} \\
  \hline
dental\_age &  0.386 &  0.318 & 0.472 & -0.378 &  0.249 & 0.955 &  1.017 & (-1.802, 1.962) \\ 
  Total\_mgFPerDay &  0.008 &  0.006 & 0.399 & -0.018 &  0.274 & 1.279 &  1.355 & (-1.842, 3.536) \\ 
  SugarAddedBeverageOzPerDay &  0.004 &  0.156 & 0.437 & -0.540 &  0.463 & 1.245 &  1.459 & (-3.364, 2.432) \\ 
  BrushingFrequencyPerDay & -0.006 & -0.009 & 0.360 &  0.057 & -0.268 & 1.720 &  1.792 & (-4.217, 2.231) \\ 
  Avg\_homeppm & -0.444 & -0.370 & 0.675 &  0.074 & -0.054 & 0.564 &  0.566 & (-0.941, 1.343) \\ 
  Prop\_DentAppt & -0.065 & -0.018 & 0.417 &  0.382 & -0.308 & 1.445 &  1.540 & (-2.621, 3.035) \\ 
  Prop\_FluorideTreatment &  0.221 &  0.039 & 0.401 & -1.036 &  1.080 & 1.550 &  2.716 & (-3.167, 3.326) \\ 
  Tooth8 & -0.123 & -0.107 & 0.745 &  0.095 &  0.058 & 1.374 &  1.378 & (-1.246, 2.670) \\ 
  Tooth9 & -0.069 & -0.129 & 0.602 &  3.952 & -3.895 & 1.172 & 16.346 & (-2.137, 2.163) \\ 
  Tooth10 &  0.093 &  0.141 & 0.473 & -0.726 &  0.487 & 0.943 &  1.181 & (-2.358, 1.006) \\ 
  ZoneM & -0.276 & -0.378 & 0.743 &  0.914 & -0.780 & 0.860 &  1.469 & (-1.428, 2.181) \\ 
  ZoneI & -0.948 & -0.352 & 0.879 &  0.163 & -0.185 & 0.719 &  0.753 & (-1.270, 1.679) \\ 
  ZoneE & -1.457 & -0.364 & 0.891 &  0.024 & -0.095 & 0.726 &  0.735 & (-1.220, 1.568) \\ 
   \hline
\end{tabular}
}
\end{subtable}
\begin{subtable}{\linewidth}
\centering
\caption{Model C.1.1.3 (age 17)}
\scalebox{0.55}{
\begin{tabular}{rrrrrrrrl}
Variable & Estimate & \makecell{Standardized\\Estimate} & \makecell{SE\\(Standardized\\Estimate)} & \makecell{James-Stein\\Estimator} & \makecell{Bias\\(James-Stein\\Estimator)} & \makecell{SE\\(James-Stein\\Estimator)} & \makecell{MSE\\(James-Stein\\Estimator)} & \makecell{95\% CI\\(James-Stein\\Estimator)} \\
  \hline
dental\_age &  1.034 &  0.678 & 0.450 & -0.804 &  0.331 & 0.915 &  1.024 & (-2.643, 0.579) \\ 
  Total\_mgFPerDay & -0.203 & -0.592 & 0.491 &  1.749 & -1.270 & 1.193 &  2.806 & (-1.542, 3.243) \\ 
  SugarAddedBeverageOzPerDay &  0.021 &  0.580 & 0.666 & -2.016 &  1.753 & 1.272 &  4.344 & (-3.701, 2.235) \\ 
  BrushingFrequencyPerDay &  0.257 &  0.365 & 0.532 & -2.314 &  2.254 & 1.258 &  6.340 & (-3.060, 2.593) \\ 
  Avg\_homeppm & -1.031 & -1.003 & 0.636 &  0.202 & -0.336 & 0.608 &  0.721 & (-1.408, 1.147) \\ 
  Prop\_DentAppt & -4.630 & -0.260 & 0.542 &  5.469 & -5.462 & 1.415 & 31.254 & (-2.684, 2.684) \\ 
  Prop\_FluorideTreatment &  2.698 &  0.169 & 0.583 & -4.495 &  4.629 & 1.226 & 22.649 & (-2.646, 2.455) \\ 
  Tooth8 & -0.134 & -0.069 & 0.764 &  0.061 &  0.001 & 1.249 &  1.249 & (-2.127, 3.995) \\ 
  Tooth9 &  0.063 &  0.043 & 0.623 & -1.326 &  1.266 & 1.740 &  3.343 & (-1.539, 2.567) \\ 
  Tooth10 & -0.286 & -0.213 & 0.454 &  1.100 & -1.251 & 1.122 &  2.688 & (-3.295, 1.576) \\ 
  ZoneM & -0.473 & -0.338 & 0.667 &  0.818 & -0.724 & 0.960 &  1.484 & (-1.514, 2.694) \\ 
  ZoneI & -2.355 & -0.914 & 0.915 &  0.424 & -0.707 & 0.852 &  1.352 & (-2.360, 1.261) \\ 
  ZoneE & -3.209 & -1.133 & 0.933 &  0.074 & -0.440 & 0.822 &  1.015 & (-2.136, 1.084) \\ 
   \hline
\end{tabular}
}
\end{subtable}
\begin{subtable}{\linewidth}
\centering
\caption{Model C.1.1.4 (age 23)}
\scalebox{0.55}{
\begin{tabular}{rrrrrrrrl}
Variable & Estimate & \makecell{Standardized\\Estimate} & \makecell{SE\\(Standardized\\Estimate)} & \makecell{James-Stein\\Estimator} & \makecell{Bias\\(James-Stein\\Estimator)} & \makecell{SE\\(James-Stein\\Estimator)} & \makecell{MSE\\(James-Stein\\Estimator)} & \makecell{95\% CI\\(James-Stein\\Estimator)} \\
  \hline
dental\_age &  0.031 &  0.018 & 0.710 & -0.021 &  0.400 & 0.835 & 0.995 & (-0.752, 2.564) \\ 
  Total\_mgFPerDay & -0.014 & -0.017 & 0.965 &  0.051 & -0.631 & 1.700 & 2.098 & (-3.332, 1.403) \\ 
  SugarAddedBeverageOzPerDay & -0.001 & -0.017 & 0.768 &  0.059 & -0.485 & 1.260 & 1.495 & (-2.930, 1.240) \\ 
  BrushingFrequencyPerDay & -0.006 & -0.012 & 1.204 &  0.079 & -0.561 & 1.468 & 1.782 & (-3.838, 1.500) \\ 
  Avg\_homeppm & -0.038 & -0.015 & 0.979 &  0.003 & -0.075 & 0.630 & 0.636 & (-2.023, 0.993) \\ 
  Prop\_DentAppt & -0.013 & -0.016 & 0.812 &  0.330 & -0.619 & 1.590 & 1.974 & (-4.800, 2.738) \\ 
  Prop\_FluorideTreatment &  0.025 &  0.019 & 0.887 & -0.502 &  0.833 & 1.562 & 2.256 & (-1.662, 3.407) \\ 
  Tooth8 & -0.021 & -0.016 & 1.287 &  0.014 & -0.227 & 1.253 & 1.305 & (-3.124, 1.956) \\ 
  Tooth9 & -0.019 & -0.016 & 1.168 &  0.499 & -0.934 & 1.348 & 2.220 & (-3.599, 1.420) \\ 
  Tooth10 &  0.003 &  0.023 & 0.412 & -0.119 &  0.249 & 1.128 & 1.190 & (-1.476, 2.475) \\ 
  ZoneM & -0.041 & -0.029 & 1.539 &  0.069 & -0.360 & 1.428 & 1.557 & (-4.375, 1.671) \\ 
  ZoneI & -0.112 & -0.020 & 1.982 &  0.009 & -0.454 & 1.783 & 1.989 & (-6.693, 0.852) \\ 
  ZoneE & -0.148 & -0.019 & 1.996 &  0.001 & -0.425 & 1.808 & 1.989 & (-6.511, 0.890) \\ 
   \hline
\end{tabular}
}
\end{subtable}
\begin{tablenotes}[para,flushleft]
\scriptsize
\item Superscripts $*+$ and $*-$ denote significant protective and risk effects at the 5\% significance level, respectively.
\end{tablenotes}
\end{threeparttable}
\end{table}

\addtocounter{table}{-1} 

\begin{table}[H]
\centering
\caption{ Severity estimates from models C.2.2.1-C.2.2.4, the combined models with the exchangeable and exchangeable presence and severity cluster correlation structures respectively. }
\label{table:comb:sev:exchangeable}
\begin{threeparttable}
\centering
\begin{subtable}{\linewidth}
\centering
\caption{Model C.2.2.1 (age 9)}
\scalebox{0.55}{
\begin{tabular}{rrrrrrrrl}
Variable & Estimate & \makecell{Standardized\\Estimate} & \makecell{SE\\(Standardized\\Estimate)} & \makecell{James-Stein\\Estimator} & \makecell{Bias\\(James-Stein\\Estimator)} & \makecell{SE\\(James-Stein\\Estimator)} & \makecell{MSE\\(James-Stein\\Estimator)} & \makecell{95\% CI\\(James-Stein\\Estimator)} \\
  \hline
dental\_age & -0.208 & -0.595 & 0.972 &  0.736 & -0.767 & 1.031 &  1.619 & (-2.069, 2.094) \\ 
  Total\_mgFPerDay & -0.065 & -0.395 & 1.276 &  0.919 & -1.104 & 1.625 &  2.843 & (-3.308, 2.615) \\ 
  SugarAddedBeverageOzPerDay & -0.004 & -0.294 & 1.101 &  0.929 & -0.926 & 1.492 &  2.349 & (-4.613, 2.462) \\ 
  BrushingFrequencyPerDay & -0.049 & -0.374 & 0.968 &  2.194 & -2.073 & 1.608 &  5.905 & (-3.082, 3.561) \\ 
  Avg\_homeppm & -0.452 & -0.722 & 2.011 &  0.192 & -1.085 & 2.231 &  3.409 & (-5.595, 2.015) \\ 
  Prop\_DentAppt & -0.132 & -0.151 & 1.006 &  2.637 & -2.809 & 1.570 &  9.463 & (-4.085, 2.986) \\ 
  Prop\_FluorideTreatment &  0.267 &  0.206 & 1.335 & -3.716 &  4.013 & 1.805 & 17.907 & (-2.389, 4.967) \\ 
  Tooth8 & -0.158 & -1.029 & 2.357 &  0.781 & -1.798 & 2.711 &  5.944 & (-8.321, 2.468) \\ 
  Tooth9 & -0.096 & -0.213 & 2.293 &  3.335 & -3.972 & 2.374 & 18.154 & (-6.152, 2.136) \\ 
  Tooth10 &  0.255 &  0.509 & 1.410 & -2.440 &  2.628 & 2.249 &  9.153 & (-3.428, 4.896) \\ 
  ZoneM & -0.582 & -0.573 & 2.392 &  1.800 & -2.665 & 2.795 &  9.898 & (-8.756, 2.660) \\ 
  ZoneI & -1.196 & -0.638 & 2.831 &  0.373 & -1.649 & 3.060 &  5.780 & (-7.644, 2.004) \\ 
  ZoneE & -1.490 & -0.679 & 2.772 &  0.103 & -1.438 & 2.720 &  4.787 & (-7.942, 1.978) \\ 
   \hline
\end{tabular}
}
\end{subtable}
\begin{subtable}{\linewidth}
\centering
\caption{Model C.2.2.2 (age 13)}
\scalebox{0.55}{
\begin{tabular}{rrrrrrrrl}
Variable & Estimate & \makecell{Standardized\\Estimate} & \makecell{SE\\(Standardized\\Estimate)} & \makecell{James-Stein\\Estimator} & \makecell{Bias\\(James-Stein\\Estimator)} & \makecell{SE\\(James-Stein\\Estimator)} & \makecell{MSE\\(James-Stein\\Estimator)} & \makecell{95\% CI\\(James-Stein\\Estimator)} \\
  \hline
dental\_age &  0.379 &  0.097 & 2.076 & -0.120 &  0.501 & 2.208 & 2.459 & (-2.421, 6.014) \\ 
  Total\_mgFPerDay &  0.008 &  0.004 & 1.190 & -0.009 &  0.074 & 1.565 & 1.570 & (-2.880, 3.439) \\ 
  SugarAddedBeverageOzPerDay &  0.004 &  0.089 & 1.500 & -0.281 &  0.367 & 1.639 & 1.774 & (-3.738, 3.493) \\ 
  BrushingFrequencyPerDay & -0.005 & -0.005 & 1.125 &  0.032 & -0.187 & 1.775 & 1.810 & (-4.455, 3.028) \\ 
  Avg\_homeppm & -0.438 & -0.103 & 2.871 &  0.027 & -0.833 & 2.997 & 3.691 & (-7.703, 2.042) \\ 
  Prop\_DentAppt & -0.051 & -0.010 & 1.538 &  0.178 & -0.304 & 1.939 & 2.032 & (-4.982, 3.830) \\ 
  Prop\_FluorideTreatment &  0.201 &  0.025 & 1.506 & -0.459 &  0.415 & 1.926 & 2.098 & (-4.065, 4.153) \\ 
  Tooth8 & -0.119 & -0.056 & 2.177 &  0.043 &  0.005 & 2.380 & 2.380 & (-4.985, 6.939) \\ 
  Tooth9 & -0.067 & -0.057 & 1.978 &  0.897 & -0.683 & 2.136 & 2.602 & (-3.515, 3.624) \\ 
  Tooth10 &  0.093 &  0.065 & 2.053 & -0.314 &  0.691 & 1.653 & 2.131 & (-2.630, 4.378) \\ 
  ZoneM & -0.270 & -0.096 & 2.446 &  0.302 & -0.734 & 2.202 & 2.741 & (-5.531, 2.843) \\ 
  ZoneI & -0.926 & -0.095 & 4.780 &  0.056 & -0.921 & 5.147 & 5.996 & (-7.619, 6.992) \\ 
  ZoneE & -1.424 & -0.097 & 4.481 &  0.015 & -0.879 & 4.716 & 5.489 & (-8.564, 5.737) \\ 
   \hline
\end{tabular}
}
\end{subtable}
\begin{subtable}{\linewidth}
\centering
\caption{Model C.2.2.3 (age 17)}
\scalebox{0.55}{
\begin{tabular}{rrrrrrrrl}
Variable & Estimate & \makecell{Standardized\\Estimate} & \makecell{SE\\(Standardized\\Estimate)} & \makecell{James-Stein\\Estimator} & \makecell{Bias\\(James-Stein\\Estimator)} & \makecell{SE\\(James-Stein\\Estimator)} & \makecell{MSE\\(James-Stein\\Estimator)} & \makecell{95\% CI\\(James-Stein\\Estimator)} \\
  \hline
dental\_age &  1.034 &  0.668 & 0.451 & -0.827 &  0.704 & 0.837 &  1.333 & (-2.210, 0.951) \\ 
  Total\_mgFPerDay & -0.203 & -0.587 & 0.485 &  1.366 & -1.112 & 1.175 &  2.411 & (-1.513, 3.478) \\ 
  SugarAddedBeverageOzPerDay &  0.021 &  0.557 & 0.652 & -1.760 &  1.752 & 1.178 &  4.247 & (-2.588, 2.360) \\ 
  BrushingFrequencyPerDay &  0.259 &  0.348 & 0.539 & -2.039 &  2.033 & 1.097 &  5.232 & (-1.692, 1.570) \\ 
  Avg\_homeppm & -1.036 & -0.990 & 0.657 &  0.263 & -0.569 & 0.703 &  1.026 & (-1.989, 0.814) \\ 
  Prop\_DentAppt & -4.611 & -0.257 & 0.554 &  4.502 & -4.432 & 1.293 & 20.937 & (-2.199, 2.662) \\ 
  Prop\_FluorideTreatment &  2.718 &  0.170 & 0.593 & -3.067 &  3.089 & 1.137 & 10.680 & (-2.438, 1.815) \\ 
  Tooth8 & -0.130 & -0.066 & 0.768 &  0.050 & -0.344 & 0.842 &  0.961 & (-2.164, 1.567) \\ 
  Tooth9 &  0.061 &  0.042 & 0.639 & -0.655 &  0.372 & 1.721 &  1.860 & (-2.025, 1.436) \\ 
  Tooth10 & -0.287 & -0.211 & 0.516 &  1.014 & -1.103 & 1.047 &  2.264 & (-3.008, 1.690) \\ 
  ZoneM & -0.469 & -0.330 & 0.690 &  1.037 & -1.121 & 0.955 &  2.211 & (-2.163, 1.836) \\ 
  ZoneI & -2.357 & -0.874 & 0.965 &  0.511 & -0.992 & 0.992 &  1.977 & (-3.414, 1.121) \\ 
  ZoneE & -3.207 & -1.088 & 0.981 &  0.165 & -0.720 & 0.969 &  1.488 & (-3.355, 0.984) \\ 
   \hline
\end{tabular}
}
\end{subtable}
\begin{subtable}{\linewidth}
\centering
\caption{Model C.2.2.4 (age 23)}
\scalebox{0.55}{
\begin{tabular}{rrrrrrrrl}
Variable & Estimate & \makecell{Standardized\\Estimate} & \makecell{SE\\(Standardized\\Estimate)} & \makecell{James-Stein\\Estimator} & \makecell{Bias\\(James-Stein\\Estimator)} & \makecell{SE\\(James-Stein\\Estimator)} & \makecell{MSE\\(James-Stein\\Estimator)} & \makecell{95\% CI\\(James-Stein\\Estimator)} \\
  \hline
dental\_age & -0.184 & -0.291 & 0.867 &  0.360 & -0.138 & 0.855 &  0.874 & (-0.870, 2.658) \\ 
  Total\_mgFPerDay &  0.087 &  0.317 & 2.232 & -0.739 &  0.385 & 1.289 &  1.438 & (-3.407, 1.378) \\ 
  SugarAddedBeverageOzPerDay &  0.004 &  0.277 & 0.832 & -0.875 &  0.593 & 1.301 &  1.653 & (-3.719, 1.356) \\ 
  BrushingFrequencyPerDay &  0.045 &  0.174 & 1.245 & -1.022 &  0.679 & 1.441 &  1.902 & (-3.616, 1.785) \\ 
  Avg\_homeppm &  0.251 &  0.261 & 1.082 & -0.070 &  0.071 & 0.711 &  0.716 & (-1.977, 1.216) \\ 
  Prop\_DentAppt &  0.070 &  0.138 & 0.843 & -2.422 &  2.229 & 1.368 &  6.338 & (-3.801, 2.965) \\ 
  Prop\_FluorideTreatment & -0.122 & -0.182 & 0.930 &  3.278 & -2.928 & 1.494 & 10.069 & (-1.252, 2.136) \\ 
  Tooth8 &  0.130 &  0.264 & 1.371 & -0.200 &  0.065 & 1.286 &  1.291 & (-2.749, 2.103) \\ 
  Tooth9 &  0.118 &  0.264 & 1.244 & -4.130 &  3.866 & 1.359 & 16.307 & (-3.422, 1.470) \\ 
  Tooth10 & -0.015 & -0.193 & 0.593 &  0.927 & -0.788 & 0.993 &  1.614 & (-1.195, 1.828) \\ 
  ZoneM &  0.148 &  0.191 & 1.583 & -0.599 &  0.309 & 1.475 &  1.571 & (-5.347, 2.003) \\ 
  ZoneI &  0.592 &  0.285 & 2.019 & -0.167 & -0.251 & 1.812 &  1.875 & (-7.019, 1.118) \\ 
  ZoneE &  0.806 &  0.286 & 2.020 & -0.043 & -0.360 & 1.844 &  1.973 & (-7.109, 1.043) \\ 
   \hline
\end{tabular}
}
\end{subtable}
\begin{tablenotes}[para,flushleft]
\scriptsize
\item Superscripts $*+$ and $*-$ denote significant protective and risk effects at the 5\% significance level, respectively.
\end{tablenotes}
\end{threeparttable}
\end{table}

\addtocounter{table}{-1} 

\begin{table}[H]
\centering
\caption{ Severity estimates from models C.3.3.1-C.3.3.4, the combined models with the ar1 and ar1 presence and severity cluster correlation structures respectively. }
\label{table:comb:sev:ar1}
\begin{threeparttable}
\centering
\begin{subtable}{\linewidth}
\centering
\caption{Model C.3.3.1 (age 9)}
\scalebox{0.55}{
\begin{tabular}{rrrrrrrrl}
Variable & Estimate & \makecell{Standardized\\Estimate} & \makecell{SE\\(Standardized\\Estimate)} & \makecell{James-Stein\\Estimator} & \makecell{Bias\\(James-Stein\\Estimator)} & \makecell{SE\\(James-Stein\\Estimator)} & \makecell{MSE\\(James-Stein\\Estimator)} & \makecell{95\% CI\\(James-Stein\\Estimator)} \\
  \hline
dental\_age & -0.248 & -0.203 & 0.420 &   9.097 &  -7.462 & 5.247 &   60.930 & ( -8.896, 11.097) \\ 
  Total\_mgFPerDay & -0.141 & -0.145 & 0.381 &  12.520 & -12.601 & 7.599 &  166.373 & (-19.403, 13.661) \\ 
  SugarAddedBeverageOzPerDay & -0.008 & -0.163 & 0.408 &  11.242 &  -9.614 & 6.744 &   99.168 & (-10.707, 10.743) \\ 
  BrushingFrequencyPerDay & -0.078 & -0.056 & 0.388 &  35.180 & -35.366 & 7.217 & 1257.966 & (-12.986, 15.485) \\ 
  Avg\_homeppm & -0.541 & -0.474 & 0.571 &   3.706 &  -1.114 & 3.789 &    5.030 & ( -3.381, 11.910) \\ 
  Prop\_DentAppt & -0.184 & -0.062 & 0.349 &  27.089 & -24.993 & 6.410 &  631.068 & ( -9.282, 12.898) \\ 
  Prop\_FluorideTreatment &  0.148 &  0.033 & 0.382 & -17.602 &  14.913 & 7.404 &  229.810 & (-16.912,  7.142) \\ 
  Tooth8 & -0.368 & -0.172 & 0.508 &  10.507 &  -8.361 & 9.799 &   79.699 & (-21.341, 17.715) \\ 
  Tooth9 & -0.256 & -0.101 & 0.471 &  16.535 & -15.620 & 9.671 &  253.657 & (-18.613, 18.576) \\ 
  Tooth10 &  0.120 &  0.088 & 0.522 & -15.883 &  13.312 & 5.100 &  182.303 & (-12.613,  7.873) \\ 
  ZoneM & -0.671 & -0.329 & 0.759 &   5.490 &  -2.574 & 4.718 &   11.345 & ( -4.993, 12.846) \\ 
  ZoneI & -1.537 & -0.584 & 0.851 &   2.799 &  -0.996 & 3.379 &    4.371 & ( -4.291,  9.107) \\ 
  ZoneE & -1.944 & -0.878 & 0.700 &   1.391 &   1.609 & 4.516 &    7.104 & ( -2.148, 10.705) \\ 
   \hline
\end{tabular}
}
\end{subtable}
\begin{subtable}{\linewidth}
\centering
\caption{Model C.3.3.2 (age 13)}
\scalebox{0.55}{
\begin{tabular}{rrrrrrrrl}
Variable & Estimate & \makecell{Standardized\\Estimate} & \makecell{SE\\(Standardized\\Estimate)} & \makecell{James-Stein\\Estimator} & \makecell{Bias\\(James-Stein\\Estimator)} & \makecell{SE\\(James-Stein\\Estimator)} & \makecell{MSE\\(James-Stein\\Estimator)} & \makecell{95\% CI\\(James-Stein\\Estimator)} \\
  \hline
dental\_age & -0.033 &  0.000 & 0.080 &  0.017 & -1.046 & 5.982 & 7.077 & (-23.175,  4.511) \\ 
  Total\_mgFPerDay & -0.002 &  0.000 & 0.089 &  0.021 &  0.581 & 4.920 & 5.257 & ( -6.624, 13.331) \\ 
  SugarAddedBeverageOzPerDay & -0.001 & -0.001 & 0.072 &  0.096 & -0.805 & 6.375 & 7.024 & (-12.727,  5.541) \\ 
  BrushingFrequencyPerDay &  0.006 &  0.001 & 0.084 & -0.553 &  0.716 & 3.853 & 4.366 & ( -9.121,  8.249) \\ 
  Avg\_homeppm &  0.056 &  0.001 & 0.145 & -0.005 &  0.843 & 3.440 & 4.151 & ( -1.278, 10.366) \\ 
  Prop\_DentAppt &  0.006 &  0.000 & 0.076 & -0.132 &  0.521 & 7.733 & 8.004 & ( -8.400, 21.022) \\ 
  Prop\_FluorideTreatment &  0.004 &  0.000 & 0.074 & -0.107 &  0.911 & 4.680 & 5.510 & ( -4.415,  7.488) \\ 
  Tooth8 &  0.025 &  0.001 & 0.131 & -0.047 &  0.834 & 6.359 & 7.054 & ( -5.110, 14.338) \\ 
  Tooth9 &  0.014 &  0.001 & 0.097 & -0.084 &  0.129 & 7.214 & 7.230 & ( -8.894,  8.417) \\ 
  Tooth10 & -0.008 &  0.000 & 0.070 &  0.081 & -0.521 & 5.614 & 5.886 & ( -6.494,  5.353) \\ 
  ZoneM & -5.183 & -0.055 & 0.089 &  0.913 & -0.443 & 3.597 & 3.793 & ( -3.924,  3.921) \\ 
  ZoneI &  0.223 &  0.001 & 0.088 & -0.006 & -0.752 & 3.800 & 4.365 & ( -9.837,  2.087) \\ 
  ZoneE &  0.248 &  0.001 & 0.177 & -0.001 &  0.578 & 3.587 & 3.922 & ( -2.391, 12.573) \\ 
   \hline
\end{tabular}
}
\end{subtable}
\begin{subtable}{\linewidth}
\centering
\caption{Model C.3.3.3 (age 17)}
\scalebox{0.55}{
\begin{tabular}{rrrrrrrrl}
Variable & Estimate & \makecell{Standardized\\Estimate} & \makecell{SE\\(Standardized\\Estimate)} & \makecell{James-Stein\\Estimator} & \makecell{Bias\\(James-Stein\\Estimator)} & \makecell{SE\\(James-Stein\\Estimator)} & \makecell{MSE\\(James-Stein\\Estimator)} & \makecell{95\% CI\\(James-Stein\\Estimator)} \\
  \hline
dental\_age & -2.449 & -0.048 & 0.142 &   2.165 & -1.554 & 4.682 &   7.097 & ( -9.584, 13.140) \\ 
  Total\_mgFPerDay &  0.513 &  0.043 & 0.136 &  -3.675 &  2.605 & 5.490 &  12.278 & (-16.772,  6.237) \\ 
  SugarAddedBeverageOzPerDay & -0.056 & -0.043 & 0.141 &   2.963 & -2.401 & 5.524 &  11.288 & (-12.463, 14.758) \\ 
  BrushingFrequencyPerDay &  0.005 &  0.001 & 0.154 &  -0.853 &  0.818 & 5.396 &   6.066 & ( -9.114, 11.250) \\ 
  Avg\_homeppm &  3.093 &  0.047 & 0.126 &  -0.369 &  0.097 & 4.046 &   4.055 & (-11.161,  5.497) \\ 
  Prop\_DentAppt &  0.181 &  0.023 & 0.119 & -10.195 &  9.302 & 7.762 &  94.297 & (-23.207,  8.072) \\ 
  Prop\_FluorideTreatment &  5.189 &  0.050 & 0.122 & -26.752 & 27.418 & 6.189 & 757.912 & (-12.940, 12.216) \\ 
  Tooth8 &  1.585 &  0.050 & 0.116 &  -3.046 &  2.398 & 6.751 &  12.502 & (-17.508, 10.249) \\ 
  Tooth9 &  1.013 &  0.043 & 0.122 &  -7.070 &  6.305 & 7.395 &  47.149 & (-20.669, 12.704) \\ 
  Tooth10 & -0.492 & -0.057 & 0.109 &  10.259 & -9.353 & 4.346 &  91.821 & ( -6.633, 15.355) \\ 
  ZoneM &  2.044 &  0.043 & 0.124 &  -0.716 & -0.306 & 2.734 &   2.827 & ( -8.561,  0.902) \\ 
  ZoneI &  7.986 &  0.063 & 0.141 &  -0.303 & -0.367 & 3.065 &   3.200 & (-10.487,  3.526) \\ 
  ZoneE & 11.193 &  0.049 & 0.151 &  -0.078 & -0.507 & 4.231 &   4.488 & (-11.807,  3.735) \\ 
   \hline
\end{tabular}
}
\end{subtable}
\begin{subtable}{\linewidth}
\centering
\caption{Model C.3.3.4 (age 23)}
\scalebox{0.55}{
\begin{tabular}{rrrrrrrrl}
Variable & Estimate & \makecell{Standardized\\Estimate} & \makecell{SE\\(Standardized\\Estimate)} & \makecell{James-Stein\\Estimator} & \makecell{Bias\\(James-Stein\\Estimator)} & \makecell{SE\\(James-Stein\\Estimator)} & \makecell{MSE\\(James-Stein\\Estimator)} & \makecell{95\% CI\\(James-Stein\\Estimator)} \\
  \hline
dental\_age &  1.192 &  0.009 & 0.250 & -0.382 &  0.380 & 5.295 &  5.440 & (-12.144,  6.846) \\ 
  Total\_mgFPerDay & -0.554 & -0.009 & 0.366 &  0.751 & -1.503 & 8.601 & 10.859 & (-17.752, 11.800) \\ 
  SugarAddedBeverageOzPerDay & -0.022 & -0.010 & 0.280 &  0.662 & -0.643 & 8.378 &  8.791 & ( -9.255,  3.933) \\ 
  BrushingFrequencyPerDay & -0.247 & -0.007 & 0.211 &  4.649 & -5.875 & 4.020 & 38.531 & (-11.861,  5.180) \\ 
  Avg\_homeppm & -1.469 & -0.010 & 0.267 &  0.075 & -0.549 & 2.308 &  2.610 & ( -7.146,  3.016) \\ 
  Prop\_DentAppt & -0.571 & -0.014 & 0.251 &  6.153 & -4.663 & 7.908 & 29.655 & ( -9.927, 23.179) \\ 
  Prop\_FluorideTreatment &  1.033 &  0.013 & 0.184 & -6.752 &  6.357 & 5.109 & 45.522 & (-11.056, 13.022) \\ 
  Tooth8 & -0.804 & -0.009 & 0.326 &  0.543 & -1.106 & 5.402 &  6.625 & (-11.224, 12.685) \\ 
  Tooth9 & -0.751 & -0.009 & 0.363 &  1.472 & -1.354 & 8.627 & 10.459 & (-10.177, 22.301) \\ 
  Tooth10 &  0.098 &  0.008 & 0.153 & -1.385 &  2.081 & 4.092 &  8.424 & ( -5.759,  7.918) \\ 
  ZoneM & -1.520 & -0.005 & 0.349 &  0.078 &  1.419 & 3.930 &  5.943 & ( -1.138,  8.981) \\ 
  ZoneI & -4.240 & -0.009 & 0.487 &  0.042 & -0.252 & 2.765 &  2.829 & ( -4.792,  3.702) \\ 
  ZoneE & -5.585 & -0.023 & 0.480 &  0.037 & -0.313 & 2.600 &  2.698 & ( -4.398,  5.853) \\ 
   \hline
\end{tabular}
}
\end{subtable}
\begin{tablenotes}[para,flushleft]
\scriptsize
\item Superscripts $*+$ and $*-$ denote significant protective and risk effects at the 5\% significance level, respectively.
\end{tablenotes}
\end{threeparttable}
\end{table}

\section{Rank Aggregation Results}



\begin{table}[H]
\centering
\caption{ Rank aggregation over different working correlation structures for the separate presence models at different ages, and overall across all ages. }
\label{ch4:table:sep:pres:rankAggreg}
\begin{tabular}{rccccccl}
Age & Rank 1 & Rank 2 & Rank 3 & Rank 4 \\
  \hline
Age 9 & A.1 & A.2 & A.4 & A.3 \\ 
  Age 13 & A.3 & A.4 & A.1 & A.2 \\ 
  Age 17 & A.4 & A.1 & A.2 & A.3 \\ 
  Age 23 & A.3 & A.2 & A.1 & A.4 \\ 
  Overall & A.3 & A.1 & A.2 & A.4 \\ 
   \hline
\end{tabular}
\begin{tablenotes}
\scriptsize
\item A.$c$ denotes the model. ``A'' indicates the separate presence model, $c$ indicates the working correlation structure: $c=1$ for \\ independence, $c=2$ for exchangeable, $c=3$ for AR(1), $c=4$ for jackknifed. Lower ranks correspond to lower SE of the \\ James-Stein adjusted standardized estimate.
\end{tablenotes}
\end{table}


\begin{table}[H]
\centering
\caption{ Rank aggregation over different working correlation structures for the separate severity models at different ages, and overall across all ages. }
\label{ch4:table:sep:sev:rankAggreg}
\begin{tabular}{rccccccl}
Age & Rank 1 & Rank 2 & Rank 3 & Rank 4 \\
  \hline
Age 9 & B.4 & B.2 & B.1 & B.3 \\ 
  Age 13 & B.3 & B.4 & B.1 & B.2 \\ 
  Age 17 & B.4 & B.2 & B.1 & B.3 \\ 
  Age 23 & B.4 & B.1 & B.2 & B.3 \\ 
  Overall & B.4 & B.2 & B.1 & B.3 \\ 
   \hline
\end{tabular}
\begin{tablenotes}
\scriptsize
\item B.$c$ denotes the model. ``B'' indicates the separate severity model, $c$ indicates the working correlation structure: $c=1$ for \\ independence, $c=2$ for exchangeable, $c=3$ for AR(1), $c=4$ for jackknifed. Lower ranks correspond to lower SE of the \\ James-Stein adjusted standardized estimate.
\end{tablenotes}
\end{table}


\begin{table}[H]
\centering
\caption{ Rank aggregation over different working correlation structures for the presence models from the combined modeling at different ages, and overall across all ages. }
\begin{tabular}{rccccccl}
\label{ch4:table:comb:pres:rankAggreg}
Age & Rank 1 & Rank 2 & Rank 3 & Rank 4 \\
  \hline
Age 9 & C.4.4 & C.1.1 & C.3.3 & C.2.2 \\ 
  Age 13 & C.4.4 & C.1.1 & C.3.3 & C.2.2 \\ 
  Age 17 & C.4.4 & C.1.1 & C.2.2 & C.3.3 \\ 
  Age 23 & C.4.4 & C.1.1 & C.2.2 & C.3.3 \\ 
  Overall & C.4.4 & C.1.1 & C.3.3 & C.2.2 \\ 
   \hline
\end{tabular}
\begin{tablenotes}
\scriptsize
\item C.$c_p$.$c_s$ denotes the model. ``C'' indicates the combined model, $c_p$ and $c_s$ denote the working correlation structures for \\presence and severity components respectively: $1$ for independence, $2$ for exchangeable, $3$ for AR(1), $4$ for jackknifed. \\Lower ranks correspond to lower SE of the James-Stein adjusted standardized estimate.
\end{tablenotes}
\end{table}


\begin{table}[H]
\centering
\caption{ Rank aggregation over different working correlation structures for the severity models from the combined modeling at different ages, and overall across all ages. }
\label{ch4:table:comb:sev:rankAggreg}
\begin{tabular}{rccccccl}
Age & Rank 1 & Rank 2 & Rank 3 & Rank 4 \\
  \hline
Age 9 & C.4.4 & C.1.1 & C.2.2 & C.3.3 \\ 
  Age 13 & C.4.4 & C.1.1 & C.2.2 & C.3.3 \\ 
  Age 17 & C.4.4 & C.2.2 & C.1.1 & C.3.3 \\ 
  Age 23 & C.4.4 & C.1.1 & C.2.2 & C.3.3 \\ 
  Overall & C.4.4 & C.1.1 & C.2.2 & C.3.3 \\ 
   \hline
\end{tabular}
\begin{tablenotes}
\scriptsize
\item C.$c_p$.$c_s$ denotes the model. ``C'' indicates the combined model, $c_p$ and $c_s$ denote the working correlation structures for \\presence and severity components respectively: $1$ for independence, $2$ for exchangeable, $3$ for AR(1), $4$ for jackknifed. \\Lower ranks correspond to lower SE of the James-Stein adjusted standardized estimate.
\end{tablenotes}
\end{table}

\section{Additional Simulation Results}

\subsection{Properties of Direct Estimates}



\begin{table}[H]
\centering
\caption{ Simulation results for age 23 data generated with independence and exchangeable correlation structures for presence and severity respectively - properties of direct estimates. }
\label{ch4:table:sim:raw:independence}
\begin{threeparttable}
\centering
\begin{subtable}{\linewidth}
\centering
\caption{Presence model A$_s$.1.4}
\scalebox{0.65}{
\begin{tabular}{l|cccc|cccc|cccc}
{} & \multicolumn{4}{c|}{\textbf{N=30}} & \multicolumn{4}{c|}{\textbf{N=50}} & \multicolumn{4}{c}{\textbf{N=200}} \\ \hline
Variable & Estimate & Bias & SE & MSE & Estimate & Bias & SE & MSE & Estimate & Bias & SE & MSE \\ 
  \hline
dental\_age & -0.446 & -0.145 & 0.614 & 0.398 & -0.228 & 0.073 & 0.254 & 0.070 & -0.325 & -0.024 & 0.113 & 0.013 \\ 
  Avg\_homeppm & -0.691 & -0.008 & 0.401 & 0.161 & -0.700 & -0.018 & 0.230 & 0.053 & -0.672 & 0.010 & 0.130 & 0.017 \\ 
  Tooth8 & -0.272 & -0.053 & 0.341 & 0.119 & -0.233 & -0.014 & 0.240 & 0.058 & -0.231 & -0.012 & 0.130 & 0.017 \\ 
  Tooth9 & -0.151 & -0.061 & 0.337 & 0.117 & -0.128 & -0.038 & 0.256 & 0.067 & -0.134 & -0.045 & 0.124 & 0.017 \\ 
  Tooth10 & 0.376 & -0.046 & 0.348 & 0.123 & 0.384 & -0.037 & 0.247 & 0.063 & 0.382 & -0.039 & 0.143 & 0.022 \\ 
  ZoneM & -0.904 & -0.135 & 0.497 & 0.266 & -0.831 & -0.062 & 0.317 & 0.104 & -0.868 & -0.099 & 0.168 & 0.038 \\ 
  ZoneI & -1.798 & -0.051 & 0.483 & 0.236 & -1.742 & 0.005 & 0.324 & 0.105 & -1.758 & -0.012 & 0.158 & 0.025 \\ 
  ZoneE & -2.277 & -0.121 & 0.500 & 0.265 & -2.207 & -0.051 & 0.284 & 0.083 & -2.181 & -0.024 & 0.156 & 0.025 \\ 
   \hline
\end{tabular}
}
\end{subtable}
\begin{subtable}{\linewidth}
\centering
\caption{Severity model B$_s$.2.4}
\scalebox{0.65}{
\begin{tabular}{l|cccc|cccc|cccc}
{} & \multicolumn{4}{c|}{\textbf{N=30}} & \multicolumn{4}{c|}{\textbf{N=50}} & \multicolumn{4}{c}{\textbf{N=200}} \\ \hline
Variable & Estimate & Bias & SE & MSE & Estimate & Bias & SE & MSE & Estimate & Bias & SE & MSE \\ 
  \hline
dental\_age & -0.658 & -0.550 & 7.079 & 50.413 & 0.622 & 0.730 & 7.260 & 53.240 & -0.237 & -0.129 & 3.104 & 9.654 \\ 
  Avg\_homeppm & -0.740 & -0.511 & 0.765 & 0.847 & -0.693 & -0.464 & 0.505 & 0.470 & -0.562 & -0.333 & 0.269 & 0.183 \\ 
  Tooth8 & -0.128 & -0.114 & 0.591 & 0.362 & -0.254 & -0.240 & 0.437 & 0.248 & -0.211 & -0.197 & 0.406 & 0.204 \\ 
  Tooth9 & -0.051 & -0.143 & 0.592 & 0.371 & -0.207 & -0.299 & 0.439 & 0.283 & -0.113 & -0.205 & 0.191 & 0.079 \\ 
  Tooth10 & 0.514 & 0.289 & 0.709 & 0.587 & 0.339 & 0.114 & 0.605 & 0.379 & 0.370 & 0.145 & 0.388 & 0.172 \\ 
  ZoneM & -6.380 & -6.256 & 14.666 & 254.234 & -4.036 & -3.912 & 10.460 & 124.711 & -0.822 & -0.698 & 0.586 & 0.831 \\ 
  ZoneI & -7.741 & -7.205 & 14.152 & 252.179 & -4.944 & -4.407 & 10.696 & 133.837 & -1.597 & -1.061 & 0.616 & 1.505 \\ 
  ZoneE & -8.257 & -7.625 & 14.139 & 258.051 & -5.405 & -4.773 & 10.701 & 137.283 & -2.038 & -1.406 & 0.595 & 2.332 \\ 
   \hline
\end{tabular}
}
\end{subtable}
\begin{subtable}{\linewidth}
\centering
\caption{Presence piece of model C$_s$.1.2.4}
\scalebox{0.65}{
\begin{tabular}{l|cccc|cccc|cccc}
{} & \multicolumn{4}{c|}{\textbf{N=30}} & \multicolumn{4}{c|}{\textbf{N=50}} & \multicolumn{4}{c}{\textbf{N=200}} \\ \hline
Variable & Estimate & Bias & SE & MSE & Estimate & Bias & SE & MSE & Estimate & Bias & SE & MSE \\ 
  \hline
dental\_age & -0.427 & -0.126 & 0.578 & 0.350 & -0.227 & 0.074 & 0.242 & 0.064 & -0.327 & -0.026 & 0.108 & 0.012 \\ 
  Avg\_homeppm & -0.691 & -0.008 & 0.384 & 0.148 & -0.686 & -0.004 & 0.223 & 0.050 & -0.664 & 0.019 & 0.128 & 0.017 \\ 
  Tooth8 & -0.267 & -0.048 & 0.334 & 0.114 & -0.237 & -0.018 & 0.239 & 0.057 & -0.231 & -0.012 & 0.128 & 0.016 \\ 
  Tooth9 & -0.145 & -0.055 & 0.328 & 0.111 & -0.140 & -0.050 & 0.250 & 0.065 & -0.140 & -0.050 & 0.120 & 0.017 \\ 
  Tooth10 & 0.381 & -0.040 & 0.330 & 0.110 & 0.370 & -0.051 & 0.246 & 0.063 & 0.377 & -0.044 & 0.133 & 0.020 \\ 
  ZoneM & -0.899 & -0.131 & 0.484 & 0.252 & -0.837 & -0.068 & 0.309 & 0.100 & -0.874 & -0.106 & 0.163 & 0.038 \\ 
  ZoneI & -1.796 & -0.049 & 0.466 & 0.220 & -1.750 & -0.003 & 0.318 & 0.101 & -1.771 & -0.024 & 0.145 & 0.022 \\ 
  ZoneE & -2.284 & -0.127 & 0.489 & 0.255 & -2.219 & -0.062 & 0.283 & 0.084 & -2.191 & -0.035 & 0.152 & 0.024 \\ 
   \hline
\end{tabular}
}
\end{subtable}
\begin{subtable}{\linewidth}
\centering
\caption{Severity piece of model C$_s$.1.2.4}
\scalebox{0.65}{
\begin{tabular}{l|cccc|cccc|cccc}
{} & \multicolumn{4}{c|}{\textbf{N=30}} & \multicolumn{4}{c|}{\textbf{N=50}} & \multicolumn{4}{c}{\textbf{N=200}} \\ \hline
Variable & Estimate & Bias & SE & MSE & Estimate & Bias & SE & MSE & Estimate & Bias & SE & MSE \\ 
  \hline
dental\_age & -1.219 & -1.112 & 5.923 & 36.320 & -0.740 & -0.632 & 2.490 & 6.599 & -0.281 & -0.173 & 0.241 & 0.088 \\ 
  Avg\_homeppm & -3.057 & -2.829 & 6.554 & 50.956 & -1.898 & -1.669 & 4.297 & 21.248 & -0.606 & -0.377 & 0.458 & 0.352 \\ 
  Tooth8 & -0.968 & -0.954 & 3.061 & 10.281 & -0.558 & -0.544 & 2.143 & 4.890 & -0.206 & -0.192 & 0.180 & 0.069 \\ 
  Tooth9 & -0.356 & -0.448 & 3.308 & 11.143 & -0.200 & -0.292 & 1.570 & 2.549 & -0.122 & -0.214 & 0.134 & 0.064 \\ 
  Tooth10 & 1.942 & 1.717 & 4.662 & 24.682 & 1.091 & 0.866 & 2.672 & 7.889 & 0.340 & 0.115 & 0.266 & 0.084 \\ 
  ZoneM & -3.118 & -2.993 & 6.010 & 45.080 & -2.161 & -2.037 & 4.871 & 27.881 & -0.781 & -0.657 & 0.620 & 0.815 \\ 
  ZoneI & -7.674 & -7.138 & 14.997 & 275.849 & -4.590 & -4.053 & 10.245 & 121.395 & -1.592 & -1.055 & 1.168 & 2.477 \\ 
  ZoneE & -9.282 & -8.650 & 17.530 & 382.117 & -5.717 & -5.085 & 12.570 & 183.863 & -1.973 & -1.341 & 1.389 & 3.729 \\ 
   \hline
\end{tabular}
}
\end{subtable}
\end{threeparttable}
\end{table}

\addtocounter{table}{-1} 

\begin{table}[H]
\centering
\caption{ Simulation results for age 23 data generated with exchangeable and independence correlation structures for presence and severity respectively - properties of direct estimates. }
\label{ch4:table:sim:raw:exchangeable}
\begin{threeparttable}
\centering
\begin{subtable}{\linewidth}
\centering
\caption{Presence model A$_s$.2.4}
\scalebox{0.65}{
\begin{tabular}{l|cccc|cccc|cccc}
{} & \multicolumn{4}{c|}{\textbf{N=30}} & \multicolumn{4}{c|}{\textbf{N=50}} & \multicolumn{4}{c}{\textbf{N=200}} \\ \hline
Variable & Estimate & Bias & SE & MSE & Estimate & Bias & SE & MSE & Estimate & Bias & SE & MSE \\ 
  \hline
dental\_age & -0.301 & -0.000 & 0.479 & 0.230 & -0.345 & -0.044 & 0.269 & 0.074 & -0.302 & -0.001 & 0.120 & 0.014 \\ 
  Avg\_homeppm & -0.729 & -0.047 & 0.358 & 0.130 & -0.674 & 0.008 & 0.241 & 0.058 & -0.666 & 0.016 & 0.136 & 0.019 \\ 
  Tooth8 & -0.221 & -0.002 & 0.276 & 0.076 & -0.227 & -0.008 & 0.227 & 0.052 & -0.232 & -0.012 & 0.233 & 0.054 \\ 
  Tooth9 & -0.090 & -0.001 & 0.296 & 0.087 & -0.136 & -0.047 & 0.222 & 0.052 & -0.149 & -0.059 & 0.109 & 0.015 \\ 
  Tooth10 & 0.457 & 0.036 & 0.308 & 0.096 & 0.400 & -0.021 & 0.245 & 0.060 & 0.353 & -0.068 & 0.250 & 0.067 \\ 
  ZoneM & -1.416 & -0.647 & 4.431 & 20.050 & -0.911 & -0.142 & 0.333 & 0.131 & -0.863 & -0.095 & 0.417 & 0.183 \\ 
  ZoneI & -2.410 & -0.663 & 4.590 & 21.506 & -1.913 & -0.166 & 0.590 & 0.376 & -1.748 & -0.001 & 0.398 & 0.158 \\ 
  ZoneE & -2.864 & -0.707 & 4.588 & 21.550 & -2.321 & -0.165 & 0.598 & 0.385 & -2.223 & -0.066 & 0.432 & 0.191 \\ 
   \hline
\end{tabular}
}
\end{subtable}
\begin{subtable}{\linewidth}
\centering
\caption{Severity model B$_s$.1.4}
\scalebox{0.65}{
\begin{tabular}{l|cccc|cccc|cccc}
{} & \multicolumn{4}{c|}{\textbf{N=30}} & \multicolumn{4}{c|}{\textbf{N=50}} & \multicolumn{4}{c}{\textbf{N=200}} \\ \hline
Variable & Estimate & Bias & SE & MSE & Estimate & Bias & SE & MSE & Estimate & Bias & SE & MSE \\ 
  \hline
dental\_age & -1.466 & -1.358 & 5.167 & 28.538 & 0.171 & 0.279 & 3.804 & 14.547 & -0.263 & -0.155 & 0.217 & 0.071 \\ 
  Avg\_homeppm & -0.549 & -0.320 & 0.734 & 0.641 & -0.480 & -0.252 & 0.470 & 0.284 & -0.508 & -0.280 & 0.243 & 0.137 \\ 
  Tooth8 & -0.151 & -0.137 & 0.637 & 0.425 & -0.229 & -0.214 & 0.416 & 0.219 & -0.180 & -0.166 & 0.225 & 0.078 \\ 
  Tooth9 & 0.008 & -0.084 & 0.658 & 0.440 & -0.176 & -0.268 & 0.461 & 0.284 & -0.089 & -0.181 & 0.214 & 0.079 \\ 
  Tooth10 & 0.752 & 0.527 & 4.337 & 19.083 & 0.225 & 0.000 & 0.463 & 0.214 & 0.294 & 0.069 & 0.223 & 0.054 \\ 
  ZoneM & -4.116 & -3.991 & 10.818 & 132.961 & -1.822 & -1.698 & 5.342 & 31.419 & -0.689 & -0.564 & 0.419 & 0.494 \\ 
  ZoneI & -5.585 & -5.049 & 11.136 & 149.498 & -2.538 & -2.002 & 5.186 & 30.900 & -1.339 & -0.802 & 0.380 & 0.788 \\ 
  ZoneE & -5.966 & -5.334 & 11.157 & 152.931 & -2.887 & -2.255 & 5.177 & 31.885 & -1.662 & -1.030 & 0.404 & 1.224 \\ 
   \hline
\end{tabular}
}
\end{subtable}
\begin{subtable}{\linewidth}
\centering
\caption{Presence piece of model C$_s$.2.1.4}
\scalebox{0.65}{
\begin{tabular}{l|cccc|cccc|cccc}
{} & \multicolumn{4}{c|}{\textbf{N=30}} & \multicolumn{4}{c|}{\textbf{N=50}} & \multicolumn{4}{c}{\textbf{N=200}} \\ \hline
Variable & Estimate & Bias & SE & MSE & Estimate & Bias & SE & MSE & Estimate & Bias & SE & MSE \\ 
  \hline
dental\_age & -0.303 & -0.002 & 0.421 & 0.178 & -0.320 & -0.019 & 0.246 & 0.061 & -0.295 & 0.006 & 0.122 & 0.015 \\ 
  Avg\_homeppm & -0.704 & -0.021 & 0.323 & 0.105 & -0.659 & 0.024 & 0.241 & 0.059 & -0.657 & 0.026 & 0.133 & 0.018 \\ 
  Tooth8 & -0.218 & 0.001 & 0.266 & 0.071 & -0.228 & -0.008 & 0.226 & 0.051 & -0.239 & -0.019 & 0.228 & 0.052 \\ 
  Tooth9 & -0.101 & -0.011 & 0.292 & 0.086 & -0.142 & -0.052 & 0.223 & 0.053 & -0.150 & -0.060 & 0.108 & 0.015 \\ 
  Tooth10 & 0.438 & 0.017 & 0.284 & 0.081 & 0.390 & -0.031 & 0.243 & 0.060 & 0.347 & -0.074 & 0.249 & 0.067 \\ 
  ZoneM & -0.956 & -0.188 & 0.464 & 0.251 & -0.911 & -0.142 & 0.326 & 0.126 & -0.866 & -0.098 & 0.408 & 0.176 \\ 
  ZoneI & -1.916 & -0.169 & 0.747 & 0.587 & -1.904 & -0.158 & 0.582 & 0.364 & -1.749 & -0.003 & 0.393 & 0.154 \\ 
  ZoneE & -2.362 & -0.206 & 0.705 & 0.540 & -2.311 & -0.154 & 0.591 & 0.373 & -2.221 & -0.065 & 0.429 & 0.189 \\ 
   \hline
\end{tabular}
}
\end{subtable}
\begin{subtable}{\linewidth}
\centering
\caption{Severity piece of model C$_s$.2.1.4}
\scalebox{0.65}{
\begin{tabular}{l|cccc|cccc|cccc}
{} & \multicolumn{4}{c|}{\textbf{N=30}} & \multicolumn{4}{c|}{\textbf{N=50}} & \multicolumn{4}{c}{\textbf{N=200}} \\ \hline
Variable & Estimate & Bias & SE & MSE & Estimate & Bias & SE & MSE & Estimate & Bias & SE & MSE \\ 
  \hline
dental\_age & -0.791 & -0.684 & 3.814 & 15.014 & -0.407 & -0.299 & 1.108 & 1.316 & -0.239 & -0.132 & 0.141 & 0.037 \\ 
  Avg\_homeppm & -1.928 & -1.699 & 4.287 & 21.265 & -0.822 & -0.593 & 1.904 & 3.976 & -0.522 & -0.294 & 0.221 & 0.135 \\ 
  Tooth8 & -0.364 & -0.350 & 2.144 & 4.721 & -0.220 & -0.206 & 0.622 & 0.429 & -0.176 & -0.162 & 0.189 & 0.062 \\ 
  Tooth9 & 0.008 & -0.084 & 1.555 & 2.426 & -0.099 & -0.191 & 0.353 & 0.161 & -0.116 & -0.208 & 0.099 & 0.053 \\ 
  Tooth10 & 1.473 & 1.248 & 3.775 & 15.808 & 0.592 & 0.367 & 1.295 & 1.813 & 0.289 & 0.064 & 0.253 & 0.068 \\ 
  ZoneM & -2.880 & -2.756 & 6.090 & 44.680 & -1.177 & -1.053 & 2.295 & 6.374 & -0.637 & -0.513 & 0.321 & 0.366 \\ 
  ZoneI & -5.531 & -4.995 & 11.571 & 158.828 & -2.555 & -2.019 & 5.690 & 36.454 & -1.340 & -0.804 & 0.463 & 0.861 \\ 
  ZoneE & -6.882 & -6.251 & 14.277 & 242.917 & -3.090 & -2.458 & 6.615 & 49.804 & -1.710 & -1.078 & 0.573 & 1.490 \\ 
   \hline
\end{tabular}
}
\end{subtable}
\end{threeparttable}
\end{table}

\addtocounter{table}{-1} 

\begin{table}[H]
\centering
\caption{ Simulation results for age 23 data generated with exchangeable and exchangeable correlation structures for presence and severity respectively - properties of direct estimates. }
\label{ch4:table:sim:raw:exchangeable}
\begin{threeparttable}
\centering
\begin{subtable}{\linewidth}
\centering
\caption{Presence model A$_s$.2.4}
\scalebox{0.65}{
\begin{tabular}{l|cccc|cccc|cccc}
{} & \multicolumn{4}{c|}{\textbf{N=30}} & \multicolumn{4}{c|}{\textbf{N=50}} & \multicolumn{4}{c}{\textbf{N=200}} \\ \hline
Variable & Estimate & Bias & SE & MSE & Estimate & Bias & SE & MSE & Estimate & Bias & SE & MSE \\ 
  \hline
dental\_age & -0.234 & 0.067 & 0.593 & 0.356 & -0.301 & 0.000 & 0.254 & 0.064 & -0.293 & 0.008 & 0.115 & 0.013 \\ 
  Avg\_homeppm & -0.681 & 0.002 & 0.292 & 0.085 & -0.684 & -0.001 & 0.245 & 0.060 & -0.648 & 0.035 & 0.128 & 0.018 \\ 
  Tooth8 & -0.220 & -0.001 & 0.286 & 0.082 & -0.220 & -0.001 & 0.238 & 0.057 & -0.221 & -0.002 & 0.190 & 0.036 \\ 
  Tooth9 & -0.126 & -0.037 & 0.290 & 0.086 & -0.137 & -0.047 & 0.235 & 0.058 & -0.125 & -0.035 & 0.137 & 0.020 \\ 
  Tooth10 & 0.397 & -0.024 & 0.310 & 0.096 & 0.366 & -0.056 & 0.263 & 0.072 & 0.380 & -0.041 & 0.221 & 0.050 \\ 
  ZoneM & -0.897 & -0.128 & 0.507 & 0.274 & -0.846 & -0.077 & 0.362 & 0.137 & -0.855 & -0.086 & 0.270 & 0.081 \\ 
  ZoneI & -1.870 & -0.123 & 0.692 & 0.495 & -1.772 & -0.025 & 0.512 & 0.263 & -1.748 & -0.001 & 0.256 & 0.066 \\ 
  ZoneE & -2.258 & -0.102 & 0.674 & 0.465 & -2.232 & -0.075 & 0.495 & 0.251 & -2.206 & -0.050 & 0.281 & 0.081 \\ 
   \hline
\end{tabular}
}
\end{subtable}
\begin{subtable}{\linewidth}
\centering
\caption{Severity model B$_s$.2.4}
\scalebox{0.65}{
\begin{tabular}{l|cccc|cccc|cccc}
{} & \multicolumn{4}{c|}{\textbf{N=30}} & \multicolumn{4}{c|}{\textbf{N=50}} & \multicolumn{4}{c}{\textbf{N=200}} \\ \hline
Variable & Estimate & Bias & SE & MSE & Estimate & Bias & SE & MSE & Estimate & Bias & SE & MSE \\ 
  \hline
dental\_age & 0.703 & 0.810 & 9.286 & 86.885 & 0.554 & 0.662 & 4.403 & 19.822 & 0.049 & 0.157 & 0.319 & 0.126 \\ 
  Avg\_homeppm & -0.244 & -0.015 & 1.139 & 1.298 & -0.296 & -0.067 & 0.538 & 0.294 & -0.280 & -0.051 & 0.270 & 0.076 \\ 
  Tooth8 & -0.026 & -0.011 & 0.654 & 0.428 & -0.105 & -0.091 & 0.492 & 0.250 & -0.037 & -0.023 & 0.412 & 0.170 \\ 
  Tooth9 & -0.968 & -1.060 & 6.087 & 38.177 & -0.103 & -0.195 & 0.440 & 0.232 & -0.033 & -0.125 & 0.256 & 0.081 \\ 
  Tooth10 & 0.301 & 0.077 & 0.750 & 0.568 & 0.148 & -0.076 & 0.626 & 0.398 & 0.218 & -0.007 & 0.431 & 0.186 \\ 
  ZoneM & 0.800 & 0.924 & 9.154 & 84.650 & -0.814 & -0.690 & 3.906 & 15.737 & -0.342 & -0.217 & 0.637 & 0.453 \\ 
  ZoneI & 0.339 & 0.875 & 9.335 & 87.903 & -1.305 & -0.769 & 3.930 & 16.033 & -0.712 & -0.176 & 0.578 & 0.365 \\ 
  ZoneE & 0.064 & 0.696 & 9.357 & 88.035 & -1.569 & -0.938 & 3.935 & 16.366 & -0.902 & -0.271 & 0.561 & 0.388 \\ 
   \hline
\end{tabular}
}
\end{subtable}
\begin{subtable}{\linewidth}
\centering
\caption{Presence piece of model C$_s$.2.2.4}
\scalebox{0.65}{
\begin{tabular}{l|cccc|cccc|cccc}
{} & \multicolumn{4}{c|}{\textbf{N=30}} & \multicolumn{4}{c|}{\textbf{N=50}} & \multicolumn{4}{c}{\textbf{N=200}} \\ \hline
Variable & Estimate & Bias & SE & MSE & Estimate & Bias & SE & MSE & Estimate & Bias & SE & MSE \\ 
  \hline
dental\_age & -0.218 & 0.083 & 0.568 & 0.329 & -0.292 & 0.009 & 0.249 & 0.062 & -0.291 & 0.010 & 0.115 & 0.013 \\ 
  Avg\_homeppm & -0.656 & 0.026 & 0.278 & 0.078 & -0.658 & 0.025 & 0.239 & 0.058 & -0.640 & 0.042 & 0.128 & 0.018 \\ 
  Tooth8 & -0.207 & 0.012 & 0.268 & 0.072 & -0.204 & 0.016 & 0.231 & 0.054 & -0.216 & 0.003 & 0.191 & 0.037 \\ 
  Tooth9 & -0.122 & -0.033 & 0.270 & 0.074 & -0.127 & -0.038 & 0.226 & 0.052 & -0.122 & -0.032 & 0.135 & 0.019 \\ 
  Tooth10 & 0.409 & -0.012 & 0.294 & 0.087 & 0.370 & -0.052 & 0.247 & 0.064 & 0.380 & -0.041 & 0.222 & 0.051 \\ 
  ZoneM & -0.857 & -0.088 & 0.471 & 0.229 & -0.825 & -0.057 & 0.360 & 0.133 & -0.851 & -0.082 & 0.264 & 0.077 \\ 
  ZoneI & -1.812 & -0.065 & 0.669 & 0.451 & -1.743 & 0.003 & 0.517 & 0.267 & -1.738 & 0.009 & 0.255 & 0.065 \\ 
  ZoneE & -2.198 & -0.042 & 0.657 & 0.433 & -2.201 & -0.044 & 0.505 & 0.257 & -2.194 & -0.038 & 0.279 & 0.079 \\ 
   \hline
\end{tabular}
}
\end{subtable}
\begin{subtable}{\linewidth}
\centering
\caption{Severity piece of model C$_s$.2.2.4}
\scalebox{0.65}{
\begin{tabular}{l|cccc|cccc|cccc}
{} & \multicolumn{4}{c|}{\textbf{N=30}} & \multicolumn{4}{c|}{\textbf{N=50}} & \multicolumn{4}{c}{\textbf{N=200}} \\ \hline
Variable & Estimate & Bias & SE & MSE & Estimate & Bias & SE & MSE & Estimate & Bias & SE & MSE \\ 
  \hline
dental\_age & 0.271 & 0.379 & 2.252 & 5.216 & -0.123 & -0.016 & 0.365 & 0.133 & -0.096 & 0.012 & 0.100 & 0.010 \\ 
  Avg\_homeppm & 0.075 & 0.304 & 3.820 & 14.684 & -0.424 & -0.195 & 1.571 & 2.506 & -0.210 & 0.019 & 0.189 & 0.036 \\ 
  Tooth8 & -0.064 & -0.050 & 1.687 & 2.849 & -0.126 & -0.112 & 0.702 & 0.506 & -0.091 & -0.077 & 0.104 & 0.017 \\ 
  Tooth9 & -0.013 & -0.105 & 0.885 & 0.795 & -0.064 & -0.156 & 0.224 & 0.075 & -0.047 & -0.139 & 0.067 & 0.024 \\ 
  Tooth10 & -0.262 & -0.486 & 3.337 & 11.369 & 0.228 & 0.003 & 0.446 & 0.199 & 0.100 & -0.124 & 0.136 & 0.034 \\ 
  ZoneM & 0.494 & 0.618 & 5.630 & 32.079 & -0.513 & -0.389 & 2.258 & 5.251 & -0.308 & -0.184 & 0.289 & 0.118 \\ 
  ZoneI & 0.037 & 0.573 & 9.047 & 82.186 & -1.125 & -0.589 & 3.876 & 15.371 & -0.604 & -0.068 & 0.566 & 0.325 \\ 
  ZoneE & 0.204 & 0.836 & 11.324 & 128.937 & -1.418 & -0.786 & 4.947 & 25.090 & -0.770 & -0.138 & 0.699 & 0.507 \\ 
   \hline
\end{tabular}
}
\end{subtable}
\end{threeparttable}
\end{table}

\subsection{Properties of Standardized Estimates}


\addtocounter{table}{-1} 

\begin{table}[H]
\centering
\caption{ Simulation results for age 23 data generated with independence and exchangeable correlation structures for presence and severity respectively - properties of standardized estimates arising from the presence model A$_s$.1.4. }
\label{ch4:table:sim:std:pres:independence}
\begin{threeparttable}
\centering
\begin{subtable}{\linewidth}
\centering
\caption{N=30}
\scalebox{0.65}{
\begin{tabular}{lcccc|cccc}
Variable & \makecell{Standardized\\Estimate} & \makecell{Bias\\(Standardized\\Estimate)} & \makecell{SE\\(Standardized\\Estimate)} & \makecell{MSE\\(Standardized\\Estimate)} & \makecell{Standardized\\James-Stein\\Estimate} & \makecell{Bias\\(Standardized\\James-Stein\\Estimate)} & \makecell{SE\\(Standardized\\James-Stein\\Estimate)} & \makecell{MSE\\(Standardized\\James-Stein\\Estimate)} \\ 
  \hline
Dental\_age & -0.198 & -0.011 & 0.26 & 0.068 & -0.142 & 0.045 & 0.233 & 0.056 \\ 
  Avg\_homeppm & -0.365 & 0.017 & 0.203 & 0.042 & -0.328 & 0.054 & 0.195 & 0.041 \\ 
  Tooth8 & -0.145 & -0.021 & 0.186 & 0.035 & -0.092 & 0.033 & 0.200 & 0.041 \\ 
  Tooth9 & -0.081 & -0.030 & 0.187 & 0.036 & -0.048 & 0.002 & 0.237 & 0.056 \\ 
  Tooth10 & 0.197 & -0.025 & 0.181 & 0.034 & 0.154 & -0.068 & 0.198 & 0.044 \\ 
  ZoneM & -0.336 & -0.023 & 0.169 & 0.029 & -0.293 & 0.021 & 0.176 & 0.032 \\ 
  ZoneI & -0.714 & 0.047 & 0.157 & 0.027 & -0.69 & 0.071 & 0.161 & 0.031 \\ 
  ZoneE & -0.917 & 0.045 & 0.163 & 0.029 & -0.898 & 0.064 & 0.165 & 0.031 \\ 
  \hline Average MSE &  &  &  & 0.038 &  &  &  & 0.042 \\ 
   \hline
\end{tabular}
}
\end{subtable}
\begin{subtable}{\linewidth}
\centering
\caption{N=50}
\scalebox{0.65}{
\begin{tabular}{lcccc|cccc}
Variable & \makecell{Standardized\\Estimate} & \makecell{Bias\\(Standardized\\Estimate)} & \makecell{SE\\(Standardized\\Estimate)} & \makecell{MSE\\(Standardized\\Estimate)} & \makecell{Standardized\\James-Stein\\Estimate} & \makecell{Bias\\(Standardized\\James-Stein\\Estimate)} & \makecell{SE\\(Standardized\\James-Stein\\Estimate)} & \makecell{MSE\\(Standardized\\James-Stein\\Estimate)} \\ 
  \hline
Dental\_age & -0.133 & 0.055 & 0.159 & 0.080 & -0.085 & 0.103 & 0.142 & 0.031 \\ 
  Avg\_homeppm & -0.402 & -0.020 & 0.156 & 0.025 & -0.377 & 0.005 & 0.152 & 0.023 \\ 
  Tooth8 & -0.133 & -0.009 & 0.136 & 0.019 & -0.087 & 0.038 & 0.151 & 0.024 \\ 
  Tooth9 & -0.073 & -0.022 & 0.141 & 0.020 & -0.062 & -0.012 & 0.171 & 0.029 \\ 
  Tooth10 & 0.197 & -0.025 & 0.128 & 0.017 & 0.145 & -0.077 & 0.150 & 0.080 \\ 
  ZoneM & -0.323 & -0.010 & 0.114 & 0.013 & -0.29 & 0.023 & 0.127 & 0.017 \\ 
  ZoneI & -0.731 & 0.030 & 0.123 & 0.016 & -0.717 & 0.044 & 0.125 & 0.017 \\ 
  ZoneE & -0.938 & 0.024 & 0.120 & 0.015 & -0.928 & 0.035 & 0.121 & 0.016 \\ 
  \hline Average MSE &  &  &  & 0.019 &  &  &  & 0.023 \\ 
   \hline
\end{tabular}
}
\end{subtable}
\begin{subtable}{\linewidth}
\centering
\caption{N=200}
\scalebox{0.65}{
\begin{tabular}{lcccc|cccc}
Variable & \makecell{Standardized\\Estimate} & \makecell{Bias\\(Standardized\\Estimate)} & \makecell{SE\\(Standardized\\Estimate)} & \makecell{MSE\\(Standardized\\Estimate)} & \makecell{Standardized\\James-Stein\\Estimate} & \makecell{Bias\\(Standardized\\James-Stein\\Estimate)} & \makecell{SE\\(Standardized\\James-Stein\\Estimate)} & \makecell{MSE\\(Standardized\\James-Stein\\Estimate)} \\ 
  \hline
Dental\_age & -0.218 & -0.031 & 0.104 & 0.012 & -0.199 & -0.011 & 0.103 & 0.011 \\ 
  Avg\_homeppm & -0.386 & -0.004 & 0.075 & 0.006 & -0.381 & 0.001 & 0.075 & 0.006 \\ 
  Tooth8 & -0.131 & -0.006 & 0.074 & 0.006 & -0.109 & 0.016 & 0.087 & 0.008 \\ 
  Tooth9 & -0.075 & -0.025 & 0.070 & 0.006 & -0.059 & -0.008 & 0.068 & 0.005 \\ 
  Tooth10 & 0.202 & -0.020 & 0.076 & 0.006 & 0.189 & -0.033 & 0.076 & 0.007 \\ 
  ZoneM & -0.349 & -0.035 & 0.061 & 0.005 & -0.341 & -0.080 & 0.062 & 0.005 \\ 
  ZoneI & -0.759 & 0.003 & 0.063 & 0.004 & -0.755 & 0.006 & 0.063 & 0.004 \\ 
  ZoneE & -0.953 & 0.009 & 0.064 & 0.004 & -0.951 & 0.011 & 0.064 & 0.004 \\ 
  \hline Average MSE &  &  &  & 0.006 &  &  &  & 0.006 \\ 
   \hline
\end{tabular}
}
\end{subtable}
\end{threeparttable}
\end{table}

\addtocounter{table}{-1} 

\begin{table}[H]
\centering
\caption{ Simulation results for age 23 data generated with exchangeable and independence correlation structures for presence and severity respectively - properties of standardized estimates arising from the presence model A$_s$.2.4. }
\label{ch4:table:sim:std:pres:exchangeable}
\begin{threeparttable}
\centering
\begin{subtable}{\linewidth}
\centering
\caption{N=30}
\scalebox{0.65}{
\begin{tabular}{lcccc|cccc}
Variable & \makecell{Standardized\\Estimate} & \makecell{Bias\\(Standardized\\Estimate)} & \makecell{SE\\(Standardized\\Estimate)} & \makecell{MSE\\(Standardized\\Estimate)} & \makecell{Standardized\\James-Stein\\Estimate} & \makecell{Bias\\(Standardized\\James-Stein\\Estimate)} & \makecell{SE\\(Standardized\\James-Stein\\Estimate)} & \makecell{MSE\\(Standardized\\James-Stein\\Estimate)} \\ 
  \hline
Dental\_age & -0.201 & -0.014 & 0.262 & 0.069 & -0.153 & 0.035 & 0.247 & 0.062 \\ 
  Avg\_homeppm & -0.418 & -0.036 & 0.204 & 0.043 & -0.379 & 0.003 & 0.200& 0.040 \\ 
  Tooth8 & -0.121 & 0.004 & 0.153 & 0.023 & -0.060 & 0.065 & 0.179 & 0.036 \\ 
  Tooth9 & -0.051 & -0.001 & 0.167 & 0.080 & -0.038 & 0.012 & 0.192 & 0.037 \\ 
  Tooth10 & 0.226 & 0.004 & 0.162 & 0.026 & 0.150 & -0.071 & 0.186 & 0.040 \\ 
  ZoneM & -0.334 & -0.021 & 0.156 & 0.025 & -0.249 & 0.065 & 0.285 & 0.085 \\ 
  ZoneI & -0.705 & 0.056 & 0.258 & 0.070 & -0.667 & 0.094 & 0.309 & 0.105 \\ 
  ZoneE & -0.898 & 0.064 & 0.267 & 0.075 & -0.869 & 0.093 & 0.309 & 0.104 \\ 
  \hline Average MSE &  &  &  & 0.045 &  &  &  & 0.064 \\ 
   \hline
\end{tabular}
}
\end{subtable}
\begin{subtable}{\linewidth}
\centering
\caption{N=50}
\scalebox{0.65}{
\begin{tabular}{lcccc|cccc}
Variable & \makecell{Standardized\\Estimate} & \makecell{Bias\\(Standardized\\Estimate)} & \makecell{SE\\(Standardized\\Estimate)} & \makecell{MSE\\(Standardized\\Estimate)} & \makecell{Standardized\\James-Stein\\Estimate} & \makecell{Bias\\(Standardized\\James-Stein\\Estimate)} & \makecell{SE\\(Standardized\\James-Stein\\Estimate)} & \makecell{MSE\\(Standardized\\James-Stein\\Estimate)} \\ 
  \hline
Dental\_age & -0.226 & -0.039 & 0.221 & 0.051 & -0.178 & 0.009 & 0.217 & 0.047 \\ 
  Avg\_homeppm & -0.406 & -0.024 & 0.159 & 0.026 & -0.384 & -0.002 & 0.157 & 0.025 \\ 
  Tooth8 & -0.124 & 0.001 & 0.124 & 0.015 & -0.085 & 0.040 & 0.178 & 0.033 \\ 
  Tooth9 & -0.071 & -0.021 & 0.120 & 0.015 & -0.054 & -0.004 & 0.168 & 0.080 \\ 
  Tooth10 & 0.198 & -0.024 & 0.116 & 0.014 & 0.149 & -0.073 & 0.120 & 0.020 \\ 
  ZoneM & -0.333 & -0.020 & 0.103 & 0.011 & -0.299 & 0.014 & 0.116 & 0.014 \\ 
  ZoneI & -0.761 & 0.000 & 0.186 & 0.035 & -0.747 & 0.014 & 0.189 & 0.036 \\ 
  ZoneE & -0.949 & 0.013 & 0.191 & 0.037 & -0.939 & 0.024 & 0.193 & 0.038 \\ 
  \hline Average MSE &  &  &  & 0.025 &  &  &  & 0.030 \\ 
   \hline
\end{tabular}
}
\end{subtable}
\begin{subtable}{\linewidth}
\centering
\caption{N=200}
\scalebox{0.65}{
\begin{tabular}{lcccc|cccc}
Variable & \makecell{Standardized\\Estimate} & \makecell{Bias\\(Standardized\\Estimate)} & \makecell{SE\\(Standardized\\Estimate)} & \makecell{MSE\\(Standardized\\Estimate)} & \makecell{Standardized\\James-Stein\\Estimate} & \makecell{Bias\\(Standardized\\James-Stein\\Estimate)} & \makecell{SE\\(Standardized\\James-Stein\\Estimate)} & \makecell{MSE\\(Standardized\\James-Stein\\Estimate)} \\ 
  \hline
Dental\_age & -0.197 & -0.010 & 0.098 & 0.010 & -0.179 & 0.009 & 0.098 & 0.010 \\ 
  Avg\_homeppm & -0.384 & -0.002 & 0.088 & 0.008 & -0.379 & 0.003 & 0.087 & 0.008 \\ 
  Tooth8 & -0.129 & -0.004 & 0.129 & 0.017 & -0.122 & 0.003 & 0.127 & 0.016 \\ 
  Tooth9 & -0.082 & -0.032 & 0.063 & 0.005 & -0.054 & -0.004 & 0.070 & 0.005 \\ 
  Tooth10 & 0.184 & -0.038 & 0.132 & 0.019 & 0.172 & -0.050 & 0.132 & 0.020 \\ 
  ZoneM & -0.335 & -0.021 & 0.139 & 0.020 & -0.327 & -0.013 & 0.141 & 0.020 \\ 
  ZoneI & -0.739 & 0.022 & 0.128 & 0.017 & -0.736 & 0.026 & 0.129 & 0.017 \\ 
  ZoneE & -0.957 & 0.005 & 0.118 & 0.014 & -0.955 & 0.008 & 0.118 & 0.014 \\ 
  \hline Average MSE &  &  &  & 0.014 &  &  &  & 0.014 \\ 
   \hline
\end{tabular}
}
\end{subtable}
\end{threeparttable}
\end{table}

\addtocounter{table}{-1} 

\begin{table}[H]
\centering
\caption{ Simulation results for age 23 data generated with exchangeable and exchangeable correlation structures for presence and severity respectively - properties of standardized estimates arising from the presence model A$_s$.2.4. }
\label{ch4:table:sim:std:pres:exchangeable}
\begin{threeparttable}
\centering
\begin{subtable}{\linewidth}
\centering
\caption{N=30}
\scalebox{0.65}{
\begin{tabular}{lcccc|cccc}
Variable & \makecell{Standardized\\Estimate} & \makecell{Bias\\(Standardized\\Estimate)} & \makecell{SE\\(Standardized\\Estimate)} & \makecell{MSE\\(Standardized\\Estimate)} & \makecell{Standardized\\James-Stein\\Estimate} & \makecell{Bias\\(Standardized\\James-Stein\\Estimate)} & \makecell{SE\\(Standardized\\James-Stein\\Estimate)} & \makecell{MSE\\(Standardized\\James-Stein\\Estimate)} \\ 
  \hline
Dental\_age & -0.138 & 0.050 & 0.266 & 0.073 & -0.093 & 0.095 & 0.236 & 0.065 \\ 
  Avg\_homeppm & -0.378 & 0.004 & 0.195 & 0.038 & -0.337 & 0.045 & 0.190 & 0.038 \\ 
  Tooth8 & -0.126 & -0.002 & 0.160 & 0.026 & -0.039 & 0.085 & 0.186 & 0.042 \\ 
  Tooth9 & -0.069 & -0.018 & 0.163 & 0.027 & -0.046 & 0.005 & 0.217 & 0.047 \\ 
  Tooth10 & 0.197 & -0.025 & 0.151 & 0.023 & 0.121 & -0.101 & 0.175 & 0.041 \\ 
  ZoneM & -0.311 & 0.003 & 0.152 & 0.023 & -0.258 & 0.055 & 0.161 & 0.029 \\ 
  ZoneI & -0.728 & 0.034 & 0.212 & 0.046 & -0.703 & 0.058 & 0.216 & 0.050 \\ 
  ZoneE & -0.897 & 0.065 & 0.230 & 0.057 & -0.878 & 0.084 & 0.232 & 0.061 \\ 
  \hline Average MSE &  &  &  & 0.039 &  &  &  & 0.047 \\ 
   \hline
\end{tabular}
}
\end{subtable}
\begin{subtable}{\linewidth}
\centering
\caption{N=50}
\scalebox{0.65}{
\begin{tabular}{lcccc|cccc}
Variable & \makecell{Standardized\\Estimate} & \makecell{Bias\\(Standardized\\Estimate)} & \makecell{SE\\(Standardized\\Estimate)} & \makecell{MSE\\(Standardized\\Estimate)} & \makecell{Standardized\\James-Stein\\Estimate} & \makecell{Bias\\(Standardized\\James-Stein\\Estimate)} & \makecell{SE\\(Standardized\\James-Stein\\Estimate)} & \makecell{MSE\\(Standardized\\James-Stein\\Estimate)} \\ 
  \hline
Dental\_age & -0.173 & 0.014 & 0.176 & 0.031 & -0.120 & 0.068 & 0.168 & 0.033 \\ 
  Avg\_homeppm & -0.399 & -0.017 & 0.173 & 0.030 & -0.375 & 0.007 & 0.174 & 0.030 \\ 
  Tooth8 & -0.121 & 0.003 & 0.132 & 0.018 & -0.071 & 0.054 & 0.161 & 0.029 \\ 
  Tooth9 & -0.079 & -0.080 & 0.134 & 0.019 & -0.075 & -0.024 & 0.181 & 0.033 \\ 
  Tooth10 & 0.188 & -0.034 & 0.135 & 0.019 & 0.152 & -0.070 & 0.146 & 0.026 \\ 
  ZoneM & -0.329 & -0.016 & 0.124 & 0.016 & -0.276 & 0.037 & 0.246 & 0.062 \\ 
  ZoneI & -0.725 & 0.037 & 0.163 & 0.080 & -0.710 & 0.051 & 0.167 & 0.030 \\ 
  ZoneE & -0.940 & 0.023 & 0.147 & 0.022 & -0.929 & 0.033 & 0.148 & 0.023 \\ 
  \hline Average MSE &  &  &  & 0.023 &  &  &  & 0.033 \\ 
   \hline
\end{tabular}
}
\end{subtable}
\begin{subtable}{\linewidth}
\centering
\caption{N=200}
\scalebox{0.65}{
\begin{tabular}{lcccc|cccc}
Variable & \makecell{Standardized\\Estimate} & \makecell{Bias\\(Standardized\\Estimate)} & \makecell{SE\\(Standardized\\Estimate)} & \makecell{MSE\\(Standardized\\Estimate)} & \makecell{Standardized\\James-Stein\\Estimate} & \makecell{Bias\\(Standardized\\James-Stein\\Estimate)} & \makecell{SE\\(Standardized\\James-Stein\\Estimate)} & \makecell{MSE\\(Standardized\\James-Stein\\Estimate)} \\ 
  \hline
Dental\_age & -0.187 & 0.001 & 0.099 & 0.010 & -0.167 & 0.021 & 0.098 & 0.010 \\ 
  Avg\_homeppm & -0.374 & 0.008 & 0.081 & 0.007 & -0.368 & 0.014 & 0.080 & 0.007 \\ 
  Tooth8 & -0.124 & 0.001 & 0.106 & 0.011 & -0.109 & 0.015 & 0.109 & 0.012 \\ 
  Tooth9 & -0.069 & -0.018 & 0.076 & 0.006 & -0.053 & -0.003 & 0.072 & 0.005 \\ 
  Tooth10 & 0.198 & -0.024 & 0.115 & 0.014 & 0.186 & -0.036 & 0.117 & 0.015 \\ 
  ZoneM & -0.339 & -0.026 & 0.100 & 0.011 & -0.331 & -0.017 & 0.103 & 0.011 \\ 
  ZoneI & -0.750 & 0.011 & 0.087 & 0.008 & -0.747 & 0.014 & 0.088 & 0.008 \\ 
  ZoneE & -0.961 & 0.001 & 0.082 & 0.007 & -0.959 & 0.004 & 0.082 & 0.007 \\ 
  \hline Average MSE &  &  &  & 0.009 &  &  &  & 0.009 \\ 
   \hline
\end{tabular}
}
\end{subtable}
\end{threeparttable}
\end{table}



\addtocounter{table}{-1} 

\begin{table}[H]
\centering
\caption{ Simulation results for age 23 data generated with independence and exchangeable correlation structures for presence and severity respectively - properties of standardized estimates arising from the severity model B$_s$.2.4. }
\label{ch4:table:sim:std:sev:exchangeable}
\begin{threeparttable}
\centering
\begin{subtable}{\linewidth}
\centering
\caption{N=30}
\scalebox{0.65}{
\begin{tabular}{lcccc|cccc}
Variable & \makecell{Standardized\\Estimate} & \makecell{Bias\\(Standardized\\Estimate)} & \makecell{SE\\(Standardized\\Estimate)} & \makecell{MSE\\(Standardized\\Estimate)} & \makecell{Standardized\\James-Stein\\Estimate} & \makecell{Bias\\(Standardized\\James-Stein\\Estimate)} & \makecell{SE\\(Standardized\\James-Stein\\Estimate)} & \makecell{MSE\\(Standardized\\James-Stein\\Estimate)} \\ 
  \hline
Dental\_age & -0.008 & 0.033 & 0.029 & 0.002 & -0.022 & 0.018 & 1.359 & 1.846 \\ 
  Avg\_homeppm & -0.187 & -0.118 & 0.187 & 0.049 & -0.141 & -0.072 & 0.160 & 0.031 \\ 
  Tooth8 & -0.080 & -0.024 & 0.149 & 0.023 & 0.015 & 0.019 & 0.201 & 0.041 \\ 
  Tooth9 & -0.011 & -0.038 & 0.147 & 0.023 & -0.001 & -0.029 & 0.169 & 0.029 \\ 
  Tooth10 & 0.120 & 0.060 & 0.161 & 0.030 & 0.065 & 0.005 & 0.190 & 0.036 \\ 
  ZoneM & -0.078 & -0.055 & 0.165 & 0.030 & 0.078 & 0.101 & 1.529 & 2.347 \\ 
  ZoneI & -0.157 & -0.046 & 0.174 & 0.032 & 0.035 & 0.147 & 0.521 & 0.293 \\ 
  ZoneE & -0.195 & -0.060 & 0.188 & 0.039 & -0.018 & 0.117 & 0.477 & 0.241 \\ 
  \hline Average MSE &  &  &  & 0.080 &  &  &  & 0.608 \\ 
   \hline
\end{tabular}
}
\end{subtable}
\begin{subtable}{\linewidth}
\centering
\caption{N=50}
\scalebox{0.65}{
\begin{tabular}{lcccc|cccc}
Variable & \makecell{Standardized\\Estimate} & \makecell{Bias\\(Standardized\\Estimate)} & \makecell{SE\\(Standardized\\Estimate)} & \makecell{MSE\\(Standardized\\Estimate)} & \makecell{Standardized\\James-Stein\\Estimate} & \makecell{Bias\\(Standardized\\James-Stein\\Estimate)} & \makecell{SE\\(Standardized\\James-Stein\\Estimate)} & \makecell{MSE\\(Standardized\\James-Stein\\Estimate)} \\ 
  \hline
Dental\_age & -0.003 & 0.037 & 0.032 & 0.002 & 0.132 & 0.172 & 1.334 & 1.808 \\ 
  Avg\_homeppm & -0.196 & -0.128 & 0.136 & 0.035 & -0.157 & -0.088 & 0.124 & 0.023 \\ 
  Tooth8 & -0.075 & -0.071 & 0.124 & 0.020 & -0.034 & -0.030 & 0.113 & 0.014 \\ 
  Tooth9 & -0.060 & -0.088 & 0.122 & 0.023 & -0.038 & -0.066 & 0.149 & 0.026 \\ 
  Tooth10 & 0.078 & 0.018 & 0.141 & 0.020 & 0.055 & -0.005 & 0.135 & 0.018 \\ 
  ZoneM & -0.086 & -0.062 & 0.096 & 0.013 & 0.094 & 0.118 & 0.613 & 0.389 \\ 
  ZoneI & -0.194 & -0.082 & 0.132 & 0.024 & -0.105 & 0.007 & 0.277 & 0.077 \\ 
  ZoneE & -0.249 & -0.114 & 0.145 & 0.034 & -0.176 & -0.041 & 0.266 & 0.072 \\ 
  \hline Average MSE &  &  &  & 0.021 &  &  &  & 0.303 \\ 
   \hline
\end{tabular}
}
\end{subtable}
\begin{subtable}{\linewidth}
\centering
\caption{N=200}
\scalebox{0.65}{
\begin{tabular}{lcccc|cccc}
Variable & \makecell{Standardized\\Estimate} & \makecell{Bias\\(Standardized\\Estimate)} & \makecell{SE\\(Standardized\\Estimate)} & \makecell{MSE\\(Standardized\\Estimate)} & \makecell{Standardized\\James-Stein\\Estimate} & \makecell{Bias\\(Standardized\\James-Stein\\Estimate)} & \makecell{SE\\(Standardized\\James-Stein\\Estimate)} & \makecell{MSE\\(Standardized\\James-Stein\\Estimate)} \\ 
  \hline
Dental\_age & -0.013 & 0.027 & 0.037 & 0.002 & 0.102 & 0.142 & 1.093 & 1.214 \\ 
  Avg\_homeppm & -0.166 & -0.098 & 0.080 & 0.016 & -0.157 & -0.088 & 0.078 & 0.014 \\ 
  Tooth8 & -0.069 & -0.065 & 0.131 & 0.021 & -0.063 & -0.058 & 0.120 & 0.018 \\ 
  Tooth9 & -0.035 & -0.063 & 0.059 & 0.007 & -0.033 & -0.060 & 0.085 & 0.011 \\ 
  Tooth10 & 0.103 & 0.043 & 0.108 & 0.013 & 0.079 & 0.019 & 0.114 & 0.013 \\ 
  ZoneM & -0.123 & -0.100 & 0.078 & 0.016 & -0.107 & -0.083 & 0.079 & 0.013 \\ 
  ZoneI & -0.266 & -0.154 & 0.080 & 0.030 & -0.256 & -0.144 & 0.083 & 0.080 \\ 
  ZoneE & -0.345 & -0.210 & 0.075 & 0.050 & -0.338 & -0.203 & 0.076 & 0.047 \\ 
  \hline Average MSE &  &  &  & 0.019 &  &  &  & 0.170 \\ 
   \hline
\end{tabular}
}
\end{subtable}
\end{threeparttable}
\end{table}

\addtocounter{table}{-1} 

\begin{table}[H]
\centering
\caption{ Simulation results for age 23 data generated with exchangeable and independence correlation structures for presence and severity respectively - properties of standardized estimates arising from the severity model B$_s$.1.4. }
\label{ch4:table:sim:std:sev:independence}
\begin{threeparttable}
\centering
\begin{subtable}{\linewidth}
\centering
\caption{N=30}
\scalebox{0.65}{
\begin{tabular}{lcccc|cccc}
Variable & \makecell{Standardized\\Estimate} & \makecell{Bias\\(Standardized\\Estimate)} & \makecell{SE\\(Standardized\\Estimate)} & \makecell{MSE\\(Standardized\\Estimate)} & \makecell{Standardized\\James-Stein\\Estimate} & \makecell{Bias\\(Standardized\\James-Stein\\Estimate)} & \makecell{SE\\(Standardized\\James-Stein\\Estimate)} & \makecell{MSE\\(Standardized\\James-Stein\\Estimate)} \\ 
  \hline
Dental\_age & -0.038 & 0.002 & 0.122 & 0.015 & 0.153 & 0.194 & 1.090 & 1.225 \\ 
  Avg\_homeppm & -0.126 & -0.058 & 0.166 & 0.031 & -0.078 & -0.010 & 0.124 & 0.016 \\ 
  Tooth8 & -0.038 & -0.034 & 0.173 & 0.031 & 0.004 & 0.008 & 0.184 & 0.034 \\ 
  Tooth9 & -0.002 & -0.030 & 0.166 & 0.080 & 0.009 & -0.019 & 0.147 & 0.022 \\ 
  Tooth10 & 0.102 & 0.042 & 0.330 & 0.111 & 0.098 & 0.038 & 0.330 & 0.110 \\ 
  ZoneM & -0.071 & -0.047 & 0.154 & 0.026 & -0.080 & -0.056 & 0.713 & 0.511 \\ 
  ZoneI & -0.153 & -0.041 & 0.252 & 0.065 & 0.161 & 0.272 & 1.006 & 1.087 \\ 
  ZoneE & -0.188 & -0.053 & 0.245 & 0.063 & -0.061 & 0.075 & 0.432 & 0.192 \\ 
  \hline Average MSE &  &  &  & 0.046 &  &  &  & 0.400 \\ 
   \hline
\end{tabular}
}
\end{subtable}
\begin{subtable}{\linewidth}
\centering
\caption{N=50}
\scalebox{0.65}{
\begin{tabular}{lcccc|cccc}
Variable & \makecell{Standardized\\Estimate} & \makecell{Bias\\(Standardized\\Estimate)} & \makecell{SE\\(Standardized\\Estimate)} & \makecell{MSE\\(Standardized\\Estimate)} & \makecell{Standardized\\James-Stein\\Estimate} & \makecell{Bias\\(Standardized\\James-Stein\\Estimate)} & \makecell{SE\\(Standardized\\James-Stein\\Estimate)} & \makecell{MSE\\(Standardized\\James-Stein\\Estimate)} \\ 
  \hline
Dental\_age & -0.019 & 0.021 & 0.065 & 0.005 & 0.051 & 0.092 & 0.744 & 0.562 \\ 
  Avg\_homeppm & -0.131 & -0.063 & 0.131 & 0.021 & -0.093 & -0.024 & 0.108 & 0.012 \\ 
  Tooth8 & -0.068 & -0.064 & 0.123 & 0.019 & -0.040 & -0.036 & 0.111 & 0.014 \\ 
  Tooth9 & -0.050 & -0.077 & 0.135 & 0.024 & -0.027 & -0.054 & 0.124 & 0.018 \\ 
  Tooth10 & 0.063 & 0.003 & 0.126 & 0.016 & 0.048 & -0.012 & 0.141 & 0.020 \\ 
  ZoneM & -0.064 & -0.041 & 0.147 & 0.023 & 0.008 & 0.031 & 0.382 & 0.147 \\ 
  ZoneI & -0.167 & -0.056 & 0.158 & 0.080 & -0.084 & 0.027 & 0.233 & 0.055 \\ 
  ZoneE & -0.219 & -0.084 & 0.162 & 0.033 & -0.138 & -0.003 & 0.250 & 0.063 \\ 
  \hline Average MSE &  &  &  & 0.021 &  &  &  & 0.111 \\ 
   \hline
\end{tabular}
}
\end{subtable}
\begin{subtable}{\linewidth}
\centering
\caption{N=200}
\scalebox{0.65}{
\begin{tabular}{lcccc|cccc}
Variable & \makecell{Standardized\\Estimate} & \makecell{Bias\\(Standardized\\Estimate)} & \makecell{SE\\(Standardized\\Estimate)} & \makecell{MSE\\(Standardized\\Estimate)} & \makecell{Standardized\\James-Stein\\Estimate} & \makecell{Bias\\(Standardized\\James-Stein\\Estimate)} & \makecell{SE\\(Standardized\\James-Stein\\Estimate)} & \makecell{MSE\\(Standardized\\James-Stein\\Estimate)} \\ 
  \hline
Dental\_age & -0.074 & -0.033 & 0.072 & 0.006 & -0.044 & -0.004 & 0.087 & 0.008 \\ 
  Avg\_homeppm & -0.154 & -0.085 & 0.075 & 0.013 & -0.142 & -0.073 & 0.073 & 0.011 \\ 
  Tooth8 & -0.057 & -0.053 & 0.071 & 0.008 & -0.043 & -0.039 & 0.070 & 0.006 \\ 
  Tooth9 & -0.027 & -0.055 & 0.066 & 0.007 & -0.015 & -0.043 & 0.077 & 0.008 \\ 
  Tooth10 & 0.080 & 0.020 & 0.060 & 0.004 & 0.052 & -0.008 & 0.076 & 0.006 \\ 
  ZoneM & -0.113 & -0.089 & 0.065 & 0.012 & -0.089 & -0.066 & 0.069 & 0.009 \\ 
  ZoneI & -0.241 & -0.129 & 0.064 & 0.021 & -0.230 & -0.118 & 0.065 & 0.018 \\ 
  ZoneE & -0.303 & -0.168 & 0.071 & 0.033 & -0.295 & -0.159 & 0.072 & 0.031 \\ 
  \hline Average MSE &  &  &  & 0.013 &  &  &  & 0.012 \\ 
   \hline
\end{tabular}
}
\end{subtable}
\end{threeparttable}
\end{table}

\addtocounter{table}{-1} 

\begin{table}[H]
\centering
\caption{ Simulation results for age 23 data generated with exchangeable and exchangeable correlation structures for presence and severity respectively - properties of standardized estimates arising from the severity model B$_s$.2.4. }
\label{ch4:table:sim:std:sev:exchangeable}
\begin{threeparttable}
\centering
\begin{subtable}{\linewidth}
\centering
\caption{N=30}
\scalebox{0.65}{
\begin{tabular}{lcccc|cccc}
Variable & \makecell{Standardized\\Estimate} & \makecell{Bias\\(Standardized\\Estimate)} & \makecell{SE\\(Standardized\\Estimate)} & \makecell{MSE\\(Standardized\\Estimate)} & \makecell{Standardized\\James-Stein\\Estimate} & \makecell{Bias\\(Standardized\\James-Stein\\Estimate)} & \makecell{SE\\(Standardized\\James-Stein\\Estimate)} & \makecell{MSE\\(Standardized\\James-Stein\\Estimate)} \\ 
  \hline
Dental\_age & 0.026 & 0.067 & 0.076 & 0.010 & -0.257 & -0.217 & 1.273 & 1.666 \\ 
  Avg\_homeppm & -0.070 & -0.001 & 0.182 & 0.033 & -0.056 & 0.013 & 0.141 & 0.020 \\ 
  Tooth8 & 0.004 & 0.008 & 0.121 & 0.015 & 0.061 & 0.065 & 0.613 & 0.38 \\ 
  Tooth9 & -0.050 & -0.077 & 0.429 & 0.190 & -0.015 & -0.042 & 0.594 & 0.355 \\ 
  Tooth10 & 0.065 & 0.005 & 0.120 & 0.014 & 0.015 & -0.045 & 0.184 & 0.036 \\ 
  ZoneM & -0.042 & -0.018 & 0.156 & 0.025 & -0.092 & -0.068 & 0.855 & 0.736 \\ 
  ZoneI & -0.098 & 0.014 & 0.171 & 0.029 & 0.162 & 0.274 & 1.826 & 3.408 \\ 
  ZoneE & -0.124 & 0.011 & 0.188 & 0.035 & -0.202 & -0.066 & 1.497 & 2.244 \\ 
  \hline Average MSE &  &  &  & 0.044 &  &  &  & 1.106 \\ 
   \hline
\end{tabular}
}
\end{subtable}
\begin{subtable}{\linewidth}
\centering
\caption{N=50}
\scalebox{0.65}{
\begin{tabular}{lcccc|cccc}
Variable & \makecell{Standardized\\Estimate} & \makecell{Bias\\(Standardized\\Estimate)} & \makecell{SE\\(Standardized\\Estimate)} & \makecell{MSE\\(Standardized\\Estimate)} & \makecell{Standardized\\James-Stein\\Estimate} & \makecell{Bias\\(Standardized\\James-Stein\\Estimate)} & \makecell{SE\\(Standardized\\James-Stein\\Estimate)} & \makecell{MSE\\(Standardized\\James-Stein\\Estimate)} \\ 
  \hline
Dental\_age & 0.025 & 0.066 & 0.103 & 0.015 & -0.071 & -0.031 & 0.925 & 0.857 \\ 
  Avg\_homeppm & -0.075 & -0.006 & 0.127 & 0.016 & -0.038 & 0.030 & 0.097 & 0.010 \\ 
  Tooth8 & -0.026 & -0.021 & 0.113 & 0.013 & 0.000 & 0.004 & 0.113 & 0.013 \\ 
  Tooth9 & -0.024 & -0.052 & 0.101 & 0.013 & -0.006 & -0.033 & 0.103 & 0.012 \\ 
  Tooth10 & 0.036 & -0.024 & 0.126 & 0.017 & 0.025 & -0.035 & 0.108 & 0.013 \\ 
  ZoneM & -0.065 & -0.041 & 0.135 & 0.020 & -0.050 & -0.026 & 0.118 & 0.015 \\ 
  ZoneI & -0.150 & -0.039 & 0.162 & 0.080 & -0.124 & -0.012 & 0.186 & 0.035 \\ 
  ZoneE & -0.202 & -0.067 & 0.163 & 0.031 & -0.153 & -0.017 & 0.205 & 0.042 \\ 
  \hline Average MSE &  &  &  & 0.019 &  &  &  & 0.125 \\ 
   \hline
\end{tabular}
}
\end{subtable}
\begin{subtable}{\linewidth}
\centering
\caption{N=200}
\scalebox{0.65}{
\begin{tabular}{lcccc|cccc}
Variable & \makecell{Standardized\\Estimate} & \makecell{Bias\\(Standardized\\Estimate)} & \makecell{SE\\(Standardized\\Estimate)} & \makecell{MSE\\(Standardized\\Estimate)} & \makecell{Standardized\\James-Stein\\Estimate} & \makecell{Bias\\(Standardized\\James-Stein\\Estimate)} & \makecell{SE\\(Standardized\\James-Stein\\Estimate)} & \makecell{MSE\\(Standardized\\James-Stein\\Estimate)} \\ 
  \hline
Dental\_age & 0.001 & 0.042 & 0.056 & 0.005 & -0.004 & 0.036 & 0.051 & 0.004 \\ 
  Avg\_homeppm & -0.075 & -0.007 & 0.074 & 0.006 & -0.063 & 0.006 & 0.064 & 0.004 \\ 
  Tooth8 & -0.010 & -0.006 & 0.107 & 0.011 & -0.011 & -0.006 & 0.098 & 0.010 \\ 
  Tooth9 & -0.009 & -0.036 & 0.065 & 0.005 & -0.007 & -0.034 & 0.057 & 0.004 \\ 
  Tooth10 & 0.052 & -0.008 & 0.103 & 0.011 & 0.040 & -0.020 & 0.092 & 0.009 \\ 
  ZoneM & -0.061 & -0.038 & 0.114 & 0.014 & -0.045 & -0.021 & 0.113 & 0.013 \\ 
  ZoneI & -0.141 & -0.029 & 0.113 & 0.014 & -0.122 & -0.010 & 0.119 & 0.014 \\ 
  ZoneE & -0.183 & -0.048 & 0.113 & 0.015 & -0.167 & -0.032 & 0.120 & 0.015 \\ 
  \hline Average MSE &  &  &  & 0.010 &  &  &  & 0.009 \\ 
   \hline
\end{tabular}
}
\end{subtable}
\end{threeparttable}
\end{table}



\addtocounter{table}{-1} 

\begin{table}[H]
\centering
\caption{ Simulation results for age 23 data generated with independence and exchangeable correlation structures for presence and severity respectively - properties of standardized estimates arising from the presence piece of the combined model C$_s$.1.2.4. }
\label{ch4:table:sim:std:comb:pres:independence}
\begin{threeparttable}
\centering
\begin{subtable}{\linewidth}
\centering
\caption{N=30}
\scalebox{0.65}{
\begin{tabular}{lcccc|cccc}
Variable & \makecell{Standardized\\Estimate} & \makecell{Bias\\(Standardized\\Estimate)} & \makecell{SE\\(Standardized\\Estimate)} & \makecell{MSE\\(Standardized\\Estimate)} & \makecell{Standardized\\James-Stein\\Estimate} & \makecell{Bias\\(Standardized\\James-Stein\\Estimate)} & \makecell{SE\\(Standardized\\James-Stein\\Estimate)} & \makecell{MSE\\(Standardized\\James-Stein\\Estimate)} \\ 
  \hline
Dental\_age & 1.837 & 2.024 & 1.453 & 6.209 & 1.823 & 2.01 & 1.453 & 6.154 \\ 
  Avg\_homeppm & 4.226 & 4.608 & 7.081 & 71.378 & 4.217 & 4.599 & 7.085 & 71.342 \\ 
  Tooth8 & 4.560 & 4.684 & 7.680 & 80.925 & 4.551 & 4.676 & 7.683 & 80.896 \\ 
  Tooth9 & -0.236 & -0.186 & 0.319 & 0.136 & -0.205 & -0.155 & 0.298 & 0.113 \\ 
  Tooth10 & -0.362 & -0.584 & 0.212 & 0.386 & -0.321 & -0.543 & 0.203 & 0.336 \\ 
  ZoneM & -0.107 & 0.207 & 0.143 & 0.063 & -0.056 & 0.257 & 0.208 & 0.110 \\ 
  ZoneI & -0.062 & 0.699 & 0.132 & 0.506 & -0.021 & 0.740 & 0.216 & 0.594 \\ 
  ZoneE & 0.159 & 1.121 & 0.149 & 1.279 & 0.102 & 1.065 & 0.176 & 1.164 \\ 
  \hline Average MSE &  &  &  & 20.110 &  &  &  & 20.089 \\ 
   \hline
\end{tabular}
}
\end{subtable}
\begin{subtable}{\linewidth}
\centering
\caption{N=50}
\scalebox{0.65}{
\begin{tabular}{lcccc|cccc}
Variable & \makecell{Standardized\\Estimate} & \makecell{Bias\\(Standardized\\Estimate)} & \makecell{SE\\(Standardized\\Estimate)} & \makecell{MSE\\(Standardized\\Estimate)} & \makecell{Standardized\\James-Stein\\Estimate} & \makecell{Bias\\(Standardized\\James-Stein\\Estimate)} & \makecell{SE\\(Standardized\\James-Stein\\Estimate)} & \makecell{MSE\\(Standardized\\James-Stein\\Estimate)} \\ 
  \hline
Dental\_age & 1.833 & 2.021 & 1.004 & 5.091 & 1.825 & 2.013 & 1.003 & 5.059 \\ 
  Avg\_homeppm & 3.085 & 3.466 & 5.626 & 43.673 & 3.078 & 3.460 & 5.628 & 43.648 \\ 
  Tooth8 & 3.335 & 3.460 & 5.518 & 42.415 & 3.329 & 3.454 & 5.520 & 42.394 \\ 
  Tooth9 & -0.131 & -0.080 & 0.139 & 0.026 & -0.100 & -0.050 & 0.121 & 0.017 \\ 
  Tooth10 & -0.360 & -0.582 & 0.133 & 0.356 & -0.332 & -0.554 & 0.130 & 0.324 \\ 
  ZoneM & -0.095 & 0.218 & 0.098 & 0.057 & -0.034 & 0.279 & 0.159 & 0.103 \\ 
  ZoneI & -0.060 & 0.702 & 0.108 & 0.504 & -0.054 & 0.707 & 0.371 & 0.638 \\ 
  ZoneE & 0.160 & 1.122 & 0.105 & 1.271 & 0.096 & 1.058 & 0.139 & 1.139 \\ 
  \hline Average MSE &  &  &  & 11.674 &  &  &  & 11.665 \\ 
   \hline
\end{tabular}
}
\end{subtable}
\begin{subtable}{\linewidth}
\centering
\caption{N=200}
\scalebox{0.65}{
\begin{tabular}{lcccc|cccc}
Variable & \makecell{Standardized\\Estimate} & \makecell{Bias\\(Standardized\\Estimate)} & \makecell{SE\\(Standardized\\Estimate)} & \makecell{MSE\\(Standardized\\Estimate)} & \makecell{Standardized\\James-Stein\\Estimate} & \makecell{Bias\\(Standardized\\James-Stein\\Estimate)} & \makecell{SE\\(Standardized\\James-Stein\\Estimate)} & \makecell{MSE\\(Standardized\\James-Stein\\Estimate)} \\ 
  \hline
Dental\_age & 2.276 & 2.464 & 0.699 & 6.559 & 2.225 & 2.412 & 0.671 & 6.270 \\ 
  Avg\_homeppm & 1.283 & 1.665 & 0.406 & 2.937 & 1.306 & 1.688 & 0.404 & 3.012 \\ 
  Tooth8 & 2.087 & 2.212 & 1.569 & 7.354 & 2.106 & 2.230 & 1.576 & 7.457 \\ 
  Tooth9 & -0.189 & -0.139 & 0.063 & 0.023 & -0.174 & -0.123 & 0.062 & 0.019 \\ 
  Tooth10 & -0.358 & -0.580 & 0.071 & 0.342 & -0.352 & -0.574 & 0.071 & 0.335 \\ 
  ZoneM & -0.094 & 0.220 & 0.051 & 0.051 & -0.061 & 0.253 & 0.071 & 0.069 \\ 
  ZoneI & -0.061 & 0.700 & 0.051 & 0.493 & -0.038 & 0.723 & 0.057 & 0.527 \\ 
  ZoneE & 0.168 & 1.130 & 0.061 & 1.281 & 0.154 & 1.116 & 0.061 & 1.250 \\ 
  \hline Average MSE &  &  &  & 2.380 &  &  &  & 2.367 \\ 
   \hline
\end{tabular}
}
\end{subtable}
\end{threeparttable}
\end{table}

\addtocounter{table}{-1} 

\begin{table}[H]
\centering
\caption{ Simulation results for age 23 data generated with exchangeable and independence correlation structures for presence and severity respectively - properties of standardized estimates arising from the presence piece of the combined model C$_s$.2.1.4. }
\label{ch4:table:sim:std:comb:pres:exchangeable}
\begin{threeparttable}
\centering
\begin{subtable}{\linewidth}
\centering
\caption{N=30}
\scalebox{0.65}{
\begin{tabular}{lcccc|cccc}
Variable & \makecell{Standardized\\Estimate} & \makecell{Bias\\(Standardized\\Estimate)} & \makecell{SE\\(Standardized\\Estimate)} & \makecell{MSE\\(Standardized\\Estimate)} & \makecell{Standardized\\James-Stein\\Estimate} & \makecell{Bias\\(Standardized\\James-Stein\\Estimate)} & \makecell{SE\\(Standardized\\James-Stein\\Estimate)} & \makecell{MSE\\(Standardized\\James-Stein\\Estimate)} \\ 
  \hline
Dental\_age & 2.034 & 2.221 & 1.643 & 7.633 & 2.003 & 2.19 & 1.661 & 7.557 \\ 
  Avg\_homeppm & 3.309 & 3.691 & 5.846 & 47.796 & 2.895 & 3.277 & 5.479 & 40.751 \\ 
  Tooth8 & 4.662 & 4.787 & 6.042 & 59.416 & 4.237 & 4.361 & 5.647 & 50.913 \\ 
  Tooth9 & -0.164 & -0.113 & 0.222 & 0.062 & -0.138 & -0.088 & 0.201 & 0.048 \\ 
  Tooth10 & -0.356 & -0.578 & 0.176 & 0.365 & -0.307 & -0.529 & 0.174 & 0.310 \\ 
  ZoneM & -0.083 & 0.231 & 0.102 & 0.064 & 0.014 & 0.327 & 0.227 & 0.159 \\ 
  ZoneI & -0.037 & 0.724 & 0.122 & 0.539 & -0.009 & 0.752 & 0.239 & 0.623 \\ 
  ZoneE & 0.177 & 1.139 & 0.130 & 1.314 & 0.076 & 1.038 & 0.198 & 1.116 \\ 
  \hline Average MSE &  &  &  & 14.649 &  &  &  & 12.685 \\ 
   \hline
\end{tabular}
}
\end{subtable}
\begin{subtable}{\linewidth}
\centering
\caption{N=50}
\scalebox{0.65}{
\begin{tabular}{lcccc|cccc}
Variable & \makecell{Standardized\\Estimate} & \makecell{Bias\\(Standardized\\Estimate)} & \makecell{SE\\(Standardized\\Estimate)} & \makecell{MSE\\(Standardized\\Estimate)} & \makecell{Standardized\\James-Stein\\Estimate} & \makecell{Bias\\(Standardized\\James-Stein\\Estimate)} & \makecell{SE\\(Standardized\\James-Stein\\Estimate)} & \makecell{MSE\\(Standardized\\James-Stein\\Estimate)} \\ 
  \hline
Dental\_age & 2.139 & 2.326 & 1.078 & 6.572 & 2.131 & 2.318 & 1.077 & 6.536 \\ 
  Avg\_homeppm & 1.848 & 2.229 & 3.099 & 14.577 & 1.84 & 2.222 & 3.101 & 14.552 \\ 
  Tooth8 & 4.025 & 4.15 & 2.836 & 25.266 & 4.022 & 4.147 & 2.837 & 25.245 \\ 
  Tooth9 & -0.173 & -0.123 & 0.136 & 0.033 & -0.135 & -0.085 & 0.129 & 0.024 \\ 
  Tooth10 & -0.338 & -0.56 & 0.135 & 0.332 & -0.310 & -0.532 & 0.134 & 0.301 \\ 
  ZoneM & -0.090 & 0.224 & 0.087 & 0.058 & -0.043 & 0.271 & 0.318 & 0.175 \\ 
  ZoneI & -0.062 & 0.699 & 0.095 & 0.498 & -0.017 & 0.745 & 0.208 & 0.598 \\ 
  ZoneE & 0.168 & 1.131 & 0.107 & 1.29 & 0.107 & 1.069 & 0.121 & 1.158 \\ 
  \hline Average MSE &  &  &  & 6.078 &  &  &  & 6.074 \\ 
   \hline
\end{tabular}
}
\end{subtable}
\begin{subtable}{\linewidth}
\centering
\caption{N=200}
\scalebox{0.65}{
\begin{tabular}{lcccc|cccc}
Variable & \makecell{Standardized\\Estimate} & \makecell{Bias\\(Standardized\\Estimate)} & \makecell{SE\\(Standardized\\Estimate)} & \makecell{MSE\\(Standardized\\Estimate)} & \makecell{Standardized\\James-Stein\\Estimate} & \makecell{Bias\\(Standardized\\James-Stein\\Estimate)} & \makecell{SE\\(Standardized\\James-Stein\\Estimate)} & \makecell{MSE\\(Standardized\\James-Stein\\Estimate)} \\ 
  \hline
Dental\_age & 2.136 & 2.324 & 0.684 & 5.868 & 2.137 & 2.324 & 0.693 & 5.883 \\ 
  Avg\_homeppm & 1.090 & 1.472 & 0.298 & 2.257 & 1.092 & 1.473 & 0.301 & 2.261 \\ 
  Tooth8 & 3.8 & 3.924 & 0.554 & 15.708 & 3.814 & 3.939 & 0.563 & 15.83 \\ 
  Tooth9 & -0.164 & -0.113 & 0.071 & 0.018 & -0.148 & -0.098 & 0.066 & 0.014 \\ 
  Tooth10 & -0.344 & -0.566 & 0.072 & 0.326 & -0.338 & -0.56 & 0.069 & 0.318 \\ 
  ZoneM & -0.095 & 0.218 & 0.093 & 0.056 & -0.085 & 0.228 & 0.104 & 0.063 \\ 
  ZoneI & -0.063 & 0.698 & 0.048 & 0.489 & -0.031 & 0.730 & 0.068 & 0.538 \\ 
  ZoneE & 0.152 & 1.114 & 0.113 & 1.254 & 0.140 & 1.102 & 0.113 & 1.227 \\ 
  \hline Average MSE &  &  &  & 3.247 &  &  &  & 3.267 \\ 
   \hline
\end{tabular}
}
\end{subtable}
\end{threeparttable}
\end{table}

\addtocounter{table}{-1} 

\begin{table}[H]
\centering
\caption{ Simulation results for age 23 data generated with exchangeable and exchangeable correlation structures for presence and severity respectively - properties of standardized estimates arising from the presence piece of the combined model C$_s$.2.2.4. }
\label{ch4:table:sim:std:comb:pres:exchangeable}
\begin{threeparttable}
\centering
\begin{subtable}{\linewidth}
\centering
\caption{N=30}
\scalebox{0.65}{
\begin{tabular}{lcccc|cccc}
Variable & \makecell{Standardized\\Estimate} & \makecell{Bias\\(Standardized\\Estimate)} & \makecell{SE\\(Standardized\\Estimate)} & \makecell{MSE\\(Standardized\\Estimate)} & \makecell{Standardized\\James-Stein\\Estimate} & \makecell{Bias\\(Standardized\\James-Stein\\Estimate)} & \makecell{SE\\(Standardized\\James-Stein\\Estimate)} & \makecell{MSE\\(Standardized\\James-Stein\\Estimate)} \\ 
  \hline
Dental\_age & 1.883 & 2.07 & 1.552 & 6.696 & 1.874 & 2.062 & 1.56 & 6.686 \\ 
  Avg\_homeppm & -1.23 & -0.848 & 6.029 & 37.062 & -1.229 & -0.847 & 6.057 & 37.407 \\ 
  Tooth8 & -0.587 & -0.463 & 5.596 & 31.528 & -0.604 & -0.479 & 5.619 & 31.807 \\ 
  Tooth9 & -0.126 & -0.075 & 0.316 & 0.106 & -0.101 & -0.050 & 0.291 & 0.087 \\ 
  Tooth10 & -0.339 & -0.561 & 0.143 & 0.335 & -0.291 & -0.513 & 0.138 & 0.282 \\ 
  ZoneM & -0.078 & 0.235 & 0.106 & 0.067 & 0.050 & 0.363 & 0.264 & 0.202 \\ 
  ZoneI & -0.055 & 0.706 & 0.119 & 0.513 & -0.016 & 0.745 & 0.224 & 0.605 \\ 
  ZoneE & 0.174 & 1.136 & 0.131 & 1.308 & 0.073 & 1.036 & 0.184 & 1.106 \\ 
  \hline Average MSE &  &  &  & 9.702 &  &  &  & 9.773 \\ 
   \hline
\end{tabular}
}
\end{subtable}
\begin{subtable}{\linewidth}
\centering
\caption{N=50}
\scalebox{0.65}{
\begin{tabular}{lcccc|cccc}
Variable & \makecell{Standardized\\Estimate} & \makecell{Bias\\(Standardized\\Estimate)} & \makecell{SE\\(Standardized\\Estimate)} & \makecell{MSE\\(Standardized\\Estimate)} & \makecell{Standardized\\James-Stein\\Estimate} & \makecell{Bias\\(Standardized\\James-Stein\\Estimate)} & \makecell{SE\\(Standardized\\James-Stein\\Estimate)} & \makecell{MSE\\(Standardized\\James-Stein\\Estimate)} \\ 
  \hline
Dental\_age & 1.922 & 2.110 & 1.115 & 5.694 & 1.914 & 2.102 & 1.115 & 5.66 \\ 
  Avg\_homeppm & -0.031 & 0.351 & 1.732 & 3.122 & -0.027 & 0.355 & 1.727 & 3.109 \\ 
  Tooth8 & 1.013 & 1.137 & 2.666 & 8.401 & 1.012 & 1.137 & 2.663 & 8.384 \\ 
  Tooth9 & -0.167 & -0.117 & 0.142 & 0.034 & -0.138 & -0.087 & 0.129 & 0.024 \\ 
  Tooth10 & -0.347 & -0.568 & 0.130 & 0.340 & -0.318 & -0.54 & 0.128 & 0.308 \\ 
  ZoneM & -0.085 & 0.228 & 0.099 & 0.062 & -0.009 & 0.304 & 0.234 & 0.147 \\ 
  ZoneI & -0.055 & 0.706 & 0.102 & 0.509 & -0.051 & 0.71 & 0.206 & 0.546 \\ 
  ZoneE & 0.162 & 1.124 & 0.114 & 1.277 & 0.111 & 1.074 & 0.135 & 1.171 \\ 
  \hline Average MSE &  &  &  & 2.430 &  &  &  & 2.419 \\ 
   \hline
\end{tabular}
}
\end{subtable}
\begin{subtable}{\linewidth}
\centering
\caption{N=200}
\scalebox{0.65}{
\begin{tabular}{lcccc|cccc}
Variable & \makecell{Standardized\\Estimate} & \makecell{Bias\\(Standardized\\Estimate)} & \makecell{SE\\(Standardized\\Estimate)} & \makecell{MSE\\(Standardized\\Estimate)} & \makecell{Standardized\\James-Stein\\Estimate} & \makecell{Bias\\(Standardized\\James-Stein\\Estimate)} & \makecell{SE\\(Standardized\\James-Stein\\Estimate)} & \makecell{MSE\\(Standardized\\James-Stein\\Estimate)} \\ 
  \hline
Dental\_age & 2.077 & 2.264 & 0.671 & 5.576 & 2.091 & 2.279 & 0.674 & 5.646 \\ 
  Avg\_homeppm & -0.278 & 0.104 & 0.327 & 0.118 & -0.273 & 0.109 & 0.326 & 0.118 \\ 
  Tooth8 & 2.22 & 2.344 & 1.438 & 7.565 & 2.223 & 2.348 & 1.460 & 7.643 \\ 
  Tooth9 & -0.163 & -0.113 & 0.065 & 0.017 & -0.148 & -0.098 & 0.063 & 0.014 \\ 
  Tooth10 & -0.339 & -0.561 & 0.071 & 0.32 & -0.333 & -0.555 & 0.072 & 0.313 \\ 
  ZoneM & -0.087 & 0.227 & 0.077 & 0.057 & -0.066 & 0.247 & 0.099 & 0.071 \\ 
  ZoneI & -0.052 & 0.709 & 0.058 & 0.506 & -0.033 & 0.728 & 0.071 & 0.535 \\ 
  ZoneE & 0.167 & 1.129 & 0.099 & 1.285 & 0.154 & 1.117 & 0.099 & 1.257 \\ 
  \hline Average MSE &  &  &  & 1.930 &  &  &  & 1.950 \\ 
   \hline
\end{tabular}
}
\end{subtable}
\end{threeparttable}
\end{table}



\addtocounter{table}{-1} 

\begin{table}[H]
\centering
\caption{ Simulation results for age 23 data generated with independence and exchangeable correlation structures for presence and severity respectively - properties of standardized estimates arising from the severity piece of the combined model C$_s$.1.2.4. }
\label{ch4:table:sim:std:comb:sev:exchangeable}
\begin{threeparttable}
\centering
\begin{subtable}{\linewidth}
\centering
\caption{N=30}
\scalebox{0.65}{
\begin{tabular}{lcccc|cccc}
Variable & \makecell{Standardized\\Estimate} & \makecell{Bias\\(Standardized\\Estimate)} & \makecell{SE\\(Standardized\\Estimate)} & \makecell{MSE\\(Standardized\\Estimate)} & \makecell{Standardized\\James-Stein\\Estimate} & \makecell{Bias\\(Standardized\\James-Stein\\Estimate)} & \makecell{SE\\(Standardized\\James-Stein\\Estimate)} & \makecell{MSE\\(Standardized\\James-Stein\\Estimate)} \\ 
  \hline
Dental\_age & 1.986 & 2.027 & 5.02 & 29.305 & 1.971 & 2.011 & 5.021 & 29.256 \\ 
  Avg\_homeppm & 2.32 & 2.389 & 5.574 & 36.778 & 2.309 & 2.378 & 5.575 & 36.739 \\ 
  Tooth8 & -0.163 & -0.158 & 0.361 & 0.156 & -0.140 & -0.136 & 0.333 & 0.129 \\ 
  Tooth9 & -0.302 & -0.330 & 0.32 & 0.211 & -0.272 & -0.299 & 0.295 & 0.177 \\ 
  Tooth10 & -0.102 & -0.162 & 0.187 & 0.061 & -0.071 & -0.131 & 0.184 & 0.051 \\ 
  ZoneM & -0.043 & -0.019 & 0.134 & 0.018 & -0.012 & 0.011 & 0.221 & 0.049 \\ 
  ZoneI & 0.120 & 0.232 & 0.170 & 0.082 & 0.082 & 0.193 & 0.176 & 0.068 \\ 
  ZoneE & -0.267 & -0.132 & 0.300 & 0.107 & -0.232 & -0.097 & 0.281 & 0.089 \\ 
  \hline Average MSE &  &  &  & 8.340 &  &  &  & 8.320 \\ 
   \hline
\end{tabular}
}
\end{subtable}
\begin{subtable}{\linewidth}
\centering
\caption{N=50}
\scalebox{0.65}{
\begin{tabular}{lcccc|cccc}
Variable & \makecell{Standardized\\Estimate} & \makecell{Bias\\(Standardized\\Estimate)} & \makecell{SE\\(Standardized\\Estimate)} & \makecell{MSE\\(Standardized\\Estimate)} & \makecell{Standardized\\James-Stein\\Estimate} & \makecell{Bias\\(Standardized\\James-Stein\\Estimate)} & \makecell{SE\\(Standardized\\James-Stein\\Estimate)} & \makecell{MSE\\(Standardized\\James-Stein\\Estimate)} \\ 
  \hline
Dental\_age & 1.499 & 1.540 & 1.437 & 4.434 & 1.489 & 1.530 & 1.437 & 4.406 \\ 
  Avg\_homeppm & 2.383 & 2.452 & 3.999 & 22.001 & 2.378 & 2.447 & 3.999 & 21.978 \\ 
  Tooth8 & -0.111 & -0.107 & 0.153 & 0.035 & -0.086 & -0.082 & 0.128 & 0.023 \\ 
  Tooth9 & -0.336 & -0.363 & 0.228 & 0.184 & -0.313 & -0.340 & 0.216 & 0.162 \\ 
  Tooth10 & -0.101 & -0.161 & 0.142 & 0.046 & -0.061 & -0.121 & 0.153 & 0.038 \\ 
  ZoneM & -0.048 & -0.024 & 0.104 & 0.011 & -0.057 & -0.034 & 0.402 & 0.163 \\ 
  ZoneI & 0.132 & 0.243 & 0.137 & 0.078 & 0.073 & 0.185 & 0.151 & 0.057 \\ 
  ZoneE & -0.310 & -0.174 & 0.215 & 0.077 & -0.287 & -0.151 & 0.206 & 0.065 \\ 
  \hline Average MSE &  &  &  & 3.358 &  &  &  & 3.362 \\ 
   \hline
\end{tabular}
}
\end{subtable}
\begin{subtable}{\linewidth}
\centering
\caption{N=200}
\scalebox{0.65}{
\begin{tabular}{lcccc|cccc}
Variable & \makecell{Standardized\\Estimate} & \makecell{Bias\\(Standardized\\Estimate)} & \makecell{SE\\(Standardized\\Estimate)} & \makecell{MSE\\(Standardized\\Estimate)} & \makecell{Standardized\\James-Stein\\Estimate} & \makecell{Bias\\(Standardized\\James-Stein\\Estimate)} & \makecell{SE\\(Standardized\\James-Stein\\Estimate)} & \makecell{MSE\\(Standardized\\James-Stein\\Estimate)} \\ 
  \hline
Dental\_age & 1.690 & 1.731 & 0.675 & 3.451 & 1.655 & 1.695 & 0.665 & 3.316 \\ 
  Avg\_homeppm & 2.487 & 2.555 & 2.165 & 11.22 & 2.522 & 2.59 & 2.222 & 11.648 \\ 
  Tooth8 & -0.178 & -0.174 & 0.088 & 0.038 & -0.163 & -0.159 & 0.086 & 0.032 \\ 
  Tooth9 & -0.369 & -0.396 & 0.129 & 0.174 & -0.362 & -0.39 & 0.131 & 0.169 \\ 
  Tooth10 & -0.120 & -0.180 & 0.076 & 0.038 & -0.095 & -0.155 & 0.087 & 0.032 \\ 
  ZoneM & -0.055 & -0.031 & 0.051 & 0.004 & -0.032 & -0.008 & 0.057 & 0.003 \\ 
  ZoneI & 0.158 & 0.270 & 0.077 & 0.079 & 0.144 & 0.255 & 0.075 & 0.071 \\ 
  ZoneE & -0.369 & -0.233 & 0.136 & 0.073 & -0.361 & -0.225 & 0.140 & 0.070 \\ 
  \hline Average MSE &  &  &  & 1.885 &  &  &  & 1.918 \\ 
   \hline
\end{tabular}
}
\end{subtable}
\end{threeparttable}
\end{table}

\addtocounter{table}{-1} 

\begin{table}[H]
\centering
\caption{ Simulation results for age 23 data generated with exchangeable and independence correlation structures for presence and severity respectively - properties of standardized estimates arising from the severity piece of the combined model C$_s$.2.1.4. }
\label{ch4:table:sim:std:comb:sev:independence}
\begin{threeparttable}
\centering
\begin{subtable}{\linewidth}
\centering
\caption{N=30}
\scalebox{0.65}{
\begin{tabular}{lcccc|cccc}
Variable & \makecell{Standardized\\Estimate} & \makecell{Bias\\(Standardized\\Estimate)} & \makecell{SE\\(Standardized\\Estimate)} & \makecell{MSE\\(Standardized\\Estimate)} & \makecell{Standardized\\James-Stein\\Estimate} & \makecell{Bias\\(Standardized\\James-Stein\\Estimate)} & \makecell{SE\\(Standardized\\James-Stein\\Estimate)} & \makecell{MSE\\(Standardized\\James-Stein\\Estimate)} \\ 
  \hline
Dental\_age & 4.560 & 4.601 & 28.744 & 847.407 & 4.648 & 4.689 & 29.200 & 874.616 \\ 
  Avg\_homeppm & 8.241 & 8.310 & 29.978 & 967.730 & 8.463 & 8.531 & 30.431 & 998.829 \\ 
  Tooth8 & -0.131 & -0.127 & 0.264 & 0.086 & -0.112 & -0.108 & 0.239 & 0.069 \\ 
  Tooth9 & -0.284 & -0.312 & 0.329 & 0.205 & -0.251 & -0.278 & 0.305 & 0.170 \\ 
  Tooth10 & -0.076 & -0.136 & 0.153 & 0.042 & -0.080 & -0.088 & 0.177 & 0.039 \\ 
  ZoneM & -0.080 & -0.004 & 0.118 & 0.014 & -0.013 & 0.010 & 0.246 & 0.061 \\ 
  ZoneI & 0.139 & 0.250 & 0.170 & 0.092 & 0.066 & 0.177 & 0.203 & 0.073 \\ 
  ZoneE & -0.287 & -0.152 & 0.273 & 0.097 & -0.251 & -0.115 & 0.261 & 0.082 \\ 
  \hline Average MSE &  &  &  & 226.959 &  &  &  & 234.242 \\ 
   \hline
\end{tabular}
}
\end{subtable}
\begin{subtable}{\linewidth}
\centering
\caption{N=50}
\scalebox{0.65}{
\begin{tabular}{lcccc|cccc}
Variable & \makecell{Standardized\\Estimate} & \makecell{Bias\\(Standardized\\Estimate)} & \makecell{SE\\(Standardized\\Estimate)} & \makecell{MSE\\(Standardized\\Estimate)} & \makecell{Standardized\\James-Stein\\Estimate} & \makecell{Bias\\(Standardized\\James-Stein\\Estimate)} & \makecell{SE\\(Standardized\\James-Stein\\Estimate)} & \makecell{MSE\\(Standardized\\James-Stein\\Estimate)} \\ 
  \hline
Dental\_age & 2.189 & 2.230 & 4.139 & 22.100 & 2.180 & 2.221 & 4.140 & 22.068 \\ 
  Avg\_homeppm & 14.269 & 14.338 & 71.694 & 5345.578 & 14.267 & 14.336 & 71.694 & 5345.557 \\ 
  Tooth8 & -0.169 & -0.165 & 0.150 & 0.049 & -0.133 & -0.129 & 0.139 & 0.036 \\ 
  Tooth9 & -0.322 & -0.350 & 0.200& 0.162 & -0.297 & -0.325 & 0.189 & 0.141 \\ 
  Tooth10 & -0.108 & -0.168 & 0.125 & 0.044 & -0.078 & -0.138 & 0.195 & 0.057 \\ 
  ZoneM & -0.053 & -0.029 & 0.090 & 0.009 & -0.003 & 0.021 & 0.228 & 0.052 \\ 
  ZoneI & 0.151 & 0.263 & 0.124 & 0.084 & 0.092 & 0.204 & 0.128 & 0.058 \\ 
  ZoneE & -0.341 & -0.206 & 0.190 & 0.078 & -0.314 & -0.179 & 0.185 & 0.066 \\ 
  \hline Average MSE &  &  &  & 671.013 &  &  &  & 671.004 \\ 
   \hline
\end{tabular}
}
\end{subtable}
\begin{subtable}{\linewidth}
\centering
\caption{N=200}
\scalebox{0.65}{
\begin{tabular}{lcccc|cccc}
Variable & \makecell{Standardized\\Estimate} & \makecell{Bias\\(Standardized\\Estimate)} & \makecell{SE\\(Standardized\\Estimate)} & \makecell{MSE\\(Standardized\\Estimate)} & \makecell{Standardized\\James-Stein\\Estimate} & \makecell{Bias\\(Standardized\\James-Stein\\Estimate)} & \makecell{SE\\(Standardized\\James-Stein\\Estimate)} & \makecell{MSE\\(Standardized\\James-Stein\\Estimate)} \\ 
  \hline
Dental\_age & 1.680 & 1.720 & 0.740 & 3.507 & 1.681 & 1.721 & 0.753 & 3.529 \\ 
  Avg\_homeppm & 5.835 & 5.904 & 3.962 & 50.549 & 5.884 & 5.953 & 4.071 & 52.012 \\ 
  Tooth8 & -0.168 & -0.164 & 0.070 & 0.032 & -0.154 & -0.150 & 0.066 & 0.027 \\ 
  Tooth9 & -0.363 & -0.391 & 0.082 & 0.159 & -0.357 & -0.385 & 0.083 & 0.155 \\ 
  Tooth10 & -0.126 & -0.186 & 0.123 & 0.050 & -0.117 & -0.177 & 0.122 & 0.046 \\ 
  ZoneM & -0.060 & -0.036 & 0.046 & 0.003 & -0.025 & -0.001 & 0.068 & 0.005 \\ 
  ZoneI & 0.151 & 0.262 & 0.113 & 0.082 & 0.138 & 0.250 & 0.113 & 0.075 \\ 
  ZoneE & -0.361 & -0.225 & 0.138 & 0.070 & -0.357 & -0.222 & 0.142 & 0.069 \\ 
  \hline Average MSE &  &  &  & 6.806 &  &  &  & 6.99 \\ 
   \hline
\end{tabular}
}
\end{subtable}
\end{threeparttable}
\end{table}

\addtocounter{table}{-1} 

\begin{table}[H]
\centering
\caption{ Simulation results for age 23 data generated with exchangeable and exchangeable correlation structures for presence and severity respectively - properties of standardized estimates arising from the severity piece of the combined model C$_s$.2.2.4. }
\label{ch4:table:sim:std:comb:sev:exchangeable}
\begin{threeparttable}
\centering
\begin{subtable}{\linewidth}
\centering
\caption{N=30}
\scalebox{0.65}{
\begin{tabular}{lcccc|cccc}
Variable & \makecell{Standardized\\Estimate} & \makecell{Bias\\(Standardized\\Estimate)} & \makecell{SE\\(Standardized\\Estimate)} & \makecell{MSE\\(Standardized\\Estimate)} & \makecell{Standardized\\James-Stein\\Estimate} & \makecell{Bias\\(Standardized\\James-Stein\\Estimate)} & \makecell{SE\\(Standardized\\James-Stein\\Estimate)} & \makecell{MSE\\(Standardized\\James-Stein\\Estimate)} \\ 
  \hline
Dental\_age & -3.858 & -3.818 & 20.943 & 453.169 & -3.869 & -3.829 & 21.048 & 457.673 \\ 
  Avg\_homeppm & 0.879 & 0.948 & 29.421 & 866.505 & 0.821 & 0.89 & 29.566 & 874.931 \\ 
  Tooth8 & -0.006 & -0.002 & 0.337 & 0.113 & -0.006 & -0.002 & 0.303 & 0.092 \\ 
  Tooth9 & -0.101 & -0.129 & 0.39 & 0.168 & -0.089 & -0.117 & 0.347 & 0.134 \\ 
  Tooth10 & -0.042 & -0.103 & 0.174 & 0.041 & -0.025 & -0.085 & 0.199 & 0.047 \\ 
  ZoneM & -0.007 & 0.016 & 0.114 & 0.013 & -0.008 & 0.016 & 0.232 & 0.054 \\ 
  ZoneI & 0.045 & 0.157 & 0.213 & 0.070 & 0.001 & 0.113 & 0.202 & 0.054 \\ 
  ZoneE & -0.093 & 0.042 & 0.364 & 0.134 & -0.080 & 0.056 & 0.323 & 0.107 \\ 
  \hline Average MSE &  &  &  & 165.027 &  &  &  & 166.637 \\ 
   \hline
\end{tabular}
}
\end{subtable}
\begin{subtable}{\linewidth}
\centering
\caption{N=50}
\scalebox{0.65}{
\begin{tabular}{lcccc|cccc}
Variable & \makecell{Standardized\\Estimate} & \makecell{Bias\\(Standardized\\Estimate)} & \makecell{SE\\(Standardized\\Estimate)} & \makecell{MSE\\(Standardized\\Estimate)} & \makecell{Standardized\\James-Stein\\Estimate} & \makecell{Bias\\(Standardized\\James-Stein\\Estimate)} & \makecell{SE\\(Standardized\\James-Stein\\Estimate)} & \makecell{MSE\\(Standardized\\James-Stein\\Estimate)} \\ 
  \hline
Dental\_age & -1.63 & -1.589 & 3.919 & 17.885 & -1.629 & -1.589 & 3.917 & 17.867 \\ 
  Avg\_homeppm & 1.464 & 1.533 & 7.367 & 56.618 & 1.463 & 1.532 & 7.366 & 56.604 \\ 
  Tooth8 & -0.095 & -0.091 & 0.197 & 0.047 & -0.078 & -0.074 & 0.170 & 0.034 \\ 
  Tooth9 & -0.224 & -0.251 & 0.332 & 0.173 & -0.207 & -0.234 & 0.313 & 0.153 \\ 
  Tooth10 & -0.050 & -0.110 & 0.154 & 0.036 & -0.015 & -0.075 & 0.184 & 0.039 \\ 
  ZoneM & -0.023 & 0.001 & 0.111 & 0.012 & -0.017 & 0.007 & 0.231 & 0.053 \\ 
  ZoneI & 0.116 & 0.228 & 0.152 & 0.075 & 0.079 & 0.191 & 0.148 & 0.058 \\ 
  ZoneE & -0.202 & -0.067 & 0.310 & 0.100 & -0.175 & -0.040 & 0.305 & 0.095 \\ 
  \hline Average MSE &  &  &  & 9.368 &  &  &  & 9.363 \\ 
   \hline
\end{tabular}
}
\end{subtable}
\begin{subtable}{\linewidth}
\centering
\caption{N=200}
\scalebox{0.65}{
\begin{tabular}{lcccc|cccc}
Variable & \makecell{Standardized\\Estimate} & \makecell{Bias\\(Standardized\\Estimate)} & \makecell{SE\\(Standardized\\Estimate)} & \makecell{MSE\\(Standardized\\Estimate)} & \makecell{Standardized\\James-Stein\\Estimate} & \makecell{Bias\\(Standardized\\James-Stein\\Estimate)} & \makecell{SE\\(Standardized\\James-Stein\\Estimate)} & \makecell{MSE\\(Standardized\\James-Stein\\Estimate)} \\ 
  \hline
Dental\_age & -5.250 & -5.210 & 12.667 & 187.585 & -5.355 & -5.314 & 12.845 & 193.242 \\ 
  Avg\_homeppm & 14.351 & 14.420 & 29.798 & 1095.879 & 14.524 & 14.593 & 30.243 & 1127.598 \\ 
  Tooth8 & -0.117 & -0.113 & 0.135 & 0.031 & -0.109 & -0.105 & 0.122 & 0.026 \\ 
  Tooth9 & -0.254 & -0.282 & 0.266 & 0.150 & -0.254 & -0.282 & 0.259 & 0.147 \\ 
  Tooth10 & -0.098 & -0.158 & 0.117 & 0.039 & -0.093 & -0.153 & 0.114 & 0.036 \\ 
  ZoneM & -0.038 & -0.015 & 0.062 & 0.004 & -0.017 & 0.007 & 0.075 & 0.006 \\ 
  ZoneI & 0.096 & 0.208 & 0.166 & 0.071 & 0.091 & 0.203 & 0.157 & 0.066 \\ 
  ZoneE & -0.274 & -0.138 & 0.264 & 0.089 & -0.270 & -0.134 & 0.255 & 0.083 \\ 
  \hline Average MSE &  &  &  & 160.481 &  &  &  & 165.15 \\ 
   \hline
\end{tabular}
}
\end{subtable}
\end{threeparttable}
\end{table}
